\documentclass[
PhD, 
DIC=off, 
12pt,
bibtotoc, 
liststotoc, 
idxtotoc, 
onehalfspacing, 
]
{icldt}

\usepackage[
            hyperref 
            ]{ihformat}

\begin{document}

\college{Queen Mary, }
\department{Physics and Astronomy}
\supervisor{Reza Tavakol}
\title{Observable predictions of generalised inflationary scenarios}
\author{Joseph Elliston}
\hypersetup{pdftitle={Mr},pdfauthor={Joseph Elliston}}

\declaration{%
I hereby certify that this thesis, which is approximately 47,000 words in length,
has been written by me; that it is the record of the work carried out by me at the
School of Physics and Astronomy, Queen Mary University of London, and that it has not been submitted in any previous application for a higher degree.
\\
The work in this thesis has been completed in collaboration with 
Laila Alabidi, 
Antonio de Felice,
Ian Huston,
David Mulryne,
David Seery,
Reza Tavakol and
Shinji Tsujikawa, and has been published in the following papers:
\begin{itemize}
\item J. Elliston, D. Seery and R. Tavakol. \newline
JCAP {\bf 11} (2012) 060 
\item J. Elliston, L. Alabidi, I. Huston, D. Mulryne and R. Tavakol. \newline
JCAP {\bf 09} (2012) 001
\item J. Elliston, D. Mulryne, D. Seery and R. Tavakol.\newline
Int.J.Mod.Phys. {\bf A26} (2011) 3821-3832
\item J. Elliston, D. Mulryne, D. Seery and R. Tavakol.\newline
JCAP {\bf 11} (2011) 005 
\item A. de Felice, S. Tsujikawa, J. Elliston and R. Tavakol.\newline
JCAP {\bf 08} (2011) 021 
\end{itemize}
With the exception of running the {\sc cosmomc} code used in chapter~\ref{ch:singlefield}, I have made a major contribution to all the original research presented in this thesis.
}

\maketitle

\chapter*{Abstract}
\label{ch:abstract}
\section*{}
\singlespacing

Inflation is an early period of accelerated cosmic expansion, thought to be sourced by high energy physics. A key task today is to use the influx of increasingly precise observational data to constrain the plethora of inflationary models suggested by fundamental theories of interactions. This requires a robust theoretical framework for quantifying the predictions of such models; helping to develop such a framework is the aim of this thesis.\\

We begin by providing the first complete quantization of subhorizon perturbations for the well-motivated class of multi-field inflationary models that possess a non-trivial field metric. In particular, the implications for the bispectrum of the Cosmic Microwave Background (\cmbr) are potentially very exciting. The subsequent evolution of perturbations in the superhorizon epoch is then considered, via a covariant extension of the transport formalism. We demonstrate appropriate matching between the subhorizon and superhorizon calculations. \\

With the aim of developing intuition about the relation between inflationary dynamics and the evolution of cosmic observables, we investigate analytic approximations of superhorizon perturbation evolution. The validity of these analytic results is contingent on reaching a state of adiabaticity which we discuss and illustrate in depth. We then apply our analytic methods to elucidate the types of inflationary dynamics that lead to an enhanced \cmbr~non-Gaussianity, both in its bispectrum and trispectrum. In addition to deriving a number of new simple relations between the non-Gaussianity parameters, we explain dynamically how and why different shapes of inflationary potential lead to particular observational signals. \\

In addition to multiple scalar fields, candidate theories of high energy physics include many possible modifications to the Einstein--Hilbert action. We consider the observational viability of single field chaotic inflation with additional corrections as motivated by low energy effective string theory. These new ingredients allow for consistency of chaotic inflationary models that are otherwise in tension with observational data.
\clearpage{\pagestyle{empty}\cleardoublepage}

\chapter*{Acknowledgements}
\label{ch:acknowledgements}

I am indebted to my supervisor Reza Tavakol for his tireless support and encouragement. His knowledge and experience is inspiring, as is his kind personality. My thanks also go to David Mulryne and David Seery for their guidance and support through our work together. I am also grateful to my other collaborators Antonio de Felice, Laila Alabidi, Ian Huston and Shinji Tsujikawa with whom it has been a pleasure and a privilege to work.

My PhD studies would not have been nearly as fun if it were not for the other members of my research group, present and past. In particular I would like to thank Adam, Alex, Ellie, Ian, Laila and Nathan for being great people to share an office with. I am also indebted to Karim and Tim for their leading role in making our group so fun and social. In addition, my thanks go to the rest of the relativity and cosmology group and all of the students in the Astronomy Unit.

I wish to thank my parents and brothers for their continual love and support. Thanks also to my friends and football team who have helped preserve some of my sanity during the creation of this thesis. 

For so very many reasons, my most sincere thanks go to Zo\"{e}. This thesis is dedicated to you.

\clearpage{\pagestyle{empty}\cleardoublepage}

\setcounter{tocdepth}{1}
\tableofcontents
\clearpage{\pagestyle{empty}\cleardoublepage}
\listoffigures
\clearpage{\pagestyle{empty}\cleardoublepage}

\onehalfspacing

%
\chapter{Introduction}
\label{ch:introduction}

\begin{addmargin}[0.05\textwidth]{0.05\textwidth}
We begin in \S\ref{sec:lcdm} by summarising the observational motivations and theoretical assumptions that underpin the prevailing \lcdm~model of our Universe. \S\ref{sec:needinflation} then describes why this picture is thought to be incomplete without the addition of an early epoch of accelerated expansion called {\it inflation}. We then proceed, in \S\ref{sec:simplest}, to demonstrate how inflation may be achieved by simple phenomenological models. Finally, \S\ref{sec:beyondsimplest} describes the observational and theoretical arguments that motivate the consideration of more general inflationary scenarios.
\end{addmargin}

\begin{center}
\partialhrule
\end{center}
\vspace{-3em}
\begin{quote}
\list{}{\leftmargin 0cm \rightmargin\leftmargin} \endlist
\begin{center}
{\it ``Confidence is what you have before you understand the problem.''}
\flushright{---Woody Allen.}
\end{center}
\end{quote}
\vspace{-1em}
\begin{center}
\partialhrule
\end{center}

\sec{The $\Lambda$\cdm~model}
\label{sec:lcdm}

Cosmology aims to understand the dynamics of our Universe: past, present and future. This task presents significant observational, theoretical and computational challenges and for these reasons it is not currently possible to provide a full definitive answer. The combination of recent high-resolution data and improved theoretical understanding has led to the \lcdm~($\Lambda$ Cold Dark Matter) model of our Universe emerging as the widely accepted standard. 

The \lcdm~model is based on three fundamental assumptions: 
\begin{enumerate}
\item The Universe is homogeneous and isotropic on large scales.
\item Gravity is described by the Einstein field equations of general relativity.
\item The content of the Universe is described by a combination of standard model particles, dark matter and a cosmological constant.
\end{enumerate}
These assumptions are far from trivial and their validity remains a matter of intense debate. In \S\S\ref{sec:smooth}--\ref{sec:stuff}, we take these three assumptions and discuss some of the key observational evidence that supports them.

\ssec{Homogeneity and isotropy}
\label{sec:smooth}

Homogeneity and isotropy describe invariance under spatial translations and rotations respectively. The constraints of homogeneity and isotropy prescribe the spacetime metric $g_{\mu \nu}$ to be of the Friedmann--Lema\^{i}tre--Robertson--Walker \frwl~form \cite{Weinberg:2008zzc}
\be
\label{eq:frwl}
\d s^2 = -\d t^2 + a^2(t) \left[ \frac{\d r^2}{1 - k r^2} + r^2 \d \theta^2 + r^2 \sin ^2 \theta \, \d \phi^2 \right],
\ee
where the scale factor $a(t)$ is a dimensionless function of coordinate time $t$ and $\{r,\theta, \phi \}$ are comoving spatial coordinates. The parameter $k = \{+1,0,-1\}$ represents positive, flat and negative spatial curvatures respectively. A growing scale factor $a(t)$ produces isotropic expansion of our Universe, which was found observationally in the 1920s by a combination of contributions from Knut Lundmark~\cite{2012arXiv1212.1359S}, Georges Lema\^{i}tre~\cite{Lemaitre} and Edwin Hubble~\cite{Hubble:1929ig}.\footnote{Hubble's recession velocities were measured by Vesto Slipher.} 

\sssec{Observational evidence for homogeneity and isotropy}

Isotropy may be directly tested by observing our Universe. The distances involved are large and are often quoted in terms of the redshift $z$ defined as
$1 + z \equiv a_0 / a(t)$, where $a_0$ and $a(t)$ are evaluated at the present time and the time of photon emission, respectively. 

\begin{figure}[h]
\begin{center}
\includegraphics[width=1\textwidth]{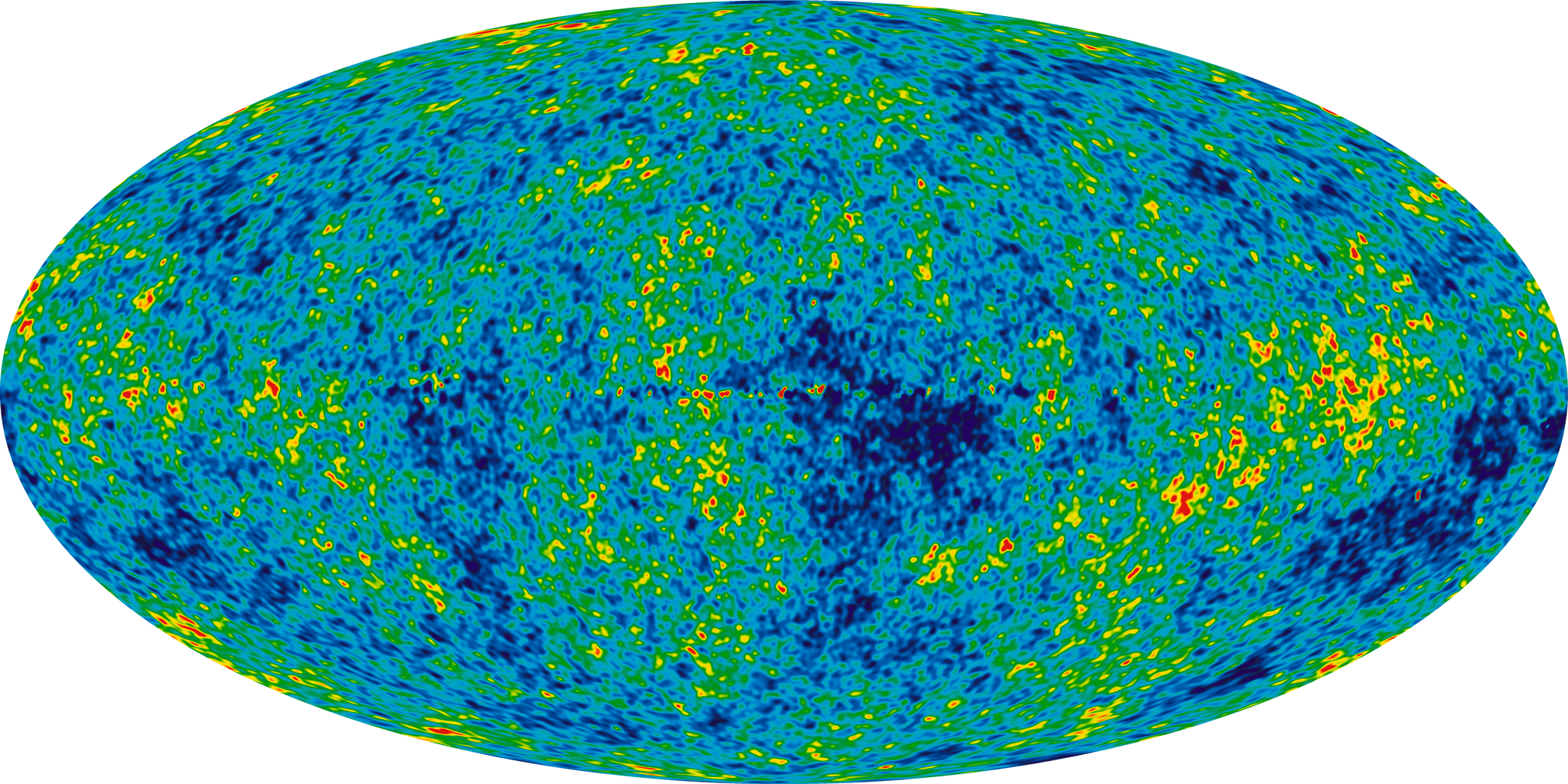}
\caption[Full-sky \cmbr~map from \wmap]{Full-sky \cmbr~map from \wmap. The monopole and dipole have been removed, as has the galactic foreground contamination. Red/blue shows temperature anisotropy of $+/- 200 ~\mu {\rm K}$ on a linear scale. Image courtesy of the \wmap~Science Team.\label{fig:wmap}}
\end{center}
\end{figure}

At high redshifts, $z \sim 1000$, cosmological data comes from the Cosmic Microwave Background (\cmbr). This was first studied in detail by the Cosmic Background Explorer (\cobe) satellite \cite{cobe}. Under a spherical harmonic decomposition, the \cmbr~has a dominant monopole with a temperature of $(2.728 \pm 0.004)\,{\rm K}$ at $95\%$ Confidence Limit (\cl). There is also a dipole arising from the Doppler shift induced by the Earth's peculiar velocity. After these are subtracted, the anisotropy remains. Following \cobe, the \wmap~mission~\cite{Larson:2010gs,Komatsu:2010fb,Hinshaw:2012fq} has imaged the \cmbr~with higher precision and provided an anisotropy map as shown in figure~\ref{fig:wmap}. The anisotropy corresponds to temperature variations of one part in $10^5$ relative to the monopole, demonstrating isotropy to high precision. 

\begin{figure}[h]
\begin{center}
\includegraphics[width=0.8\textwidth]{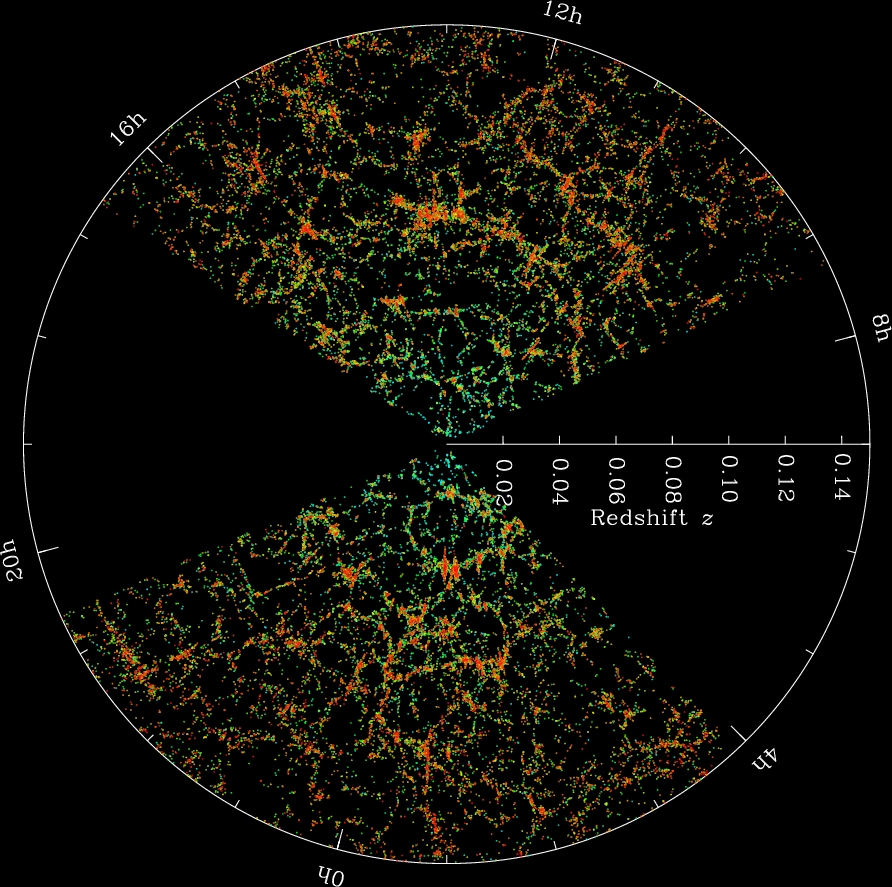}
\caption[\sdss~galaxy survey]{A two-dimensional slice through the \sdss~galaxy map with Earth at the center. Points represent galaxies, each typically containing $10^{11}$ stars. Galaxies containing older stars are coloured increasingly red. The black regions were not mapped due to the galactic foreground. Image courtesy of M. Blanton and the Sloan Digital Sky Survey.
\label{fig:sdss}}
\end{center}
\end{figure}

At lower redshifts, $z \sim 0.5-2$, isotropy is seen in spectroscopic observations of the Large Scale Structure (\lss) of galaxies and galaxy clusters. Two of the most comprehensive galaxy surveys are the Two-degree-Field Galaxy Redshift Survey ({\sc 2df})~\cite{Colless:2001gk,Cole:2005sx} and the Sloan Digital Sky Survey (\sdss)~\cite{Paris:2012iw}. A two-dimensional ({\sc 2d}) slice through \sdss~data is illustrated in figure~\ref{fig:sdss} which clearly illustrates small-scale structure in the form of filaments of enhanced galaxy population. However, after smoothing on some large scale, this provides excellent evidence for large-scale isotropy. 

\lss~and \cmbr~data provide a clear observational justification for the assumption of large-scale isotropy. Large-scale homogeneity, on the other hand, is challenging to justify on the basis of spectroscopic observations since these probe uniform redshift hypersurfaces within our past lightcone, whereas homogeneity is defined on spatial hypersurfaces~\cite{Maartens:2011yx}. The standard justification of the assumption of large-scale homogeneity appeals to the Copernican Principle: If we do not live in a special location in which the \cmbr~is isotropic by chance, then the \cmbr~must be isotropic for all freely-falling observers. In this case the Ehlers-Geren-Sachs theorem~\cite{Ehlers:1966ad} prescribes that the Universe is both homogeneous and isotropic.\footnote{The Ehlers-Geren-Sachs theorem formally only applies for a purely isotropic \cmbr~which is clearly only approximately true. The theorem was extended by Maartens et al.~\cite{Maartens:1995hh} to apply to the case with a small anisotropy.} 

The large smoothing scale applied to \lss~observations---usually referred to as the {\it homogeneity scale}---is typically taken to be $100\,{\rm Mpc}$,\footnote{`Mpc' is a standard astronomical unit of length, ${\rm Mpc} = 3.09 \times 10^{22} {\rm ~m}$.} although recent work has suggested that it could be much larger~\cite{Clowes:2012pn,Park:2012dn}. A homogeneity scale of $100~{\rm Mpc}$ corresponds to a redshift variation $\Delta z \approx 0.023$ which corresponds to roughly 16\% of the radial extent of the \lss~map shown in figure~\ref{fig:sdss}.

We also note that an area of active research considers the consequences of violating the assumptions of homogeneity and isotropy which can lead to modified cosmic dynamics~\cite{Bolejko:2011jc,AndrzejKrasinski:1997zz,Clarkson:2010uz}.

\ssec{General Relativity}

The \lcdm~model incorporates gravity as described by Einstein's general theory of relativity. The gravitational field equations may be derived from varying the Einstein--Hilbert action
\be
\label{eq:ehaction}
S_{\rm EH} = \frac{1}{2} \Mpl^2 \int \! \d^4 x \, \sqrt{-g} \, (R - 2 \Lambda)\,,
\ee
where $R$ is the Ricci scalar corresponding to the spacetime metric $g_{\mu \nu}$, $g=\det(g_{\mu \nu})$ and $\Lambda$ is the cosmological constant which we shall discuss in \S\ref{sec:stuff}. Demanding that $S_{\rm EH}$ has zero variation with respect to $g_{\mu \nu}$, one obtains the left hand side (\lhs) of the gravitational field equations
\be
R^{\mu \nu} - \frac{1}{2} g^{\mu \nu}R - g^{\mu \nu} \Lambda = \Mpl^{-2} \, T^{\mu \nu},
\label{eq:einsteinFE}
\ee
where $R^{\mu \nu}$ is the Ricci curvature tensor derived from $g^{\mu \nu}$. The spacetime curvature on the \lhs~of eq.~\eqref{eq:einsteinFE} is coupled to the matter content of the Universe through the symmetric stress-energy tensor $T^{\mu \nu}$ appearing on the right hand side (\rhs). We note that both sides of the Einstein field equations are covariantly constant.

There are a multitude of suggested alternatives to Einstein's theory of general relativity which arise from changing the form of the Einstein--Hilbert action \eqref{eq:ehaction}. For a thorough review we refer the reader to Clifton et al.~\cite{Clifton:2011jh}.

\ssec{The content of the Universe}
\label{sec:stuff}

The stress-energy tensor $T^{\mu \nu}$ may be decomposed into an energy density $T^{00}$, momentum density $T^{0i}$ and stress $T^{ij}$. The \lcdm~model provides a very simple description of the content of the Universe in terms of perfect fluids which are characterised only by their energy density $\rho$ and isotropic pressure $p$. Isotropy prescribes an absence of anisotropic stress, such that $T^{ij} = p \delta^{ij}$. If we pick a frame in which the 3-space velocity $v^i$ is small, but non-zero, then the momentum density may be written as $T^{0i}=(\rho + p)v^i$ to first order in $v^i$. At this same order, the energy density is simply $T^{00}=\rho$. Together, these conditions combine to yield a general form for the energy momentum tensor as
\be
T^{\mu \nu} = (\rho + p)u^\mu u^\nu + p g^{\mu \nu},
\label{eq:tmunu}
\ee
where $u^\mu$ is the fluid 4-velocity. By covariance, this expression is valid in any inertial frame. 

The continuity equation follows from the covariant constancy of the energy-momentum tensor, $\nabla_\mu T^{\mu \nu}=0$. Each non-interacting particle species `A', with density $\rho_{\rm A}$ and pressure $p_{\rm A}$, obeys a continuity equation as
\be
\dot \rho_{\rm A} + 3 H (\rho_{\rm A} + p_{\rm A}) = 0,
\label{eq:continuity}
\ee
where $H \equiv \dot a / a$ defines the Hubble parameter. For perfect fluids it is convenient to define the equation of state parameter $\omega_{\rm A}$ such that $p_{\rm A} = \omega_{\rm A} \rho_{\rm A}$. The value of $\omega_{\rm A}$ depends on the fluid and in general will vary with time. 

The fluids present in the \lcdm~model are baryons $(\omega_{\rm b}=0)$, relativistic photons and neutrinos $(\omega_{\rm rel}=1/3)$, Cold Dark Matter (\cdm) $(\omega_\textsc{cdm}=0)$ and dark energy $(\omega_\Lambda=-1)$. The dynamics of these homogeneous and isotropic fluids follow from the continuity equation~\eqref{eq:continuity} which integrates to give $\{\rho_{\rm b},\rho_\textsc{cdm} \} \propto a^{-3}$ and $\rho_{\rm rel} \propto a^{-4}$. 

\cdm~does not interact electromagnetically and so is not directly detectable by observation of photons; as such the nature of dark matter is unknown and is the subject of much debate~\cite{Bertone:2004pz}. Motivations for the existence of \cdm~are {\it indirect} and include: The inferred rapidity of structure formation~\cite{Davis:1985rj}, non-Keplerian galaxy rotation curves and Baryon Acoustic Oscillations (\bao)~\cite{Bassett:2009mm}. The \bao~arise in the Universe around the time of photon decoupling, when competition between gravitational collapse and pressure lead to the generation of acoustic waves. Measurements of \bao, such as by {\sc sdss}~\cite{Paris:2012iw}, provide a key constraint on the \lcdm~model.

The \lcdm~model provides a simple realisation of the dark energy component by identifying it with the cosmological constant appearing in the Einstein--Hilbert action~\eqref{eq:ehaction}. 

\ssec{Dynamics of the \lcdm~universe}
\label{sec:quantifyinglcdm}

Substituting the energy momentum tensor~\eqref{eq:tmunu} and the \frwl~metric \eqref{eq:frwl} into the gravitational field equations \eqref{eq:einsteinFE}, we obtain the {\it Friedmann} and {\it Raychaudhuri} equations for $a(t)$ respectively as
\begin{align}
\label{eq:general-friedmann}
\frac{\dot a^2}{a^2} &= \frac{\rho}{3\Mpl^2} + \frac{\Lambda}{3}  - \frac{k}{a^2},\\
\frac{\ddot a}{a} &= -\frac{1}{6\Mpl^2} ( \rho+ 3p) + \frac{\Lambda}{3}.
\label{eq:acceleration}
\end{align}

The Raychaudhuri equation tells us that the \lcdm~universe, in the absence of the $\Lambda$ term, would be decelerating. This is at odds with observational data from surveys of type-1A supernovae (\snia)~\cite{Perlmutter:1997zf,Riess:1998cb} that have led to the conclusion that the Universe is accelerating in its expansion. This acceleration has since been independently corroborated by \cmbr~data from high multipole measurements using the South Pole Telescope (\spt)~\cite{Story:2012wx} which possess much higher sensitivity to weak gravitational lensing by large-scale structure at low redshifts. The \lcdm~model achieves this acceleration through a small positive value of $\Lambda$.

We now present observational bounds on some of the key \lcdm~model parameters. Dividing the Friedmann equation \eqref{eq:general-friedmann} by the critical density $\rho_{\rm crit} = 3 \Mpl^2 H^2$ yields
\be
\Omega_{\rm tot} = \Omega_{\rm b} + \Omega_\textsc{cdm} + \Omega_{\rm rel} + \Omega_\Lambda = 1- \Omega_{\rm k} ,
\label{eq:omegaeqn}
\ee
where the density parameters are defined as $\Omega_{\rm A} = \rho_{\rm A} / \rho_{\rm crit}$, except for $\Omega_{\rm \Lambda} = \Lambda / 3 H^2 $ and $\Omega_{\rm k} = -k / H^2 a^2 $. The latest \wmap~9-year results \cite{Hinshaw:2012fq}, combined with data from \bao~\cite{Percival:2009xn} and the Hubble Space Telescope (\hst) \cite{Riess:2011yx}, give $100 \, \Omega_{\rm b} \, h^2 = 2.266 \pm 0.043$, $\Omega_\textsc{cdm} \, h^2 = 0.1157 \pm 0.0023$ and $\Omega_\Lambda = 0.712 \pm 0.010$ at $68\%$ \cl, where $h$ is the dimensionless present value of the Hubble parameter defined as 
\be
H_0 = 100 \, h \,{\rm Km~s}^{-1} {\rm ~Mpc}^{-1} = (3.247 \times 10^{-18} s^{-1})\,h.
\ee
The current best constraint on this value is $h = 0.6933 \pm 0.0088$ at $68\%$~\cl~\cite{Hinshaw:2012fq}. 
We note that the Friedmann, Raychaudhuri and continuity equations \eqref{eq:general-friedmann}, \eqref{eq:acceleration} and \eqref{eq:continuity} together inform us that the Hubble rate monotonically decreases with time. Note that we have neglected the present density of relativistic species. This is much smaller than the other components and may be calculated using the tools developed in \S\ref{sec:appendix_quantifying_inflation} to be $\Omega_{{\rm rel},0} = (4.13 \pm 0.11) \times 10^{-5}$. 

\begin{figure}[h]
\begin{center}
\includegraphics[width=0.5\textwidth]{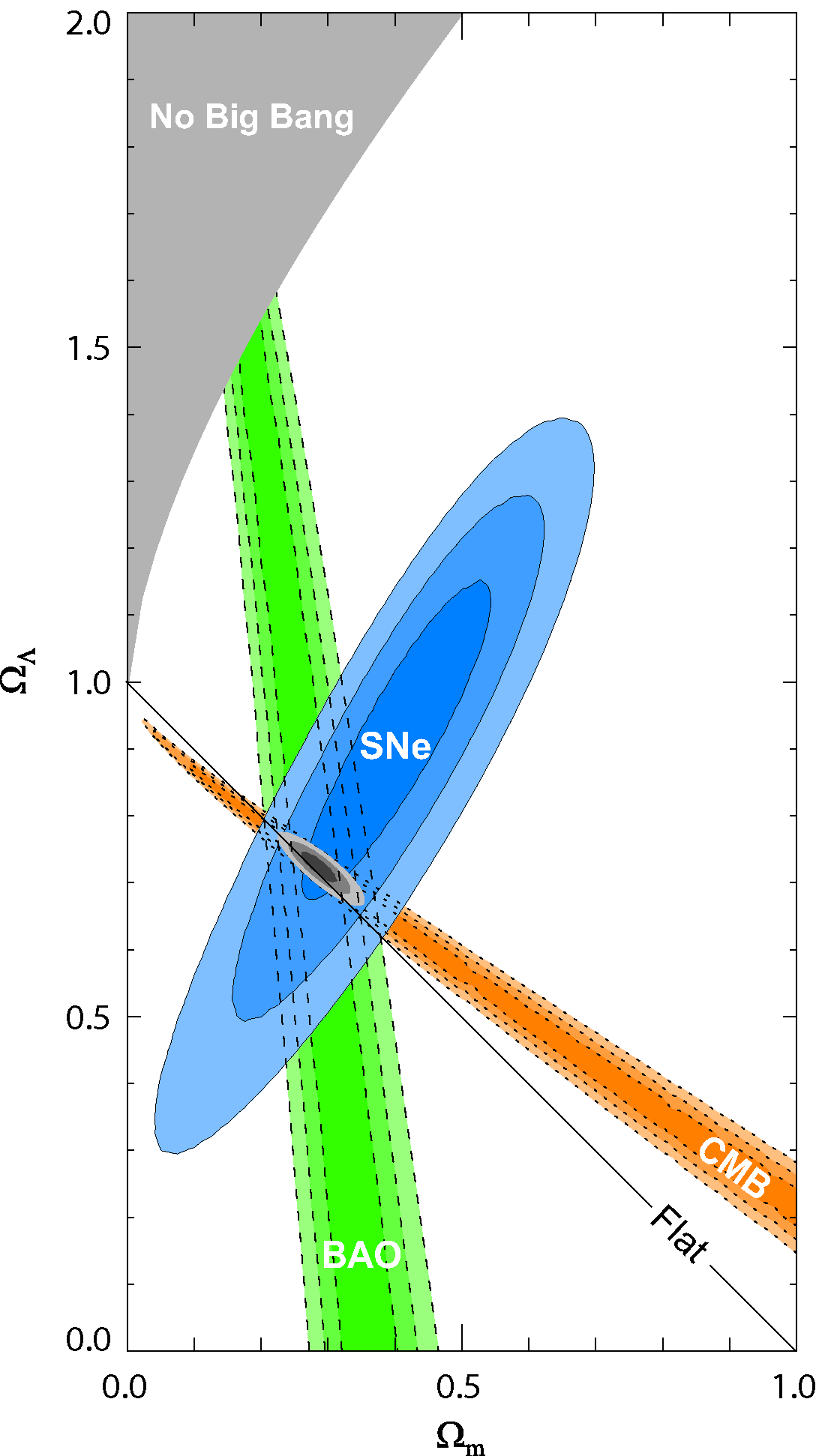}
\caption[Constraints on cosmic density parameters]{Observational constraints on the cosmic density parameters $\Omega_\Lambda$ and $\Omega_{\rm m} = \Omega_\textsc{cdm} + \Omega_{\rm b}$ from \cmbr, \bao~and~\snia. Image courtesy of the Supernova Cosmology Project~{\rm \cite{Kowalski:2008ez}}.\label{fig:triangle}}
\end{center}
\end{figure}

These data are illustrated in figure~\ref{fig:triangle}, which demonstrates that $\Omega_{\rm tot}$ is observationally consistent with unity, representing a flat spatial geometry.

\sec{The need for inflation}
\label{sec:needinflation}

Thus far we have discussed the \lcdm~model as an effective model of our Universe. There are, however, serious issues with the \lcdm~model that may be averted by adding an initial epoch of accelerated expansion. This epoch, which was originally proposed in the 1980s by a number of independent authors~\cite{Guth:1980zm,Starobinsky:1980te,Sato:1980yn}, is called {\it inflation}. In order to explain these issues, we must first introduce some theoretical concepts.

\ssec{Theoretical prerequisites}
\label{sec:theoretical}

\paragraph{Conformal time.} The spatially flat $(k=0)$ \frwl~metric can be put into conformally flat form by introducing a conformal time coordinate
\be
\eta = \int \frac{\d t}{a}.
\ee
For inflationary dynamics we choose integration limits between $\infty$ and the time of measurement $t$, such that $-\infty < \eta < 0$. However, for a universe filled with dust or radiation then the appropriate integration limits are from $0$ to $t$ which lead to a positive definite range for $\eta$. The spatially flat \frwl~metric then takes the form
\be
\label{eq:conformalfrwl}
\d s^2  = a(\eta)^2 \big(-\d \eta^2 + \d x^2 + \d y^2 + \d z^2 \big).
\ee

\paragraph{Particle horizon.} The {\it particle horizon} is the maximal distance that light may have travelled since the start of the Universe where $t=0$. Assuming causality, spacetime points that are separated by more than their particle horizon have never interacted. If the Universe is flat, light follows null geodesics of the conformal \frwl~metric \eqref{eq:conformalfrwl} and the comoving particle horizon distance $d_{\rm ph}$ is simply the change in the conformal time
\be
d_{\rm ph} = \eta(t) - \eta(0).
\ee
As an example, if the Universe is presumed to be spatially flat and contain only radiation, then one can integrate the Friedmann equation \eqref{eq:general-friedmann} to obtain $\eta = (aH)^{-1}$. Thus the particle horizon equals the comoving Hubble length. Allowing for pressureless matter only modifies this conclusion by a $\O(1)$ factor.

\paragraph{Inflation.} The acceleration of the \frwl~universe is precisely correlated with a shrinking comoving particle horizon size 
\be
\label{eq:inflationcorrespondance}
\ddot a > 0 \quad \Leftrightarrow \quad \frac{\d}{\d t} (aH)^{-1} <0.
\ee
Under such accelerating conditions, the \frwl~universe is described as {\it inflating}. It proves useful to define the positive definite parameter $\epH$ as
\be
\label{eq:epH}
\epH = -\frac{\dot H}{H^2},
\ee
such that inflation corresponds to the regime $\epH <1$. We note that eq.~\eqref{eq:inflationcorrespondance} informs us that a decelerating universe has a growing comoving Hubble length $(aH)^{-1}$.  

\paragraph{Efolds.} The expansion of the Universe between times $t_1$ and $t_2$ is often written in terms of the dimensionless number of {\it efolds} $N$. Defining the scale factor and Hubble rate at time $t_1$ as $a_1$ and $H_1$ respectively (and similarly for $t_2$, $a_2$ and $H_2$), $N$ is defined as
\be
\label{eq:N}
N = \ln \left( \frac{a_2}{a_1} \right) = \int_{t_1}^{t_2} \! H \, \d t.
\ee
An alternative definition of efolds relates the comoving horizon size $(a H)^{-1}$ between two different times and may be defined as
\be
\label{eq:Nbar}
\bar N = \ln \left(\frac{a_2 H_2}{a_1 H_1} \right).
\ee
Throughout this thesis, our use of `efolds' shall refer to $N$, rather than $\bar N$.

\ssec{How inflation resolves problems with the \lcdm~model}
\label{sec:bbnprobs}

\sssec{Horizon Problem}
In \S\ref{sec:theoretical} we calculated that the early phases of the \lcdm~model (which may be reasonably approximated as radiation-dominated) have a particle horizon of order the comoving horizon size. Hence the particle horizon at the time of \cmbr~last scattering is roughly $205$ Mpc, which corresponds to an angular separation of roughly one degree as measured today~\cite{Lyth:2009zz}. Thus the \lcdm~model prescribes that the \cmbr~should be comprised of about 40,000 causally disconnected regions.

The Horizon Problem is how to reconcile the uniformity of the \cmbr~with the causal separation of its component parts. A complementary description of the Horizon Problem is to ask why the \cmbr~has a near-perfect black-body spectrum \cite{cobe} which strongly implies a system in thermal equilibrium. To achieve the observed \cmbr~isotropy within the \lcdm~model requires fine-tuning of the initial conditions of each of the causally disconnected patches. 

This problem is averted if one presumes an initial inflationary epoch because this will reduce the comoving horizon size via eq.~\eqref{eq:inflationcorrespondance}. Given sufficient inflation, the comoving horizon size reduces to encompass a small homogeneous portion of the particle horizon. If inflation then ends and is followed by the standard \lcdm~decelerating expansion, we will {\it expect} the near-isotropy of the \cmbr~as a {\it requirement} for our model to be valid.

\sssec{Flatness Problem}
To understand this problem in the context of the \lcdm~model let us take a non-flat model with $k = \pm 1$ and approximate the \lcdm~universe to be radiation dominated. Taking the ratio of the Friedmann equation \eqref{eq:omegaeqn} between an early time $t=t_e$ and today, where $t=t_0$, we find
\be
\frac{1-\Omega_{{\rm tot},e}}{1-\Omega_{{\rm tot},0}} = \frac{t_e}{t_0}.
\label{eq:flatness_explained}
\ee
We may take the age of the Universe to be approximately 13.7 Gyr \cite{Lyth:2009zz} and $t_e \approx 1$s to be the early epoch of Big Bang Nucleosynthesis ({\sc bbn}). {\sc bbn} is the process by which the light elements such as hydrogen, helium, lithium and deuterium are produced in the early Universe, which is in excellent agreement with observed abundances found in clouds of interstellar gas~\cite{Coc:2003ce}. Between {\sc bbn} and today, eq.~\eqref{eq:flatness_explained} gives
\be
|1-\Omega_{{\rm tot},e}| \sim 10^{-17} |1-\Omega_{{\rm tot},0}|.
\ee
We therefore see that, since the Universe is close to flatness today, it must have been incredibly fine-tuned towards flatness at earlier times. In this respect the Flatness Problem is similar to the Horizon Problem, in that the \lcdm~model provides no means for choosing such delicate initial conditions. 

It is easy to see how inflation resolves the Flatness Problem by reconsidering 
the ratio of the Friedmann equation under the assumption that our Universe undergoes an early period of quasi-exponential expansion in some time interval $t_i < t < t_e$. In this case the Hubble rate $H$ remains roughly constant and one finds
\be
\frac{1-\Omega_i}{1-\Omega_e} \approx \frac{a_e^2}{a_i^2}. 
\ee
As inflation proceeds, the scale factor grows quasi-exponentially and so $a_e \ggg a_i$. For a generic initial density $\Omega_i \neq 1$, this entails that the density parameter $\Omega_e$ is driven incredibly close to unity and so the Universe evolves very close to flatness at time $t_e$. Given enough inflation, this behaviour is sufficient to ensure that the Universe that we see today is spatially flat. In fact, the near-spatial flatness of our Universe may be regarded as an important {\it consequence} of inflation.

\sssec{Relic particle production}
In extended models of high energy particle physics it is possible to generate high-mass particles in the early Universe, such as magnetic monopoles \cite{'tHooft:1974qc}. If their mass is very large then they will become non-relativistic whilst the remainder of the Universe is sufficiently hot to be relativistic. As shown in \S\ref{sec:stuff}, the energy density of non-relativistic species decays more slowly and so even a small initial population of such heavy particles will quickly dominate the Universe, which is inconsistent with what we observe.

Inflation solves this problem by diluting the relic particles to sufficiently small densities that any such particle production would not conflict with observations. One then requires that the particles are not reproduced at the end of inflation, in the period called {\it reheating}, which may be achieved by having the reheating temperature sufficiently low.

\ssec{Quantifying the amount of observable inflation}
\label{sec:quantifying_inflation}

We have discussed qualitatively how the horizon, flatness and relic particle problems can be resolved by admitting a period of accelerated expansion. We now quantify the amount of inflation required. The standard answer follows from the procedure outlined in Liddle and Leach~\cite{Liddle:2003as}. Instead, we follow a thermodynamic argument which avoids presuming an instantaneous transition between radiation and matter epochs. For simplicity, we presume instantaneous reheating at a temperature $T_{\rm reh}$ which defines the time of transition between inflation and decelerated expansion.

In \S\ref{sec:theoretical} we showed that, in all but the most recent phase of the post-inflationary Universe, the comoving Hubble length $(a H)^{-1}$ is growing. Thus the present Hubble length $(a_0 H_0)^{-1}$ is much greater than it was at the end of inflation, $(a_{\rm reh} H_{\rm reh})^{-1}$. Under the assumption that the present comoving horizon size was well inside the particle horizon at some earlier time, it follows that the present comoving horizon size crossed the horizon at some point during inflation, which we denote with a label `$*$'.\footnote{In the body of this thesis we shall use $*$ to denote the time at which a given Fourier mode $k$ crossed the horizon. For contact with observational data this is conventionally taken to be the \wmap~pivot scale $k = 0.002 \, {\rm Mpc}^{-1}$, which corresponds to a spherical multipole $l \simeq 30$.} We therefore have
\be
\label{eq:efoldsequate}
\frac{(aH)_0}{(aH)_*} = 1 = \frac{(aH)_0}{(aH)_{\rm reh}} 
e^{\bar{N}_{\rm inf}},
\ee
where 
\be
e^{\bar{N}_{\rm inf}} = \frac{(aH)_{\rm reh}}{(aH)_*},
\ee
is the number of efolds of {\it observable inflation}. During inflation with $\epH \ll 1$, the Hubble parameter does not vary appreciably. Therefore, $\bar N_{\rm inf} \approx \ln (a_{\rm reh}/a_*) = N_{\rm inf}$ and it is conventional to calculate $N_{\rm inf}$ rather than $\bar{N}_{\rm inf}$. Using eq.~\eqref{eq:efoldsequate}, we calculate that the required amount of inflation is therefore equal to the ratios of comoving Hubble lengths after inflation as
\be
\label{eq:Ninf}
N_{\rm inf} = \ln \left(\frac{a_{\rm reh}}{a_0}\right) + \frac{1}{2}\ln \left(\frac{H^2_{\rm reh}}{H^2_0} \right).
\ee
We note that the total number of inflationary efolds may be significantly larger, but that only the last $N_{\rm inf}$ efolds correspond to scales of observable interest. 

In \S\ref{sec:appendix_quantifying_inflation} we show how we can evaluate the two parts of eq.~\eqref{eq:Ninf} separately. We find
\be
N_{\rm inf} = \frac{1}{2}\ln \Omega_{{\rm rel},0} 
+ \frac{1}{6} \ln \left( \frac{g_{\rm reh}}{g_0}\right) 
+ \ln \left( \frac{T_{\rm reh}}{T_0} \right),
\ee
where $g_{\rm reh}$ is the particle degeneracy factor and $g_0=43/11$ (see \S\ref{sec:appendix_quantifying_inflation} for details). Taking the typical value $T_{\rm reh} = 10^{16}~{\rm GeV}$ \cite{Lyth:2009zz}, compared with $T_0 = 2.35 \times 10^{-13}~{\rm GeV}$ \cite{cobe}, and further approximating $g_{\rm reh} = 100$, one finds $N_{\rm inf} = 61.1$. Whist this is relatively insensitive to variations of $g_{\rm reh}$, it is sensitive to the choice of $T_{\rm reh}$, which one may argue could be different by many orders of magnitude \cite{Liddle:2003as}. In conclusion, a necessary condition for a given inflationary model to describe our Universe is that it provides $\sim 60$ efolds of inflation. 

\ssec{Cosmic perturbations from inflation}

We have seen that inflation is capable of resolving the Horizon, Flatness and Relic Particle Problems for which it was originally designed. However, its biggest success has been its prediction of the anisotropy that has since been observed in the \cmbr. At present, the whole-sky anisotropy of the \cmbr~has been most precisely observed by the \wmap~satellite~\cite{Larson:2010gs,Komatsu:2010fb,Hinshaw:2012fq}. Recently, this data has been augmented by measurements of the small-scale anisotropy as measured by the \spt~\cite{Keisler:2011aw,Story:2012wx} and the Atacama Cosmology Telescope ({\sc act})~\cite{Fowler:2010cy,Das:2011ak}. In the near future, the full-sky anisotropy will be observed with increased precision by the Planck mission~\cite{Planck:2006aa}. 

Full-sky \cmbr~data does not provide the only observational constraints on inflation. Observations of \bao~in \lss~surveys have been very successful in constraining the parameters of the \lcdm~model~\cite{Bassett:2009mm}. In addition, future constraints are expected from {\sc 3d} observations of 21cm emissions~\cite{Morales:2009gs} and \cmbr~polarisation observations such as {\sc spider}~\cite{Crill:2008rd}.

The statistics of the anisotropy provides a powerful constraint on the dynamics of inflation. We shall defer a thorough discussion to chapter \ref{ch:formalisms} where we shall properly discuss observable quantities and their evolution. Nevertheless, we emphasise that a model of inflation is not judged purely on the basis of generating sufficient background expansion; it is also necessary to ensure that the model produces observably consistent perturbation statistics.

We now describe one of the simplest phenomenological inflationary models---where the expansion is riven by a single, canonical and minimally coupled scalar field---for the dual purposes of orientation and introducing important concepts.

\sec{The simplest inflationary scenarios}
\label{sec:simplest}

The simplest and most commonly studied models of inflation employ a single canonical scalar field $\vp$ that is minimally coupled to the Ricci curvature (see \cite{Lidsey:1995np,Lyth:1998xn,Linde:2005ht,Bassett:2005xm} for reviews). A particularly simple realisation of this scenario is provided by the chaotic inflationary model of Linde~\cite{Linde:1983gd}, a more general version of which we shall study in chapter~\ref{ch:singlefield}. The scalar field $\vp$ is known as the {\it inflaton} and it is considered to dominate the energy density of the Universe during the early epoch of accelerated expansion. 

The dynamics follow from varying the action 
\be
\label{eq:single_field_action}
S = \frac{1}{2} \int \! \d^4 x \, \sqrt{-g} \, 
\Big( \Mpl^2 R + X 
- V(\vp) \Big).
\ee
which combines the theory of gravity in the form of the Einstein--Hilbert action~\eqref{eq:ehaction} with the Lagrangian ${\cal L}_\vp$ for the inflationary field. The Lagrangian ${\cal L}_\vp$ is the usual combination of kinetic energy $X = -g^{\mu \nu} \partial_\mu \vp \partial_\nu \vp/2$ and the potential energy $V(\vp)$. From a particle physics perspective, the potential $V(\vp)$ defines the self-interaction of inflaton particles. 

This action is varied with respect to a background {\frwl} metric and a homogeneous scalar evolution $\phi(t)$---full details are given in chapter~\ref{ch:subhorizon}. One finds background equations of motion of the form
\begin{align}
3 \Mpl^2 H^2 &= \dot \phi^2/2 + V(\phi) , \\
2 \Mpl^2 \dot H &= -\dot \phi^2, \\
\label{eq:singleKG}
\ddot \phi &= -3 H \dot \phi - V'(\phi),
\end{align}
where $V'(\phi) = \partial V(\phi)/\partial \phi$. The parameter $\epH$ becomes
\be
\label{eq:epHsingle}
\epH = \frac{ 3 \dot \phi^2}{\dot \phi^2 + 2 V} ,
\ee
which yields inflation in the limit that the potential energy dominates over the kinetic energy, $V(\phi) \gg \dot \phi^2/2$. 

\ssec{The slow-roll approximation}
\label{sec:intro-slow-roll}

The majority of viable candidate models of inflation satisfy the constraints of {\it slow-roll} for at least some of their evolution. Slow-roll implies that the kinetic energy is much smaller than the potential energy. In this regime the Friedmann and Klein--Gordon equations may be approximated as
\begin{align}
\label{eq:friedmannSR}
3 \Mpl^2 H^2 &\simeq V(\phi), \\
\label{eq:KGSR}
3 H \dot \phi &\simeq -V'(\phi).
\end{align}
These simplifications are crucial for facilitating further analytic progress, since they reduce the scalar field degrees of freedom from $\{\phi,\dot \phi\}$ to merely $\{\phi\}$. Such reduction of the dynamical phase space is formally the limit of convergence to a lower-dimensional hyperspace---referred to in the literature as the {\it inflationary attractor}~\cite{Lyth:2009zz}.

Slow-roll may be formalised by demanding the smallness of a series of {\it potential slow-roll parameters} that describe the shape of the inflationary potential. The first such parameter is 
\be
\label{eq:sf-ep}
\ep = \frac{\Mpl^2}{2} \left(\frac{V'}{V}\right)^2.
\ee
In the limit that $\ep \ll 1$ then $\ep \simeq \epH$ and so the scale factor will obey quasi-exponential expansion $a(t) \propto e^{H t}$, with the Hubble rate $H$ approximately constant. From eq.~\eqref{eq:sf-ep}, the constraint on the inflationary potential for this to occur is that it has sufficient magnitude and flatness.

The condition $\ep \ll 1$ describes the criterion for inflation at a particular field value. By itself, this is insufficient, since we need inflation to endure for $\sim 60$ efolds and in this time the field value will evolve in time. One may guarantee the enduring smallness of $\ep$ by making a temporal Taylor expansion of the inflationary potential and demanding that the series of coefficients produced are sufficiently small. The first such expansion term is
\be
\label{eq:sf-eta}
\frac{\d \ep}{\d N} \simeq 2 \ep (2 \ep - \eta) \ll 1 , \qquad
\eta = \Mpl^2 \frac{V''}{V},
\ee
where we have presumed that the slowly-rolling inflaton evolves monotonically from positive values towards zero, such that $\dot \phi \simeq - \Mpl H \sqrt{2 \ep}$. The condition that $\ep$ varies slowly corresponds to stipulating $\eta \ll 1$ and so we refer to $\eta$ as the second slow-roll parameter.\footnote{Note that we use $\eta$ both as a slow-roll parameter and to denote conformal time, but it should be clear from the context which is the appropriate meaning.} Within the slow-roll approximation, both $\ep$ and $\eta$ are typically considered to have comparable magnitudes. Higher-order slow-roll parameters may be derived analogously as discussed in ref.~\cite{Liddle:1994dx}. 

We also note that one can define a complementary set of {\it Hubble slow-roll} parameters derived around the variation of the Hubble rate rather than the potential~\cite{Liddle:1994dx}. The first such parameter is $\epH$. The Hubble slow-roll and the potential slow-roll parameters are precisely related, but the potential slow-roll parameters provide a more direct means for discerning how the shape of the potential affects the dynamics of inflation. Since developing such intuition will be a focus of this thesis, in the following we shall use the potential slow-roll parameters.

\ssec{Types of inflation}

Inflationary models fall into two broad categories: {\it large-field inflation} and {\it small-field inflation}. Large-field inflation involves scalar fields that evolve over super-Planckian field values, $|\Delta \vp| > \Mpl$. The best known example is chaotic inflation where the inflationary potential is of power-law form $V(\vp) = V_0 \vp^p$, for positive constants $V_0$ and $p$, and this can generate inflation in the regime $|\vp| \gg \Mpl$. Such models are also interesting because they typically lead to the generation of gravitational waves~\cite{Lyth:1996im}.

On the other hand, small field inflation involves scalar fields that evolve over sub-Planckian field values, $|\Delta \vp| < \Mpl$. Small field models often arise as the effect of spontaneous symmetry breaking, such as in Hybrid inflation~\cite{Mulryne:2011ni}. In either class of these inflationary models, inflation typically ends when the slow-roll conditions are violated, after which reheating occurs. 

\ssec{Reheating}

Reheating is the phase of the Universe's evolution between the inflationary epoch and the hot dense state described by the early phases of the \lcdm~model. The dynamics of reheating is complex, uncertain and the subject of much study (see e.g. refs.~\cite{Allahverdi:2010xz,Bassett:2005xm} for reviews). Models fall into two broad categories, the first simply known as `reheating' and the second as `preheating'.

The standard reheating mechanism is based on the single-body decay of the inflaton into standard model particles. This model views the inflaton field $\vp$ as a collection of scalar particles, each of which has equal probability of decaying into some other species. For example, decay into a fermion $\psi$ may arise through an interaction term $h \vp \psi \bar \psi$ appearing the the Lagrangian, where $h$ is some coupling constant. This may be most simply modelled alongside the inflationary dynamics by addition of a perturbative decay term $\Gamma \dot \vp$ in the equations of motion (for example, see ref.~\cite{Leung:2012ve}), where $\Gamma$ is typically of order the inflaton mass.

Preheating generates standard model particles through non-perturbative particle production during the oscillations of the inflaton field $\vp$. The simplest example involves a potential of the quadratic form $V = m^2 \vp^2 / 2$ which, after inflation has ended, exhibits underdamped oscillations as $\vp(t) = [\Mpl/(\sqrt{3 \pi} m t)] \sin (m t)$~\cite{Bassett:2005xm}. We may now consider a coupling of the inflaton which, for the sake of simplicity, we take to be with another scalar $\chi$. If such a coupling takes the form $g^2 \vp^2 \chi^2 / 2$ for some dimensionless coupling constant $g$, then we see that the $\chi$ field has an effective mass $m_\chi^{\rm eff} = g^2 \vp^2$. Crucially, this becomes zero at the minimum of each $\vp$ oscillation. For suitable parameters that can produce instability in the perturbation evolution equations, resonance may then occur, leading to exponential growth of the perturbations $\delta \chi$ which corresponds to explosive particle production.

\ssec{Virtues of the simplest inflationary models}

Many of the inflationary models that were first studied had only one dynamically relevant scalar field. This initial choice is not without reason, indeed such models have redeeming features including~\cite{Cicoli:2012cy}:
\begin{itemize}
\item {\it Simplicity} -- whilst the degree of simplicity possessed by a theory need have no correlation to its accuracy in characterising physical law, there are clear pragmatic advantages.
\item {\it Effectiveness} -- not only do such models resolve the known issues with the \lcdm~model, they have been widely shown to generate observably consistent perturbation statistics.
\item {\it Predictivity} -- the predicted statistics of different inflationary models are not wholly degenerate, providing the means for differentiating between candidate models.
\item {\it Robustness} -- the conservation of the curvature perturbation shortly after horizon exit (as discussed in chapter \ref{ch:formalisms}) guarantees that the predictions of single field inflationary models are easily related to current observations, despite our uncertainty regarding the physics operating at intermediate times. 
\end{itemize}

Why then should we consider more complex models? There are two compelling reasons: The first arises from the growing strength of observational data, particularly the imminent possibility of a detection of non-Gaussianity which would provide new ways to constrain inflationary models. The second is from our desire to situate inflation within theories of high energy physics which motivates more complex inflationary scenarios. We now discuss these possibilities.

\sec{Generalised inflationary scenarios}
\label{sec:beyondsimplest}

Improvements in cosmological data will inevitably lead to tighter constraints on all of the cosmological parameters. This will, in turn, restrict the parameter space of viable inflationary models. A potentially exciting example lies with \cmbr~non-Gaussianity. With current data, the primordial \cmbr~perturbations are compatible with a Gaussian random distribution. This is the generic prediction of the simplest models of inflation, as first shown by Maldacena in 2002~\cite{Maldacena:2002vr}, under the joint assumptions that inflation is driven by a single field with canonical kinetic energy and evolving under global slow-roll conditions.

However, there is a possibility that new data, such as that from Planck~\cite{Planck:2006aa}, will detect a {\it non-Gaussian} signal. By Maldacena's theorem, such a detection would {\it require} us to consider inflationary models beyond the simple scenarios he considered, and furthermore, consider how these more complex models may produce observable non-Gaussianity. At the same time, understanding the non-Gaussian predictions of different inflationary models may in principle allow us to constrain candidate theories of fundamental interactions. 

To ensure that the full implications of future cosmological data are appreciated requires a firm theoretical framework that links inflationary dynamics and observables such as non-Gaussianity. This thesis provides such a study: Working with a broad class of non-canonical multi-field inflationary models, chapters \ref{ch:subhorizon} and \ref{ch:formalisms} derive new theoretical tools that enhance our ability to track the evolution of non-Gaussianity in both the subhorizon and superhorizon regimes. This includes analytic approximations of the superhorizon evolution of non-Gaussianity, the validity of which is then discussed and investigated in chapter \ref{ch:adiabatic}. Chapter \ref{ch:heatmaps} then uses these analytic results to draw new conclusions, demonstrating those conditions under which the non-Gaussianity may evolve above the detection threshold, and also developing important intuition to explain why this happens.

\ssec{Inflation from fundamental theory}
\label{sec:fundamental_theory}

Our ultimate goal is to realise inflation as a product of fundamental laws of physical interactions. The physics operating at the high energy scales of inflation are not known to us and it is interesting to consider how situating inflation within high energy theory may modify the simplest phenomenological inflationary scenarios. Modifications may include multiple fields, non-canonical kinetic energy, non-minimal coupling and a compilation of terms arising from effective string theory corrections. We now very briefly consider these possibilities and their motivation.

\sssec{Multi-field inflation}
\label{sec:multiple_fields}

The energy scales of inflation are vastly higher than those to which the Standard Model applies. In this untested regime, string theory is currently our best-developed framework for exploring issues such as quantum gravity. The result of years of work has led to the consensus opinion that multi-field dynamics are to be expected~\cite{Quevedo:2002xw,Linde:2005dd,McAllister:2007bg}. This motivates our study multi-field inflation in chapters \ref{ch:subhorizon}--\ref{ch:heatmaps}. 

Furthermore, it has proven challenging for fundamental theories of interactions to generate inflatons with a mass sufficiently smaller than the Hubble rate~\cite{Cicoli:2011zz}---a necessary prerequisite for observable consistency. When the mass is of the order of the Hubble rate, the slow-roll expansion becomes non-perturbative and this is known as the $\eta$-problem. Realisations of inflation nevertheless exist~\cite{Copeland:1994vg}, by employing a range of procedures to effect a reduced mass. Importantly, the act of reducing the inflaton mass often leads to activating couplings to other fields which may affect cosmological evolution~\cite{Cicoli:2012cy,Mulryne:2011ja}. The dynamics of these additional fields must therefore be followed in the context of a multi-field inflationary model.

\sssec{Non-canonicality}

Non-canonicality refers to the presence of a non-standard kinetic energy term in the action. For the simplest models of single field inflation discussed in \S\ref{sec:simplest}, the kinetic energy assumed the {\it canonical} form $ {\cal L}_\vp \supseteq X$ where $X = - \frac{1}{2} g^{\mu \nu} \partial_\mu \vp \partial_\nu \vp$. For multi-field inflation, we choose {\it canonical} to mean $ {\cal L}_\vp \supseteq X$ where $X = - \frac{1}{2} \delta_{IJ} g^{\mu \nu} \partial_\mu \vp^I \partial_\nu \vp^J$. 

Assuming a smooth potential, canonical inflationary models have been shown to generate unobservable non-Gaussianity at the time of horizon crossing \cite{Seery:2005gb,Seery:2006vu,Seery:2008ax,Chen:2010xka}. However, if one considers {\it non-canonical} kinetic terms, then a large non-Gaussianity can develop. 
A typical non-canonical Lagrangian is a complex function of the kinetic energy $X$, such as for Dirac-Born-Infeld (\dbi) inflation. When this may effect a reduced sound speed for perturbations, it can significantly enhance the non-Gaussian signal \cite{Alishahiha:2004eh,Seery:2005wm,Chen:2006nt,Cheung:2007st,
Arroja:2008yy,Langlois:2008qf,RenauxPetel:2011uk}. We discuss a specific case of this at greater length in chapter~\ref{ch:singlefield} when we consider Galileon corrections to chaotic inflation.

However, other non-canonical models naturally arise. High energy theories such as string theory and supergravity generically involve kinetic energy terms of the form 
\begin{equation}
\label{eq:X_non-canonical}
	X = - \frac{1}{2} G_{IJ}(\vp^K) g^{\mu \nu}
	\partial_\mu \vp^I \partial_\nu \vp^J ,
\end{equation}
where $G_{IJ}(\vp^K)$ is an arbitrary, symmetric function of the fields $\vp^I$. The simplest example is the non-linear $\sigma$-model of
Gell-Mann and L\'{e}vy, originally introduced to describe spin-0 mesons~\cite{GellMann:1960np}. A second example is K\"{a}hler moduli inflation, where the metric arises as the Hessian of the K\"{a}hler potential $K$ as $G_{IJ} \propto \partial_I \partial_J K$ \cite{Conlon:2005jm}. In chapters~\ref{ch:subhorizon} and~\ref{ch:formalisms} we shall derive the necessary new theoretical framework required to evolve perturbations for non-canonical models with kinetic energies of the form given in eq.~\eqref{eq:X_non-canonical}. 

\sssec{Non-minimal coupling}
\label{sec:nonminimal}

The simple single field inflationary model described in eq.~\eqref{eq:single_field_action} is {\it minimally coupled}, in that the Ricci scalar $R$ is not directly multiplied by any other fields. {\it Non-minimally coupled models} incorporate such a coupling. A simple example exists within the standard model of particle physics via Higgs inflation: The Higgs scalar $\vp$ can support a viable inflationary phase as a result of a non-minimal coupling to the Ricci scalar of the form $(1-\zeta \vp^2 / \Mpl^2)R$~\cite{Bezrukov:2007ep}. We shall discuss this scenario in greater depth in \S\ref{sec:single-minimally}.

In the context of multi-scalar field inflation, a non-minimal coupling manifests as an action such as 
\be
	S = \int \d^4 x \, \sqrt{- g} \left[
		\frac{1}{2} \Mpl^2 \, f(\vp^K) R
		- \frac{1}{2} G_{IJ} g^{\mu \nu}
	\partial_\mu \vp^I \partial_\nu \vp^J
		- V(\vp^K)
	\right] .
\ee
Such a scenario may arise naturally in the low-energy limit of higher-dimensional theories including supergravity, string theory and Kaluza--Klein models \cite{Rainer:1996gw,Gunther:1997ft,Appelquist:1987nr}, or as counterterms in curved spacetime \cite{Birrell:1982ix,Buchbinder:1992rb}. To be general, let us assume that $f(\vp^K)$ is an arbitrary function of the fields $\vp^I$, although we require it to be positive definite. This action defines the {\it Jordan frame}, where the Ricci scalar is coupled to the inflaton fields. 

One can recover the {\it Einstein frame} action (in which the Ricci scalar is only minimally coupled to the scalar fields) via a conformal redefinition of the spacetime metric~\cite{Kaiser:2010ps}. Denoting quantities in the Einstein frame with a circumflex, the appropriate conformal redefinition is 
\begin{equation}
	\hat g_{\mu\nu} = f g_{\mu\nu} .
\end{equation}
The Einstein frame action is
\begin{equation}
	\label{eq:einstein-action}
	\hat S = \int \d^4 x \, \sqrt{- \hat g} \left[
		\frac{1}{2} \Mpl^2 \hat R
		- \frac{1}{2} \hat G_{IJ} 
	\hat g^{\mu \nu} \partial_\mu \vp^I \partial_\nu \vp^J - \hat V
	\right] ,
\end{equation}
where
$\hat V = V/f^2$ is the potential and
\be
\hat G_{IJ} = \frac{1}{f} G_{IJ}
		+ \frac{3}{2} \Mpl^2 \frac{f_{,I} f_{,J}}{f^2} .
\ee
We note that cosmological observables are unaltered by this procedure~\cite{Kaiser:1995nv,Flanagan:2004bz,Chiba:2008ia,Gong:2011qe}. However, quantities that are not directly observable may be altered~\cite{White:2012ya}.

The Einstein-frame kinetic energy is of the same form as eq.~\eqref{eq:X_non-canonical} and so our analysis of such non-canonical models in chapters~\ref{ch:subhorizon} and \ref{ch:formalisms} also applies to these non-minimally coupled scenarios.

\sssec{Effective string theory corrections}
\label{sec:alphaprime}

Since string theory is the leading candidate theory for the physics at high energies, it is natural to ask if and how simple phenomenological models of inflation may be situated within effective theories motivated by string theory. In chapter~\ref{ch:singlefield} we perform such an analysis for what is arguably the simplest inflationary model of all: chaotic inflation. 

If chaotic inflation is derived from compactification of a higher dimensional string theory action to yield the {\sc 4d} effective action~\cite{Hertzberg:2007ke}, one expects terms appearing at next order in the Regge parameter $\alpha'$~\cite{Metsaev,DeFelice:2011zh,DeFelice:2011jm} which may lead to much richer inflationary dynamics. This encourages us to consider models possessing a number of such $\alpha'$ terms. We have already discussed some of these, including non-minimal coupling of the inflaton $\vp$ to the Ricci scalar and non-canonical kinetic terms. 

In chapter \ref{ch:singlefield} we also consider the effect of higher derivative quantum gravity terms such as the Gauss-Bonnet term $\G \equiv R^2 -4 R_{\alpha \beta}R^{\alpha \beta} + R_{\alpha \beta \gamma \delta}R^{\alpha \beta \gamma \delta}$ and non-linear field interactions such as the Galileon term $J(\vp,X) \Box \vp$.\footnote{`$\square$' denotes the d'Alembertian operator $\square = g^{\mu \nu} \partial_\mu \partial_\nu$.} We take the Galileon coupling $J(\vp,X)$ to be a general differential function of the field $\vp$ and its kinetic energy $X=-(1/2)g^{\mu\nu}\partial_{\mu}\vp\partial_{\nu}\vp$.\footnote{In ref.~\cite{DeFelice:2011jm} the Galileon term was denoted $G(\vp,X)$. We have changed $G \to J$ to avoid confusion with the field metric $G_{IJ}(\vp^K)$.} This is a generalisation of interaction $X \square \vp$ that appears in the Dvali-Gabadadze-Porrati ({\sc dgp}) braneworld model~\cite{Dvali:2000hr,Deffayet:2010qz,Pujolas:2011he}. Furthermore, the term of the form $\xi(\vp) X \square \vp$ also appears as an $\alpha'$ correction in low-energy effective string theory~\cite{Metsaev,Cartier:2001is}. 

\ssec{Thesis aims and structure}

We are thus motivated to consider two broad generalised inflationary scenarios: multi-field inflation and generalised single field inflation. This thesis is structured chronologically in cosmological terms: Chapter~\ref{ch:subhorizon} will quantify the subhorizon perturbations for multi-field inflationary models with a non-trivial field metric. This then sets the initial conditions for chapter~\ref{ch:formalisms}, where we demonstrate how to evolve these perturbations in the superhorizon epoch and make contact with observations. We also show that, in certain simple scenarios, it is possible to analytically approximate the superhorizon evolution of perturbations. Chapter \ref{ch:adiabatic} then provides some context to the superhorizon evolution of perturbations which must be followed until such a time as they become conserved. We will discuss and illustrate how such conservation may occur. We then proceed in chapter~\ref{ch:heatmaps} to use our analytic results to develop intuition about the relationship between inflationary dynamics and the evolution of cosmic observables. Combined, these chapters represent a thorough study of the theory and phenomenology of a number of important aspects of multi-field inflation. As discussed above, we are also motivated to study generalised single field inflation which we consider in chapter~\ref{ch:singlefield} where we consider an range of generalisations to chaotic inflation. We conclude in chapter~\ref{ch:conclusions}. Appendix \ref{ch:appendices} contains some supporting calculations.

To assist reading, we begin each chapter with a short abstract that summarises the material included and how this contributes to the overall aims of this thesis. 

\sec{Conventions}
\label{sec:conventions}

We use natural units with $c = \hbar = k_B = 1$ such that $\Mpl = (8 \pi G)^{-1/2} = 2.4 \times 10^{18} \, {\rm GeV}$. The spacetime metric has signature $(-1,1,1,1)$, Greek indices run over the four spacetime coordinates $\{0,1,2,3\}$ and lowercase Roman indices run over the purely spatial coordinates $\{1,2,3\}$. Partial and covariant derivatives with respect to spacetime indices may be respectively denoted by a comma and semi-colon, such that for vector components $A^\alpha$
\be
A^\alpha_{;\beta} = A^\alpha_{,\beta} + \Gamma^\alpha_{\beta \gamma} A^\gamma ,
\ee
where $\Gamma^\alpha_{\beta \gamma}$ are the Christoffel symbols of the second kind corresponding to the metric $g_{\alpha \beta}$ as
\be
\Gamma^\alpha_{\beta \gamma} = \frac{1}{2} g^{\alpha \delta}
\Big( 
\partial_{\gamma} g_{\beta \delta}+
\partial_{\beta} g_{\gamma \delta}-
\partial_{\delta} g_{\beta \gamma}
\Big).
\ee 
The Riemann curvature tensor is then constructed as
\be
{R^\alpha}_{\beta \gamma \delta} = \Gamma^\alpha_{\beta \gamma,\delta} - \Gamma^\alpha_{\beta \delta,\gamma}
+ \Gamma^\alpha_{\delta \ep} \Gamma^\ep_{\beta \gamma} - \Gamma^\alpha_{\gamma \ep} \Gamma^\ep_{\beta \delta}.
\ee
Partial derivatives with respect to spacetime coordinates may be abbreviated as 
$\partial A / \partial x^\mu = \partial_\mu A = A_{,\mu}$. 

We also take derivatives of functions of scalar fields. This involves scalar field indices which are usually denoted by capitalised Roman letters (exceptions to this rule are explained where they occur). These derivatives may be partial derivatives $\partial_I$, or covariant $\grad_I$ with respect to the field metric $G_{IJ}$, as detailed in chapter \ref{ch:subhorizon}. It is useful to define intrinsic derivatives on the space of scalar fields $\vp^I$ with respect to coordinate time $t$, conformal time $\eta$ and efolds $N$ as
\begin{equation}
	\Dt = \frac{\d \phi^I}{\d t} \grad_I ,
	\qquad
	\Deta = \frac{\d \phi^I}{\d \eta} \grad_I ,
	\qquad
	\DN = \frac{\d \phi^I}{\d N} \grad_I .
\end{equation}

Finally, we employ a Fourier convention such that
\be
Q_{\vect{k}} = \int \! \d^3 \vect{x} ~ Q(\vect{x}) e^{-i \vect{k} \cdot \vect{x}}.
\ee

\clearpage{\pagestyle{empty}\cleardoublepage}
\chapter{Subhorizon perturbations}
\label{ch:subhorizon}
\begin{addmargin}[0.05\textwidth]{0.05\textwidth}
As discussed in \S\ref{sec:beyondsimplest}, there is good motivation to consider general multi-field inflationary models with non-canonical kinetic terms as defined by the action
\begin{equation}
	\label{eq:mf-action}
	S = \frac{1}{2} \int \d^4 x \, \sqrt{-g} \left[ \Mpl^2 R -
	G_{IJ}(\vp^K) g^{\mu \nu} \partial_\mu \vp^I \partial_\nu \vp^J - 2 V(\vp^K) \right] .
\end{equation}
In this chapter we follow Elliston et al.~\cite{Elliston:2012ab} which provided the first full computation of the quantized perturbation statistics for the subhorizon epoch of such a scenario, without restriction on the functional form of $G_{IJ}(\vp^K)$ or $V(\vp^K)$. In \S\S\ref{sec:intro_pert}--\ref{sec:matter-perturbations} we introduce metric and covariant matter perturbations. These are then combined in \S\ref{sec:perts} to yield covariant perturbations up to third order which are then quantized in \S\ref{sec:quantisation} to obtain the two and three-point functions of the field perturbations at horizon exit. As we shall show, the non-canonical nature of the action \eqref{eq:mf-action} has profound implications for the three-point function. The results derived also serve as the initial conditions for the superhorizon evolution discussed in subsequent chapters.
\end{addmargin}

\begin{center}
\partialhrule
\end{center}
\vspace{-3em}
\begin{quote}
\list{}{\leftmargin -0.5cm \rightmargin\leftmargin} \endlist
\begin{center}
{\it ``Clouds are not spheres, mountains are not cones, coastlines are not circles, and bark is not smooth, nor does lightning travel in a straight line.''}
\flushright{---Beno\^{i}t Mandelbrot, 1982.}
\end{center}
\end{quote}
\vspace{-1em}
\begin{center}
\partialhrule
\end{center}

\sec{Introduction to perturbations}
\label{sec:intro_pert}

In chapter \ref{ch:introduction} we discussed the background dynamics of a perfectly homogeneous and isotropic Universe. In order to make contact with \cmbr~observations it is necessary to consider inhomogeneous perturbations by using the methods of cosmological perturbation theory~\cite{Malik:2008im,Malik:2008yp,Mukhanov:1990me}. The inhomogeneous Universe contains perturbations both of the matter content and also of the spacetime metric. 

One of the complexities of cosmological perturbation theory is that there is no unique prescription with which to describe the perturbations, due to the lack of a preferred coordinate system. A coordinate system is built by foliating the 4{\sc d} spacetime with spatial hypersurfaces, each with a different constant time $t$. This $3+1$ split is often referred to as the {\it slicing} and {\it threading} of spacetime~\cite{Misner:1974qy}. One is free to choose the threading and slicing, and this alters the description of any inhomogeneity~\cite{Lyth:2009zz}. Since the choice of coordinate system cannot have physical implications, it is clear that our splitting of physical quantities into a background and a perturbation introduces spurious gauge dependencies~\cite{Malik:2012dr}. 

One may solve this problem by working with gauge-invariant variables. These are constructed from combinations of perturbations that are invariant under changes of the threading and slicing. That is not to say that the choice of gauge invariant variables is unique; one is free to pick between a number of possible gauges and this decision is usually made for pragmatic reasons. 

We now consider the metric and matter perturbations in turn. For the metric perturbations we begin with the Arnowitt-Deser-Misner (\adm) metric~\cite{Arnowitt:1962hi} and then expand this around a background \frwl~metric. This is one of the standard calculation procedures for multi-field inflation 
(for examples see refs.~\cite{Seery:2005gb,Langlois:2008mn,Gao:2008dt,Langlois:2008qf}). The matter perturbations involve splitting the scalar fields $\vp^I$ into background and perturbed pieces for which we employ the recent covariant perturbation scheme of Gong and Tanaka~\cite{Gong:2011uw}.

\sec{Metric Perturbations}
\label{sec:metric_perts}

A general spacetime metric may be written in terms of \adm~variables~\cite{Arnowitt:1962hi} as
\begin{equation}
	\label{eq:ADM}
	\d s^2 = - L^2 \, \d t^2
		+ h_{ij}(\d x^i + N^i \,\d t)(\d x^j + N^j \,\d t) ,
\end{equation}
where $h_{ij}$ is the metric of the 3-slices, $L$ is the lapse function and $N^i$ is the shift vector. Spatial indices are raised and lowered using the 3-metric $h_{ij}$, which has determinant $h$. The contravariant metric components are given by
\be
g^{00} = -\frac{1}{L^2} \,, \qquad g^{0i} = \frac{N^i}{L^2} \,, \qquad g^{ij} = h^{ij} - \frac{N^i N^j}{L^2} .
\ee
Substituting this metric into the action \eqref{eq:mf-action} gives~\cite{Langlois:2008mn}
\be
	S = \frac{1}{2} \int \d^4 x \, \sqrt{h} \left\{
		\Mpl^2
		\left[
			L R^{(3)} + \frac{1}{L} (E_{ij}E^{ij}-E^2)
		\right]
		+ 2 L p
	\right\} ,
\label{eq:ADM-action}
\ee
where $p$ is the pressure, $p = -g^{\mu \nu} G_{IJ} \partial_\mu \vp^I \partial_\nu \vp^J/2 - V$, and $R^{(3)}$ is the Ricci scalar built from the 3-metric $h_{ij}$.  $E_{ij}$ is proportional to the extrinsic curvature of spatial slices and is given by
\begin{equation}
	E_{ij} = \frac{1}{2} \left( \dot h_{ij} - N_{i|j} - N_{j|i} \right),
\end{equation}
where a vertical bar denotes the covariant derivative corresponding to the 3-metric $h_{ij}$. Finally, $E$ is the trace ${E^i}_i$.

Metric perturbations may be scalar, vector or tensor in nature~\cite{Lyth:2009zz}. At linear order these types of perturbation are decoupled from one another and so may be treated independently. We shall ignore vector perturbations since they are not sourced at linear order during inflation and are attenuated during the expansion~\cite{Baumann:2009ds}. It is therefore consistent to decompose the shift vector into scalar and vector components as $N^i \equiv \partial^i \vartheta + \beta^i$ where $\partial_i \beta^i=0$, and retain only the scalar part.\footnote{Note that in ref.~\cite{Elliston:2012ab} the vector component $\beta^i$ was kept, but the perturbation equations constrain it to be null.} The tensor perturbations shall be discussed independently in \S\ref{sec:tensormodes}. We now focus on the scalar perturbations that are the dominant source of the \cmbr~anisotropy. The lapse $L$ and scalar shift $\vartheta$ appear in~\eqref{eq:ADM-action} without time derivatives. Consequently their variation leads to constraint equations rather than propagating modes, which may be substituted back into the second order action to eliminate the metric perturbations.

We consider the \adm~metric as a perturbation about a background \frwl~universe. It is therefore necessary to remove redundant gauge modes which, for the reasons to be described in \S\ref{sec:ir-safety}, we achieve by specialising to the spatially flat gauge where $h_{ij} = a^2 \delta_{ij}$. The remaining metric perturbations $L$ and $\vartheta$ can then be expanded as
\begin{align}
	L &= 1 + \alpha_1 + \cdots , \nonumber \\
	\vartheta &= \vartheta_1 + \cdots .
\end{align}
To expand the action~\eqref{eq:mf-action} to third-order in field perturbations we only need consider the linear order metric components $\alpha_1$ and $\vartheta_1$ since the second order metric perturbations cancel by virtue of the equations of motion~\cite{Maldacena:2002vr,Chen:2006nt,Gong:2011uw}. 

\subsection{Gauge choice and infrared safety}
\label{sec:ir-safety}

Now that we have defined our metric, we are in a position to explain our choice of the flat gauge rather than the popular $\zeta$ gauge. 

\subsubsection{The $\zeta$ gauge}

A common choice of gauge for single field inflation is the uniform-field foliation $\delta \vp=0$. This leaves a scalar metric mode $\zeta$ which measures the perturbation of the volume element as $\zeta = (1/6) \delta \ln \det h$. The disadvantage of this gauge is that the calculations are long~\cite{Maldacena:2002vr}. In order to get the action into a suitable form, multiple partial integrations are required which generate boundary terms which must be retained~\cite{Adshead:2011bw,Arroja:2011yj,Burrage:2011hd}. 

The advantage of the $\zeta$ gauge for single field inflation is that $\zeta$ becomes classically time-independent after horizon crossing~\cite{Bardeen:1983qw,Lyth:1984gv, Rigopoulos:2003ak,Lyth:2004gb,Weinberg:2004kf,Langlois:2005ii,Langlois:2005qp}. This result can also be seen in the (tree-level) quantum calculation of the correlation functions of $\zeta$ \cite{Weinberg:2008mc}: evolution of $\zeta$ after horizon crossing would appear in the form of divergent vertex integrals appearing in the correlations functions~\cite{Zaldarriaga:2003my,Seery:2007wf,Seery:2010kh}. Such divergence would indicate that interactions continue arbitrarily far into the future, creating what it classified as {\it infrared dynamics}.

Consequently, the great merit of the $\zeta$-gauge in single field inflationary models is that the $n$-point functions of $\zeta$ cease to evolve in the superhorizon epoch. Thus calculations of the perturbations are dominated by contributions occurring near horizon crossing and are decoupled from the infrared dynamics of the theory. 

However, when faced with a multi-field model the situation becomes more complex. $\zeta$ is now able to evolve in the superhorizon epoch and is dependent on infrared dynamics which must be followed (see, for examples, refs.~\cite{Salopek:1988qh, Polarski:1994rz, Sasaki:1995aw,GarciaBellido:1995qq, Salopek:1995vw, 
Langlois:1999dw, Gordon:2000hv, Langlois:2010vx, 
Mulryne:2009kh, Mulryne:2010rp,Dias:2011xy,Seery:2012vj}).
This nullifies the main advantage of this gauge and so motivates an alternative choice.

\subsubsection{The flat gauge}

The flat gauge foliates spacetime with spatial hypersurfaces which carry a flat metric, leaving perturbations in the scalar field values. Usefully, the lowest-order perturbed actions are simpler to compute in this gauge~\cite{Seery:2005gb}. However, the field perturbations are sensitive to infrared dynamics. This can be seen from the quantum calculation where the integrals which define its correlation functions receive contributions from all times, not just those near horizon crossing~\cite{Falk:1992sf,Stewart:1993bc,
Nakamura:1996da,Zaldarriaga:2003my,Seery:2008qj}.
An important consequence of this is that terms in the action that are subleading in slow-roll may, after sufficient superhorizon evolution, lead to important evolutionary effects that must be followed. 

Seery and Lidsey~\cite{Seery:2005gb} proposed a solution to the problem of infrared dynamics by evaluating each $n$-point function in the spatially flat gauge a few efolds after horizon crossing. Restricting the duration of superhorizon evolution in this way prevents enhancements of the subleading terms from contaminating the lowest-order slow-roll prediction. Furthermore, one benefits from the computational simplicity of the flat gauge. The subsequent infrared evolution of these correlation functions must then be accounted for by other means. For canonical multi-field models, the non-linear separate universe or `$\delta N$' formalism introduced by Lyth and Rodr\'{i}guez~\cite{Lyth:2005fi} can be used for this purpose. In the presence of a non-trivial field metric $G_{IJ}(\vp^K)$, we require a suitable generalisation of the standard $\delta N$ formalism. We shall develop such a generalisation in chapter~\ref{ch:formalisms}.

The conclusion of this discussion is that we will exploit the computational simplicity of the flat gauge. The subhorizon calculation that will occupy the remainder of this chapter will generate infrared divergent vertex integrals. We must then be careful to ensure that our procedure for evolving the superhorizon perturbations, as described in chapter~\ref{ch:formalisms}, exactly matches these growing modes.

\sec{Matter Perturbations}
\label{sec:matter-perturbations}

The previous section focussed on the perturbations of the spacetime metric. In this section we shall discuss the matter perturbations. For the present purposes, inflation is driven by multiple scalar fields and so the {\it matter} in question is simply the scalar fields $\vp^I$. The presence of the non-trivial field metric $G_{IJ}(\vp^K)$ complicates the procedure of taking perturbations. We first motivate a suitable perturbation scheme that is covariant in field space, before applying it to the action \eqref{eq:mf-action}.

\sssec{Covariant perturbations}
\label{sec:covariant-perts}

When tasked with perturbing the action \eqref{eq:mf-action} one is perfectly entitled to employ standard non-covariant perturbation theory and expand fields about a homogeneous background $\phi^I(t)$ as $\vp^I(t,\bm x) = \phi^I(t) + \delta \vp^I(t,\bm x)$. Under this prescription it is necessary to treat $G_{IJ}(\vp^K)$ simply as an additional function, and so perturbations naturally invoke new terms proportional to the derivatives such as $G_{IJ,K}$. Refs.~\cite{Langlois:2008qf,RenauxPetel:2011uk} used such a procedure in their study of \dbi~inflation, providing expressions for the quantized perturbations, valid in the limit that the sound speed is very small. The action \eqref{eq:mf-action} has a sound speed of unity and so a more complete analysis is required.

We shall demonstrate that it is {\it economical} to interpret the fields $\vp^I$ as living on a curved manifold with {\it field metric} $G_{IJ}$, as first suggested by Sasaki and Stewart~\cite{Sasaki:1995aw}. In this interpretation, the field metric raises and lowers tangent-space indices. A well-defined inverse metric requires that $\det G_{IJ}$ is non-zero in the domain of application. Early work on this problem was performed by Groot Nibbelink and van Tent~\cite{GrootNibbelink:2001qt}, although this work was limited to linear perturbations and did not employ to covariant techniques discussed in the next section.

Local inhomogeneity is produced by a finite coordinate displacement $\delta \vp^I$ that does not lie in the tangent space of the field manifold at $\phi^I$. As a result, $\delta \vp^I$ is not tensorial, a fact that is not desirable within a covariant framework. Recently, Gong and Tanaka resolved this issue by providing a covariant perturbation scheme~\cite{Gong:2011uw}\footnote{A similar argument later appeared by Saffin~\cite{Saffin:2012et}.} and perturbing the \dbi~action of Langlois et al.~\cite{Langlois:2008qf} up to third-order. The action~\eqref{eq:mf-action} is a specific case of that considered in ref.~\cite{Langlois:2008qf}, and so the first step of our calculation is implicitly contained in the analysis of ref.~\cite{Gong:2011uw}. By performing covariant perturbations with respect to the field metric, the lengthy terms involving explicit derivatives of the field metric are thankfully absent. Rather, these are replaced with quantities related to the field space curvature which allow for a more intuitive interpretation. 

\sssec{Gong and Tanaka's covariant methodology}

\begin{figure}[h]
\begin{center}
\includegraphics[width=0.8\textwidth]{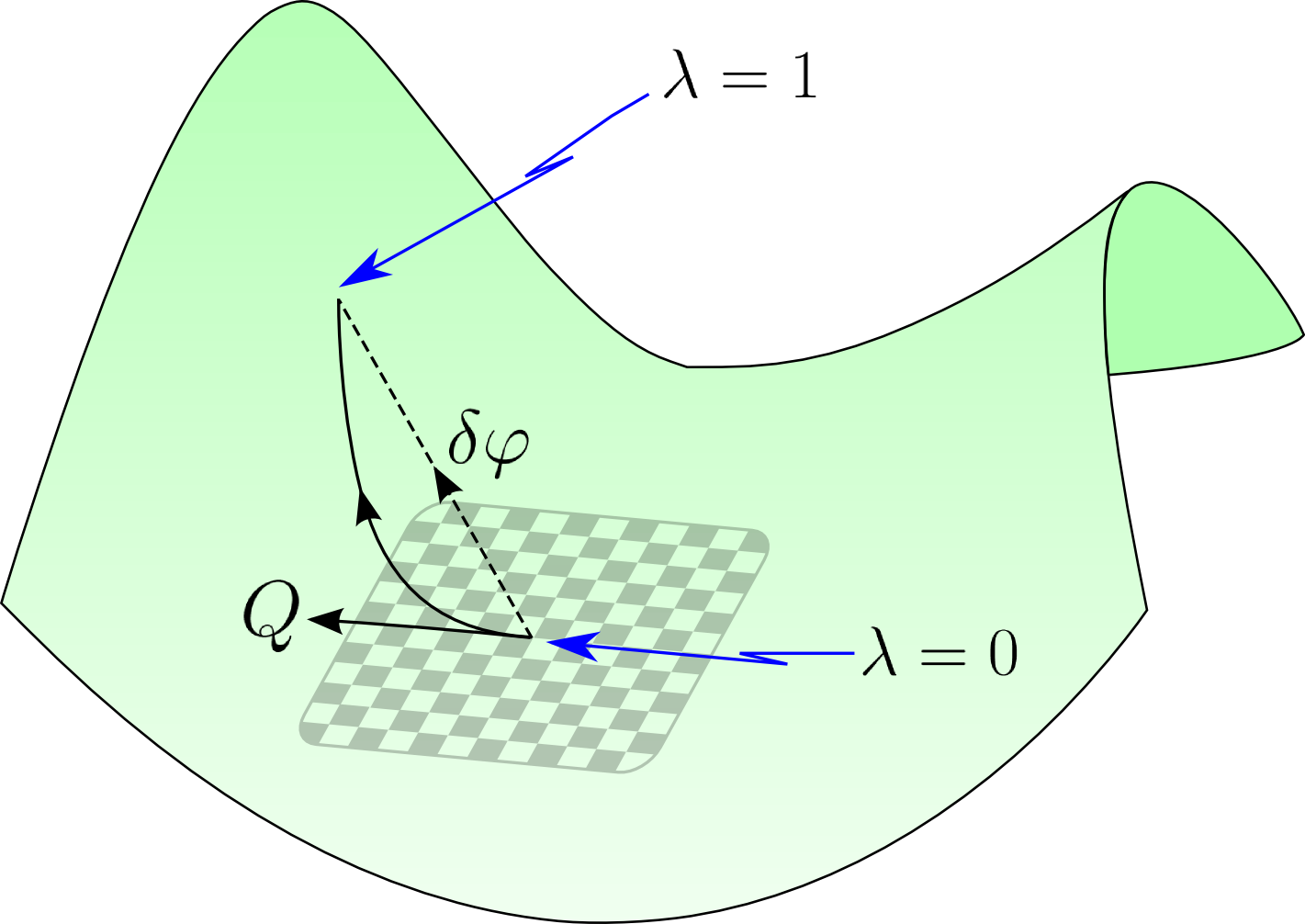}
\caption[Covariant perturbations]{The scalar fields are defined on a curved manifold (shown schematically in green), with $\lambda=0$ defining the background field location and $\lambda=1$ the perturbed point. The standard perturbation $\delta \vp$ (dashed line) does not in general lie in the tangent space at the background point (shown as cross-hatching). We may define a geodesic (curved solid line) linking the two points $\lambda=\{0,1\}$. To obtain a description that obeys correct tensorial transformation properties such that the perturbations are manifestly covariant, perturbations are described via the geodesic tangent vector $Q^I$.
\label{fig:GT}}
\end{center}
\end{figure}

The field-space metric is assumed to be smoothly differentiable and the points $\phi^I$ and $\phi^I + \delta \vp^I$ are presumed to be linked by a unique geodesic, labelled by a parameter $\lambda$. The normalization is set so that $\lambda = 0$ corresponds to the background coordinate $\phi^I$ and $\lambda = 1$ corresponds to the perturbed coordinate $\phi^I + \delta \vp^I$, as shown in figure~\ref{fig:GT}. One may then expand the displacement $\delta \vp^I$ via a Taylor series along this geodesic
\begin{equation}
	\delta \vp^I
	\equiv 
	\left. \frac{\d \vp^I}{\d \lambda} \right|_{\lambda=0} + 
	\frac{1}{2!} \left. \frac{\d^2 \vp^I}{\d \lambda^2} \right|_{\lambda=0} + 
	\frac{1}{3!} \left. \frac{\d^3 \vp^I}{\d \lambda^3} \right|_{\lambda=0} +
	\cdots ,
	\label{eq:perturbation-expansion}
\end{equation}
which we note is independent of our choice of normalisation of $\lambda$. This expansion is subject to the geodesic equation
\begin{equation}
	\Dl^2 \vp^I =
	\frac{\d^2 \vp^I}{\d \lambda^2}
	+ \Gamma^I_{JK} \frac{\d \vp^J}{\d \lambda} \frac{\d \vp^K}{\d \lambda}
	= 0 ,
	\label{eq:geodesic}
\end{equation}
where $\Dl \equiv (\d \vp^I/\d \lambda) \grad_I$. It is also useful to define $Q^I \equiv \d \vp^I / \d \lambda |_{\lambda = 0}$. Using eq.~\eqref{eq:geodesic}, the Taylor expansion \eqref{eq:perturbation-expansion} may then be rewritten as a power series in $Q^I$, yielding
\begin{equation}
	\label{eq:Q_expansion}
	\delta\vp^I =
	Q^I
	-\frac{1}{2!} \Gamma^I_{JK} Q^J Q^K
	+ \frac{1}{3!} \left( \Gamma^I_{LM} \Gamma^M_{JK} - \Gamma^I_{JK,L} \right)
		Q^J Q^K Q^L
	+ \cdots ,
\end{equation}
where the coefficients $\Gamma^I_{JK}$,
$\Gamma^I_{JK,L}$, \ldots are evaluated to background order, i.e.~at $\lambda = 0$. For the standard calculation where the field-space is Euclidean, only the first term is non-vanishing and so $\delta \vp^I = Q^I$. Our normalization of $\lambda$ was chosen to achieve this correspondence. 

\sssec{Covariant Taylor expansions}

A crucial point is that $Q^I$ transforms as a vector in the tangent space at $\phi^I$. Consequently, the perturbative expansion of the action \eqref{eq:mf-action} will be manifestly covariant if expressed in powers of $Q^I$. As a general example, let us expand an arbitrary tensor $\F_{I \cdots J}$. It is necessary to prescribe which indices belong to which tangent space, so we prime indices transforming in the tangent space at the perturbed position $\vp^I$ and leave un-primed those indices transforming in the tangent space at the background position $\phi^I$. The two are then straightforwardly related as
\begin{equation}
	\F_{I' \cdots J'} = {\G_{I'}}^I \cdots {\G_{J'}}^J
		\left(
		\left. \F_{I \cdots J} \right|_{\lambda=0}
		+ \left. \Dl \F_{I \cdots J} \right|_{\lambda=0}
		+ \frac{1}{2!} \left. \Dl^2 \F_{I \cdots J} \right|_{\lambda=0}
		+ \cdots
	\right) ,
	\label{eq:covariant-expansion}
\end{equation}
where ${\G_{I'}}^{I}$ is the parallel propagator,
which expresses parallel transport along the geodesic
connecting $\phi^I$ with $\phi^I + \delta\vp^I$.
For details, see Poisson, Pound and Vega~\cite{lrr-2011-7}.

In the action~\eqref{eq:mf-action}, the covariant expansion is only required for expanding the pressure $p$. Since this is a field-space scalar, the perturbation expansion is rather simpler and may be achieved by applying successive $\Dl$ operators as
\begin{align}
\label{eq:P_expansion}
p &= \left. p \right|_{\lambda =0}
+ \left. \Dl X \right|_{\lambda =0} - V_{,I} Q^I
+ \, \frac{1}{2!} \!\! \left. \Dl^2 X \right|_{\lambda =0} - \frac{1}{2!} V_{;IJ} Q^I Q^J
\nonumber \\
& \qquad + \, \frac{1}{3!} \!\! \left. \Dl^3 X \right|_{\lambda =0} - \frac{1}{3!}V_{;IJK} Q^I Q^J Q^K
+ \O(Q^4) ,
\end{align}
where $X = -\frac{1}{2}	G_{IJ} g^{\mu \nu} \partial_\mu \vp^I \partial_\nu \vp^J$.

\sssec{Useful results}

The covariant notation may be extended to define $\D_\mu$ as the spacetime derivative covariant with respect to the field metric. Hence, for a vector $U^I$ we have
\be
\D_\mu U^I  = \partial_\mu U^I + \Gamma^I_{JK} \partial_\mu \vp^J U^K .
\label{eq:Dmu}
\ee
Taking $U^I = \d \vp^I / \d \lambda$ in eq.~\eqref{eq:Dmu} and evaluating at $\lambda=0$ gives the useful result
\be
\Dl \partial_\mu \phi^I = \partial_\mu Q^I + \Gamma^I_{JK} \partial_\mu \phi^J Q^K = \D_\mu Q^I . \label{eq:covariant-mu}
\ee
It is also useful to note that eq.~\eqref{eq:covariant-mu} implies that $\D_i Q^I = \partial_i Q^I$ for spatial indices $i$.
Commutators of covariant derivatives are concisely given in terms of the Riemann tensor $R_{IJKL}$ corresponding to the field metric $G_{IJ}$ as
\be
[\Dl , \D_\mu] U_I = R_{IJKL} \partial_\mu \vp^L \frac{\d \vp^J}{\d \lambda} U^K , \\
\ee
for some vector $U_I$. For the specific case $U_I = Q_I$, and evaluating at $\lambda=0$, this yields another useful result
\be
\label{eq:riemann}
\Dl \D_\mu Q_I = R_{IJKL} \partial_\mu \phi^L Q^J Q^K ,
\ee
by virtue of the geodesic equation~\eqref{eq:geodesic} whereby $\Dl Q^I =0$.
Finally, derivatives of the Riemann tensor, when evaluated at $\lambda=0$, are simply given by
\be
\label{eq:riemann_deriv}
\left. \Dl R_{IJKL} \right|_{\lambda=0} = R_{IJKL;M} Q^M .
\ee
It is to be understood that the Riemann tensor components and its derivatives that appear in such expressions are all evaluated along the background trajectory where $\lambda=0$. Applying eqs. \eqref{eq:covariant-mu}, \eqref{eq:riemann} and \eqref{eq:riemann_deriv} we find
\bea
\left. \Dl X \right|_{\lambda =0} \!\!\!&=&\!\!\! - g^{\mu \nu} \partial_\mu \phi_I \D_\nu Q^I , \nonumber \\
\left. \Dl^2 X \right|_{\lambda =0} \!\!\!&=&\!\!\! - g^{\mu \nu} \left(
R_{KIJL} \partial_\mu \phi^K \partial_\nu \phi^L Q^I Q^J + \D_\mu Q_I \D_\nu Q^I 
\right) , \nonumber \\
\left. \Dl^3 X \right|_{\lambda =0} \!\!\!&=&\!\!\! - g^{\mu \nu} \left(
R_{MIJL;K} \partial_\mu \phi^L \partial_\nu \phi^M Q^I Q^J Q^K 
+ 4 R_{KIJL} \partial_\mu \phi^L Q^I Q^J \D_\nu Q^K \right) , ~~~~~~~~~
\label{eq:X-derivs}
\eea
where it is then necessary to substitute in the \adm~metric for $g^{\mu \nu}$. For example, $\left. \Dl X \right|_{\lambda =0}$ appears in the linear order action $S_{(1)}$ as a term $\dot \phi_I \Dt Q^I$. It also appears in the second order action $S_{(2)}$ as two terms: $-2 \alpha_1 \dot \phi_I \Dt Q^I$ and $-\partial^i \vartheta \dot \phi_I \partial_i Q^I$. 

\sec{Combined perturbations}
\label{sec:perts}

We now combine the matter and metric perturbations from \S\S\ref{sec:metric_perts} and~\ref{sec:matter-perturbations} and construct the perturbed actions. We will proceed order-by-order.

\ssec{First order}
\label{sec:first_order}

At linear order we find
\begin{equation}
	S_{(1)}
	= \int \d^4 x \left\{
		a^3 \left[
			3\Mpl^2 H^2
			- \frac{1}{2} \dot \phi_I \dot \phi^I - V
		\right] \alpha_1 
		- \left[
			\Dt (a^3 \dot \phi_I)
			+ a^3 V_{,I} \right]
		Q^I
	\right\} ,
\end{equation}
where we have integrated by parts and removed total spatial derivatives.
The background field equations follow after varying this action with respect to $\alpha_1$ and $Q^I$,
\begin{align}
	3 \Mpl^2 H^2 & =
		\frac{1}{2} \dot \phi_I \dot \phi^I + V
	\label{eq:friedmann} ,
	\\
	\Dt \dot{\phi}_I + 3H \dot \phi_I
		& = - V_{,I} .
	\label{eq:KG}
\end{align}
Acting $\Dt$ on the Friedmann eq.~\eqref{eq:friedmann} and eliminating $\Dt \dot \phi$ by using the Klein-Gordon eq.~\eqref{eq:KG}, we obtain the Raychaudhuri equation
\be
2 \Mpl^2 \dot H = - \dot \phi_I \dot \phi^I .
\ee
We therefore see that, at background level, the effect of the curved field space is only evident through the covariant time derivative in the Klein-Gordon equation ~\eqref{eq:KG}.

\sssec{Slow-roll}

The slow-roll regime in curved field-space was discussed by Sasaki and Stewart~\cite{Sasaki:1995aw} and later by Nakamura and Stewart~\cite{Nakamura:1996da}. Extending the discussion in \S\ref{sec:intro-slow-roll}, this requires $\epH$ to be sufficiently small and slowly varying such that
\be
\eta_{\rm H} \equiv \frac{\d \ln \epH}{\d N}
		= \frac{2}{H} \frac{\dot{\phi}^I
		\Dt \dot{\phi}_I}{\dot{\phi}^J \dot{\phi}_J}
		+ 2 \epH \ll 1 .
\ee
Geometrically, $\eta_{\rm H} \ll 1$ constrains the acceleration tangent to the phase space velocity to be small. These conditions are satisfied for fields obeying the slow-roll equation
\be
	3 H \dot \phi_I + V_{,I} \simeq 0,
	\label{eq:sr_eom}
\ee
if the potential is sufficiently flat. Detailed conditions are given in refs.~\cite{Sasaki:1995aw,Nakamura:1996da}. 

\sssec{Beyond slow-roll}

We presume that the background theory realises an era of inflation with $\epH \ll 1$ in which $H$ varies smoothly during horizon exit. This enables us to generate a reasonably general formalism, but the assumption of slow-roll is not mandatory. For example, it is also interesting to consider cases involving potential {\it features} that violate slow-roll, which requires a dedicated analysis~\cite{Adshead:2011bw,Adshead:2011jq}.

The presumption of a globally near-flat potential is also not necessary. One may admit large potential gradients orthogonal to the inflationary trajectory via the inclusion of heavy fields~\cite{Tolley:2009fg,Achucarro:2010jv,Achucarro:2010da,
Avgoustidis:2012yc}. 
When we compute the two and three-point functions we will only consider field-space directions which are light during horizon exit, implicitly ignoring any such heavy fields.\footnote{It is for this reason that, when employing slow-roll ordering to the terms in the second and third-order actions, we quote powers of $\dot{\phi}^I/H$ to emphasize that individual components of this vector are not necessarily of order $\epsilon^{1/2}$.} In simple models this is acceptable because large masses rapidly drive any fluctuations to extinction. In more complex models it has been suggested that modest corrections can occur where the phase-space flow drives power from massive modes into the curvature perturbation before
decay~\cite{Chen:2009zp,Chen:2012ge,McAllister:2012am,Gao:2012uq}.
To capture these effects would require an extension of the formalism derived in this chapter.

\ssec{Second order}

Expanding the action to second order, performing multiple partial integrations and removing total derivatives, we find
\begin{align}
	S_{(2)} =
	\frac{1}{2} \int \d^4 x \, a^3 \Big\{
	& \alpha_1 \Big[
		-6 \Mpl^2 H^2 \alpha_1
		+ \dot \phi_I \dot \phi^I \alpha_1
		- 2 \dot \phi_I \Dt Q^I
		- 2  V_{,I}Q^I
	\Big] \nonumber
	\\
	&
	\mbox{} \qquad 
	- \frac{2}{a^2} \partial^2 \vartheta_1 \Big[
		2 \Mpl^2 H \alpha_1
		- \dot \phi_I Q^I
	\Big] + R_{KIJL} \dot \phi^K \dot \phi^L Q^I Q^J \nonumber
	\\
	&
	\mbox{} \qquad
	+ \Dt Q_I \Dt Q^I
	- \frac{1}{a^2} \partial_i Q_I \partial_i Q^I
	- V_{;IJ}Q^I Q^J
	\Big\} .
	\label{eq:2nd_order}
\end{align}
Repeated covariant indices denote summation with the Dirac delta function $\delta^{ij}$ rather than the 3-space metric $h^{ij}$. We employ the Laplacian operator $\partial^2 = \partial_i \partial_i$ such that it defines the Laplacian in the Euclidean metric of the Fourier components $Q^I_{\vect{k}}$. The energy and momentum constraints can be respectively obtained by varying the action with respect to $\alpha_1$ and $\vartheta_1$. We find
\bea
-2 \Mpl^2 \frac{H}{a^2} \partial^2 \vartheta_1
	&=&
		6 \Mpl^2 H^2 \alpha_1 -\alpha_1 \dot \phi_I \dot \phi^I 
		+ \dot \phi_I  \Dt Q^I
		+ V_{,I} Q^I ,
	\label{eq:energy-constraint}\\
	2 \Mpl^2 H \alpha_1
	&=&
		\dot \phi_I Q^I .
		 \label{eq:momentum-constraint}
\eea
Eqs.~\eqref{eq:energy-constraint} and~\eqref{eq:momentum-constraint} can now be employed to eliminate the metric perturbations $\alpha_1$ and $\vartheta_1$ in $S_{(2)}$ to obtain
\begin{equation}
	\label{eq:S2}
	S_{(2)} =
	\frac{1}{2} \int \d^4 x \, a^3 \Big\{
		\Dt Q_I \Dt Q^I
		- \frac{1}{a^2} \partial_i Q_I \partial_i Q^I
		- \M_{IJ} Q^I Q^J
	\Big\} ,
\end{equation}
where $\M_{IJ}$ is the symmetric mass matrix which satisfies
\begin{equation}
	\label{eq:mass}
	\M_{IJ} = V_{;IJ} - R_{LIJM} \dot \phi^L \dot \phi^M
	- \frac{1}{\Mpl^2 a^3} \Dt
	\left( \frac{a^3}{H} \dot \phi_I \dot \phi_J \right) .
\end{equation}
This result was first obtained by Sasaki and Stewart~\cite{Sasaki:1995aw} 
(see also Nakamura and Stewart~\cite{Nakamura:1996da} and Gong and Stewart~\cite{Gong:2001he,Gong:2002cx}).
The second order action is almost identical to the canonical case where $G_{IJ} = \delta_{IJ}$, except for the presence of covariant derivatives and the term in $\M_{IJ}$ involving the Riemann tensor. The Riemann term in the mass matrix alters the effective mass of modes orthogonal to the field velocity which will affect the way that these modes evolve and couple.

\ssec{Third order}
\label{sec:third order}

The third order action is
\begin{align}
\label{eq:s3-full}
S_{(3)} &=
	\frac{1}{2} \int \d^4 x \, a^3
	\bigg\{
		6 \Mpl^2 H^2 \alpha_1^3 
		+ 4 \Mpl^2 \frac{H}{a^2} \alpha_1^2 \partial^2 \vartheta_1 
		- \frac{\Mpl^2 \alpha_1}{a^4} \Big(
			\partial_i \partial_j \vartheta_1 \partial_i \partial_j \vartheta_1
			- \partial^2 \vartheta_1 \partial^2 \vartheta_1
		\Big)
		\nonumber \\ & \quad
		- \dot \phi_I \dot \phi^I \alpha_1^3
		+ 2 \alpha_1^2\dot \phi_I \Dt Q^I
		+ \frac{2}{a^2} \alpha_1 \dot{\phi}_I
			\partial_i \vartheta_1 \partial_i Q^I
		- \alpha_1 R^{L(IJ)M}\dot \phi_L \dot \phi_M Q_I Q_J
		\nonumber \\ & \quad
		- \alpha_1 \Big(
			\Dt Q_I \Dt Q^I
			+ \frac{1}{a^2} \partial_i Q_I \partial_i Q^I
		\Big)
		- \frac{2}{a^2} \partial_i \vartheta_1 \Dt Q_I \partial_i Q^I
		+ \frac{4}{3} R_{I(JK)L}\dot \phi^L \Dt Q^I Q^J Q^K 
		\nonumber \\ & \quad
		+ \frac{1}{3} R_{(I|LM|J;K)}\dot \phi^L \dot \phi^M Q^I Q^J Q^K
		- \frac{1}{3} V_{;(IJK)} Q^I Q^J Q^K
		- V_{;(IJ)} \alpha_1 Q^I Q^J
	\bigg\} .
\end{align}
We have indicated the symmetric index combinations\footnote{Our symmetrization conventions are $2 A_{(IJ)} = A_{IJ} + A_{JI}$ and	$6 A_{(IJK)} = A_{IJK} + \{ \text{5 perms} \}$. Vertical bars delimit indices excluded from symmetrization.} picked out by each product of the $Q^I$. The metric perturbations $\alpha_1$ and $\vartheta_1$ can be eliminated using eqs.~\eqref{eq:energy-constraint} and \eqref{eq:momentum-constraint}. The resulting expression is exact and does not invoke an expansion in powers of slow-roll. However, as we shall see in \S\ref{sec:quantisation}, it is necessary to presume slow-roll in order to integrate the contributions to the three point function. Furthermore, it is these leading order terms that will produce the dominant contributions to the three-point function. Isolating the lowest-order slow-roll contributions and rewriting in terms of conformal time, we find
\begin{align}
\label{eq:s3sr}
	S_{(3)} \supseteq \int \d^3 x \, \d \eta \, \bigg\{
		&- \frac{a^2}{4 \Mpl^2 H} \dot \phi^I Q_I \Deta Q^J \Deta Q_J 
		- \frac{a^2}{4 \Mpl^2 H} \dot \phi^I Q_I \partial_i Q^J \partial_i Q_J
		\nonumber \\  
		&+ \frac{a^2}{2 \Mpl^2 H} 
		\dot \phi^I \partial_i \partial^{-2} \Deta Q_I
		\partial_i Q_J \Deta Q^J 
		+ \frac{2 a^3}{3} R_{I(JK)L}\dot \phi^L \Deta Q^I Q^J Q^K 
		\nonumber \\  
		&+ \frac{a^4}{6} R_{(I|LM|J;K)}\dot \phi^L \dot \phi^M Q^I Q^J Q^K
	\bigg\} .
\end{align}
The first two lines of eq.~\eqref{eq:s3sr} are identical to the result of the canonical $(G_{IJ}=\delta_{IJ})$ calculation~\cite{Seery:2005gb} with the partial derivatives promoted to covariant derivatives. These terms do not produce infrared divergences at lowest order in slow-roll. Subleading corrections to these terms enter at $\O(\dot{\phi}/H)^3$ and hence are negligible around horizon crossing. However, these terms are time-dependent and so exhibit important infrared behaviour at later times. The third line of eq.~\eqref{eq:s3sr} includes explicitly new terms proportional to the Riemann tensor. Such terms were first calculated, but not quantized, by Gong and Tanaka~\cite{Gong:2011uw} and are a new feature associated with the curvature of field space. In \S\ref{sec:quantisation} we show that these terms produce infrared divergences; this is particularly significant since these divergences arise at leading order in slow-roll. 

To track this time-dependence as clearly as possible we have elected to retain terms up to $\O(\dot{\phi}/H)^2$. This procedure is consistent because eq.~\eqref{eq:mass} demonstrates that next-order corrections in the two-point statistics enter at $\O(\dot{\phi}/H)^3$ and are thus neglected at requisite order. One finds the same result for corrections to the scale factor and Hubble rate. There is, however, one correction term at $\O(\dot{\phi}/H)^2$ which arises from the time-dependence of the term proportional to $R_{I(JK)L}$
[See eqs.~\eqref{eq:riemann_3pf_expansion} and~\eqref{eq:3pfv6}]. In order to maintain consistency we shall include this term in our quantization. Additionally, by retaining terms consistently to $\O(\dot{\phi}/H)^2$, we shall be able to demonstrate in chapter~\ref{ch:formalisms} that the quantum and classical time-dependence of the perturbations match at both leading and subleading orders.

\sec{Quantization}
\label{sec:quantisation}

The previous section derived the perturbed actions $S_{(2)}$ and $S_{(3)}$. Upon quantization these respectively will provide us with the two and three-point correlators of the field perturbations $\langle Q^I_{\vect{k}_1} Q^J_{\vect{k}_2} \rangle$ and $\langle Q^I_{\vect{k}_1} Q^J_{\vect{k}_2} Q^K_{\vect{k}_3}\rangle$. Obtaining these is the principal aim of this chapter.

We shall frequently refer to the two-point correlator $\langle Q^I_{\vect{k}_1} Q^J_{\vect{k}_2} \rangle$ as the {\it propagator} since it has precisely this interpretation in terms of Feynman diagrams. The calculation of the propagator and all higher $n$-point functions follows from the `in--in' formulation of quantum field theory. Details may be found in refs.~\cite{Maldacena:2002vr,Weinberg:2005vy,Seery:2007we,Chen:2010xka,Koyama:2010xj}.
For the present calculation, the curved field-space generates an extra complication because we must be careful to ensure that each $n$-point function satisfies the correct tensorial transformation properties. We find that this may be enforced by parallel transport of the tangent-space bases along the direction of the inflationary phase space flow.

\ssec{Two-point statistics}

The two-point statistics have previously been calculated in the presence of a non-trivial field metric. At lowest order in slow-roll this was calculated by Sasaki and Stewart~\cite{Sasaki:1995aw}. Next-leading-order corrections were included by Gong and Stewart~\cite{Gong:2001he}. Neither of these references employed the covariant perturbation scheme of Gong and Tanaka~\cite{Gong:2011uw}, but the calculation follows almost identically and so we shall only summarise the key steps.

We begin with the second order action \eqref{eq:S2}. After moving to conformal time $\eta$ and integrating by parts this takes the form
\be
\label{eq:s2conformal}
S_{(2)} =
	-\frac{1}{2} \int \d^3 x \, \d \eta \, a^2 Q^I \Big\{
		G_{IJ} \left(
		\D_\eta^2 Q^J 
		+ 2 \frac{a'}{a} \D_\eta Q^J
		- \partial^2 Q^J \right)
		+ a^2 \M_{IJ} Q^J
	\Big\} ,
\ee
where a prime denotes differentiation with respect to conformal time. One may then move to Fourier space perturbations $Q^I_{\vect{k}}$ and define $u^I_{\vect{k}} = a Q^I_{\vect{k}}$ and so obtain the Sasaki--Mukhanov equation
\be
\label{eq:desitterM}
\D_\eta^2 u^I_{\vect{k}} + \left( \Big[k^2 - \frac{a''}{a} \Big] \delta^I_J + a^2 M^I_J \right) u^J_{\vect{k}} = 0.
\ee
One may write 
\be
\frac{a''}{a} = 2 a^2 H^2 \Big(1-\frac{1}{2} \epH \Big) ,
\ee
and we note that the mass matrix $\M_{IJ} = \O(\epH)$. Working to leading order in slow-roll we find\footnote{Perturbative mass corrections can be considered~\cite{Gong:2001he}. As argued at the end of \S\ref{sec:third order}, such corrections appear beyond the order of our bispectrum calculation and so we may safely ignore them. Note, however, that we only ignore such couplings between modes for the brief period around horizon exit. After horizon crossing such couplings are definitely not negligible, causing flow of power from isocurvature perturbations to the adiabatic mode. This is accounted for in chapter \ref{ch:formalisms} by the covariant transport formalism.}
\be
\D_\eta^2 u^I_{\vect{k}} + \left( k^2 - \frac{2}{\eta^2} \right) u^I_{\vect{k}} = 0.
\ee
The mode variable $u^I_{\vect{k}}$ may then be quantized by writing in terms of raising and lowering operators
\be
\label{eq:makeoperator}
u^I_{\vect{k}}(\eta) \rightarrow 
u(\eta) {\hat a}^I_{\vect{k}} + 
u^*(\eta) {\hat a}^{\dag I}_{\vect{-k}} ,
\ee
where $u^*$ is the complex conjugate of $u$. Provided that these are normalised as
\be
u_{\vect{k}}^*u_{\vect{k}}' - {u_{\vect{k}}^*}'u_{\vect{k}} = -i,
\ee
then the creation and annihilation operators satisfy
\be
\label{eq:commutation}
\Big[{\hat a}^I_{\vect{k}_1} , {\hat a}^{\dag J'}_{\vect{k}_2} \Big]
= (2 \pi)^3 \delta(\vect{k}_1 - \vect{k}_2) \trajp^{IJ'}.
\ee
In this commutation relation we have introduced the bitensor $\trajp^{IJ'}$ which defines parallel transport along the direction of phase space flow rather than along geodesics. The transport is between the unprimed tangent space at $\eta_1$ and the primed tangent space at $\eta_2$. This bitensor solves the equation $\Deta \trajp^{IJ'} = 0$ which prescribes a formal solution as a path ordered exponential,
\begin{equation}
	\trajp^{IJ'}
	= \pathorder
		\exp \bigg(
			-\int_{\eta_1}^{\eta_2} \d \eta
			\;
			{\Gamma^{I''}}_{K''L''} \frac{\d \phi^{K''}}{\d \tau}
		\bigg)
		G^{L'J'} ,
	\label{eq:pi}
\end{equation}
where the integral is computed along the phase space trajectory traversed by the inflationary dynamics and the symbol `$\pathorder$' denotes path ordering. Note that in the limit $\eta_2 \rightarrow \eta_1$ we obtain $\trajp^{IJ'} \rightarrow G^{IJ}$.

One may enquire about the motivation for the appearance of the parallel propagator in eq.~\eqref{eq:commutation}. Gong and Stewart~\cite{Gong:2002cx}, working in the limit of an equal-time commutator, used the field metric $G^{IJ}$ in place of the parallel propagator. Eq.~\eqref{eq:commutation} recovers this result if the operators on the {\lhs} are taken at equal times, since then their tangent spaces are equivalent and $\trajp^{IJ'} \to G^{IJ}$. 

However, we require a commutator where the creation and annihilation operators may be taken at different times (this will become evident in the calculation of the three-point function when one operator will be taken at the time of measurement and one operator will be summed over past times). The free Minkowski vacuum $|0\rangle$ is inert under time evolution, but the operators ${\hat a}_{\vect{k}}^I$ will evolve under parallel propagation over the non-trivial metric $G^{IJ}$. It is for this reason that we insert the factor of the trajectory propagator.

An alternative but equivalent argument may be made in terms of the derivation of the propagator as a Green's function of the Klein-Gordon operator $\Delta_{IJ}$. This operator is defined from $S_{(2)}$ which yields the equation of motion for $Q^J$ as $\Delta_{IJ} Q^J = 0$. Schematically, the propagator $\langle Q^I Q^{J'} \rangle$ obeys $\Delta_{IK} \langle Q^I(\eta,\vect{x}) Q^{J'}(\sigma,\vect{y}) \rangle \propto \trajp_K^{J'} \delta(\eta-\sigma) \delta(\vect{x}-\vect{y})$, where the non-time-equality is explicit. From this argument it is clear that the trajectory propagator relates field perturbations defined at different times $\eta$ and $\sigma$. 

The parallel propagator factorises the index structure of the Sasaki--Mukhanov equation \eqref{eq:desitterM} leaving the mode functions $u(\eta)$ obeying a scalar mode equation 
\be
\label{eq:desittermuchanov}
u''(\eta) + \Big( k^2 - \frac{2}{\eta^2} \Big) u(\eta) = 0,
\ee
where we have used the fact that $\D_\eta^2 u(\eta) = u''(\eta)$, since $u(\eta)$ is a field space scalar. Eq.~\eqref{eq:desittermuchanov} is precisely the standard mode equation for single field inflation.

Its initial condition arises in the deep subhorizon where $\eta \to -\infty$ and the {\frwl} spacetime is approximately Minkowski. The Bunch-Davies initial vacuum condition then follows as
\be
u(\eta) = \frac{1}{\sqrt{2 k}} e^{-i k \eta}.
\ee
We now use this as a boundary condition for eq.~\eqref{eq:desittermuchanov} to find the solution
\be
\label{eq:modefunction}
u(\eta) = \frac{1}{\sqrt{2 k}} \Big( 1 - \frac{i}{k \eta}\Big) 
e^{-i k \eta}.
\ee
One thus finds the power spectrum of $Q^I_{\vect{k}}$ as
\be
\langle \timeorder Q^I_{\vect{k}_1}(\eta_1)Q^{J'}_{\vect{k}_2}(\eta_2) \rangle
= (2 \pi)^3 \delta(\vect{k}_1 + \vect{k}_2) \trajp^{IJ'}
\frac{u(\eta_1)u^*(\eta_2)}{a^2},
\label{eq:pspectrumone}
\ee
where we have presumed $\eta_1 > \eta_2$ and `$\timeorder$' denotes time ordering. The solution for $\eta_1 < \eta_2$ is identical with $\eta_1$ and $\eta_2$ interchanged. Substituting eq.~\eqref{eq:modefunction} into eq.~\eqref{eq:pspectrumone} we arrive at our final result
\begin{equation}	
	\!\!\! \langle
		\timeorder
		Q^I_{\vect{k}_1}(\eta_1)
		Q^{J'}_{\vect{k}_2}(\eta_2)
	\rangle
	\simeq
	(2\pi)^3 \delta(\vect{k}_1 + \vect{k}_2)
	\trajp^{IJ'}
	\frac{H_\ast^2}{2k^3}
	\times
	\left\{ \! \! \! \begin{array}{l@{\hspace{5mm}}l}
		(1 - \im k \eta_1) (1 + \im k \eta_2) \e{\im k (\eta_1 - \eta_2)}, &
		\eta_1 < \eta_2 \\
		(1 + \im k \eta_1) (1 - \im k \eta_2) \e{\im k (\eta_2 - \eta_1)}, &
		\eta_2 < \eta_1
	\end{array} \right. 
	\label{eq:2pf-pp}
\end{equation}
where $k =|\vect{k}_1| = |\vect{k}_2|$ and this result is valid within a few efolds of the horizon crossing times $|k\eta_1| = 1$ and $|k \eta_2| = 1$. The label `$*$' denotes evaluation precisely at the time of horizon crossing. We note that the trajectory propagator ensures that the \rhs~of eq.~\eqref{eq:2pf-pp} has the correct bitensorial transformation properties.

Taking the equal-time limit of eq.~\eqref{eq:2pf-pp} and evaluating soon after horizon crossing, the two-point function becomes
\begin{equation}
	\langle Q^I_{\vect{k}_1} Q^J_{\vect{k}_2} \rangle
	\simeq
	(2\pi)^3 \delta(\vect{k}_1 + \vect{k}_2) G^{IJ}
	P_Q(k) \,,
	\label{eq:power-spectrum}
\end{equation}
where $P_Q(k)$ is the {\it power spectrum} of $Q^I_{\vect{k}}$ defined as
\be
P_Q(k) \equiv \frac{H_*^2}{2k^3} \,.
\ee
One may also define the dimensionless power spectrum
\be
\label{eq:PQdimless}
\P_Q (k) = \frac{k^3}{2 \pi^2} P_Q (k) = \left(\frac{H_*}{2 \pi} \right)^2.
\ee 
For this estimate to be valid we require that all of the non-growing power-law terms in~\eqref{eq:2pf-pp} have frozen out at constant values~\cite{Nalson:2011gc}. However, $Q^I$ has infrared dynamics (see \S\ref{sec:ir-safety}) which manifest as growing terms at subleading order~\cite{Stewart:1993bc,Nakamura:1996da} and so~\eqref{eq:power-spectrum} becomes untrustworthy soon after horizon crossing. Consequently, this result remains valid only for a very narrow range of efolds.

\subsection{Tensor spectrum}
\label{sec:tensormodes}

The calculation for the tensor modes is considerably simpler than their scalar counterparts. Ignoring the independent scalar perturbations, the tensor perturbations are defined from the conformal metric~\cite{Baumann:2009ds,Lyth:2009zz} as
\be
\d s^2 = a^2(\eta) \left[ - \d \eta^2 +  \left(\delta_{ij} + h_{ij} \right) \d x^i \d x^j \right] ,
\ee
where $h_{ij}$ denotes the perturbed metric which obeys $h^{ij}_{,i} = h^{i}_{i} = 0$. The second order action $S_{(2)}$ for the tensor perturbation takes the form~\cite{Baumann:2009ds}
\be
\label{eq:s2tensor1}
S_{(2)} =
	-\frac{\Mpl^2}{8} \int \d^3 x \, \d \eta \, a^2 h^{ij} \left[
		h_{ij}'' + 2 \frac{a'}{a} h_{ij}'
		- \partial^2 h_{ij} \right] ,
\ee
which is simply a rescaled version of the scalar action \eqref{eq:s2conformal} without the mass term. One may then move to Fourier space as
\be
\label{eq:tensorfourier}
h_{ij}(\vect{x},\eta) = 
\int \frac{\d^3 k}{(2 \pi)^3} \sum_{s=+,\times} e^s_{ij} h^s_{\vect{k}}(\eta) e^{i\vect{k}\cdot \vect{x}} + ~{\rm c.c.} \,,
\ee
where we have used the two eigenmodes of the spatial Laplacian $e_{ij}^{(+,\times)}$ and $\vect{k}$ is the comoving wavenumber. We write `$+ ~{\rm c.c.}$' to denote the addition of the complex conjugate. If we choose $\vect{k}$ to be aligned with the $z$-axis then
\be
e_{ij}^+ = \begin{pmatrix}
1&0&0\\
0&-1&0\\
0&0&0
\end{pmatrix}, \qquad
e_{ij}^\times = \begin{pmatrix}
0&1&0\\
1&0&0\\
0&0&0
\end{pmatrix}.
\ee

Noting that $e_{ii} = k^i e_{ij} = 0$ and $e^s_{ij}(k) e^{s'}_{ij}(k) = 2 \delta^{s s'}$, the action $S_{(2)}$ in eq.~\eqref{eq:s2tensor1} becomes
\be
S_{(2)} = -\frac{\Mpl^2}{4} \sum_s \int \d^3 \vect{k} \, \d \eta \, a^2 h^s_{\vect{k}} \left[
		{h^s_{\vect{k}}}'' + 2 \frac{a'}{a} {h^s_{\vect{k}}}'
		+ k^2 h^s_{\vect{k}} \right] .
\ee
Defining $v^s_{\vect{k}} = a\Mpl h^s_{\vect{k}}/2$ we obtain the Sasaki--Mukhanov equation
\be
\label{eq:desitterMtensor}
{v^s_{\vect{k}}}'' + \left( k^2 - \frac{a''}{a} \right) v^s_{\vect{k}} = 0.
\ee
As before, one may write 
\be
\frac{a''}{a} = \frac{2}{\eta^2} + \O( \epH).
\ee
We see that the two polarisations of the gravitational waves behave as renormalised massless scalar fields under the identifications
\be
Q^I_{\vect{k}} \to \frac{\Mpl}{2} h^s_{\vect{k}}, \qquad
u^I_{\vect{k}} \to v^s_{\vect{k}}.
\ee
The dimensionless power spectrum $\P_h(k)$ for each of these tensor fluctuations is thus
\be
\P_h(k) = \frac{4}{\Mpl^2} \left( \frac{H_*}{2 \pi} \right)^2,
\ee 
where we have evaluated around the time of horizon exit. Summing over both polarisations yields a total tensor spectrum as
\be
\label{eq:Ptdimless}
\P_t(k) = \frac{8}{\Mpl^2} \left( \frac{H_*}{2 \pi} \right)^2.
\ee
The lack of mass corrections or a non-trivial field space makes the calculation considerably simpler. 

\ssec{Three-point statistics}
\label{sec:third-order}

There are two popular and equivalent methods for deriving the three-point function $\langle Q^I Q^J Q^K \rangle$: One follows from the path integral formalism described pedagogically in ref.~\cite{Seery:2005wm} in the context of single field inflation and subsequently extended to the multi-field case~\cite{Seery:2005gb}. The second approach uses the Hamiltonian formulation of quantum mechanics and was the method used by Maldacena~\cite{Maldacena:2002vr}. We shall present the latter method since it makes some of the calculational steps more transparent. Using this method, the full bispectrum in the presence of an arbitrary non-trivial field metric was first computed by Elliston et al.~\cite{Elliston:2012ab}, which we now review.

Our calculation is similar to standard scattering calculations, where one computes transition amplitudes between an `input' state $| {\rm in} \rangle$ and an `output' state $\langle {\rm out} |$ as $\langle {\rm out} |S| {\rm in} \rangle$ where $S$ is the scattering matrix. However, in our case we want to calculate an expectation value of an operator $\F$ at a given time, which is schematically $\langle {\rm in} |\F| {\rm in} \rangle$. This structure gives the formalism of our calculation its name: the `{\it in--in formalism}'. 

We work in the interaction picture such that the Hamiltonian is split into two components as $H = H_0 + H_{\rm int}$. The dominant term is $H_0$ which contains perturbations only up to second order in $Q^I$ and represents the `free-theory' calculation. It is $H_0$ that generates the time dependence of operators such as $Q^I Q^J Q^K$. Meanwhile, $H_{\rm int}$ contains higher order perturbations such as those arising in $S_{(3)}$ and it is $H_{\rm int}$ that evolves the states.

The interacting vacuum $| \Omega \rangle$ at some time $t$ may be calculated in terms of the free-theory vacuum $| 0 \rangle$ at some early time $t_0$, provided that we take $t_0$ to be sufficiently early. This follows from standard theory~\cite{Peskin:1995ev} as
\be
| \Omega \rangle = \timeorder \exp \left( 
-i \int_{t_0(1-i \delta)}^{t(1-i \delta)} H_{\rm int} (t') \, \d t'
\right) |0 \rangle.
\ee
The contour of integration is rotated infinitesimally as $t' \to t' (1-i \delta)$ in order that the interacting vacuum state may be defined. 

Our goal is to calculate the correlator of perturbations $\langle \Omega | Q^I(t) Q^J(t) Q^K(t) | \Omega \rangle$. Using the above result and expanding the exponents to first order (which is sufficient for the present calculation) one finds
\be
\label{eq:qqq1}
\langle \Omega | Q^I(t) Q^J(t) Q^K(t) | \Omega \rangle = 
-i \int_{t_0(1-i \delta)}^{t(1-i \delta)} \d t' 
\langle Q^I(t) Q^J(t) Q^K(t) H_{\rm int} (t') \rangle + ~{\rm c.c.}
\ee
At the level of our computation we are neglecting loop corrections. If such terms are computed then they will be suppressed by factors of $Q \sim H/\Mpl$ which is very small. 

We are now in a position to calculate the three-point function $\langle Q^I(t) Q^J(t) Q^K(t) \rangle$ by adding up the contributions from each of the terms in $S_{(3)}$. We shall split the terms into two types, those that do not produce log-divergences, and those that do. For both classes, we shall explicitly illustrate one calculation, the method of which then applies analogously to the other contributions. 

\sssec{Convergent contributions}

We shall begin with the first term in eq.~\eqref{eq:s3sr} 
\be
S_{(3)} \supseteq \int \d^3 x \, \d \eta \, \bigg\{- \frac{a^2}{4 \Mpl^2 H} \dot \phi_I G_{JK} Q^I \D_\eta Q^J \D_\eta Q^K \bigg\},
\ee
which produces a contribution to the three-point function as
\bea
\label{eq:3ptcorrelator1}
&&\langle \Omega | Q^I(\tau,\vect{y_1}) Q^J(\tau,\vect{y_2}) Q^K(\tau,\vect{y_3}) | \Omega \rangle
= - i \int_{\cal C} \d \eta \, \d^3 x \frac{a^2}{4 \Mpl^2 H} \dot \phi_{L'} G_{M'N'} \times \nonumber \\
&& \qquad \langle Q^I(\tau,\vect{y_1})Q^{L'}(\eta,\vect{x}) \rangle
\D_\eta \langle Q^J(\tau,\vect{y_2})Q^{M'}(\eta,\vect{x}) \rangle
\D_\eta \langle Q^K(\tau,\vect{y_3})Q^{N'}(\eta,\vect{x}) \rangle \nonumber \\
&& \qquad + ~{\rm perms.} + ~{\rm c.c.}
\eea
The permutations arise from Wick's theorem, since there are six ways of contracting pairs of external and internal Feynman lines. $\C$ denotes the integration contour as previously described. Since we are computing quantities over a range of times, the field perturbations are naturally defined in a range of tangent spaces. We label tangent-space indices at the time of calculation, $\tau$, with unprimed capitalised labels $I$, $J$, $K$. Indices associated with the integration variable $\eta$ are primed and capitalised viz $I'$, $J'$, $K'$.

One may now move to Fourier space. The variable $Q^I(\tau,\vect{y}_1)$ is then written in terms of the Fourier mode $Q^I_{\vect{k}_1}$ and so on. Eq.~\eqref{eq:3ptcorrelator1} contains six separate perturbations $Q^I, Q^J, \dots, Q^{N'}$ which correspond to Fourier modes $\vect{k}_1, \vect{k}_2, \dots, \vect{k}_6$. We now substitute for the propagators using eq.~\eqref{eq:2pf-pp}, where we set $\eta_1 \to \eta$ as the time at which the interaction occurs, and $\eta_2 \to \tau$ as the time at which the field perturbations are calculated. The propagator factors introduce delta functions of the form $\delta(\vect{k}_1 + \vect{k}_4)$ which allow us to eliminate $\vect{k}_4,\vect{k}_5,\vect{k}_6$. Performing the integration over the spatial coordinates we obtain
\begin{align}
\label{eq:3ptcorrelator2}
\langle Q^I_{\vect{k}_1} Q^J_{\vect{k}_2} Q^K_{\vect{k}_3} \rangle
&= - i (2 \pi)^3 \delta(\vect{k}_1 + \vect{k}_2 + \vect{k}_3)
\trajp^I_{I'}\trajp^J_{J'}\trajp^K_{K'}
\prod_{i=1,2,3} \! \bigg(\frac{1 + i k_i \tau}{2 k_i^3} \bigg) \times \nonumber \\
& \qquad \int_{-\infty(1-i\delta)}^{\tau(1-i\delta)} \d \eta \, \frac{a^2 H^4}{4 \Mpl^2} \frac{\dot \phi^{I'}}{H} G^{J'K'} (1-ik_1 \eta) k_2^2 \eta \,k_3^2 \eta \,\e{i k_t (\eta - \tau)}
\nonumber \\
& \raisebox{-3mm}{\qquad + ~{\rm c.c.} + {\rm perms}.}
\end{align}
We have introduced $k_t = k_1 + k_2 + k_3$. Note that the factors of the trajectory propagator $\trajp^I_{I'}$ ensure that eq.~\eqref{eq:3ptcorrelator2} has the correct tensorial transformation properties. 

To perform the integral in eq.~\eqref{eq:3ptcorrelator2} we must specify the time dependence of the field velocity $\dot \phi^I$ and the scale factor $a$ (and thus the Hubble rate $H$). In the absence of a specific model, this is not in general possible. However, the exponent in the integrand is highly oscillatory in the deep subhorizon and so it is only the times near to horizon exit that will significantly contribute to the result. We may parametrise our ignorance of the inflationary model by expanding in terms of slow-roll. Firstly, we may substitute $a = -(H \eta)^{-1} + \O(\epH)$. Secondly, we temporally Taylor expand the factors of $\dot \phi^I$ and $H$ about their values at the time of horizon exit of a particular Fourier mode $k_*$. Tangent-space indices at this time are labelled with unprimed lower-case indices $i$, $j$, $k$. The Taylor expansion proceeds via eq.~\eqref{eq:covariant-expansion} with the trajectory propagator in place of the parallel propagator. The zeroth order term is simply a constant, augmented by the appropriate factors of the trajectory propagator. Slow-roll corrections may be incorporated perturbatively, although they appear beyond requisite order for the present calculation~\cite{Dias:2012qy} and so are ignored. 

We shall find that there are two types of terms contributing to the three-point function: {\it growing modes} and {\it frozen modes}. The frozen modes are terms that have a vanishing or constant contribution to the three-point function at times a few efolds after horizon exit. The asymptotic value of these contributions is obtained by taking the limit $\tau \to \infty$. The growing modes behave differently, sourcing contributions to the three-point function at times significantly after horizon exit. These will manifest as terms proportional to powers of the number of superhorizon efolds $N$. We will truncate the evolution of such terms near the time of horizon exit when $N$ is small, and leave the superhorizon evolution to be tracked using the methods developed in chapter \ref{ch:formalisms}. 

It proves convenient to extract a number of common factors such that the three-point function may be written in terms of the bitensor $A^{ijk}(N)$ as
\begin{equation}
	\langle
		Q^I_{\vect{k}_1}
		Q^J_{\vect{k}_2}
		Q^K_{\vect{k}_3}
	\rangle
	=
	(2\pi)^3 \delta(\vect{k}_1 + \vect{k}_2 + \vect{k}_3)
	\frac{H_*^4}{4 k_1^3 k_2^3 k_3^3}
	{\trajp^I}_i {\trajp^J}_j {\trajp^K}_k
	A^{ijk}(N) .
	\label{eq:field_3pf}
\end{equation}
We start our efolding clock at the time of horizon crossing for $k_*$ and so $N_* = 0$. The three-point function $\langle Q^I_{\vect{k}_1} Q^J_{\vect{k}_2} Q^K_{\vect{k}_3} \rangle$ is a rank-three tensor evaluated in the tangent space at a slightly different time $N$. This leaves $A^{ijk}(N)$ as a bitensorial quantity, transforming as a rank-three tensor at $N_*$ but as a scalar at $N$. Eq.~\eqref{eq:3ptcorrelator2} then becomes
\be
A^{ijk} \supseteq - \frac{i}{8} \frac{\dot \phi_*^i}{H_*} G^{jk} k_2^2 k_3^2
\, \lim_{\tau \to 0} \,
\prod_{l=1}^3 \! (1+ik_l \tau) \e{-i k_t \tau} 
\! \int_{-\infty(1-i\delta)}^{\tau(1-i\delta)} \! \d \eta \, (1-ik_1 \eta) \e{i k_t \eta}
+ ~{\rm c.c.} + {\rm perms}.
\ee
The contour integral can be calculated and so one obtains the final result
\be
\label{eq:3ptcorrelator3}
A^{ijk} \supseteq - \frac{1}{2} \frac{\dot \phi_*^i}{H_*} G^{jk} k_2^2 k_3^2 \Big( \frac{1}{k_t} + \frac{k_1}{k_t^2}\Big) + {\rm cyclic},
\ee
where a number of terms have dropped out following the complex conjugation and `+ cyclic' sums over the two permutations generated
by simultaneous exchange of $\{i$, $\vect{k}_1\}$, $\{j$, $\vect{k}_2\}$ and $\{k$, $\vect{k}_3\}$.

This result is valid at lowest order in slow-roll and one can see that eq.~\eqref{eq:3ptcorrelator3} does not contain any explicit time-dependent terms. Physically this means that such a bispectrum contribution does not contain growing modes at lowest order in slow-roll. Rather, it contains frozen modes and eq.~\eqref{eq:3ptcorrelator3} represents the asymptotic limit in which these have all reached constant values. Note that this is not precisely the time of horizon crossing~\cite{Nalson:2011gc}. Interesting infrared dynamics occur when one considers the subleading slow-roll corrections to eq.~\eqref{eq:3ptcorrelator3}, but as discussed in \S\ref{sec:ir-safety}, we may ignore these provided we constrain our attention to the few efolds immediately after horizon exit.

We may now consider the contributions arising from the other two terms in $S_{(3)}$ that are simply the covariantized versions of those present in flat field-space. The calculation is analogous to that above and the three results together are
\begin{itemize}
\item $ \displaystyle{-\frac{a^2}{4 \Mpl^2 H} \dot \phi^I Q_I 
\Deta Q^J \Deta Q_J} $ \\
\begin{equation}
	\label{eq:3pfv1}
	A^{ijk} \supseteq  
	-\frac{1}{2 \Mpl^2} \frac{\dot \phi^i_*}{H_*} G^{jk}_* k_2^2 k_3^2 
	\Bigg( \frac{1}{k_t} + \frac{k_1}{k_t^2}\Bigg) + \text{cyclic} ,
\end{equation}
\item $ \displaystyle{-\frac{a^2}{4 \Mpl^2 H} \dot \phi^I Q_I \partial_i Q^J \partial_i Q_J} $ \\
\begin{equation}
	\label{eq:3pfv2}
	A^{ijk} \supseteq  
	\frac{1}{2 \Mpl^2} \frac{\dot \phi^i_*}{H_*} G^{jk}_*
	(\vect{k}_2 \cdot \vect{k}_3)
	\Bigg( k_t - \frac{\kappa^2}{k_t} - \frac{k_1 k_2 k_3}{k_t^2} \Bigg)
 	+ \text{cyclic} ,
\end{equation}
\item $ \displaystyle{\frac{a^2}{2 \Mpl^2 H} \dot \phi^I \partial_i \partial^{-2} \Deta Q_I \partial_i Q^J \Deta Q_J} $ \\
\begin{equation}
	\label{eq:3pfv3}
	A^{ijk} \supseteq  
	\frac{1}{2\Mpl^2} \frac{\dot \phi^i_*}{H_*} G^{jk}_* \Bigg[
	(\vect{k}_1 \cdot \vect{k}_2) k_3^2
	\Bigg( \frac{1}{k_t} + \frac{k_2}{k_t^2} \Bigg)
	+(\vect{k}_1 \cdot \vect{k}_3) k_2^2
	\Bigg( \frac{1}{k_t} + \frac{k_3}{k_t^2} \Bigg)
	\Bigg] + \text{cyclic} ,
\end{equation}
\end{itemize}
where $\kappa^2 = \sum_{i < j} k_i k_j$. Elliston et al.~\cite{Elliston:2012ab} showed that these three contributions can be combined to produce the more elegant result
\begin{equation}
	A^{ijk} \supseteq 
	\frac{1}{\Mpl^2} \frac{\dot \phi^i_*}{H_*} G^{jk}_* 
	\Bigg[
		-2 \frac{k_2^2 k_3^2}{k_t}
		+ \frac{1}{2} k_1 (\vect{k}_2 \cdot \vect{k}_3)
	\Bigg] + \text{cyclic} .
\end{equation}

\sssec{Log-divergent contributions}

We now consider the remaining terms in $S_{(3)}$ that involve the Riemann tensor. The quantization procedure is identical apart from the final integration. As an example, let us consider the bispectrum contribution arising from the term
\be
S_{(3)} \supseteq \int \d^3 x \, \d \eta \, \bigg\{\frac{2 a^3}{3} R^{I(JK)L}\dot \phi_L \Deta Q_I Q_J Q_K \bigg\}.
\ee
Following the previous methodology one finds
\begin{align}
A^{ijk} &\supseteq - \frac{i}{3} \Mpl^2 \frac{\dot \phi_*^l}{H_*} {R^{ijk}}_l k_1^2
\, \lim_{\tau \to 0} \, (1+ik_t \tau + \O(\tau^2)) \e{-i k_t \tau} \times \nonumber \\
& \qquad \int_{-\infty(1-i\delta)}^{\tau(1-i\delta)} \! \d \eta \, \frac{1}{\eta^2} (1-ik_2 \eta)(1-ik_3 \eta) \e{i k_t \eta}
+ ~{\rm c.c.} + {\rm perms}.
\label{eq:madras1}
\end{align}
The integral in eq.~\eqref{eq:madras1} may be performed by parts, starting with the most divergent pieces, as
\begin{align}
\hspace{-2em} {\rm Integral} &= i \int_{-\infty(1-i\delta)}^{\tau(1-i\delta)} \! \d \eta \, \frac{1}{\eta^2} (1-ik_2 \eta)(1-ik_3 \eta) \e{i k_t \eta} \nonumber \\
&= \frac{-i e^{i k_t \tau}}{\tau} - \int_{-\infty(1-i\delta)}^{\tau(1-i\delta)} \! \d \eta \, \left( \frac{k_1}{\eta} + i k_2 k_3 \right) \e{i k_t \eta} \nonumber \\
&= \frac{-i e^{i k_t \tau}}{\tau} - k_1 \ln (k_t \tau) e^{i k_t \tau} - \frac{k_2 k_3}{k_t} e^{i k_t \tau} + i k_1 \int_{-\infty(1-i\delta)}^{k_t \tau(1-i\delta)} \! \d x \, \ln (-x) e^{ix}. \label{eq:mascheroni_int}
\end{align}
The final integral is convergent as $k_t \tau \to 0$ and so we take this limit:
\begin{align}
i \int_{-\infty}^{0} \! \d x \, \ln (-x) e^{ix} &=
\int_{0}^{\infty} \! \d w \, \ln (-iw) e^{-w} \nonumber \\ &=
\int_{0}^{\infty} \! \d w \, \left( \ln w - \frac{i \pi}{2} \right) e^{-w} \nonumber \\ &
= -\left(\gamma_{\rm E} + \frac{i \pi}{2} \right),
\end{align}
where $\EulerGamma \approx 0.577$ is the Euler--Mascheroni constant. We therefore find the integral in eq.~\eqref{eq:mascheroni_int} as 
\be
{\rm Integral} = e^{i k_t \tau} \left[ 
\frac{-i}{\tau} - k_1 \Big(	N + \ln \frac{k_t}{k_\ast} \Big) - \frac{k_2 k_3}{k_t}
\right] -k_1 \left( \EulerGamma + \frac{i \pi}{2} \right).
\ee
Combining this with eq.~\eqref{eq:madras1}, one finds that the terms which diverge polynomially as $\tau \to 0$ all cancel under the complex conjugation. One is then left with convergent and log-divergent terms that form the final result~\cite{Elliston:2012ab}
\be
A^{ijk} \supseteq
	\frac{4}{3} R^{i(jk)m}_* \frac{\dot \phi_m^*}{H_*} 
	\Bigg[
		k_1^3 \bigg( \EulerGamma -N + \ln \frac{k_t}{k_*} \bigg)
		- k_t k_1^2 + \frac{k_1^2 k_2 k_3}{k_t}
	\Bigg] + \text{cyclic} ,
\ee
The remaining log-divergence is explicit in the term proportional to $N = - \ln |k_\ast \tau|$ and so there is explicit infrared dynamics at leading order. It is such terms that ensure that the three-point function result as calculated in this section is not valid in the limit $N \gg 1$. 

A similar calculation finds the contribution from the other term $S_{(3)}$ as
\begin{itemize}
\item $ \displaystyle{\frac{a^4}{6} R^{(I|LM|J;K)}\dot \phi_L \dot \phi_M Q_I Q_J Q_K} $ \\
\begin{equation}
	\label{eq:3pfv5}
	A^{ijk} \supseteq  
	\frac{1}{3} R^{(i|mn|j;k)}_\ast
		\frac{\dot{\phi}_m^\ast}{H_\ast}
		\frac{\dot{\phi}_n^\ast}{H_\ast}
		\left[
			k_1^3 \Big(
				N - \ln \frac{k_t}{k_\ast} - \EulerGamma - \frac{1}{3}
			\Big) 
			+ \frac{4}{9} k_t^3
			- k_t \kappa^2
		\right] + \text{cyclic} .
\end{equation}
\end{itemize}

In addition, since we are working to $\O(\dot \phi / H)^2$, there is one final contribution arising from the Taylor expansion of the Riemann tensor 
\begin{equation}
\label{eq:riemann_3pf_expansion}
	R^{I'(J'K')L'} \frac{\dot{\phi}_{L'}}{H}
	=
	{\trajp^{I'}}_i
	{\trajp^{J'}}_j
	{\trajp^{K'}}_k
	\bigg(
		R^{i(jk)l} \frac{\dot{\phi}_l}{H}
		+ \DN
			R^{i(jk)l} \frac{\dot{\phi}_l}{H}
			N
		+
		\cdots
	\bigg)_\ast .
\end{equation}
This provides us with one additional contribution to the three-point function that appears at requisite order in slow-roll:
\begin{itemize}
\item $ \displaystyle{\frac{2 a^3}{3H} R^{I(JK)L;M}\dot \phi_L \dot \phi_M N \Deta Q_I Q_J Q_K} $ \\
\begin{align}
	\label{eq:3pfv6}
	A^{ijk} \supseteq 
	\frac{2}{3} R^{i(jk)m;n}_* \frac{\dot \phi_m^*}{H_*}
	\frac{\dot \phi_n^*}{H_*} 
	\Bigg[ &
		-k_1^3 N^2 + k_1^3 \bigg(
			\EulerGamma^2 - \frac{\pi^2}{12}
			+ \ln \frac{k_t}{k_*} \Big(
				2\EulerGamma + \ln \frac{k_t}{k_*}
			\Big) \bigg)
	\nonumber \\ & \mbox{}
		- 2 k_t k_1^2 \Big(
			\ln \frac{k_t}{k_*} + \EulerGamma - 1
		\Big)
		+ 2\frac{k_1^2 k_2 k_3}{k_t}\Big(
			\ln \frac{k_t}{k_*} + \EulerGamma
		\Big)
	\Bigg]
	\nonumber \\ & \mbox{} + \text{cyclic} .
\end{align}
\end{itemize}

\sssec{Final result}

We may now simply combine the convergent and log-divergent pieces together to obtain the full bispectrum at horizon crossing:
\bea
	A^{ijk}(N) &=& 
		\frac{1}{\Mpl^2} \frac{\dot{\phi}^i_\ast}{H_\ast} G^{jk}_\ast
		\bigg(
			- 2 \frac{k_2^2 k_3^2}{k_t}
			+
			\frac{k_1}{2} \vect{k}_2 \cdot \vect{k}_3
		\bigg)
		\nonumber \\ && 
		+ \frac{4}{3} R^{i(jk)m}_\ast \frac{\dot{\phi}_m^\ast}{H_\ast}
		\left[
			k_1^3 \Big(
				\EulerGamma - N + \ln \frac{k_t}{k_\ast}
			\Big)
			- k_1^2 k_t
			+ \frac{k_1^2 k_2 k_3}{k_t}
		\right]
		\nonumber \\ &&
		+ \frac{1}{3} R^{(i|mn|j;k)}_\ast
		\frac{\dot{\phi}_m^\ast}{H_\ast}
		\frac{\dot{\phi}_n^\ast}{H_\ast}
		\left[
			k_1^3 \Big(
				N - \ln \frac{k_t}{k_\ast} - \EulerGamma - \frac{1}{3}
			\Big) 
			+ \frac{4}{9} k_t^3
			- k_t \kappa^2
		\right]
		\nonumber \\ &&
		- \frac{4}{3} R^{i(jk)m;n}_\ast
		\frac{\dot{\phi}_m^\ast}{H_\ast}
		\frac{\dot{\phi}_n^\ast}{H_\ast}
		\Bigg[
			\frac{k_1^3}{2} \Big(
				N^2
				- \EulerGamma^2 + \frac{\pi^2}{12}
				- \Big[
					2 \EulerGamma + \ln \frac{k_t}{k_\ast}
				\Big]
				\ln \frac{k_t}{k_\ast}
			\Big)
			\nonumber \\ && \hspace{36mm} 
			+ k_1^2 k_t \Big(
				\ln \frac{k_t}{k_\ast} + \EulerGamma - 1
			\Big)
			- \frac{k_1^2 k_2 k_3}{k_t}
			\Big(
				\EulerGamma + \ln \frac{k_t}{k_\ast}
			\Big)
		\Bigg]
		\nonumber \\ &&
		+ \text{cyclic} .
	\label{eq:bispectrum}
\eea

This result constitutes one of the principal results of this thesis. The first line of eq.~\eqref{eq:bispectrum} reduces to the standard result of Seery and Lidsey~\cite{Seery:2005gb} for multi-scalar field inflation in the presence of a trivial field metric. The remaining terms in eq.~\eqref{eq:bispectrum} are new and mediated by the field space curvature tensor. The added complexity provides exciting prospects for bispectrum phenomenology at horizon crossing. Perhaps the most exciting aspect is the presence of time-dependence in the bispectrum at lowest order in slow-roll, suggesting that it is possible for a field space metric to mediate significant bispectrum evolution. 

\sssec{Isolating the growing modes}
\label{sec:isolatedgrowingmodes}

The time-dependent terms in~\eqref{eq:bispectrum} (those involving $N$) are divergent in the late-time limit $\tau \rightarrow 0$. It is these terms that generate the infrared dynamics described in~\S\ref{sec:ir-safety} by sourcing time evolution after horizon
exit~\cite{Zaldarriaga:2003my,Seery:2007wf,Seery:2010kh}, which ensures that the expressions derived in this chapter are only valid for a few efolds after horizon crossing. In chapter~\ref{ch:formalisms} we shall show that the divergences calculated in this chapter may be properly accounted for by a covariant version of the `separate universe'
method~\cite{Lyth:2005fi,Seery:2012vj}. To demonstrate the agreement between this superhorizon formalism and the quantization calculation in this chapter, it is necessary to demonstrate that both calculations have the same time-dependence at horizon crossing. At this point it is therefore useful to isolate those terms in $A^{ijk}$ which are linear or quadratic in diverging logarithms as
\bea
	A^{ijk}_{\text{1-$\log$}} &=&
		\frac{1}{3} N
		k_1^3
		\bigg(
			R^{(i|mn|j;k)}
			\frac{\dot{\phi}_{m}}{H}
			\frac{\dot{\phi}_{n}}{H}
			-
			4 {R^{i(jk)m}}
			\frac{\dot{\phi}_m}{H}
		\bigg)_*
		+
		\text{cyclic} ,
	\label{eq:bispectrum-single-log} \\
	A^{ijk}_{\text{2-$\log$}} &=&
		-\frac{4}{6} N^2 k_1^3
		\bigg(		
		R^{i(jk)m;n}
		\frac{\dot{\phi}_m}{H}
		\frac{\dot{\phi}_n}{H}
		\bigg)_*
	+ \text{cyclic} .
	\label{eq:bispectrum-double-log}
\eea

\sssec{Summary}
This chapter followed Elliston et al.~\cite{Elliston:2012ab} where we provided the first full computation of the three-point function $\langle	Q^I_{\vect{k}_1} Q^J_{\vect{k}_2} Q^K_{\vect{k}_3} \rangle$ for non-canonical models of the form~\eqref{eq:mf-action}. Because we have chosen to employ the covariant perturbation scheme of Gong and Tanaka~\cite{Gong:2011uw}, one must be careful to ensure that all expressions have the correct tensorial transformation properties. We have achieved this through the introduction of the trajectory propagator $\trajp^{IJ'}$. The three-point function is written in eq.~\eqref{eq:bispectrum}, which generalises the canonical result of Seery and Lidsey~\cite{Seery:2005gb}.

The non-trivial field metric does not alter the power spectrum at leading order. However, it has a notable effect on the bispectrum through the introduction of new terms in $S_{(3)}$ mediated by the field space curvature. One effect of this is to produce richer frozen modes; these may be observable if the curvature tensor takes sufficiently large values. A second effect is the presence of new time-dependent growing modes that appear at lowest order in slow-roll. This is in contrast to the canonical scenario where such time-dependence enters with $\O(\dot \phi / H)^3$ slow-roll suppression. Consequently there is an expectation that such models may lead to the generation of observable local-shape non-Gaussianity, which may then be constrained by future observational data.

The results of this chapter do not in themselves represent observables. Rather, they represent stochastic initial conditions for the subsequent superhorizon evolution that we consider in chapter~\ref{ch:formalisms}. 
\clearpage{\pagestyle{empty}\cleardoublepage}
\chapter{Superhorizon formalisms}
\label{ch:formalisms}

\begin{addmargin}[0.05\textwidth]{0.05\textwidth}
Chapter \ref{ch:subhorizon} calculated covariant quantum perturbations around the time of horizon exit. The aim of this chapter is to develop the necessary mathematical formalism to enable these perturbations to be tracked in the subsequent superhorizon epoch and then linked with observationally relevant quantities. Observational quantities, discussed in \S\ref{sec:obs}, are derived from the $n$-point correlators of the curvature perturbation on uniform density hypersurfaces, $\zeta$. In \S\ref{sec:deltaN} we review the $\delta N$ formalism and show how this formally allows $\zeta$ to be computed in terms of field perturbations at horizon exit. To obtain usable formulae for the evolution of covariant perturbations, \S\ref{sec:transport} follows Elliston et al.~\cite{Elliston:2012ab} by employing and extending the transport formalism of Mulryne, Seery and Wesley~\cite{Mulryne:2009kh,Mulryne:2010rp,Mulryne:2013uka}. 
$\zeta$ is then derived from the covariant gauge transformation defined in \S\ref{sec:gaugetransformation}.
We then derive analytic solutions to the superhorizon evolution of perturbations in \S\ref{sec:analytics} so that we can develop intuition about the relationship between inflationary dynamics and the evolution of cosmic observables. This follows Elliston et al.~\cite{Elliston:2012wm}, deriving compact expressions for $\delta N$ coefficients for two-field canonical inflation which lead to new expressions for the local bispectrum and trispectrum. The analysis of these analytic results is deferred to chapters \ref{ch:adiabatic} and \ref{ch:heatmaps}.
\end{addmargin}

\begin{center}
\partialhrule
\end{center}
\vspace{-3em}
\begin{quote}
\list{}{\leftmargin 1cm \rightmargin\leftmargin} \endlist
\begin{center}
{\it ``Philosophy is written in this grand book---I mean the Universe---which stands continually open to our gaze, but it cannot be understood unless one first learns to comprehend the language and interpret the characters in which it is written. It is written in the language of mathematics, and its characters are triangles, circles, and other geometrical figures, without which it is humanly impossible to understand a single word of it; without these,\newline one is wandering around in a dark labyrinth.''}
\flushright{---Galileo Galilei, The Assayer, 1622.}
\end{center}
\end{quote}
\vspace{-1em}
\begin{center}
\partialhrule
\end{center}
\newpage

\sec{Observable statistics of inflation}
\label{sec:obs}

Observational quantities are derived from the $n$-point correlators of $\zeta$, the curvature perturbation on uniform density hypersurfaces, which we now define.

\subsection{The uniform density curvature perturbation $\zeta$}

Working with the~\adm~metric in the flat gauge, at linear order the curvature perturbation on uniform density hypersurfaces takes the form
\be
\label{eq:zeta_defn}
\zeta = - \frac{H}{\dot \rho} \delta \rho  = \frac{1}{3} \frac{\delta \rho}{\rho + p},
\ee
where $H(t)$, $\rho(t)$ and $p(t)$ are background quantities whereas $\zeta(\vect{x},t)$ and $\delta \rho(\vect{x},t)$ are inhomogeneous perturbations. If we had not specialised to the flat gauge then eq.~\eqref{eq:zeta_defn} would include another term.\footnote{General expressions can be found in Malik and~Wands~\cite{Malik:2008im}.} This definition of $\zeta$ is {\it gauge invariant} to linear order, and may be suitably redefined to ensure that gauge invariance holds to second order~\cite{Malik:2003mv,Malik:2005cy}.

At this point we can state a very useful result called {\it the $\delta N$ formula}. Noting that the density perturbation may be written
$\delta \rho(\vect{x},t) = -\dot \rho(t) \delta t(\vect{x},t)$, we find that $\zeta$ may be written as~\cite{Wands:2000dp}
\be
\zeta(\vect{x},t) = \delta N(\vect{x},t),
\label{eq:zetadeltaN}
\ee
where we have employed the sign convention of Lyth and Rodr\'{i}guez~\cite{Lyth:2005fi}. This result has also been extended beyond linear order~\cite{Lyth:2004gb}.
Eq.~\eqref{eq:zetadeltaN} applies between a flat hypersurface of constant efolding number and the uniform density hypersurface on which $\zeta$ is defined. As the Universe evolves, spatially-separated regions will undergo different expansion histories before reaching this hypersurface. The average expansion is absorbed into the background number of efolds $N(t)$, leaving the perturbation $\delta N(\vect{x},t)$ with zero mean. 

Physically, $\zeta$ describes the spatial curvature of surfaces of uniform density. At the time of horizon re-entry one can consider $\zeta$ as the primordial perturbation. This is subsequently evolved via transfer functions to describe the subhorizon perturbations that we observe~\cite{Dodelson:2003ft}. 

In chapter \ref{ch:subhorizon} we derived the two and three-point correlators of the Fourier field perturbations $Q^I_{\vect{k}}$. These can be related to the Fourier components $\zeta_{\vect{k}}$ via either the {\it $\delta N$ formalism} or the {\it transport formalism}, which we respectively discuss in \S\ref{sec:deltaN} and \S\ref{sec:transport}. 

\sssec{The comoving curvature perturbation $\R$}

It is useful at this point to define a second gauge-invariant perturbation variable which we shall use in chapter~\ref{ch:singlefield}. In the flat gauge at linear order, the {\it comoving curvature perturbation} takes the form~\cite{Baumann:2009ds}
\be
\R = - \frac{H}{\rho + p} \delta q ,
\ee
where $\partial_i \delta q$ is the scalar part of the 3-momentum density $T_i^0 = \partial_i \delta q$. The perturbation $\R(\vect{x},t)$ denotes the spatial curvature on comoving hypersurfaces. For inflation driven by a single scalar field, the two curvature perturbations $\zeta$ and $\R$ are approximately equal~\cite{Baumann:2009ds}, up to a minus sign. This approximate equality also holds in the superhorizon regime, both for single field and multi-field inflation.

\subsection{Two-point statistics of $\zeta$}

We saw in chapter \ref{ch:subhorizon} that the perturbations in the \cmbr~have a quantum origin. It is therefore only meaningful to ask {\it statistical} questions about the perturbations that we observe. The most basic statistical information is provided by the $2$-point correlator of $\zeta$. 

\sssec{Power spectrum}

The power spectrum $P_\zeta(k)$ is defined as the two-point correlator of the Fourier modes $\zeta_{\vect{k}}$ as
\be
\label{eq:powerspectrum}
\langle \zeta_{\vect{k}_1} \zeta_{\vect{k}_2} \rangle
\equiv (2 \pi)^3 \delta(\vect{k}_1 + \vect{k}_2) P_{\zeta} (k),
\ee
where $\langle \cdots \rangle$ denotes the ensemble average and $k$ is the common magnitude of $\vect{k}_1$ and $\vect{k}_2$. If the volume is sufficiently large then the ensemble average equates to a volume average by the ergodic theorem~\cite{Lyth:2009zz}. We may also define a dimensionless power spectrum
\be
\P_\zeta (k) \equiv \frac{k^3}{2 \pi^2} P_\zeta (k).
\ee 
The dimensionless power spectrum varies weakly with the comoving scale $k$, so it is conventional to fix a particular {\it pivot scale}. The best data at present is a combination of \wmap~9-year, \bao~and \hst~data~\cite{Hinshaw:2012fq} which uses a pivot scale $k_0 = 0.002 ~{\rm \Mpl}^{-1}$ and finds
\be
\label{eq:Pzeta_data}
\P_\zeta \big|_{k_0} = \big(2.427^{+0.078}_{-0.079} \big) \times 10^{-9} ~{\rm at}~ 68\%~ \textsc{cl}.
\ee

\sssec{Spectral index}

The spectral index $n_\zeta (k)$ measures the scale-dependence of the power spectrum as
\be
\label{eq:nsdef}
n_\zeta -1 \equiv \frac{\d \ln \P_\zeta (k)}{\d \ln k}.
\ee
The factor of unity is conventional, since the power spectrum is observationally found to be nearly scale invariant and so the above definition implies $n_\zeta \approx 1$. Current data~\cite{Hinshaw:2012fq} favours a {\it red-tilted} spectrum as
\be
n_\zeta \big|_{k_0} = 0.971 \pm 0.010 ~{\rm at}~ 68\%~ \textsc{cl}.
\ee
This places a strong constraint on many models of inflation, although there is often sufficient parameter freedom to ensure observational compatibility. This will be increasingly tough as new data increasingly constrains the observational parameters of cosmology. In particular, Planck is forecast to improve constraints on the spectral index by up to a factor of $5$~\cite{Burigana:2010hg}.

\sssec{Spectral index running}

One may continue differentiating the power spectrum to gain more information about its scale-dependence. The {\it running} of the spectral index is defined as
\be
\alpha_\zeta \equiv \frac{\d \ln n_\zeta (k)}{\d \ln k}.
\ee
The constraint on $\alpha_\zeta (k)$ varies significantly depending on whether the tensor modes (discussed in \S\ref{sec:tensor}) are jointly constrained. Presuming an absence of tensor modes, the running is constrained as~\cite{Hinshaw:2012fq}
\be
\alpha_\zeta \big|_{k_0}= -0.023 \pm 0.011 ~{\rm at}~ 68\%~ \textsc{cl},
\ee
although this is insufficiently precise to infer inconsistency with a power-law spectrum which has constant $n_\zeta$ and correspondingly zero $\alpha_\zeta$.

\subsection{Tensor modes}
\label{sec:tensor}

The tensor power spectrum $\P_t(k)$ near horizon crossing was calculated in eq.~\eqref{eq:Ptdimless}. The tensor modes obey the equation of motion derived from $S_{(2)}$ in eq.~\eqref{eq:s2tensor1} as
\be
\label{eq:tensoreom}
\ddot h_{ij} + 3 H \dot h_{ij} + \frac{k^2}{a^2} h_{ij} = 0.
\ee
In the superhorizon regime the final term in \eqref{eq:tensoreom} is negligible, leading to superhorizon evolution that admits a constant solution. The tensor modes are therefore frozen during the superhorizon evolution, until the point of re-entry, after which they undergo damped oscillations that may be observable as gravitational waves. As a result of this lack of superhorizon time-evolution, the result $\P_t(k)$ in eq.~\eqref{eq:Ptdimless} holds until the time of re-entry. 

The amplitude of the tensor modes is commonly written in terms of the {\it tensor--scalar ratio} defined as
\be
r \equiv \frac{\P_t(k)}{\P_\zeta(k)}.
\ee
For a single field model, $\P_\zeta(k)$ is conserved shortly after horizon crossing and then $r$ is similarly conserved. Conversely, $r$ may vary for multi-field models due to the superhorizon evolution of $\P_\zeta(k)$.

At present we only have an upper bound on the tensor--scalar ratio $r$. The tightest constraint arises from a combination of \wmap~9-year, \bao~and \hst~data~\cite{Hinshaw:2012fq} with additional small-scale data supplied by the \spt~\cite{Keisler:2011aw} and {\sc act}~\cite{Fowler:2010cy,Das:2011ak} which produces the upper bound of
\be
r < 0.13 {\rm ~at ~95\%}~ \textsc{cl}.
\ee
This constraint is sensitive to any assumption made on the running of the scalar spectral index $\alpha_\zeta$. The value we quote is valid for the power-law assumption $\alpha_\zeta = 0$. 

\begin{figure}[h]
\begin{center}
\includegraphics[width=0.8\textwidth]{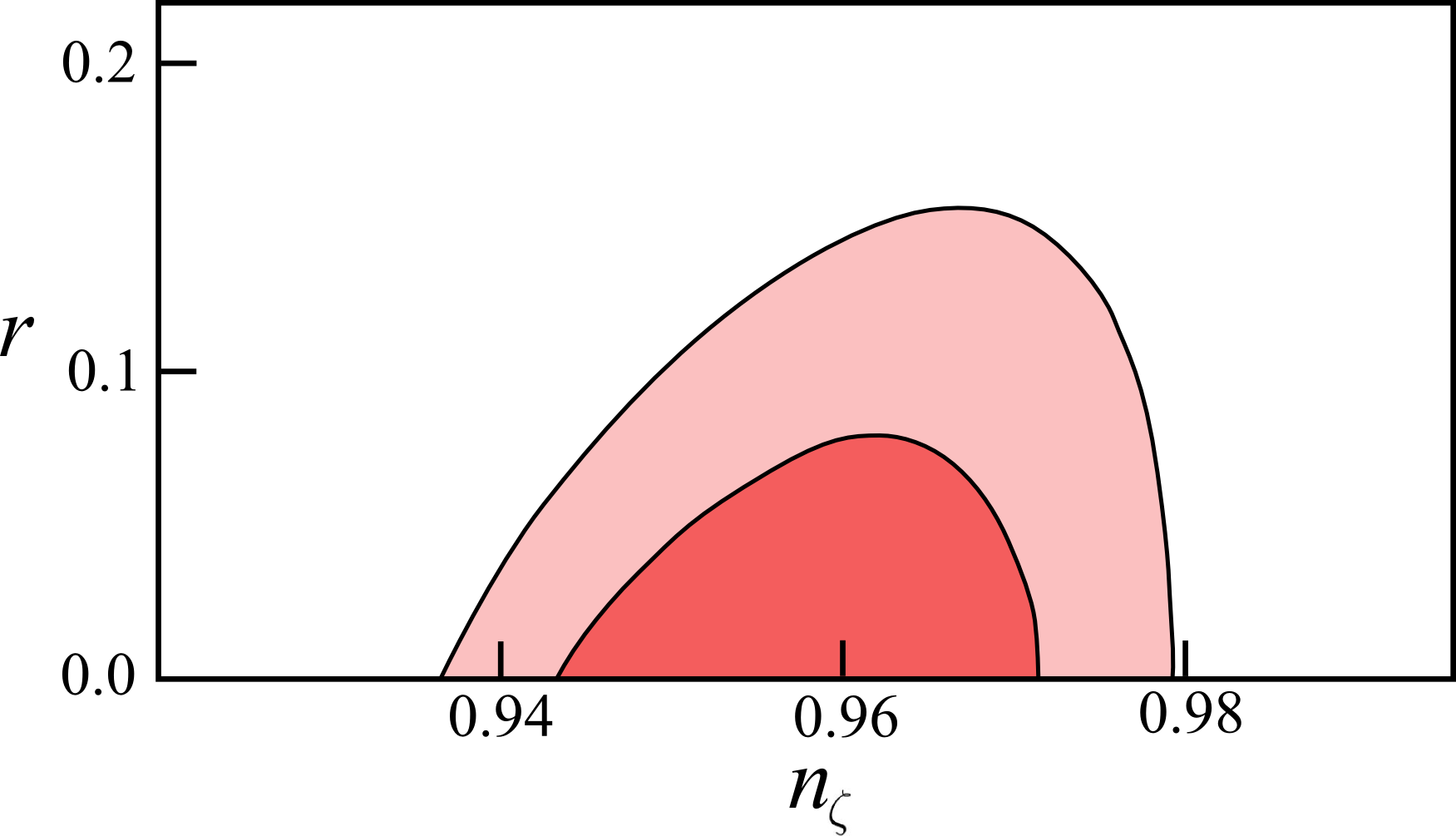}
\caption[Current constraints on $n_\zeta$ and $r$]{Current observational bounds on the tensor--scalar ratio $r$ and the spectral index $n_\zeta$ at $68\%$~{\sc cl} and $95\%$~{\sc cl} respectively. The scale-invariant Harrison-Zel'dovich spectrum with $n_\zeta=1$ is strongly ruled out. Image traced from Hinshaw et al.~\cite{Hinshaw:2012fq}.
\label{fig:constraint}}
\end{center}
\end{figure}

The upper bound on $r$ and the narrow bounds on $n_\zeta$ provide two of the most stringent current constraints on inflationary models. For this reason it is commonplace to provide a joint constraint as shown in figure~\ref{fig:constraint}. The upper bounds on $r$ will be further constrained by \cmbr~polarisation observations such as {\sc spider}~\cite{Crill:2008rd} which, in the absence of a detection, will place limits of $r < 0.03$ at 3$\sigma$ confidence. 

\sssec{Tensor spectral index}

The tensor spectral index $n_t (k)$ is defined similarly to the scalar spectral index $n_\zeta$ as
\be
\label{eq:ntdef}
n_t \equiv \frac{\d \ln \P_t (k)}{\d \ln k}.
\ee
Note that, as opposed to $n_\zeta$, it is $n_t$ not $n_t-1$ that is defined in this way. Observational data is yet to provide useful constraints on the tensor spectral index.

\subsection{Three and four-point statistics of $\zeta$}

\sssec{Bispectrum}
\label{sec:bispectrum}

The three-point statistics of $\zeta$ are written in terms of the bispectrum $B_\zeta (k_1, k_2, k_3)$ as
\be
\label{eq:Bz}
\langle	\zeta_{\vect{k}_1}
		\zeta_{\vect{k}_2}
		\zeta_{\vect{k}_3}	\rangle
\equiv(2\pi)^3 \delta(\vect{k}_1 + \vect{k}_2 + \vect{k}_3)
B_\zeta(k_1, k_2, k_3) .
\ee
The overall delta function multiplying the bispectrum constrains the wavevectors $\vect{k}_1$, $\vect{k}_2$ and $\vect{k}_3$ to form a triangle. One degree of freedom is the {\it size} of this triangle, and in this sense the bispectrum has scale-dependence much like the power spectrum $\P_\zeta(k)$. However, the bispectrum contains considerably more information because one can also vary the triangle {\it shape}. 

\begin{figure}[h]
\begin{center}
\includegraphics[width=0.6\textwidth]{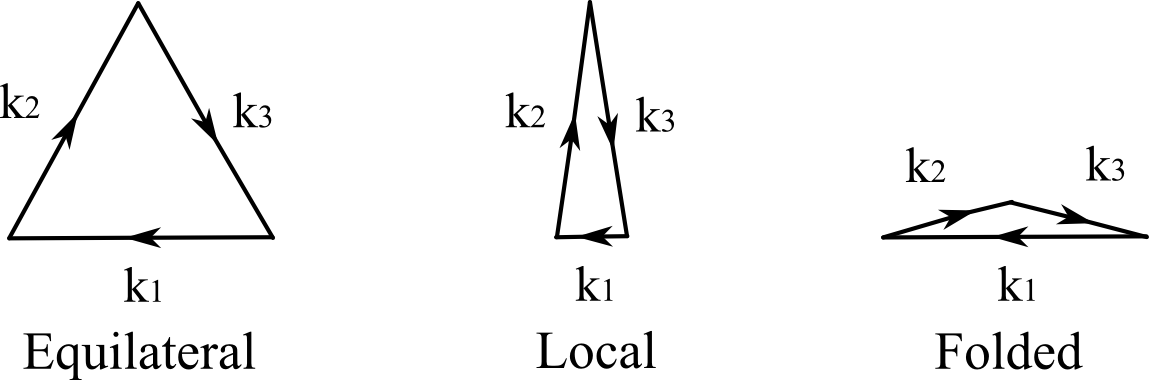}
\caption[Bispectrum shapes: equilateral, local and folded]
{The bispectrum wavenumbers $\vect{k}_1,\vect{k}_2,\vect{k}_3$ form a triangle but various shape possibilities exist. These are commonly parametrized in terms of the {\it equilateral} ($k_1=k_2=k_3$), {\it local} ($k_1 \ll k_2=k_3$) and {\it folded} ($k_1=2k_2=2k_3$) shapes as shown.
\label{fig:triangles}}
\end{center}
\end{figure}

For a fixed total scale $k_t = k_1 + k_2 + k_3$, the triangle shape is dependent on the ratios $k_2/k_1$ and $k_3/k_1$. The shape is commonly referred to in terms of three limiting cases: {\it equilateral} ($k_1=k_2=k_3$), {\it local} ($k_1 \ll k_2=k_3$) and {\it folded} ($k_1=2k_2=2k_3$). These are shown in figure~\ref{fig:triangles}. In our present state of uncertainty regarding the bispectrum, it is sufficient to consider these simple shapes. However, in light of increasingly precise data, it may become necessary to employ a complete basis decomposition such as that developed by Fergusson and Shellard~\cite{Fergusson:2008ra}. 

Physically, the bispectrum has terms of two types: First, those arising from frozen modes near horizon exit; and second, those arising from interactions between growing modes far outside the horizon. By {\it frozen} we simply mean that such modes cease to evolve after a few efolds of superhorizon evolution and that they settle to a constant value. The frozen modes can have complex dependence on the $k$-modes $k_1$, $k_2$ and $k_3$ which manifests as a bispectrum with complex shape-dependence. The second type appears only in the {\it local} combination $(k_1^3 k_2^3)^{-1}$ or its permutations. With canonical kinetic terms, applying global slow-roll conditions to the potential and assuming that only light fields contribute to $\zeta$, Lyth~and~Zaballa~\cite{Lyth:2005qj} showed that the multi-field result of Seery~and~Lidsey~\cite{Seery:2005gb} implies that only the local bispectrum shape can be observable (see also refs.~\cite{Lyth:2005qj,Vernizzi:2006ve,Lyth:2005fi}). In this chapter we shall investigate the non-canonical extension of this result.

It is commonplace to further parametrise the bispectrum in terms of the dimensionless parameter $\fnl(k_1,k_2,k_3)$ defined by
\be
\label{fnl}
B_{\zeta}(k_1,k_2,k_3) \equiv \frac{6}{5} \fnl(k_1,k_2,k_3) \big[ P_\zeta(k_1) P_\zeta(k_2) + {\rm cyclic} \big],
\ee
where `$+~{\rm cyclic}$' includes the additional terms $P_\zeta(k_2) P_\zeta(k_3)$ and $P_\zeta(k_3) P_\zeta(k_1)$. This definition may be used for any bispectrum shape. However, it is particularly useful if the bispectrum is dominated by the local shape in which case $\fnl(k_1,k_2,k_3) \to \fnl^{\rm local}$ is simply a number. Chapters \ref{ch:adiabatic}--\ref{ch:heatmaps} consider canonical multi-field inflation for which this local limit is the only observationally relevant shape. Conversely, chapter \ref{ch:singlefield} will work with a range of scenarios of generalised single field inflation for which the only bispectrum generated is of the equilateral type. 

The best bispectrum constraints to date are provided by the analysis of data from the \wmap~satellite~\cite{Komatsu:2010fb} as 
\begin{align}
-10 < &\fnl^{\rm local} < 74, \\
-214 < &\fnl^{\rm equil} < 266,\\
-410 < &\fnl^{\rm fold} < 6, 
\end{align}
where the uncertainties are at 95\%~{\sc cl}. Planck will improve on these constraints considerably, and in the absence of a detection is expected to give the bounds $|\fnl^{\rm local}| < 5$~\cite{Komatsu:2010hc}. 

We recall that the quantum computation of $\langle Q^I_{\vect{k}_1} Q^J_{\vect{k}_2} Q^K_{\vect{k}_3} \rangle$ in chapter \ref{ch:subhorizon} worked under the assumption that all of the modes $\vect{k}_i$ left the horizon at approximately the same time. This may seem at odds with the notion of a bispectrum $B_{\zeta}(k_1,k_2,k_3)$ where the $\vect{k}_i$ are not equal. However, it is not inconsistent to take $|\vect{k}_i| \neq |\vect{k}_*|$ provided that the difference is moderate, as we shall now demonstrate. 

The dominant contribution to $\langle Q^I_{\vect{k}_1} Q^J_{\vect{k}_2} Q^K_{\vect{k}_3} \rangle$ for non-log-divergent terms arises from the narrow range of efolds near to the time of horizon exit of some reference scale $\vect{k}_*$. We approximate the slow-roll parameters to equal their constant value at the horizon exit time of the mode $\vect{k}_*$; the time-dependence of the slow-roll parameters then only enters perturbatively at higher order in the slow-roll expansion. If we label $N_*$ as the time at which $\vect{k}_*$ exits the horizon, then $|\vect{k}_i| \neq |\vect{k}_*|$ exits the horizon at a time $N_* + \Delta N_i$. Provided that $\Delta N_i$ is not much larger than the range of efolds which provide the dominant contribution to $\langle Q^I_{\vect{k}_1} Q^J_{\vect{k}_2} Q^K_{\vect{k}_3} \rangle$, then the effect of $\Delta N_i \neq 0$ will manifest through the presence of new higher order corrections. Since we have neglected all higher order perturbations in our analysis, it is fully consistent to ignore these effects.

\sssec{Trispectrum}

There are two types of terms that contribute to the four-point statistics $\langle \zeta_{\vect{k}_1} \zeta_{\vect{k}_2} \zeta_{\vect{k}_3} \zeta_{\vect{k}_4} \rangle$. Firstly, there are {\it disconnected} terms that are simply related to two copies of the power spectrum and arise when any two of the $\vect{k}_i$ sum to zero. These therefore provide no new information and so we shall follow refs.~\cite{Seery:2006vu,Byrnes:2006vq} and ignore these contributions. This equates to ignoring parallelogram shapes. The remaining {\it connected} contribution $\langle \zeta_{\vect{k}_1} \zeta_{\vect{k}_2} \zeta_{\vect{k}_3} \zeta_{\vect{k}_4} \rangle_{\rm conn}$ defines the trispectrum $T_\zeta(k_1, k_2, k_3, k_4)$ as
\be
\langle	\zeta_{\vect{k}_1}
		\zeta_{\vect{k}_2}
		\zeta_{\vect{k}_3}
		\zeta_{\vect{k}_4}	\rangle_{\rm conn}
\equiv (2\pi)^3 \delta(\vect{k}_1 + \vect{k}_2 + \vect{k}_3 + \vect{k}_4)
\, T_\zeta(k_1, k_2, k_3, k_4) .
\ee
Analogously to the bispectrum, the trispectrum will in general exhibit complex shape-dependence which may be parametrised by extending the methods used for the bispectrum~\cite{Regan:2010cn}. However, the growing superhorizon modes that we study in chapters~\ref{ch:adiabatic}--\ref{ch:heatmaps} only source particular shapes. For canonical fields and a smooth potential, these superhorizon modes represent the only observable source of the trispectrum and so we can recast the trispectrum as~\cite{Byrnes:2006vq,Seery:2006js}
\begin{align}
\label{eq:tnlgnl}
T_\zeta(k_1, k_2, k_3,k_4) &= \tnl \big[ P_\zeta(k_{13}) P_\zeta(k_3) P_\zeta(k_4) + \mathrm{11~perms}\big] \nonumber\\
& \qquad+ \frac{54}{25}\gnl \big[ P_\zeta(k_2)  P_\zeta(k_3) P_\zeta(k_4) +\mathrm{3~perms} \big],
\end{align}
where $\vect{k}_{ij} = \vect{k}_i + \vect{k}_j$.

The parameters $\tnl$ and $\gnl$ are presently constrained as $-0.6 < \tnl/10^4 < 3.3$ and $-7.4 < \gnl/10^5 < 8.2$ \cite{Smidt:2010ra, Smidt:2010sv}, with ref.~\cite{Fergusson:2010gn} finding the slightly different constraint $-5.4 < \gnl/10^5 < 8.6$. In the absence of a detection, Planck is expected to give bounds of $\tnl < 560$ \cite{Komatsu:2010hc} and $|\gnl| < 1.6 \times 10^5$ \cite{Smidt:2010ra}.

If a given inflationary model predicts a magnitude of any of the parameters $\fnl$, $\tnl$ or $\gnl$ that are greater than the Planck forecast bounds, then we refer to such a prediction as {\it observationally relevant}. A non-Gaussianity that could in principle be measured by an ideal observation conservatively requires these non-Gaussianity parameters to have magnitudes greater than unity. Following the standard diction in the literature, we describe such models as producing a {\it large} non-Gaussianity.

\sec{$\delta N$ formalism}
\label{sec:deltaN}

\ssec{Separate universes}

One may follow the superhorizon evolution of cosmic perturbations by using the tools of cosmological perturbation theory. The perturbation equations of motion include {\it gradient terms} that measure the deviation from exact homogeneity. When expressed in Fourier space, these gradient terms appear as powers of $k/aH$. During inflation, $H$ is approximately constant, whereas $a$ is growing quasi-exponentially. Therefore $k/aH$ quickly becomes negligible after horizon crossing. This motivates a simplifying assumption whereby the gradient terms are neglected.\footnote{One can incorporate the gradient terms perturbatively via a gradient expansion~\cite{Lyth:2004gb}.} 

The above discussion suggests that we can consider the superhorizon Universe as a collection of independent \frwl~universes; independent by virtue of the lack of gradient terms. In reality this procedure only works down to a cut-off {\it smoothing scale} somewhat larger than the comoving horizon size. If we were to equate the smoothing scale to the \wmap~pivot scale, this implies that the observed \cmbr~is a composition of at least $40,\!000$ {\it separate universes}, although ref.~\cite{Lyth:2006gd} suggests that \cmbr~observations sample $10^6$ or more separate universes. These separate universes evolve according to the same background equations of motion, but they have different initial conditions sourced by quantum perturbations. 

This framework, known as the {\it separate universe assumption}, allows for a simpler analysis of superhorizon perturbation evolution~\cite{Starobinsky:1982ee,Bardeen:1983qw,Lyth:1984gv,
Starobinsky:1986fxa,Salopek:1990jq,Wands:2000dp,Lyth:2005fi}. 

\ssec{Separate universes in phase space}
\label{sec:separate_phase}

We shall now show that it may be intuitive to consider the separate universe picture in terms of the relevant phase space. Each separate universe defines a different phase space point that evolves independently. The task of evolving the perturbed Universe is thus reduced to the simpler task of tracking this ensemble of phase space points. This ensemble is drawn from a spacetime region of finite comoving extent. Provided that this extent is not too large then we expect the ensemble of phase space points to be narrowly distributed. We refer to this ensemble of clustered points as a bundle, since this is the shape generated in the phase space when the ensemble is allowed to evolve with time. In practice, the analysis is simplified by working in a thermodynamic limit where the bundle formally contains an infinite number of trajectories. To avoid spurious infrared problems we should demand that these reheat in the same vacuum. This bundle approach can be traced to Hawking's formulation of perturbation theory \cite{Hawking:1966qi}, and has been applied to inflation by several authors \cite{Lyth:1984gv,Starobinsky:1982ee,Starobinsky:1986fxa,Sasaki:1995aw}. 

The full phase space incorporates two dimensions per scalar field from each of the $\{\vp^I,\dot \vp^I\}$ pairs. Extending the discussion in \S\ref{sec:intro-slow-roll}, the {\it slow-roll} limit, where 
$\epH \ll 1$, causes reduction of dynamics to a submanifold on which (for example) the field velocities are unique functions of the field values. The growing mode on this submanifold satisfies $3H \dot \vp_I + V_{,I} = 0$. An explicit description in terms of trajectories on the slow-roll submanifold was given by Salopek \cite{Salopek:1995vw} and Garc\'{\i}a-Bellido and Wands \cite{GarciaBellido:1995qq}. 

There are two types of perturbation associated with the relative phase space location of separate universes, and these two types of perturbation have distinct geometrical definitions. Perturbations along the direction of phase space flow are referred to as {\it adiabatic} perturbations. Since these perturbations are defined by time evolution, they may be written in terms of the variation in efold number $\delta N$, or equivalently the curvature perturbation $\zeta$. The remaining perturbations distinguish different bundle elements. These perturbations are called {\it isocurvature} or {\it entropy} perturbations. It is useful to define a new field basis aligned with the bundle, such that $\sigma$ is the local adiabatic field and $s^I$ are local isocurvature fields~\cite{Gordon:2000hv}. Each isocurvature field $s^I$ has an equation of motion $\dot s^I = 0$ and constitutes a conserved quantity \cite{Salopek:1995vw,GarciaBellido:1995qq}. Together, these conserved quantities identify a trajectory.

\ssec{The $\delta N$ expansion}

The $\delta N$ expansion~\cite{Starobinsky:1986fxa,Sasaki:1995aw,Lyth:2005fi} provides a prescription for finding $\zeta$ in terms of the field perturbations at horizon exit and at subsequent times. It is necessary to assume that the slow roll attractor is operative at the time of horizon exit, such that the subsequent number of efolds until a later-time uniform density hypersurface can be expressed as a function of the field values at horizon exit, $N = N(\vp^I_*)$. We note that the slow-roll assumption was also required in order to compute the quantum field correlators at horizon exit and so it is fully consistent to demand slow-roll behaviour for the first few efolds of superhorizon evolution. 

In standard non-covariant perturbation theory the perturbed fields at horizon exit $\vp^I_*(\vect{x},t)$ and the unperturbed background $\phi^I_*(t)$ are related by the perturbation $\delta \vp^I_*(\vect{x},t)$. We may therefore expand $\delta N$ in terms of these perturbations as
\be
\zeta \equiv \delta N = 
N_{,I} \delta \vp_I^* + \frac{1}{2} N_{,IJ} \delta \vp_I^* \delta \vp_J^*
		+ \cdots ,
\ee
where $N_{,I}$ denotes $\partial N / \partial \phi^I_*$. We have written all indices as covariant to emphasise that they are contracted with the Dirac delta function $\delta^{IJ}$ rather than the field metric $G^{IJ}$.  Lyth~and~Rodr\'{\i}guez~\cite{Lyth:2005fi} were the first to extend this Taylor expansion to second order as required for calculating the bispectrum. 

It is no harder to work with covariant perturbations $Q^I$. The number of efolds $N$ is a field-space scalar and so it Taylor expands via eq.~\eqref{eq:covariant-expansion} as
\be
\label{eq:deltaN}
\zeta \equiv \delta N = 
N_{,i} Q^i_* + \frac{1}{2} N_{;ij} 
		Q^i_* Q^j_* + \cdots .
\ee
Note that we have used lower case Roman indices to be consistent with indices transforming in the tangent space at horizon crossing. The derivatives of $N$ are now covariant derivatives, and they are formally bitensorial quantities, transforming as rank-one or rank-two covariant tensors in the tangent space at horizon exit and with scalar dependence in the tangent space at the time of calculation. We describe these as {\it $\delta N$ coefficients}. This covariant expansion was first written down by Saffin~\cite{Saffin:2012et}. A similar expansion has already been used by Peterson~and~Tegmark~\cite{Peterson:2011yt}, although this did not incorporate covariant perturbations.

\sssec{The finite difference approach}
\label{sec:computation_techniques}

The simplest numerical procedure for calculating the (non-covariant) $\delta N$ coefficients $N_{,I}$ and $N_{,IJ}$ is the {\it finite difference} approach. This entails setting up a grid of different initial conditions at horizon crossing to represent different separate universes. Each universe is then evolved onto subsequent uniform density hypersurfaces and the time taken is recorded. Dividing the variation in the number of efolds by the relative field separation at horizon crossing, one obtains a crude but effective approximation of the linear $\delta N$ coefficients. Second order $\delta N$ coefficients are obtained by comparing neighbouring linear $\delta N$ coefficients (and so on for higher orders).

The advantage of the finite difference approach is its simplicity and that it can trivially account for non-slow-roll behaviour after horizon crossing. The disadvantage is that the numerical calculation can be delicate; having too small an initial grid leads to rapidly growing numerical error, whereas too large an initial grid fails to calculate the coefficients at the correct point in phase space. 

\sec{The covariant transport formalism}
\label{sec:transport}

The transport formalism of Mulryne, Seery and Wesley~\cite{Mulryne:2009kh,Mulryne:2010rp,Mulryne:2013uka} provides an alternative to the $\delta N$ formalism. The key idea is to generate evolution equations directly for the moments of the probability distribution of $\zeta$. Equivalently, one can consider the evolution of the {\it Jacobi fields}---the vectors $Q^I$ which define the separate universes~\cite{Yokoyama:2007uu}. Observable quantities are thus evolved via systems of ordinary differential equations which have good numerical stability. Once the field perturbations $Q^I$ are found at the time of interest, one can then compute the correlators of $\zeta$ by making an appropriate gauge transformation. In flat field-space the transport equations can be integrated to reproduce the $\delta N$ Taylor expansion~\cite{Seery:2012vj} and so these procedures are equivalent.

In curved field-space we must be cautious when comparing the relative motion of neighbouring trajectories. Analogously to general relativity, the relative motion between separate universes will be altered through the effect of geodesic deviation as mediated by field space curvature. One way to proceed is to use the finite difference approach with non-covariant perturbation theory. The background field equations are now augmented by additional terms describing the geodesic flow. This procedure is intuitively simple, but requires correlators of the form $\langle \delta \vp^I \delta \vp^J \rangle_*$. We recall that in chapter \ref{ch:subhorizon} we purposefully avoided calculating such field perturbations because the perturbed Einstein--Hilbert action included a complex array of terms proportional to derivatives of $G_{IJ}(\vp^K)$. It is therefore preferable to work with covariant perturbations $Q^I$. In Elliston et al.~\cite{Elliston:2012ab} we developed such an approach by deriving a covariant transport formalism to evolve the Jacobi fields. When promoted to curved field-space, the Jacobi approach is automatically covariant and accounts naturally for time-dependent effects generated by the Riemann curvature, including its known contribution to the effective mass-matrix. We therefore elect to use the covariant transport approach which we now derive.

\ssec{Jacobi fields}

The Jacobi method provides a simple way to implement the separate universe approach in curved field-space. Consider two separate universes, with a small relative displacement in phase space at horizon exit. This displacement is described covariantly using a tangent-space vector $Q^I$ described in~\S\ref{sec:covariant-perts}.

Each universe evolves according to the field equation
\begin{equation}
	\frac{1}{3} \DN^2 \vp^I +
	\left( 1 - \frac{\epsilon}{3} \right) \DN \vp^I = u^I ,
	\label{eq:trajectory-eq}
\end{equation}
where $u_I = - V_{,I} / 3H^2$. (Recall that $N$ in the derivative $\DN$ is not a field-space index, but the number of efolds.) Under the slow-roll approximation the acceleration term $\DN^2 \vp^I$ is negligible along each trajectory. In flat field-space this means that the change in acceleration term between neighbouring trajectories also contributes at higher-order in slow-roll. In curved field-space this is no longer true because derivatives do not commute. Therefore we must retain the acceleration term when studying how trajectories disperse.

The evolution of $Q^I$ can be determined by making a Taylor expansion of eq.~\eqref{eq:trajectory-eq} along a geodesic connecting the adjacent trajectories, as in \S\ref{sec:covariant-perts}. To describe evolution of the two and three-point functions we require this expansion up to second-order. Dropping the explicit $\Or(\epsilon)$ term, which can contribute only at higher order in the slow-roll expansion, and discarding a common factor of the parallel propagator we find
\begin{equation}
	\left( \Dl + \frac{1}{2} \Dl^2 \right)
	\left(
		\frac{1}{3}	\DN^2 \vp^I
		+
		\DN \vp^I
	\right) \bigg|_{\lambda=0}
	=
	{u^I}_{;J} Q^J
	+ \frac{1}{2} {u^I}_{;JK} Q^J Q^K .
	\label{eq:taylor-jacobi}
\end{equation}
Performing the covariant expansion using the results developed in \S\ref{sec:matter-perturbations} one finds
\begin{align}
\DN Q^I + \frac{1}{3} \DN^2 Q^I &=
\Bigg(
u^I_{;J}
+\frac{1}{3} {R^I}_{LMJ} \frac{\dot \phi^L}{H}\frac{\dot \phi^M}{H} \Bigg) Q^J
\nonumber \\ &
+ \Bigg( \frac{1}{2} u^I_{;JK}
-\frac{1}{2} {R^I}_{JKL} \frac{\dot \phi^L}{H} 
-\frac{1}{6} \left( {R^I}_{JKL;M}-{R^I}_{LMJ;K}\right) \frac{\dot \phi^L}{H}\frac{\dot \phi^M}{H} \Bigg) Q^J Q^K \nonumber \\
&-\frac{2}{3} {R^I}_{JKL} \frac{\dot \phi^L}{H} \DN Q^J Q^K ,
\label{eq:jacobi_exp_1}
\end{align}
where we have applied the Bianchi identity ${R^I}_{[JKL]}=0$ to simplify the final term. 

To ensure that the Jacobi equation is consistent with the bispectrum as calculated in eq.~\eqref{eq:bispectrum}, it is necessary to ensure that all terms in eq.~\eqref{eq:jacobi_exp_1} have the correct symmetry properties. We note that in the canonical framework, such symmetry considerations are irrelevant since the Jacobi equation is simply an expansion in terms of symmetric partial derivatives. One bispectrum term that requires symmetrization, for example, is to rewrite $R_{IJKL;M}$ in terms of $R_{(I|LM|J;K)}$. The curvature tensor terms in eq.~\eqref{eq:jacobi_exp_1} are symmetrised using the Bianchi identities, allowing us to find
\be
\bigg( R_{I(JK)L;M} - R_{ILM(J;K)}\bigg)\dot \phi^L \dot \phi^M = 
\bigg( \frac{4}{3} R_{I(JK)L;M} - R_{(I|LM|J;K)}\bigg)\dot \phi^L \dot \phi^M.
\ee
Similarly, we need to symmetrise the $u$-tensors. $u_{I;J}$ is symmetric, whereas $u_{I;JK}$ may be found as
\be
u_{I;JK} = u_{(I;JK)} - \frac{1}{3} R_{I(JK)L} \frac{\dot \phi^L}{H}
\ee
at lowest order in slow-roll. Combining these ingredients one finds
\begin{align}
\left( 1 + \frac{1}{3} \DN \right) \DN Q_I &=
\Bigg(
u_{(I;J)} +\frac{1}{3} R_{ILMJ} \frac{\dot \phi^L}{H}\frac{\dot \phi^M}{H} \Bigg) Q^J
\nonumber \\ &
+ \frac{1}{2} \Bigg( 
u_{(I;JK)} -\frac{4}{3} R_{IJKL} \frac{\dot \phi^L}{H} 
+\frac{1}{3} R_{(I|LM|J;K)} \frac{\dot \phi^L}{H}\frac{\dot \phi^M}{H} 
\Bigg) Q^J Q^K \nonumber \\
&-\frac{2}{3} R_{IJKL} \frac{\dot \phi^L}{H} \DN Q^J Q^K 
-\frac{2}{9} R_{I(JK)L;M}\frac{\dot \phi^L}{H}\frac{\dot \phi^M}{H} Q^J Q^K .
\label{eq:jacobi_exp_2}
\end{align}
The final step is to eliminate the factor of $(1 + \DN/3)$ from the \lhs~of eq.~\eqref{eq:jacobi_exp_2}. We can do this by multiplying both sides by $(1 - \DN/3)$ and then taking the leading order slow-roll contribution. Working to $\Or(\dot{\phi}/H)^2$ in the terms involving the curvature tensor and $\Or(\dot{\phi}/H)^3$ in other terms, we find that the final term in eq.~\eqref{eq:jacobi_exp_2} cancels. The penultimate term in eq.~\eqref{eq:jacobi_exp_2} is neglected at the order of our calculation.

We thus conclude that $Q^I$ evolves according to the {\it Jacobi equation}
\begin{equation}
	\label{eq:jacobi-eq}
	\DN Q^I
	= {w^I}_J Q^J + \frac{1}{2} {w^I}_{(JK)} Q^J Q^K + \cdots ,
\end{equation}
where the coefficients ${w^I}_J$ and ${w^I}_{(JK)}$ satisfy
\begin{align}
	\label{eq:jacobi-w2}
	w_{IJ} & = u_{(I;J)} 
	+ \frac{1}{3} R_{L(IJ)M}
	\frac{\dot{\phi}^L}{H}
	\frac{\dot{\phi}^M}{H} , \\
	\label{eq:jacobi-w3}
	w_{I(JK)} & = u_{(I;JK)}
		+ \frac{1}{3} \bigg(
			R_{(I|LM|J;K)} \frac{\dot{\phi}^L}{H} \frac{\dot{\phi}^M}{H}
			- 4 R_{I(JK)L} \frac{\dot{\phi}^L}{H}
		\bigg) .
\end{align}
It is possible to extend this analysis to non-slow-roll scenarios but we shall confine ourselves to the slow-roll limit. As usual, the background trajectory is denoted by $\phi^I(t)$. All curvature quantities and derivatives of $u^I$ are evaluated on this trajectory and therefore powers of slow-roll can be counted in the usual way. Because we have used the slow-roll approximation, eqs.~\eqref{eq:jacobi-w2}--\eqref{eq:jacobi-w3} are trustworthy only to lowest order in slow-roll in respect of derivatives of $u^I$, and to $\Or(\dot{\phi}/H)^2$ in terms multiplying the Riemann tensor and its derivatives. This accuracy is sufficient to make a comparison with the divergent terms isolated in~\S\ref{sec:isolatedgrowingmodes}.

Although both terms in eq.~\eqref{eq:jacobi-w2} are automatically symmetric under exchange of $IJ$, we have indicated this explicitly. However, $w_{IJK}$ is symmetric only on $JK$. This is different to the case of flat field-space, where terms involving the Riemann tensor are absent and each coefficient on the right-hand side of the Jacobi equation is always a symmetric combination of partial derivatives. When writing ${w^I}_{(JK)}$ we add brackets to emphasize this symmetry.

\ssec{Time-evolution operators}

The Jacobi equation~\eqref{eq:jacobi-eq} is a first-order differential equation, and therefore its solution can be expanded in powers of the initial conditions $Q^i_*$,%
	\footnote{The quantities ${\bigamma^I}_m$ and ${\bigamma^I}_{mn}$ were written	${\Gamma^I}_m$ and ${\Gamma^I}_{mn}$ in refs.~\cite{Seery:2012vj,Anderson:2012em}. In this paper we reserve $\Gamma$ to mean the Levi-Civita	connection compatible with the field-space metric
	$G^{IJ}$.}
\begin{equation}
	Q^I = {\bigamma^I}_i Q^i_*
		+ \frac{1}{2} {\bigamma^I}_{(ij)} Q^i_* Q^j_* + \cdots .
	\label{eq:jacobi-soln}
\end{equation}
To write eq.~\eqref{eq:jacobi-soln} we have used the index convention introduced in~\S\ref{sec:third-order}. The fluctuation $Q^I$ is evaluated at some late time $N$, and its index $I$ transforms as a contravariant vector in the tangent space at this time. Conversely, $Q^i_*$ and its index $i$ transform as a contravariant vector at an earlier time $N_*$. Like the trajectory propagator~\eqref{eq:pi}, the coefficients ${\bigamma^I}_i$ and ${\bigamma^I}_{(ij)}$ are bitensors. The initial conditions require ${\bigamma^I}_i = \delta^I_i$ and ${\bigamma^I}_{(ij)} = 0$ when $N = N_\ast$.

Eq.~\eqref{eq:jacobi-soln} solves the Jacobi equation~\eqref{eq:jacobi-eq} provided the $\bigamma$ coefficients satisfy
\begin{align}
	\DN {\bigamma^I}_i & =
		{w^I}_J {\bigamma^J}_i ,
	\label{eq:gamma2-eq}
	\\
	\DN {\bigamma^I}_{(ij)}
	& = {w^I}_J {\bigamma^J}_{(ij)}
	+ {w^I}_{(JK)} {\bigamma^J}_i {\bigamma^K}_j .
	\label{eq:gamma3-eq}
\end{align}
We describe the coefficients ${\bigamma^I}_{(i \cdots j)}$ collectively as {\it time-evolution operators}. They are covariant analogues of the coefficients $\partial \phi^I / \partial \phi^i_*$ and its higher derivatives which occur when applying the separate-universe method in flat field-space~\cite{DeWittMorette:1976up,Lewandowski:1993zq}. These could be obtained by solving for $\vp^I$ using eq.~\eqref{eq:Q_expansion} and then computing its derivatives with respect to the initial conditions, but in practice it is much easier to integrate eqs.~\eqref{eq:gamma2-eq}--\eqref{eq:gamma3-eq} directly.

\ssec{Matching to subhorizon divergences}
\label{sec:matching}

Terms in $\langle Q^I_{\vect{k}_1} Q^J_{\vect{k}_2} Q^K_{\vect{k}_3} \rangle$  involving $N$ are divergent in the late-time limit $\tau \rightarrow 0$ and are responsible for spoiling infrared safety, as described in~\S\ref{sec:ir-safety}. They generate time evolution after horizon exit~\cite{Zaldarriaga:2003my,Seery:2007wf,Seery:2010kh} and rapidly invalidate the expressions derived in chapter~\ref{ch:subhorizon}. It is to properly account for this superhorizon evolution that led us to develop the covariant transport formalism.

In this section we perform an important cross-check: We show that the divergent Riemann curvature terms in eq.~\eqref{eq:bispectrum} are correctly matched by the covariant transport method derived above. This also provides new intuition about the geometrical origin of the curvature-mediated divergences arising in the in--in calculation. 

In \S\ref{sec:isolatedgrowingmodes} we isolated the terms diverging linearly and quadratically with $N$. We now show that eqs.~\eqref{eq:jacobi-eq}--\eqref{eq:jacobi-w3} reproduce these divergences. The argument is similar to that of Zaldarriaga~\cite{Zaldarriaga:2003my}.  Solving eqs.~\eqref{eq:gamma2-eq}--\eqref{eq:gamma3-eq} perturbatively yields a power series in $N$. The lowest-order terms are
\begin{align}
	\label{eq:gamma2-soln}
	{\bigamma^I}_i & =
		{\trajp^I}_i
		+ {\trajp^I}_j \Big[{w^j}_i \Big]_\ast N
		+ \frac{1}{2} {\trajp^I}_j
			\Big[{w^j}_k {w^k}_i + \DN {w^j}_i \Big]_\ast N^2
		+ \cdots , \\
		\label{eq:gamma3-soln}
	{\bigamma^I}_{(ij)} & =
		{\trajp^I}_k \Big[ {w^k}_{(ij)} \Big]_* N
		\\ & \qquad \nonumber 
		+ \frac{1}{2} {\trajp^I}_k \Big[
			\DN {w^k}_{(ij)}
			+ {w^k}_l {w^l}_{(ij)}		
			+ {w^k}_{(il)} {w^l}_j
			+ {w^k}_{(jl)} {w^l}_i
		\Big]_\ast N^2
		+ \cdots ,
\end{align}
where $N = -\ln|k_\ast \tau|$.

Eq.~\eqref{eq:gamma2-soln} shows that the time-evolution operator ${\bigamma^I}_i$ can be understood as a modification of the trajectory propagator to include the effect of time-dependence along the inflationary trajectory in addition to parallel transport. This follows because the trajectory propagator ${\trajp^I}_i$ satisfies~\eqref{eq:gamma2-eq} with ${w^I}_J = 0$.

At linear order in $N$,
the two and three-point functions following
from eqs.~\eqref{eq:gamma2-soln}--\eqref{eq:gamma3-soln} are
\begin{align}
	\label{eq:divergent-2pf}
	\langle Q^I_{\vect{k}_1} Q^J_{\vect{k}_2} \rangle
	& \supseteq
	(2\pi)^3 \delta(\vect{k}_1 + \vect{k}_2)
	\frac{N H_*^2}{k^3}
	{\trajp^I}_{i}
	{\trajp^J}_{j}
	{w}^{ij}_\ast , \\
	\label{eq:divergent-3pf}
	\langle Q^I_{\vect{k}_1} Q^J_{\vect{k}_2} Q^K_{\vect{k}_3} \rangle
	& \supseteq
	(2\pi)^3 \delta(\vect{k}_1 + \vect{k}_2 + \vect{k}_3)
	\frac{N H_*^4}{4\prod_i k_i^3} 
		{\trajp^I}_i {\trajp^J}_j {\trajp^K}_k
		w^{i(jk)}_\ast k_1^3
		+ \text{cyclic} .
\end{align}
One may then employ eqs.~\eqref{eq:gamma2-eq} and~\eqref{eq:gamma3-eq} to substitute for $w^{ij}$ and $w^{i(jk)}$. 

At quadratic order in $N$ and lowest order in slow-roll one finds a single contribution arising from the $\DN {w^k}_{(ij)}$ term in eq.~\eqref{eq:gamma3-soln}. This gives
\begin{equation}
\label{eq:double-divergent-3pf}
\begin{split}
	\langle Q^I_{\vect{k}_1} Q^J_{\vect{k}_2} Q^K_{\vect{k}_3} \rangle
	\supseteq \mbox{}
	&
	(2\pi)^3 \delta(\vect{k}_1 + \vect{k}_2 + \vect{k}_3)
	\frac{N^2 H_\ast^4}{4\prod_i k_i^3}
	\\ & \mbox{} \times
		{\trajp^I}_i {\trajp^J}_j {\trajp^K}_k
		\bigg(
			-\frac{4}{6} R^{i(jk)l;m}
			\frac{\dot{\phi}_l}{H}
			\frac{\dot{\phi}_m}{H}
		\bigg)_* k_1^3
		+ \text{cyclic} .
\end{split}
\end{equation}

It can be checked that eq.~\eqref{eq:divergent-2pf} reproduces the divergence in the two-point function (including the term involving the Riemann tensor) found by Nakamura and Stewart~\cite{Nakamura:1996da}.%
	\footnote{In ref.~\cite{Nakamura:1996da} the factors of ${\trajp^I}_i$ were omitted.}
Comparing eqs.~\eqref{eq:divergent-3pf} and~\eqref{eq:jacobi-w3}, it can also be checked that the terms in $w^{i(jk)}$ involving the Riemann tensor reproduce the 1-$\log$ divergences in eq.~\eqref{eq:bispectrum-single-log}. It was to enable a non-trivial check of this matching that we elected to keep divergences up to $\O(\dot{\phi}/H)^2$ in the Riemann-tensor terms of eq.~\eqref{eq:s3sr}. Finally, comparing eqs.~\eqref{eq:double-divergent-3pf} and~\eqref{eq:bispectrum-double-log} it can be checked that the lowest-order double-logarithmic divergence is also correctly reproduced. At the accuracy of our present calculation it is not possible to check whether the divergences proportional to $u_{(i;jk)}$ also agree. Higher-order terms in $N$ and $\dot{\phi}/H$ could also be retained in the perturbative expansions~\eqref{eq:gamma2-soln}--\eqref{eq:gamma3-soln}, which would enable a check of matching at all orders. This has subsequently been verified to all orders in ref.~\cite{Dias:2012qy}.

\ssec{Transport equations}
\label{sec:transport-equations}

The time-evolution operators enable us to determine each $n$-point function after horizon exit. Translating the formulae of Lyth~and~Rodr\'{\i}guez~\cite{Lyth:2005fi} using eq.~\eqref{eq:jacobi-soln} we obtain
\begin{equation}
	\langle
		Q^I_{\vect{k}_1}
		Q^J_{\vect{k}_2}
	\rangle
	=
		{\bigamma^I}_i
		{\bigamma^J}_j
		\langle
			Q^i_{\vect{k}_1}
			Q^j_{\vect{k}_2}
		\rangle_*
	\label{eq:sep-univ-2pf}
\end{equation}
and
\begin{equation}
\begin{split}
	\langle
		Q^I_{\vect{k}_1}
		&
		Q^J_{\vect{k}_2}
		Q^K_{\vect{k}_3}
	\rangle
	=
		{\bigamma^I}_i
		{\bigamma^J}_j
		{\bigamma^K}_k
		\langle
			Q^i_{\vect{k}_1}
			Q^j_{\vect{k}_2}
			Q^k_{\vect{k}_3}
		\rangle_*
	\\
	& \mbox{}
		+
		{\bigamma^I}_{(ij)}
		{\bigamma^K}_{k}
		{\bigamma^J}_{l}
		\int \frac{\d^3 q}{(2\pi)^3}
		\langle
			Q^i_{\vect{k}_1 - \vect{q}}
			Q^k_{\vect{k}_2}
		\rangle_*
		\langle
			Q^j_{\vect{q}}
			Q^l_{\vect{k}_3}
		\rangle_*
		+
		\text{cyclic} ,
\end{split}
\label{eq:sep-univ-3pf}
\end{equation}
where `cyclic' denotes the two permutations of the second line in~\eqref{eq:sep-univ-3pf} obtained by exchanging $\{ I, \vect{k}_1 \}$, $\{ J, \vect{k}_2 \}$ and $\{ K, \vect{k}_3 \}$. When there is no time evolution (i.e. ${w^I}_J = 0$), eq.~\eqref{eq:sep-univ-3pf} reproduces the horizon crossing result of eq.~\eqref{eq:field_3pf}.

Up to this point we have worked in a frame derived from the field-space coordinates, but other possibilities exist. Since an $n$-point function of the $Q^m_*$ transforms as a rank-$n$ tensor in the tangent-space at time $N_\ast$, eqs.~\eqref{eq:sep-univ-2pf}--\eqref{eq:sep-univ-3pf} are manifestly covariant. As a result, we are free to select a basis for the tangent space independently at the early and late times $N_\ast$ and $N$. 

The approach given above is simple and emphasizes its similarity with familiar $\delta N$ methods, but it is also possible to write transport equations for the $n$-point functions. We write the two-point function as
\begin{equation}
	\langle
		Q^I_{\vect{k}_1}
		Q^J_{\vect{k}_2}
	\rangle
	=
	(2\pi)^3
	\delta(\vect{k}_1 + \vect{k}_2)
	\frac{\Sigma^{IJ}(k)}{2k^3} ,
\end{equation}
where $\Sigma^{IJ}(k)$ is symmetric and weakly scale-dependent. The amplitude of the local mode of the three-point function can be parametrized
\begin{equation}
	\langle
		Q^I_{\vect{k}_1}
		Q^J_{\vect{k}_2}
		Q^K_{\vect{k}_3}
	\rangle
	=
	(2\pi)^3
	\delta(\vect{k}_1 + \vect{k}_2 + \vect{k}_3)
	\left[
		\frac{\alpha^{I(JK)}}{k_2^3 k_3^3}
		+
		\frac{\alpha^{J(IK)}}{k_1^3 k_3^3}
		+
		\frac{\alpha^{K(IJ)}}{k_1^3 k_2^3}
	\right] ,
\end{equation}
where the $\alpha^{IJK}$ are formally functions of $k$ with weak scale-dependence. Direct differentiation, followed by use of the Jacobi equation~\eqref{eq:jacobi-eq}, yields 
\begin{align}
	\DN \Sigma^{IJ}
	& =
	{w^I}_L \Sigma^{LJ} + {w^J}_L \Sigma^{LI} + \cdots ,
	\label{eq:2pf-transport}
	\\
	\DN \alpha^{I(JK)}
	& =
	{w^I}_L \alpha^{L(JK)}
	+ {w^J}_L \alpha^{I(LK)}
	+ {w^K}_L \alpha^{I(JL)}
	+ {w^I}_{(LM)} \Sigma^{LJ} \Sigma^{MK}
	+ \cdots , \label{eq:3pf-transport}
\end{align}
where the omitted terms involve higher-order correlation functions and are negligible in typical inflationary theories. Following the method described in ref.~\cite{Seery:2012vj} it can be verified that eqs.~\eqref{eq:2pf-transport}--\eqref{eq:3pf-transport} reproduce eqs.~\eqref{eq:sep-univ-2pf}--\eqref{eq:sep-univ-3pf}.

\sssec{Interpretation of Riemann terms}
\label{sec:riemann-terms}

It is now possible to understand the significance of those interactions in $\langle Q^I Q^J Q^K \rangle$, as calculated in eq.~\eqref{eq:bispectrum}, which are mediated by the Riemann curvature. The formalism derived in this section makes it clear that this effect in field space is entirely analogous to geodesic deviation between freely-falling observers in curved spacetime. These new sources of time-dependence arise mathematically from tidal effects in field space due to the non-vanishing field space curvature and their physical meaning can be understood as follows: An initial perturbation $Q^i_*$ generically represents a mix of adiabatic and isocurvature fluctuations. The isocurvature fluctuations differentiate between `separate universes', and correspond to a choice of inflationary trajectory measured from the fiducial trajectory at $Q^i = 0$. As these trajectories flow over field-space their separation will be subject to the effect of geodesic deviation which will generate a new mechanism by which the bundle may dilate and/or shear. This may cause a non-linear redistribution of power between adjacent trajectories which can generate non-Gaussianity. 

The Riemann tensor is antisymmetric on its first and second pairs of indices. Since the field velocity is proportional to the bundle tangent vector, we conclude that the Riemann contribution to ${w^I}_J$ is zero when either index is aligned with the adiabatic direction. Gong~and~Tanaka emphasized that this leads only to new couplings between isocurvature modes~\cite{Gong:2011uw}. Eq.~\eqref{eq:2pf-transport} shows that these couplings influence how the isocurvature modes share power between themselves, but do not cause power to flow between the isocurvature and adiabatic directions. Such a flow must be mediated by the potential through $u_{(I;J)}$.

In the special case where the trajectory follows an exact geodesic, its tangent vector is parallel-transported proportional to itself. In this case, the adiabatic mode decouples completely and no power flows to or from it.

\ssec{Gauge transformation to the curvature perturbation}
\label{sec:gaugetransformation}

We have now developed a formalism for computing the superhorizon evolution of the correlators of the field perturbations during slow-roll inflation. For comparison with microwave background observations or galaxy surveys we must compute the $n$-point functions of the primordial curvature perturbation, $\zeta$. This entails a gauge transformation from $Q^I$ to the curvature perturbation $\zeta$ and in curved field-space this can be performed economically using a covariant extension of the method introduced in ref.~\cite{Anderson:2012em}.

We expand $N$ as a function of the density $\rho$. Taking $\Delta \rho$ to be the displacement from a point of fixed density $\rho_c$ to an arbitrary initial location, we find
\begin{equation}
	\Delta N =
		\frac{\d N}{\d \rho}
		\Delta \rho
		+
		\frac{1}{2}
		\frac{\d^2 N}{\d \rho^2}
		(\Delta \rho)^2
		+
		\cdots
		.
	\label{eq:primitive-deltaN}
\end{equation}
To determine the variation of eq.~\eqref{eq:primitive-deltaN} under a change in the initial location we expand along a geodesic, as in~\S\ref{sec:covariant-perts}, along which both $\Delta\rho$ and the differential coefficients will vary. The variation of $\Delta\rho$ satisfies
\begin{equation}
	\delta(\Delta \rho) =
		- V_{;I} Q^I - \frac{1}{2} V_{;IJ} Q^I Q^J + \cdots .
\end{equation}
Therefore, up to second order, we can express $\zeta$ as
\begin{equation}
	\zeta
		= \delta(\Delta N)
		= N_I Q^I + \frac{1}{2} N_{IJ} Q^I Q^J + \cdots .
	\label{eq:zeta-def}
\end{equation}
The coefficients $N_I$ and $N_{IJ}$ are distinct from the $\delta N$ coefficients by the lack of a comma or semicolon. $N_I$ and $N_{IJ}$ satisfy
\begin{align}
	\label{eq:N1-gauge}
	N_I & = - \frac{\d N}{\d \rho} V_{;I} , \\
	\label{eq:N2-gauge}
	N_{IJ} & = - \frac{\d N}{\d \rho} V_{;IJ}
		+ \frac{\d^2 N}{\d\rho^2} V_{;I} V_{;J}
		+ \frac{1}{\Mpl^2} \big(
			A_{I} V_{;J} + A_{J} V_{;I}
		\big) ,
\end{align}
where 
\begin{align}
	A_I & =
		\frac{V_{;I}}{V^{;J}V_{;J}}
		- \frac{2V}{(V^{;J}V_{;J})^2} V^{;K}V_{;IK} , \\
	\frac{\d N}{\d \rho} &=
		- \frac{1}{\Mpl^2} \frac{V}{V^{;I}V_{;I}} , \\
	\frac{\d^2 N}{\d \rho^2} &=
		- \frac{1}{\Mpl^2} \frac{1}{V^{;I}V_{;I}}
		+ \frac{2}{\Mpl^2} \frac{V}{(V^{;I}V_{;I})^3} V^{;J}V^{;K}V_{;JK} .
\end{align}

Eqs.~\eqref{eq:N1-gauge}--\eqref{eq:N2-gauge} are defined at a single point
in field space; they are not bilocal in the sense of the coefficients ${\bigamma^I}_m$ and ${\bigamma^I}_{(mn)}$. We can obtain analogues of these bilocal coefficients using the time evolution operators to relate the $Q^I$ to their values at horizon crossing. This yields eq.~\eqref{eq:deltaN} as
\begin{equation}
	\zeta(N)
	=
		N_{,i} Q^i_\ast
		+
		\frac{1}{2} N_{;ij} Q^i_\ast Q^j_\ast
		+
		\cdots ,
	\label{eq:covariant-deltaN}
\end{equation}
where $N_{,i}$ and $N_{;ij}$ may be calculated as
\begin{align}
	\label{eq:N1}
	N_{,i} & = N_I {\bigamma^I}_i , \\
	\label{eq:N2}
	N_{;ij} & = N_I {\bigamma^I}_{(ij)} +
		N_I N_J {\bigamma^I}_i {\bigamma^J}_j .
\end{align}
Our procedure, culminating in eq.~\eqref{eq:covariant-deltaN}, therefore agrees with the
covariant $\delta N$ expansion discussed by Saffin~\cite{Saffin:2012et}. However, we have been the first to provide a procedure for computing the covariant $\delta N$ coefficients. 

\ssec{Correlators of $\zeta$}
\label{sec:zetacorrelators}

The expansion \eqref{eq:covariant-deltaN} allows us to write the correlators of $\zeta$ in terms of the correlators of field perturbations at horizon exit. We now have all of the necessary formalism with which to make predictions for inflationary observables.

\paragraph{Power spectrum.} The power spectrum follows simply from the two-point correlator of $\zeta_{\vect{k}}$
\be
	\langle
		\zeta_{\vect{k}_1}
		\zeta_{\vect{k}_2}
	\rangle
	=
		N_{,i} N_{,j}
		\langle
			Q^i_{\vect{k}_1}
			Q^j_{\vect{k}_2}
		\rangle_* .
	\label{eq:zeta-2pf}
\ee
Substituting the two-point field correlator $\langle Q^i_{\vect{k}_1} Q^j_{\vect{k}_2} \rangle_*$ using eq.~\eqref{eq:power-spectrum}, the power spectum $\P_\zeta(k)$ follows from eq.~\eqref{eq:powerspectrum} as
\be
\P_\zeta(k) = \frac{k^3}{2\pi^2} P_\zeta(k)
	= N_{,i} N_{,j} G^{ij} \left( \frac{H^*}{2\pi} \right)^2 ,
	\label{eq:Pzeta}
\ee
where the label `$*$' on \rhs~denotes evaluation at the horizon crossing time for the mode $k$. 

\paragraph{Scalar spectral index.} The scalar spectral index defined in eq.~\eqref{eq:nsdef} follow directly from differentiating the power spectrum $\P_\zeta(k)$ as given in eq.~\eqref{eq:Pzeta}. Using the chain rule one finds
\be
n_\zeta-1 = \frac{\d \ln \P_\zeta}{\d N} \times 
\frac{\d N}{\d \ln k}.
\label{eq:nsline1}
\ee
The latter term in eq.~\eqref{eq:nsline1} may be found given we are working about the time of horizon exit of the mode $k_*$ such that $k_*=aH$. One finds
\be
\frac{\d N}{\d \ln k} = \frac{1}{1-\epH^*}.
\ee
The former term in eq.~\eqref{eq:nsline1} is calculated by taking $\d / \d N$ of the expression for $\P_\zeta$. Noting that $\P_\zeta$ is itself a field-space scalar, we have $\d \P_\zeta / \d N = \D_N \P_\zeta$. The covariant derivative of eq.~\eqref{eq:Pzeta} is then simply found and putting the pieces together to find
\be
n_\zeta-1 = \left[ 
-2 \epH^* + \frac{2}{H_*} \frac{N_{,ik}N_{,j} G^{ij} \dot \phi_*^k}{N_{,l}N_{,m}G^{lm}}
\right].
\ee
This result is valid to $\O(\dot \phi^I/H)^2$.

\paragraph{Bispectrum.} The three point correlator of $\zeta$ may be expanded as
\be
\begin{split}
	\langle
		\zeta_{\vect{k}_1}
		\zeta_{\vect{k}_2}
		\zeta_{\vect{k}_3}
	\rangle
	&=
		N_{,i} N_{,j} N_{,k}
		\langle
			Q^i_{\vect{k}_1}
			Q^j_{\vect{k}_2}
			Q^k_{\vect{k}_3}
		\rangle_* 
	\\
	& \qquad +
		N_{;ij} N_{,k} N_{,l}
		\int \frac{\d^3 q}{(2\pi)^3}
		\langle
			Q^i_{\vect{k}_1 - \vect{q}}
			Q^k_{\vect{k}_2}
		\rangle_*
		\langle
			Q^j_{\vect{q}}
			Q^l_{\vect{k}_3}
		\rangle_*
		+ \text{cyclic} ,
	\label{eq:zeta-3pf}
\end{split}
\ee
where `cyclic' indicates the usual combination of permutations, as in eq.~\eqref{eq:3ptcorrelator3}. In \S\ref{sec:bispectrum} we described how it is useful for the bispectrum to be written in terms of the dimensionless parameter $\fnl^{\rm local}$, provided that the dominant bispectrum contribution arises from the non-linear interaction of superhorizon growing modes. For this to be valid we require that the other bispectrum contributions arising from the frozen modes in eq.~\eqref{eq:bispectrum} are negligible. Depending on the field-space curvature it is possible that `non-local' contributions in eq.~\eqref{eq:bispectrum} could be enhanced, but to determine whether this happens would require an extension of the analysis in refs.~\cite{Lyth:2005qj,Vernizzi:2006ve}. On the other hand, the field metric curvature certainly modifies the evolution of the amplitude of the local shape and so it is possible that the dominant bispectrum signal may be local regardless. In this case one is able to neglect the intrinsic non-Gaussianity of the $Q^I_*$, which amounts to ignoring the first term on the \rhs~of eq.~\eqref{eq:zeta-3pf}. One then obtains a covariant analogue of the familiar $\delta N$ formula for the amplitude of the local bispectrum,
\be
	\frac{6}{5} \fnl^{\rm local} \approx
		\frac{N_{,i} N_{,j} N^{;ij}}{(N_{,k} N^{,k})^2}.
	\label{eq:fnl}
\ee
Since this formula assumes that the bispectrum generated at horizon exit is negligible then subsequent time evolution is necessary to generate an observable non-Gaussian signal.

\paragraph{Tensor--scalar ratio.} Given the expression for $\P_\zeta(k)$ in eq.~\eqref{eq:Pzeta}, and the result for $\P_t(k)$ derived in eq.~\eqref{eq:Ptdimless}, the tensor--scalar ratio follows immediately as
\be
r = \frac{8}{\Mpl^2 N_{,i} N_{,j} G^{ij}}.
\ee

\paragraph{Tensor spectral index.} We may equally derive a formula for the scale-dependence of the tensor spectrum $n_t$ as defined in eq.~\eqref{eq:ntdef}. This is considerably simpler than the scalar spectral index since it does not evolve after horizon exit. At leading order one finds
\be
n_t = -2 \epH^*.
\ee

\sec{Analytic superhorizon evolution}
\label{sec:analytics}

\ssec{Orientation}

So far we have described how the superhorizon evolution of perturbations in multi-field inflation could be computed within the $\delta N$ or transport formalisms. The non-covariant $\delta N$ coefficients $N_{,I} ,N_{,IJ} , \cdots$ may be found numerically via the finite difference method discussed in \S\ref{sec:computation_techniques}. The transport method necessitates evolution of eqs.~\eqref{eq:2pf-transport} and \eqref{eq:3pf-transport} to evolve the power spectrum and bispectrum respectively. These procedures are suitable for numerical computations but are less applicable towards our goal of generating intuition about how the dynamics of inflation affect the evolution of cosmic observables. One expects this type of intuition to arise naturally from analytic solutions of the superhorizon perturbation evolution, where such solutions exist.

The most direct method of obtaining analytic solutions for the evolution of perturbations $Q^I$ would be to integrate the Jacobi equation~\eqref{eq:jacobi-w2}. In practice it is considerably easier to compute the $\delta N$ coefficients directly, which shall be the focus of this section. It is not known how to derive analytic solutions in the presence of a general metric $G_{IJ}$ and so we henceforth specialise to the canonical multi-field scenario with $G_{IJ} = \delta_{IJ}$. Since we are assuming a canonical field metric in this section, we leave all field indices covariant and capitalized. 

Whilst the following results enable the computation of the superhorizon evolution of a range of cosmic observables, including the power spectrum $\P_\zeta$ and the spectral index $n_\zeta$, we choose to focus on non-Gaussianity both because of its known sensitivity to superhorizon dynamics and also because of the possibility of its imminent detection. The presence of multiple dynamically relevant fields is not covered by Maldacena's theorem~\cite{Maldacena:2002vr} and large non-Gaussianity have been shown to develop in a number of inflationary models due to the non-linear interaction of growing modes in the superhorizon epoch~\cite{Rigopoulos:2005us,Vernizzi:2006ve,Alabidi:2006hg,Byrnes:2008wi}. As discussed in chapter~\ref{ch:formalisms}, only the `local' shape non-Gaussianity is observable in this scenario and so we focus on this exclusively for the remainder of this chapter. To avoid redundant notation, we shall drop the label `local'. 

We do not consider alternative scenarios which may give rise to a large non-Gaussianity after horizon exit, such as the curvaton mechanism \cite{Lyth:2001nq,Moroi:2001ct} or modulated reheating \cite{Dvali:2003em,Zaldarriaga:2003my}, which rely on an inflationary seed perturbation which is then amplified by a non-inflationary mechanism.

\sssec{$\delta N$ expressions for observable parameters}

Before launching into the derivation of the $\delta N$ coefficients, we need to know how these lead to expressions for observable parameters. The power spectrum $\P_\zeta$, scalar spectral index $n_\zeta$, tensor--scalar ratio $r$, tensor spectral index $n_t$ and the bispectrum parameter $\fnl$ were derived in \S\ref{sec:zetacorrelators}. The canonical analogues of these expressions are simply found and two of these are written below. In addition to these results we now include formulae for the local-shape trispectrum parameters $\tnl$ and $\gnl$~\cite{Byrnes:2006vq,Seery:2006js}. Together, the results that we shall require for this section are
\begin{align}
\label{eq:dN_sindex}
n_\zeta -1 &= \frac{2 \dot{\phi}^*_I}{H^*}\frac{ N_{,IJ} N_{,J}}{N_{,K} N_{,K}}
-2 \epH^* , \\
\label{eq:dN_fnl}
\frac{6}{5} \fnl &= \frac{ N_{,I} N_{,J} N_{,IJ}}{\left(N_{,K} N_{,K} \right) ^2} , \\
\label{eq:dN_tnl}
\tnl &= \frac{ N_{,IJ} N_{,IK} N_{,J} N_{,K} }{\left(  N_{,L}N_{,L} \right) ^3} , \\
\label{eq:dN_gnl}
\frac{54}{25} \gnl &= \frac{ N_{,IJK} N_{,I} N_{,J} N_{,K} }{\left(  N_{,L} N_{,L} \right) ^3} ,
\end{align}
where we have assumed the field fluctuations to be approximately Gaussian at horizon crossing, which we recall is an excellent approximation for canonical fields \cite{Seery:2005gb,Seery:2006vu,Seery:2008ax}.

We will generate expressions for the $\delta N$ coefficients in terms of the potential and its derivatives. These quantities may then be written in terms of the multi-field potential slow-roll parameters 
\be
\bal{3}
\epsilon_I &= \ds{\frac{\Mpl^2}{2} \frac{V_{,I}^2}{V^2} } \,, \qquad &\epsilon &= \sum_I \ep_I , \vspace{2mm}\\
\eta_{IJ} &= \ds{{\Mpl^2} \frac{V_{,IJ}}{V}} \,, \qquad & \xi^2_{IJK} 
&= \ds{{\Mpl^3} \sqrt{2 \ep} \frac{V_{,IJK}}{V}}\,.
\eal
\label{eq:sr_parameters}
\ee
In the limit where $\ep \ll 1$ then $\ep \simeq \epH$ and the dynamics describe an accelerating cosmology. 

Since the $\delta N$ method involves comparison of background field values of different separate universes, we shall be working with $\phi_K(t)$ rather than $\vp_K(t,\vect{x})$. 

\ssec{Computing $N_{,I}$}

In \S\ref{sec:firstdeltaN} we give an alternative derivation of the linear $\delta N$ coefficients $N_{,I}$ for canonical multi-field inflation, which is fully consistent with previous results~\cite{Vernizzi:2006ve,Wang:2010si}. Rather than employing the standard analytical tricks that simplify the functional variation, we evaluate the path integrals directly. The main advantage of this method is its direct approach.

\begin{figure}[h]
\begin{center}
\includegraphics[width=0.5\textwidth]{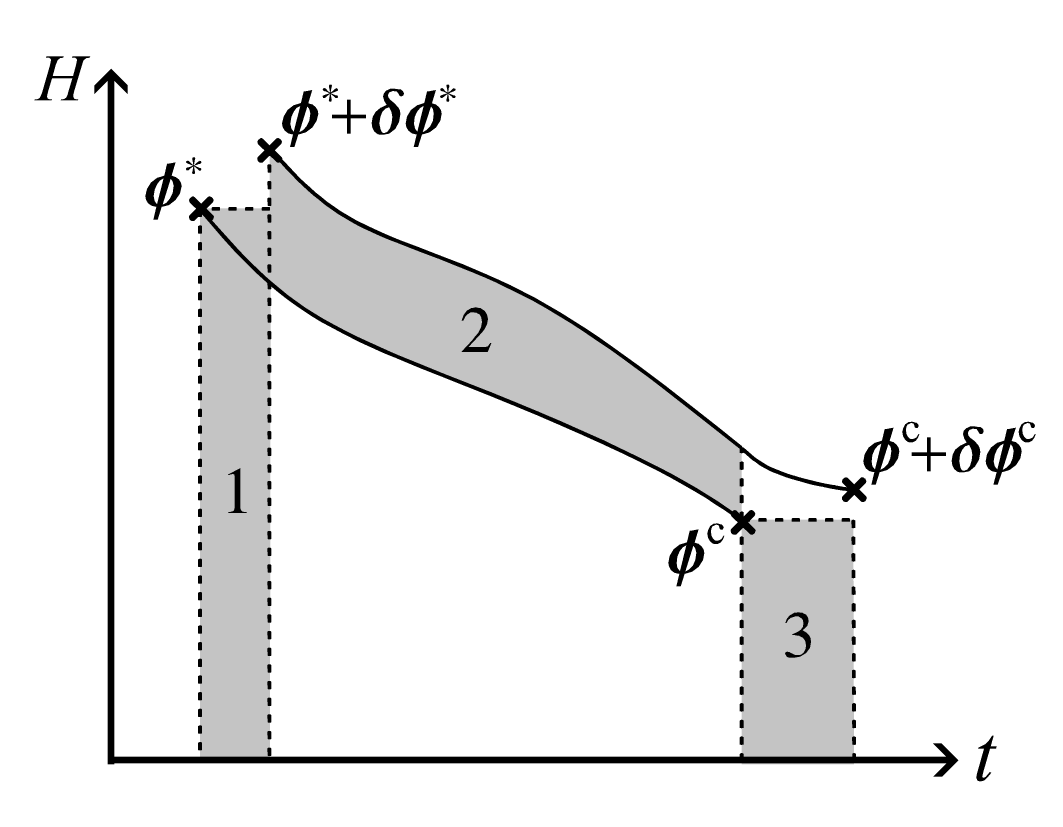}
\caption[$\delta N$ via functional variation]{
The functional integral for $N$ between horizon exit and a later-time uniform density hypersurface. In general there are three contributions, the `initial term' (region $1$) and `end term' (region $3$) arise from varying the boundaries and region $2$ is the `path term' from the variation of the intergrand. 
\label{fig:functional}}
\end{center}
\end{figure}

To obtain analytic tractability we must simplify the dynamics. One possible route that we will not consider is to take the Hubble rate to be of the separable form $H = \sum_J H_J(\phi_J)$~\cite{Battefeld:2009ym,Byrnes:2009qy}. A more popular assumption is that of {\it exact} slow-roll dynamics such that 
\be
\label{eq:exactSR}
3 H \dot \phi_I + V_{,I} = 0.
\ee
Under the additional assumption of monotonicity (chosen such that $\dot \phi_K < 0$, without loss of generality), the number of efolds $N$ can be written with the field $\phi_K$ as a time variable as
\be
N = - \int_*^c \frac{V}{V_{,K}} \, \d \phi_K, \qquad \mbox{(no sum)}.
\ee
Note that there is no summation over $K$. This integral is performed between the initial flat hypersurface `$*$' and later-time uniform density hypersurface `$c$', such that the functional variation $\delta N$ yields $\zeta$. Varying this functional with respect to the horizon exit field values will generate three terms, an {\it initial term} from varying the initial boundary, an {\it end term} from varying the final boundary and a {\it path term} from varying the integrand itself. These are shown schematically in figure~\ref{fig:functional}. The result is
\be
\label{eq:initial_expansion}
N_{,I} = \left. \frac{V}{V_{,K}} \right|_* \delta_{IK} - 
\left. \frac{V}{V_{,K}} \right|_c \frac{\partial \phi_K^c}{\partial \phi_I^*}-
\int_*^c \frac{\partial}{\partial \phi_I^*} \left( \frac{V}{V_{,K}} \right)_{\phi_K} \d \phi_K  , \qquad \mbox{(no sum)}.
\ee
where we employ the notation $\left( \right)_{\phi_K}$ to denote that the variable $\phi_k$ is being held constant under differentiation. One may be concerned about the free index `$K$' that appears only on the \rhs~of eq.~\eqref{eq:initial_expansion}. However, whilst the various terms in eq.~\eqref{eq:initial_expansion} may individually vary with $K$, their sum is necessarily invariant.

After performing the manipulations in \S\ref{sec:firstdeltaN} we show that eq.~\eqref{eq:initial_expansion} may be written as 
\begin{alignat}{2}
\label{eq:ps_Ni_hca2}
\Mpl^2 \, N_{,I} &= \frac{U_I^*}{{U_I '}^*} \delta_{IJ} - \frac{U_J^c}{{U_J'}^c} \frac{\partial \phi_J^c}{\partial \phi_I^*},\qquad & &\mbox{---}{\rm for~}V=P, \\ 
\Mpl^2 \, N_{,I} &= A\frac{U_I^*}{{U_I '}^*} - A\sum_J \frac{U_J^c}{{U_J'}^c} \frac{\partial \phi_J^c}{\partial \phi_I^*},\qquad & &\mbox{---}{\rm for~}V=S^{1/A},
\label{eq:ss_Ni_hca2}
\end{alignat}
where we have used the shorthand $U_I = U_I (\phi_I)$, $S = \sum_I U_I$ and $P = \Mpl^4 \prod_I U_I$. The derivatives appearing in these expressions are evaluated in eqs.~\eqref{eq:ps_deriv} and \eqref{eq:ss_deriv}. Higher order $\delta N$ coefficients such as $N_{,IJ}$ may be calculated by differentiation of \eqref{eq:ps_Ni_hca2} and \eqref{eq:ss_Ni_hca2}.

This method of deriving $N_{,I}$ for such separable potentials is fully consistent with previous work: Expressions for the sum-separable potential $V=S$ have appeared in refs.~\cite{GarciaBellido:1995qq, Vernizzi:2006ve, Battefeld:2006sz, Seery:2006js} and the product-separable counterparts $V = P$ in ref.~\cite{Choi:2007su}. The extension to sum separable potentials of the form $V = S^{1/A}$ for constant $A$ was first shown by Wang~\cite{Wang:2010si}. In our subsequent discussion of sum-separable potentials we shall set the constant $A = 1$.

\sec{Two-field analytic non-Gaussianity}
\label{sec:twofields}

The above results for $N_{,I}$ hold for any number of scalar fields. In order to provide a simple analytic analysis of the superhorizon evolution, we now focus on the important subclass of two-field inflation. This is the simplest scalar field inflationary scenario possessing isocurvature perturbations during slow-roll evolution and so is a good setting in which to develop intuition about how the inflationary dynamics are related to observational predictions. We label the two scalar fields $\phi_1 = \phi$ and $\phi_2 = \chi$ to reduce proliferation of unnecessary indices. 

Analytic formulae for the bispectrum for two-field inflation may be found in refs.~\cite{Lyth:2005fi,Vernizzi:2006ve,Seery:2006js,Byrnes:2006vq,
Wang:2010si,Meyers:2010rg}, which entails the use of linear and quadratic $\delta N$ coefficients. These were extended to multiple fields by Battefeld and Easther~\cite{Battefeld:2006sz}. Trispectrum formulae have appeared in ref.~\cite{Meyers:2011mm} where it is necessary to consider the cubic order $\delta N$ coefficients in order to calculate $\gnl$. Our approach in this section draws on all of this previous work. But, by performing a rotation of the field basis on {\it both} the horizon exit and uniform density hypersurfaces, we find new ways of simplifying the results.

We shall use a field basis $\{\sigma,s\}$ such that the perturbations $\d \sigma$ are aligned with the instantaneous adiabatic direction, and perpendicular isocurvature perturbations are described by $\d s$. This basis was introduced by Gordon et al.~\cite{Gordon:2000hv} and we shall follow ref.~\cite{Peterson:2010mv} in referring to this as the {\it kinematic basis}. For two-field slow-roll inflation, the kinematic basis is defined by a simple rotation of the field basis as
\be
\begin{pmatrix}
\d \sigma \\ 
\d s
\end{pmatrix}
=
\bm Y 
\cdot
\begin{pmatrix}
\d \phi \\
\d \chi
\end{pmatrix},
\ee
where $\bm Y$ is nothing more than the standard rotation matrix
\be
\label{eq:rotation}
\bm Y = 
\begin{pmatrix}
\cos \theta & \sin \theta \\
-\sin \theta & \cos \theta
\end{pmatrix} .
\ee
This therefore defines $\theta$ as the angle of instantaneous phase space velocity which will prove to be a useful physical variable with which to describe the dynamics. We emphasise that $\theta(t)$ is a dynamical variable. This means that the kinematic frame at the time of horizon exit `$*$' is different from the kinematic frame on some later uniform density hypersurface `$c$'. 

For notational ease we now drop the `$c$' label that has been attached to quantities evaluated on a later-time uniform density hypersurface. This is unambiguous since all quantities henceforth not evaluated at horizon crossing are evaluated on uniform density hypersurfaces.

To facilitate our simplifications of the $\delta N$ coefficients it is necessary to have a prescription for writing the potential slow-roll parameters in terms of the kinematic basis $\{\sigma,s\}$. We therefore introduce a suitable formal prescription: Let $\phi_I$ define the original field basis and $\tilde \phi_I$ the kinematic basis. These are related as $\tilde \phi_I = Y_{IJ} \phi_J$ where $Y_{IJ}$ are the elements of $\bm Y$. Using the shorthand $\partial_I = \partial / \partial \phi_I$ one may verify that $\tilde \partial_I = Y_{IJ} \partial_J$. From this result one finds
\begin{align}
\label{eq:rotate_eta}
\tilde \eta_{IJ} &= Y_{IK} Y_{JL} \eta_{KL} \,,\\
\label{eq:rotate_xi}
\tilde \xi^2_{IJK} &= Y_{IL} Y_{JM} Y_{KN} \xi^2_{LMN} \,.
\end{align}
We now employ these formulae for the sum and product-separable potentials. The calculations appear in \S\ref{sec:form_sum} and \S\ref{sec:form_prod} respectively.

\ssec{Formulae for sum-separable potentials}
\label{sec:form_sum}

We follow the work of Vernizzi and Wands \cite{Vernizzi:2006ve} and consider a potential with the sum-separable form $V = U(\phi) + W(\chi)$. The first derivatives of $N$ follow from the previous section and may be written as~\cite{Vernizzi:2006ve}
\begin{align}
\Mpl N_{,\phi}&= \frac{u}{\sqrt{2 \epp^*}} \,, \qquad u = \frac{U^* + Z}{V^*}\,, 
\label{eq:ss_u}\\
\Mpl N_{,\chi}&= \frac{w}{\sqrt{2 \epc^*}} \,, \qquad w = \frac{W^* - Z}{V^*}\,, 
\label{eq:ss_v}
\end{align}
where
\be
\label{eq:Z}
Z = \frac{W \epp - U \epc}{\ep}\,.
\ee
For the purposes of calculating our desired expressions for $N_{,IJ}$ and $N_{,IJK}$ we shall need to decompose $\eta_{IJ}$ and $\xi_{IJK}^2$ into the kinematic basis. In the original $\{ \phi,\chi \}$ frame the potential is sum-separable and the only non-zero values of $\eta_{IJ}$ and $\xi_{IJK}^2$ are those for which all of the indices are identical,  which allows us to use the single-index notation $\etp$ and $\xi_\phi^2$. After rotating into the kinematic basis with fields $\sigma$ and $s$, the potential generally loses it separable form and so it is necessary to use all of the indices. The three $\eta$ components are found from eq.~\eqref{eq:rotate_eta} as
\begin{align}
\eta_{\sigma \sigma} &= \frac{\epp \etp + \epc \etc}{\ep}\,, \\
\eta_{\sigma s} &= \frac{\sqrt{\epp \epc}}{\ep} (\etc - \etp)\,, \\
\eta_{ss} &= \frac{\epc \etp + \epp \etc}{\ep}\,.
\end{align}
The components of $\xi_{IJK}^2$ in the kinematic frame follow from eq.~\eqref{eq:rotate_xi} as
\begin{align}
\ep^{3/2}\, \xi^2_{\sigma \sigma \sigma} &= \epc^{3/2}\, \xic^2 + \epp^{3/2} \, \xip^2 \,, \\
\ep^{3/2}\, \xi^2_{\sigma \sigma s} &= \epc \sqrt{\epp} \, \xic^2 - \epp \sqrt{\epc}\, \xip^2 \,, \\
\ep^{3/2}\, \xi^2_{\sigma ss} &= \epp \sqrt{\epc} \, \xic^2 + \epc \sqrt{\epp}\, \xip^2\,,  \\
\ep^{3/2}\, \xi^2_{sss} &= \epp^{3/2} \, \xic^2 - \epc^{3/2}\, \xip^2 \,.
\end{align}
The second derivatives $N_{,IJ}$ are found by differentiation of eqs.~\eqref{eq:ss_u}--\eqref{eq:ss_v}, giving 
\begin{align}
\Mpl^2 N_{,\phi \phi}&= 1 - \frac{u \etp^*}{2 \epp^*} + \frac{\A}{\epp^*}\,, \\
\Mpl^2 N_{,\phi \chi}&= - \frac{\A}{\sqrt{\epp^* \epc^*}}\,, \\
\Mpl^2 N_{,\chi \chi}&= 1 - \frac{w \etc^*}{2 \epc^*} + \frac{\A}{\epc^*}\,,
\end{align}
where
\be
\frac{\Mpl}{V^*}\sqrt{\frac{\epp^*}{2}} \frac{\partial Z}{\partial \phi^*} 
= - \frac{\Mpl}{V^*}\sqrt{\frac{\epc^*}{2}} \frac{\partial Z}{\partial \chi^*} 
= \A \equiv \frac{V^2}{V_*^2} \frac{\epp \epc}{\ep^2}
	\left( \eta_{ss} - \ep \right).
\label{eq:ss_A}
\ee
Taking the next derivative we find
\begin{align}
\Mpl^3 N_{,\phi \phi \phi}&= \frac{1}{\epp^* \sqrt{2 \epp^*}}
	\left(
	- \frac{u}{2} \sqrt{\frac{\epp^*}{\ep^*}} {\xip^*}^2  
	- \epp^* \etp^*
	+ u {\etp^*}^2
	- 3 \etp^* \A + \B^2 
	\right),\\
\Mpl^3 N_{,\phi \phi \chi}&= \frac{1}{\epp^* \sqrt{2 \epc^*}}\left( \etp^* \A - \B^2 \right), \\
\Mpl^3 N_{,\phi \chi \chi}&= \frac{1}{\epc^* \sqrt{2 \epp^*}}\left( \etc^* \A + \B^2 \right), \\
\Mpl^3 N_{,\chi \chi \chi}&= \frac{1}{\epc^* \sqrt{2 \epc^*}}
	\left(
	- \frac{w}{2} \sqrt{\frac{\epc^*}{\ep^*}} {\xic^*}^2  
	- \epc^* \etc^*
	+ w {\etc^*}^2
	- 3 \etc^* \A - \B^2  
	\right), \\
\Mpl \sqrt{2 \epp^*} \frac{\partial \A}{\partial \phi_*} &\equiv -4 \epp^* \A + \B^2 , \\
\B^2 &\equiv   -\frac{V^3}{V_*^3} \frac{\sqrt{\epp \epc}^3}{\ep^3} \Big[
\xi_{sss}^2 + 2 \frac{\epp - \epc}{\sqrt{\epp \epc}} \eta_{ss} (\eta_{ss}-\ep)
- 2 \eta_{\sigma s} (\eta_{ss} + \ep)
\Big] .
\label{eq:ss_B}
\end{align}

The derivative $\partial \A / \partial \chi^*$ follows by a simple permutation of eq.~\eqref{eq:ss_B} under the joint permutations $\{u \leftrightarrow v \}$ and $\{\phi \leftrightarrow \chi \}$ which has the effect of negating $\B^2$ whilst $\A$ remains unchanged. We thus obtain expressions for inflationary observables by substituting these results into eqs.~\eqref{eq:dN_sindex}--\eqref{eq:dN_gnl}. This yields
\begin{align}
\hspace{-3em} n_\zeta-1 &= - 4 \left(\frac{u^2}{\epp^*} + \frac{w^2}{\epc^*} \right)^{-1}
&& \sneg_space \left[ 1 - \frac{u^2 \etp^*}{2 \epp^*} - \frac{w^2 \etc^*}{2 \epc^*} \right] - 2 \ep^*, \label{eq:ss_params1_ns} \displaybreak[0]\\
\fnl &= \frac{5}{6} \left(\frac{u^2}{\epp^*} + \frac{w^2}{\epc^*} \right)^{-2} 
&& \sneg_space \left[ 
	2 \left( \frac{u^2}{\epp^*} +\frac{w^2}{\epc^*} \right)
	- \frac{u^3 \etp^*}{{\epp^*}^2 }
	- \frac{w^3 \etc^*}{{\epc^*}^2 } 
	+2 \left( \frac{u}{\epp^*} - \frac{w}{\epc^*} \right)^2 \A 
\right]\!\!, \label{eq:ss_params1_fnl} \displaybreak[0]\\
\tnl &= 4 \left(\frac{u^2}{\epp^*} + \frac{w^2}{\epc^*} \right)^{-3} 
&& \sneg_space \Bigg[
	-\frac{u^3 \etp^*}{{\epp^*}^2}
	-\frac{w^3 \etc^*}{{\epc^*}^2}
	+\frac{u^4 {\etp^*}^2 }{4{\epp^*}^3 }
	+\frac{w^4 {\etc^*}^2 }{4{\epc^*}^3 } 
	+\frac{u^2}{\epp^*}
	+\frac{w^2}{\epc^*} \nonumber \\
	& && \sneg_space \left. \quad - \frac{u^2}{ {\epp^*}^2} \left( \frac{u}{\epp^*} - \frac{w}{\epc^*} \right) \etp^* \A 
	- \frac{w^2}{ {\epc^*}^2} \left( \frac{w}{\epc^*} - \frac{u}{\epp^*} \right) \etc^* \A \right. \nonumber \\
	& && \sneg_space \quad + 2 \left( \frac{u}{\epp^*} - \frac{w}{\epc^*} \right)^2 \A  
	+\left( \frac{u}{\epp^*} - \frac{w}{\epc^*}\right)^2 \left( \frac{1}{\epp^*} + \frac{1}{\epc^*}\right) \A^2
\Bigg], \label{eq:ss_params1_tnl} \displaybreak[0]\\
\gnl &= \frac{25}{27}\left(\frac{u^2}{\epp^*} + \frac{w^2}{\epc^*} \right)^{-3} 
&& \sneg_space \Bigg[
	- \frac{u^3 \etp^*}{{\epp^*}^2}
	- \frac{w^3 \etc^*}{{\epc^*}^2}
	+ \frac{u^4 {\etp^*}^2}{{\epp^*}^3}
	+ \frac{w^4 {\etc^*}^2}{{\epc^*}^3}
	- \frac{1}{2}\frac{u^4 {\xip^*}^2}{{\epp^*}^2 \sqrt{\ep^* \, \epp^*} }
	  \nonumber \\ & && \sneg_space 
	 - \frac{1}{2}\frac{w^4 {\xic^*}^2}{{\epc^*}^2 \sqrt{\ep^* \, \epc^*}}
	-3 \frac{u^2}{{\epp^*}^2} \left(\frac{u}{\epp^*} - \frac{w}{\epc^*} \right) \etp^* \A 
	\nonumber \\ & && \sneg_space
	-3 \frac{w^2}{{\epc^*}^2} \left(\frac{w}{\epc^*} - \frac{u}{\epp^*} \right) \etc^* \A 
	+ \left(\frac{u}{\epp^*} - \frac{w}{\epc^*} \right)^3 \B^2
\Bigg]. \label{eq:ss_params1_gnl}
\end{align}

Thus far we have employed the kinematic basis on the later-time hypersurface. In order to facilitate the simplification of the above expressions in chapter \ref{ch:heatmaps} we rewrite the horizon crossing slow-roll parameters in terms of their kinematic counterparts such as $\eta_{ss}^*$. Note that $\eta_{ss}$ and $\eta_{ss}^*$ are calculated in two different bases, since the kinematic basis evolves with time. There are three kinematic $\eta$ parameters and four $\xi^2$ parameters, which means that there is no unique way to write the slow-roll parameters in the kinematic basis. We shall elect to write $\etp^*$ and $\etc^*$ in terms of $\eta_{ss}^*$ and $\eta_{\sigma s}^*$ and ${\xip^*}^2$ and ${\xic^*}^2$ in terms of ${\xi_{sss}^*}^2$ and ${\xi_{\sigma ss}^*}^2$. This gives the relations
\begin{align}
\etp &= \eta_{ss} - \sqrt{\frac{\epp}{\epc}} \eta_{\sigma s} \, , \label{eq:kinematic_subs1} \\ 
\etc &= \eta_{ss} + \sqrt{\frac{\epc}{\epp}} \eta_{\sigma s} \, , \label{eq:kinematic_subs2} \\
\xip^2 &= \frac{\sqrt{\ep \epp}}{\epc}\xi_{\sigma ss}^2 - \sqrt{\frac{\ep}{\epc}} \xi_{sss}^2 \, , \label{eq:kinematic_subs3} \\
\xic^2 &= \frac{\sqrt{\ep \epc}}{\epp}\xi_{\sigma ss}^2 + \sqrt{\frac{\ep}{\epp}} \xi_{sss}^2 \, .
\label{eq:kinematic_subs4}
\end{align}
Substituting the relations~\eqref{eq:kinematic_subs1}--\eqref{eq:kinematic_subs4} into eqs.~\eqref{eq:ss_params1_ns}--\eqref{eq:ss_params1_gnl} and simplifying we obtain
\begin{align}
\hspace{-1em} n_\zeta-1 &= \left(\frac{u^2}{\epp^*} +
\frac{w^2}{\epc^*} \right)^{-1} && \lneg_space
\left[ -4 + \frac{2w-2u}{\sqrt{\epp^* \epc^*}}
\eta_{\sigma s}^* \right] + 2(\eta_{ss}^* - \ep^*)\,, \label{eq:ss_params_ns} \displaybreak[0]\\
\hspace{-1em}\frac{6}{5} \fnl &= \left(\frac{u^2}{\epp^*} + \frac{w^2}{\epc^*} \right)^{-2} && \lneg_space
\Bigg[ 
	2 \left( \frac{u^2}{\epp^*} +\frac{w^2}{\epc^*} \right)
	- \left( \frac{u^3}{{\epp^*}^2} + \frac{w^3}{{\epc^*}^2} \right) \eta_{ss}^* \nonumber  \\ & && \lneg_space
	+ \left( \frac{u^3}{\epp^*} - \frac{w^3}{\epc^*} \right) \frac{\eta_{\sigma s}^*}{\sqrt{\epp^* \epc^*}}
	+2 \left( \frac{u}{\epp^*} - \frac{w}{\epc^*} \right)^2 \A 
\Bigg], \label{eq:ss_params_fnl}\displaybreak[0]\\
\hspace{-1em}\tnl &= \left(\frac{u^2}{\epp^*} + \frac{w^2}{\epc^*} \right)^{-3} && \lneg_space
\Bigg[
	\left( \frac{u^4}{{\epp^*}^3} + \frac{w^4}{{\epc^*}^3} \right){\eta_{ss}^*}^2
	-2\left( \frac{u^4}{{\epp^*}^2} - \frac{w^4}{{\epc^*}^2} \right)\frac{\eta_{ss}^* \eta_{\sigma s}^*}{\sqrt{\epp^* \epc^*}} 
	+\left( \frac{u^4}{\epp^*} + \frac{w^4}{\epc^*} \right)\frac{{\eta_{\sigma s}^*}^2}{\epp^* \epc^*}	
	\nonumber \\ & && \lneg_space\left.
	-4 \left(\frac{u^3}{{\epp^*}^2} + \frac{w^3}{{\epc^*}^2} \right)\eta_{ss}^* 
	+ 4\left(\frac{u^3}{\epp^*} - \frac{w^3}{\epc^*} \right)\frac{\eta_{\sigma s}^*}{\sqrt{\epp^* \epc^*}} 
	+4\left(\frac{u^2}{\epp^*} + \frac{w^2}{\epc^*}\right) 
	\right. \nonumber \\ & && \lneg_space\left.	
	- 4\left(\frac{u}{\epp^*}-\frac{w}{\epc^*}\right)^2 \left( \frac{u}{\epp^*} + \frac{w}{\epc^*} \right) \eta_{ss}^* \A 
	+ 8 \left( \frac{u}{\epp^*} - \frac{w}{\epc^*}
\right)^2 \A  
	\right. \nonumber \\ & && \lneg_space
	+ 4\left(\frac{u}{\epp^*}-\frac{w}{\epc^*}\right) \left( \frac{u^2}{\epp^*} + \frac{w^2}{\epc^*} \right) \frac{\eta_{\sigma s}^* \A}{\sqrt{\epp^* \epc^*}} 
	+4\left( \frac{u}{\epp^*} - \frac{w}{\epc^*}\right)^2 \frac{\ep^* \A^2}{\epp^* \epc^*}
\Bigg], \label{eq:ss_params_tnl} \displaybreak[0]\\
\hspace{-1em}\frac{27}{25} \gnl &= \left(\frac{u^2}{\epp^*} + \frac{w^2}{\epc^*} \right)^{-3} &&\lneg_space
\Bigg[
	\left( \frac{u^4}{{\epp^*}^3} + \frac{w^4}{{\epc^*}^3} \right){\eta_{ss}^*}^2
	-2\left( \frac{u^4}{{\epp^*}^2} - \frac{w^4}{{\epc^*}^2} \right)\frac{\eta_{ss}^* \eta_{\sigma s}^*}{\sqrt{\epp^* \epc^*}} 
	+\left( \frac{u^4}{\epp^*} + \frac{w^4}{\epc^*} \right)\frac{{\eta_{\sigma s}^*}^2}{\epp^* \epc^*}
	\nonumber \\ & && \lneg_space\left.	
	- \left(\frac{u^3}{{\epp^*}^2} + \frac{w^3}{{\epc^*}^2} \right)\eta_{ss}^* 
	+\left(\frac{u^3}{\epp^*} - \frac{w^3}{\epc^*} \right)\frac{\eta_{\sigma
s}^*}{\sqrt{\epp^* \epc^*}} 
	+\frac{1}{2} \left(\frac{u^4}{{\epp^*}^2} - \frac{w^4}{{\epc^*}^2} \right)\frac{{\xi_{sss}^*}^2}{\sqrt{\epp^* \epc^*}}
	\right. \nonumber \\ & && \lneg_space\left.	
	-\frac{1}{2} \left(\frac{u^4}{\epp^*} + \frac{w^4}{\epc^*} \right)\frac{{\xi_{\sigma ss}^*}^2}{\epp^* \epc^*}
	- 3\left(\frac{u}{\epp^*}-\frac{w}{\epc^*}\right)^2 \left( \frac{u}{\epp^*} + \frac{w}{\epc^*} \right) \eta_{ss}^* \A 
	\right. \nonumber \\ & && \lneg_space
	+ 3\left(\frac{u}{\epp^*}-\frac{w}{\epc^*}\right) \left( \frac{u^2}{\epp^*} + \frac{w^2}{\epc^*} \right) \frac{\eta_{\sigma s}^* \A}{\sqrt{\epp^* \epc^*}}  + \left( \frac{u}{\epp^*} - \frac{w}{\epc^*} \right)^3 \B^2  
\Bigg] \label{eq:ss_params_gnl}
\end{align}
Whilst these expressions may appear more complex than those in eqs.~\eqref{eq:ss_params1_ns}--\eqref{eq:ss_params1_gnl}, we shall find in chapter \ref{ch:heatmaps} that this method of recasting the expressions ultimately allows for greater simplification. The product separable counterparts to these formulae may be found in \S\ref{sec:form_prod}.

\sssec{Summary} 

This chapter has developed the technology to evolve covariant perturbations in the superhorizon epoch. The covariant $\delta N$ coefficients $N_{,i}, N_{;ij}, \dots$ may be found using the covariant transport method that we have derived. Our calculation presumes slow-roll inflation, but there is no barrier to developing a non-slow-roll version of this formalism. We have then used this theoretical framework to show how to derive observational predictions from inflationary models. This framework in general requires numerical methods for its computation. However, there is an important subclass of scenarios where the evolution is analytically tractable. These are very useful since they allow us to develop intuition about the relation between inflationary dynamics and the evolution of inflationary observables. 

Such analytic formulae are then discussed. Under the assumptions of slow-roll and a separable potential, we have derived expressions that describe the evolution of non-Gaussianity. There are, however, important scenarios in which this evolution ceases, with profound implications for the predictivity of inflationary models. We shall discuss this at length in chapter \ref{ch:adiabatic}. Then, in chapter \ref{ch:heatmaps}, we shall interpret and simplify the key results of this chapter appearing in eqs.~\eqref{eq:ss_params_ns}--\eqref{eq:ss_params_gnl} and \eqref{eq:ps_params_ns}--\eqref{eq:ps_params_gnl}.
\clearpage{\pagestyle{empty}\cleardoublepage}
\chapter{Adiabaticity}
\label{ch:adiabatic}

\begin{addmargin}[0.05\textwidth]{0.05\textwidth}
One of the key goals of this thesis is to calculate observational predictions for inflationary models. However, there are intermediate phases between inflation and the time of \cmbr~decoupling and one may generally expect these to source additional evolution of the curvature perturbation. In our present state of ignorance of the physical processes occurring at these intermediate times, the only scenario in which inflationary models will generate robust predictions occurs when $\zeta$ and its statistics become conserved during or shortly after the inflationary epoch. We show in \S\ref{sec:adiabaticity} that conservation of this form arises if cosmic dynamics reach a state of {\it adiabaticity}. Following Elliston et al.~\cite{Elliston:2011dr}, \S\S\ref{sec:model1}--\ref{sec:model3} then illustrate this discussion with a range of examples covering three different adiabaticity regimes. A key emphasis of this study is its implications for the appropriate choice of analytical or numerical tools for calculating cosmic observables.
\end{addmargin}

\begin{center}
\partialhrule
\end{center}
\vspace{-3em}
\begin{quote}
\list{}{\leftmargin 2cm \rightmargin\leftmargin} \endlist
\begin{center}
{\it ``The only real valuable thing is intuition.''}
\flushright{---Albert Einstein.}
\end{center}
\end{quote}
\vspace{-1em}
\begin{center}
\partialhrule
\end{center}

\sec{Criterion for adiabaticity}
\label{sec:adiabaticity}

In \S\ref{sec:ir-safety} we discussed how multi-field models of inflation may be sensitive to infrared dynamics. We then developed the necessary formalism for tracking these dynamics in chapter~\ref{ch:formalisms}. Such sensitivity has the exciting capacity to be able to differentiate between candidate models of inflation. Unfortunately, precisely because of these desirable properties, the corresponding observables possess an added complication: they may be equally sensitive to {\it post-inflationary} dynamics, further complicating the task of extracting predictions. In principle, the statistics of the curvature perturbation should be tracked until the time of last scattering---where the microwave background anisotropy was imprinted---and in our present state of ignorance this is an impossible undertaking. Therefore, to connect the physics of inflation with observations, an important approach in practice is to rely on {\it conservation}; it is the statistics which apply at the onset of conservation which will be inherited by observable quantities.

This point of view was developed soon after multiple-field models entered the literature~\cite{GarciaBellido:1995qq}. For practical purposes we require a characterization of the conditions under which $\zeta$ cannot subsequently evolve. {\it This occurs when the isocurvature modes are attenuated}, as demonstrated by Rigopoulos and Shellard~\cite{Rigopoulos:2003ak}, Lyth, Malik and Sasaki~\cite{Lyth:2004gb} and later Langlois and Vernizzi~\cite{Langlois:2005ii,Langlois:2005qp,Langlois:2006iq,Langlois:2008vk,Langlois:2010vx} using a gradient expansion. Christopherson and Malik~\cite{Christopherson:2008ry} extended these results to models in which the Lagrangian can be an arbitrary Lorentz-invariant function of the scalar field and its first derivatives. More recently, Naruko and Sasaki~\cite{Naruko:2011zk} and Gao~\cite{Gao:2011mz} applied similar arguments to higher-derivative models which preserve second-derivative field equations, where conservation can be subtle~\cite{Khoury:2008wj,Baumann:2011dt}. Weinberg developed a different approach \cite{Weinberg:2004kr,Weinberg:2004kf,Weinberg:2008nf,Weinberg:2008si}, adapting the techniques of Goldstone's theorem to show that $\zeta$ would become massless on superhorizon scales, admitting a time-independent solution. Whether this solution is selected is a model-dependent question.

{\it When are the isocurvature modes attenuated?} The conditions under which this occurs are known as the {\it adiabatic conditions}. They require that every component of the Universe's content (such as dust and radiation) is uniquely prescribed by a single parameter, such as the local density $\rho(\vect{x},t)$. For perfect fluids, the continuity equation \eqref{eq:continuity} then uniquely prescribes the relative magnitude of each fluid's perturbation. 

Let us now consider the adiabatic conditions in terms of the relevant phase space (the dimensions of this space describe every component of the Universe's content, though we do not require any further details for the present argument). The adiabatic conditions require that all of the separate universes reside on a {\sc 1d} line with different separate universes parametrized by translations in time. Upon making such a translation to reach a nearby uniform density hypersurface,\footnote{We have assumed that the {\sc 1d} adiabatic line is nowhere tangent to the uniform density hypersurfaces. This condition is always fulfilled by slow-roll inflation where the bundle obeys gradient flow and thus flows normal to uniform density hypersurfaces. More generally, eq.~\eqref{eq:acceleration} guarantees that $\dot H <0$ even out of the slow-roll regime, such that the bundle is nowhere tangent to surfaces of constant $H$.} one generates the curvature perturbation $\zeta$. In this gauge, the adiabatic conditions prescribe the separate universes to be at a single point in the phase space. They thus all undergo identical subsequent evolution to any later uniform density hypersurface, and in this evolution $\zeta$ remains unchanged. This demonstrates that $\zeta$ is conserved under adiabatic conditions, regardless of the content of the Universe. Consequently, it is only necessary to evolve the dynamics until the onset of adiabaticity.

The simplest models of inflation have a single scalar field. As discussed in \S\ref{sec:ir-safety}, for such models the curvature perturbation $\zeta$ is conserved at times slightly beyond horizon exit~\cite{Bardeen:1983qw,Lyth:1984gv, Rigopoulos:2003ak,Lyth:2004gb,Weinberg:2004kf,Langlois:2005ii,Langlois:2005qp}. We may now understand this result in terms of the above argument: Single field models would in general have both adiabatic and isocurvature perturbations, corresponding to the variations $\{\delta \vp_I, \delta \dot \vp_I\}$. However, by virtue of the inflationary attractor~\cite{Lyth:2009zz}, these are not independent and may be written in terms of a single perturbation which defines relative time-translation between different separate universes. Up to a gauge choice, this perturbation is $\zeta$. Since only this adiabatic perturbation persists, the above argument shows that $\zeta$ will not evolve between subsequent uniform density hypersurfaces in the superhorizon regime. Consequently, the $n$-point functions $\langle \zeta^n \rangle$ are conserved for all subsequent times, regardless of the dynamics of subsequent phases.\footnote{Note that this argument only holds a few efolds after horizon crossing, since it is necessary to ensure that the non-growing subhorizon modes have frozen out at constant values, as demonstrated explicitly by Nalson et al.~\cite{Nalson:2011gc}.} 

Multi-field inflation is more complex because the isocurvature modes may persist into the superhorizon epoch and source interesting infrared dynamics. For example, turning of the separate universe bundle is known to cause the curvature perturbation to evolve~\cite{Gordon:2000hv,Rigopoulos:2005us,Vernizzi:2006ve}. However, the previous argument demonstrates that if the separate universe bundle converges to a {\sc 1d} line then an {\it adiabatic limit} is reached and cosmic observables at that time are conserved thereafter. This notion of {\it adiabaticity} has profound implications for the application of any theoretical tools which we may employ to track the dynamics. Whether these tools are analytic or computational in nature, they will be based on certain simplifying approximations. {\it If} an adiabatic limit is reached {\it before} these approximations break down, then the tools will reliably calculate the limiting value of cosmic observables. On the other hand, an adiabatic limit may not be reached until some time after the approximations have broken down. In this case we cannot expect the tools to track the complete evolution. 

A fundamentally important question therefore is: {\it When does the adiabatic regime begin?} If and how the phase space bundle may focus to a {\sc 1d}~line is a model-dependent question. A `natural' method is to have a valley shape in the inflationary potential such that neighbouring separate universe trajectories converge. Examples include Nflation and related models~\cite{Dimopoulos:2005ac,Alabidi:2005qi,
Vernizzi:2006ve,Alabidi:2006hg,Kim:2006ys,Kim:2006te,Kim:2010ud}. Alternatively, one may consider interrupting the dynamics, such as via a hybrid transition, in which case one must verify whether or not the interruption itself modifies the predictions for cosmic observables.\footnote{We calculate statistics on uniform density hypersurfaces. Should inflation end suddenly on some different hypersurface, and if this happens before any adiabatic limit is attained, then this will generate an additional source of $\zeta$ \cite{Lyth:2005qk,Huang:2009vk}.}

\ssec{Analytic adiabaticity}

In certain scenarios the evolution of the curvature perturbation can be accounted for analytically. In chapter \ref{ch:formalisms} we showed that this may be achieved with slow-roll evolution and a separable potential. Where analytic representations are faithful, they provide valuable insights into the inflationary dynamics and the corresponding observational effects. Naturally, it is necessary to consider the domain of validity of these analytic approximations. Any individual case may be easily verified by comparing to a numerical code which does not employ the slow-roll approximation after horizon exit, or require a separable potential. Whilst this is an important check, it does not develop intuition. Developing such intuition, that can then be cautiously applied in other scenarios, is a focus of this chapter.

\begin{figure}[h]
\begin{center}
\parbox{0.5\textwidth}{\includegraphics[width=0.5\textwidth]{4analytic/functional_variation}}%
\begin{minipage}{0.5\textwidth}%
\includegraphics[width=\textwidth]{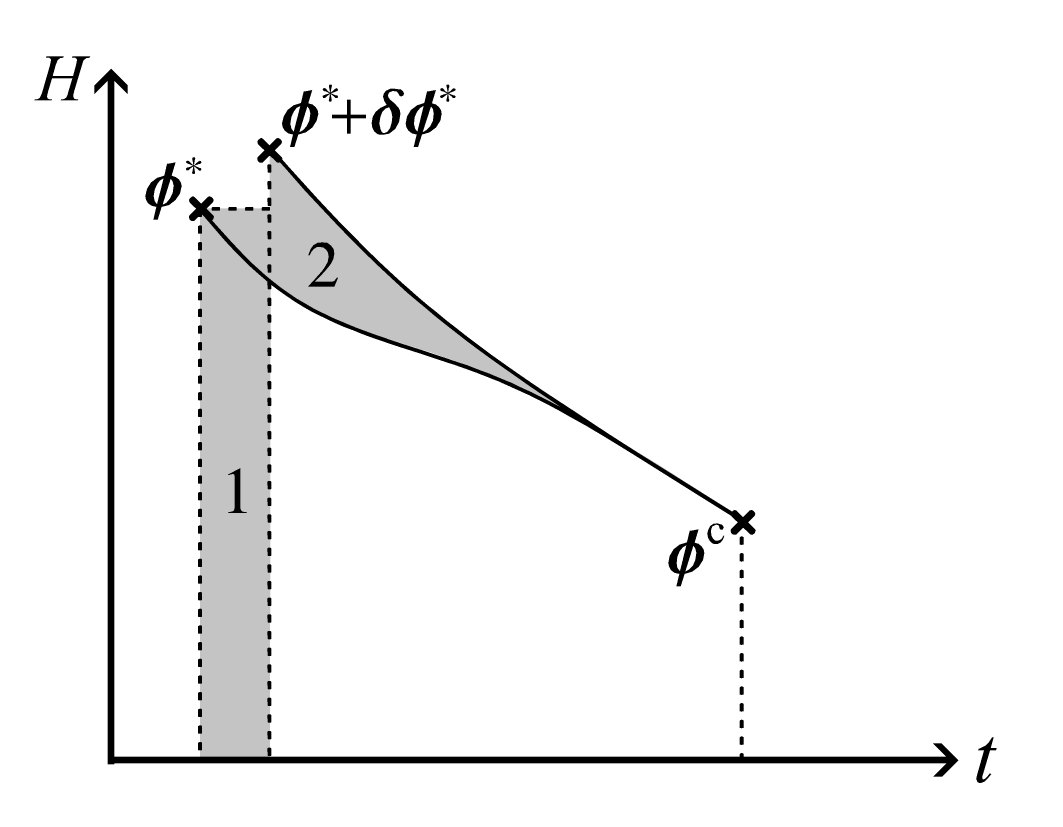}
\end{minipage}%
\caption[Functional variation and the \hca]{
The functional variation for $\delta N$ between horizon exit and a later-time uniform density hypersurface. The left pane is the general scenario shown in figure \ref{fig:functional}. The right pane shows the effect of realising an adiabatic limit at some intermediate time: There is no final boundary term and the functional variation $\delta N$ (and thus $\zeta$) becomes conserved after the onset of adiabaticity.  
\label{fig:functional2}}
\end{center}
\end{figure}

If the separate universe bundle has converged to a {\sc 1d}~line then it intersects each uniform density hypersurface at a unique point $\phi_J^c$. Making a small change of trajectory by perturbing the horizon exit conditions as $\phi_I^* \to \phi_I^* + \delta \phi_I^*$, will not change $\phi_J^c$. Therefore $\delta \phi^c_J = 0$ and we conclude that derivatives of the form $\partial \phi^c_J / \partial \phi_I^*$ are null in regions where $\zeta$ is conserved.\footnote{The condition that $\partial \phi^c_J / \partial \phi_I^* \rightarrow 0$ is a sufficient but not necessary condition for $\zeta$ to become conserved within the separate universe picture.} This has a profound effect on the functional variation of $N$ as shown in figure \ref{fig:functional2}. These principles provide a formal analytic procedure for determining the statistics of $\zeta$ in an adiabatic limit: One uses methods such as those developed in \S\ref{sec:analytics} to determine any required $n$-point correlation functions, and then imposes the requirement $\partial \phi_J^c / \partial \phi_I^* \rightarrow 0$. One must ensure that this limit is faithfully attained by the dynamics, and furthermore, that this occurs before the slow-roll approximation breaks down (after which the analytic results of \S\ref{sec:analytics} no longer apply). We now discuss in greater detail how this adiabatic limit may be attained via a focussing of trajectories during slow-roll inflation.

\sssec{Focussing of trajectories}

A typical example of a focusing region is a valley of the potential landscape, perhaps terminating in a local minimum. For $\M$ fields, this prescribes there to be $\M-1$ heavy directions with masses greater than the Hubble rate. The steep slopes cause exponential convergence, and rapidly focus the bundle to a line. In the neighbourhood of the valley floor, let us assume it is possible to choose coordinates on field space for which the potential approximately separates as
\begin{equation}
	V \approx U_\sigma(\sigma) + \sum_\alpha U_\alpha(s_\alpha)
	\approx
	U_\sigma(\sigma) + \frac{1}{2} \sum_\alpha m_\alpha^2 s_\alpha^2 ,
\label{eq:potential-valley}
\end{equation}
where $\sigma$ labels distance along the valley floor---which may be a light direction---and the $\M-1$ fields $s_\alpha$ are stabilized with masses $m_\alpha \gtrsim H$. To describe a complicated valley it may be necessary to glue several such regions together. Focusing on the particular region described by eq.~\eqref{eq:potential-valley}, we denote the field values on entry to its domain of validity as $\bar{\sigma}$ and $\bar{s}_\alpha$. These will be functions of the initial fields $\phi_i^*$ which may lie far away from the valley where eq.~\eqref{eq:potential-valley} need not be a good approximation.

If $m > 3H/2$ then the dynamics of the $s$ field are underdamped and so exhibit damped oscillations about the valley minimum. This amplitude decay is identical to that which occurs in the simpler over-damped case, which occurs for $m \lesssim H$. For brevity, let us focus on the overdamped case where the field obeys slow-roll as $3H \dot{s}_\alpha = - m_\alpha^2 s_\alpha$. After $N$ efolds from the point of entry, one finds
\begin{equation}
	s_\alpha = 
		\bar{s}_\alpha(\phi_I^*)
		\e{ - \int_0^N \eta_\alpha(N') \; \d N' } ,
	\label{eq:phi-decay}
\end{equation}
where $\eta_\alpha(N')= m_\alpha^2/3H^2$. The total number of efolds available within the valley is model-dependent. In a long valley the focusing may practically go to completion, making $s_\alpha$ effectively zero. Alternatively, if the valley rapidly terminates in a local minimum there may be insufficient time to focus the 
bundle completely.

The fields $\phi_K$ can be written as linear combinations of $\sigma$ and the $s_\alpha$, giving $\phi_K = \gamma_K \sigma + \sum_\alpha \beta^\alpha_K s_\alpha$. The $\gamma_K$ and $\beta^\alpha_K$ are constants, which depend only on the choice of separable coordinates used in eq.~\eqref{eq:potential-valley}. From this we find
\begin{equation}
\frac{\partial \phi_K^c}{\partial \phi_J^*} = \sum_\alpha \left( 
\beta_K^\alpha - \gamma_K \frac{{U_\alpha '}^c}{{U_\sigma '}^c}
\right) \frac{\partial s_\alpha^c}{\partial \phi_J^*} .
\label{eq:decay-estimate}
\end{equation}
Therefore $\partial \phi_K^c / \partial \phi^*_J$ evolves as a linear combination of derivatives $\partial s^c_\alpha / \partial \phi_J^*$.

In eq.~\eqref{eq:decay-estimate} the term proportional to $\gamma_K$ will typically decay exponentially, because $U_\alpha'^c \sim m_\alpha^2 s_\alpha$ whereas $U_\sigma'^c$ decays less rapidly. Therefore eq.~\eqref{eq:phi-decay} implies the derivatives $\partial \phi_K^c / \partial \phi_J^*$ decay at least as fast as the lightest isocurvature field. We conclude that the partial derivatives of the form $\partial \phi_K^c / \partial \phi_J^*$ decay as\footnote{The asymptotic notation $x \asymp y$ indicates that $x$ and $y$ share a common decay rate.}
\begin{equation}
	\label{eq:deriv_decay}
	\frac{\partial \phi_K^c}{\partial \phi_J^*}
	\asymp
	\e{ - \int_{0}^{N} \eta_s(N') \; \d N' }
	\sim
	\e{ - \eta_s N },
\end{equation}
where $\eta_s = \min \{ \eta_\alpha \}$ and $N$ is the same quantity as that occurring in eq.~\eqref{eq:phi-decay}. The final equality applies if $\eta_s$ is approximately constant during the focusing process.

It is not straightforward to estimate the minimum number of efolds required to make the final `$c$-terms' of eqs.~\eqref{eq:ps_Ni_hca2} and \eqref{eq:ss_Ni_hca2} negligible. Although eq.~\eqref{eq:deriv_decay} gives information concerning the decay rate, the number of efolds required to damp any contribution from the $c$-terms depends on their amplitude on entry to the valley. This is a function of each species' relative contribution to the energy density of the Universe on the initial and final slices $c$ and $\ast$, from which it does not appear straightforward to draw general conclusions. However, if we presume the isocurvature to be sufficiently heavier than the Hubble scale, the parameter $\eta_s$ will typically be much larger than unity. In these circumstances, rather less than $\Or(10)$ efoldings are usually required to accumulate a very substantial suppression of the $c$-terms.

Presuming that the potential possesses a focusing region, one might expect that the final values of quantities such as $\fnl$ would depend only on the local shape of the potential in the focusing region, which will typically be a stable parabolic minimum. If so, the asymptotic value of $\fnl$ would be universal among all potentials sharing a similarly-shaped minimum. However, this is not the case since the number of efolds $N$ is dependent on the dynamics of the entire superhorizon evolution. Hence, the asymptotic value of quantities such as $\fnl$ will generally depend on properties of the potential far from the focusing region. 

If a natural focusing region is not available, or is not selected, then one must either impose an adiabatic limit through some other mechanism or continue to evolve the perturbations indefinitely in order to obtain observationally meaningful quantities. 

\sssec{Adiabaticity and non-Gaussianity}

The concept of adiabaticity applies equally to all inflationary observables. We choose to focus on non-Gaussianity because its evolution is currently less explored and because of its known sensitivity to superhorizon dynamics. For canonical multi-field inflation, the results of chapters~\ref{ch:subhorizon} and~\ref{ch:formalisms} showed that the curvature perturbation at horizon exit has nearly Gaussian statistics (see also~\cite{Seery:2005gb,Seery:2006vu,Seery:2008ax}). During a turn, typically in the early stages, it is possible for the statistics of the curvature perturbation to become highly non-Gaussian \cite{Alabidi:2006hg,Byrnes:2008wi,Byrnes:2008zy}. If an adiabatic limit is imposed suddenly, perhaps through a hybrid transition, it is \emph{possible} that this large non-Gaussianity is subsequently conserved. This strategy has been invoked by various authors~\cite{Sasaki:2008uc,Naruko:2008sq,Byrnes:2008wi,Byrnes:2008zy,Mulryne:2011ni}. Alternatively, one may consider scenarios in which an adiabatic limit is imposed `naturally' as we have just described. This may lead to negligible non-Gaussianity as argued by Meyers and Sivanandam~\cite{Meyers:2010rg, Meyers:2011mm}, but this is not the only option. It is also possible for the non-Gaussianity to tend to a large value after a turn \cite{Kim:2010ud}, and furthermore, its evolution towards this large value may be characterised by growth of non-Gaussianity rather than decay. This broad range of evolutionary behaviour on approach to adiabaticity was first studied in detail by Elliston et al.~\cite{Elliston:2011dr}.

\ssec{Asymptotic behaviour}
\label{sec:hca}

For globally separable potentials and slow-roll evolution, $N_{,I}$ satisfies eqs.~\eqref{eq:ps_Ni_hca2} and \eqref{eq:ss_Ni_hca2}. It may happen that eq.~\eqref{eq:deriv_decay} is sufficiently strong to make the final `$c$-term' irrelevant. This limit is known as the {\it Horizon Crossing Approximation} (\hca)~\cite{Kim:2006ys,Kim:2006te,Kim:2010ud}. Thus the \hca~corresponds to the analytic realisation of the adiabatic conditions $\partial \phi^c_J / \partial \phi_I^* \to 0$. When one employs the \hca~correctly, the $\delta N$ derivatives are functions only of the field values at horizon crossing and are thus constant. When applied to observables such as $\fnl$, the \hca~yields the adiabatic value but does not calculate the prior evolution. Since the \hca~is based on the above slow-roll formulae, it is clear that a necessary condition for the \hca~to yield accurate results is that an adiabatic limit is attained before the breakdown of the slow-roll approximation. 

The expressions \eqref{eq:ss_params_fnl}--\eqref{eq:ss_params_gnl} and \eqref{eq:ps_params_fnl}--\eqref{eq:ps_params_gnl} for $\fnl$, $\tnl$ and $\gnl$ simplify considerably if one invokes the \hca. In particular, the terms involving $Z$, $\A$ and $\B^2$ (or $\A_P$ and $\B_P^2$ ) as defined in eqs. \eqref{eq:Z}, \eqref{eq:ss_A} and \eqref{eq:ss_B} (or eqs. \eqref{eq:ps_A} and \eqref{eq:ps_B}) all tend to zero. These simplifications will allow us to produce some very simple results in the analysis carried out in chapter \ref{ch:heatmaps}.

From eqs.~\eqref{eq:ps_deriv} and \eqref{eq:ss_deriv}, we see that the constraint of having a separable potential means that the adiabatic conditions $\partial \phi^c_K / \partial \phi_I^* \to 0$ can only be satisfied in a number of simple ways. Firstly, for either class of separable potential, this condition is met if the phase space velocity is aligned with one of the field axes such that only one field is evolving. Secondly, the adiabatic conditions are met if the global potential $V^c$ goes to a minimum at zero. One can see that this implies $\partial \phi^c_K / \partial \phi_I^* \to 0$ for the product-separable potential because if the fields are in the potential minimum then there is no velocity in any direction and so $\ep_K^c=0$. 

When applied to any specific potential, it is important to consider whether or not one expects the isocurvature modes to have decayed on approach to one of these limits. To illustrate this point we take two different examples:
\begin{itemize}
\item Taking a multi-field sum-separable potential, the phase space velocity may be enforced to align to an axis because all the other fields are in the minima of their potentials (in this case the potential landscape is a multi-dimensional valley), then one would expect attenuation of isocurvature modes and an adiabatic limit to be reached.
\item A modified scenario may have an additional field $\chi$ with potential $U_\chi$ that is part of the global sum-separable potential. If $U_\chi$ is very flat then the field $\chi$ is essentially frozen. The other fields may evolve to their minima and the inflationary dynamics then become aligned with one of the phase space axes, but this does {\it not} imply adiabaticity since the isocurvature present in the $\chi$ field remains. This scenario is very similar to the {\it curvaton} scenario~\cite{Lyth:2001nq,Moroi:2001ct}.
\end{itemize}

If we are careful of these issues, we can impose the \hca~by eliminating the final terms in eqs.~\eqref{eq:ps_Ni_hca2} and \eqref{eq:ss_Ni_hca2}. Further care is required if the potential $U_J$ has a minimum at some non-zero value, such as for simple models of hybrid inflation with a non-zero vacuum energy. In this case the derivative prefactor $U_J/U_J'$ will diverge as the field $\phi_J$ approaches the minimum.\footnote{If the field $\phi_J$ approaches the minimum of a potential at which $U_J=0$ then $U_J/U_J'$ tends to zero and there is no divergence.} One may side-step the issue, however, by simply allocating the constant part of the potential to the potential component $U_K$ that corresponds to a field that is not purely isocurvature and so $U_K' \neq 0$. A similar issue arises for the product separable formula~\eqref{eq:ps_Ni_hca2}. In this case the resolution is easily made by choosing the free index $J$ such that $U_J' \neq 0$.

\sssec{Large $\fnl$ after natural focusing}

We now show how it is possible to have a large non-Gaussianity in an adiabatic limit in a general class of inflationary models. We focus on the bispectrum parameter $\fnl$ for simplicity. The foregoing discussion implies that eqs.~\eqref{eq:ps_Ni_hca2} and \eqref{eq:ss_Ni_hca2} may be simplified. For sum-separable potentials one finds
\be
\label{eq:sshca}
N_{,I} = \frac{U_I^*}{{U_I '}^*} ,\qquad
N_{,IJ} = \delta_{IJ}\left(1-\frac{U_I U_I''}{{U_I '}^2} \right)_* .
\ee
The general conditions required to achieve a large $\fnl$ from these results may still be complicated. However, a relatively simple picture emerges if we assume that $N_{,I}$ is dominated by one field $\chi$~\cite{Kim:2010ud}. Therefore, $U_\chi^*/{U_\chi'}^*$ dominates the analogous terms for all other fields and $\fnl$ can be written 
\begin{equation}
\frac{6}{5} \fnl \approx - \left. \frac{U_\chi''}{U_\chi} \right|_* .
\label{eq:hilltop-fnl}
\end{equation}
In a single field model, the quantity $U_\chi''/U_\chi$ would be a slow-roll parameter and thus small. But in assisted inflation the total potential may be much larger than $U_\chi$~\cite{Liddle:1998jc,Green:1999vv}. Therefore $\eta_\chi$ can remain small, making $\chi$ light at horizon crossing and causing it to acquire quantum fluctuations by the usual mechanism, while $U_\chi''/U_\chi$ can be appreciable. We will see an example of this with the axion-quadratic model, whose bispectrum evolution is shown in figure~\ref{figAx1}. 

If several fields have comparable $N_{,I}$ then their perturbations contribute equally to $\zeta$ at the adiabatic limit and this will dilute any non-Gaussianity. Therefore the largest values of $\fnl$ will be achieved where a single field has a dominant $N_{,I}$.

For product-separable potentials, eqs.~\eqref{eq:sshca} are only modified by multiplying by a factor of $\delta_{IK}$ where '$K$' must be chosen to such that only the field $\phi_K$ is still evolving at the adiabatic limit. One then finds 
\be
\frac{6}{5} \fnl =2 \ep_K^* - \eta_{KK}^* ,
\ee
and so a large $\fnl$ would require a violation of slow-roll. We conclude that large $|\fnl|$ is not possible at the natural adiabatic limit in this class of models.

\ssec{Classification of adiabatic regimes}

Regardless of the details by which an adiabatic limit is achieved, one can identify three broad classes of behaviour:
\begin{enumerate}
\item The focussing of the bundle may occur during slow-roll inflation. If this is the case then it is not necessary to specify details of the subsequent phases such as reheating. Additionally, because the slow-roll dynamical attractor is operative, the relevant phase space is simplified.  We can then employ the analytic tools developed in \S\ref{sec:analytics} to obtain robust predictions for cosmic observables.
\item Alternatively, an adiabatic limit may be achieved during the non-slow-roll epoch, but still before the onset of subsequent phases like reheating. In this case we will find that our analytic tools give a good indication of the qualitative evolution of cosmic observables, but that we must resort to numerical methods to obtain reliable quantitative results.
\item Finally it is possible that no adiabatic limit is reached during the epoch dominated by scalar-fields. In such cases the inflationary model does not make unambiguous predictions by itself, and must be embedded in a larger scenario which determines at least the mechanism by which inflation ends and the universe reheats~\cite{Huston:2011fr}. In this case we must choose among the various scenarios for reheating and later dynamics, and the evolution of cosmological observables may depend on our choice. 
\end{enumerate}
We now consider examples to illustrate these three different scenarios, for which we will compare the analytic predictions of \S\ref{sec:analytics} with numeric results derived via the finite difference method.

\sec{Slow-roll adiabatic limit}
\label{sec:model1}

This class of inflationary models achieves an adiabatic limit before the slow-roll approximation fails. We consider two examples of this class: A model with interrupted dynamics and a model with a `natural' end to inflation.

\ssec{Transitory model with interruption}

This example begins with a simple slow-roll two-field model that produces a large transient enhancement of $\fnl$. Such models have been studied by a number of authors~\cite{Alabidi:2006hg,Byrnes:2008wi,Byrnes:2008zy} who identified that the non-Gaussianity would typically grow and decay in a transitory `spike' feature. We were the first to explain the dynamical mechanism behind this behaviour in ref.~\cite{Elliston:2011dr}, details of which appear in chapter \ref{ch:heatmaps}. A generic feature of these models is the considerable fine-tuning of initial conditions required for the spike to attain large magnitudes. In two-field models with separable potentials, the parameter combinations required to ensure these conditions were given by Byrnes et al.~\cite{Byrnes:2008wi}. Chapter \ref{ch:heatmaps} will also review, simplify and extend this analysis. In addition to this two-field dynamics, one often requires a mechanism to terminate inflation and one must tune the relevant parameters such that this interruption occurs whilst the non-Gaussianity is large. 

To be specific, let us consider a model of two-field hybrid inflation studied by Alabidi and Lyth~\cite{Alabidi:2006hg} and later by Byrnes et al.~\cite{Byrnes:2008wi,Byrnes:2008zy}. The potential for this model is given by
\begin{equation}
	\label{eq:hybrid}
	V
	=
	\frac{1}{2} m_\phi^2 \phi^2
	+ \frac{1}{2} m_\chi^2 \chi^2
	+ \frac{1}{2} \left(
		g_\phi^2 \phi^2 \psi^2
		+ g_\chi^2 \chi^2 \psi^2
	\right)
	+ \frac{1}{4} \lambda \left ( \psi^2 - v^2 \right)^2,
\end{equation}
where $\phi$ and $\chi$ are slowly-rolling fields, and $\psi$ is a waterfall field. The waterfall field $\psi$ becomes destabilized when ${g_\phi}^2 \phi^2 + {g_\chi}^2 \chi^2 = \lambda v^2$ and rolls rapidly, terminating inflation. We take the masses $m_\phi$ and $m_\chi$ to be positive, and assume $g^2_\phi/g^2_\chi = m^2_\phi/m^2_\chi$. This ensures that the waterfall occurs at a fixed energy density, making it unnecessary to account for the effect of inflation ending on different hypersurfaces~\cite{Alabidi:2006wa,Sasaki:2008uc,Naruko:2008sq,Huang:2009vk}. 

If the masses are not equal, there is a steep slope in the direction of the more massive field. The trajectories evolve along this steep direction and then turn towards the global minimum and this turning leads to a large `spike' of transitory non-Gaussianity. Later, in \S\ref{sec:shapes}, we shall discuss and explain why valley shapes lead to {\it positive} spikes in $\fnl$.

\begin{figure}[h] 
\center{\includegraphics[width = 0.7\textwidth]{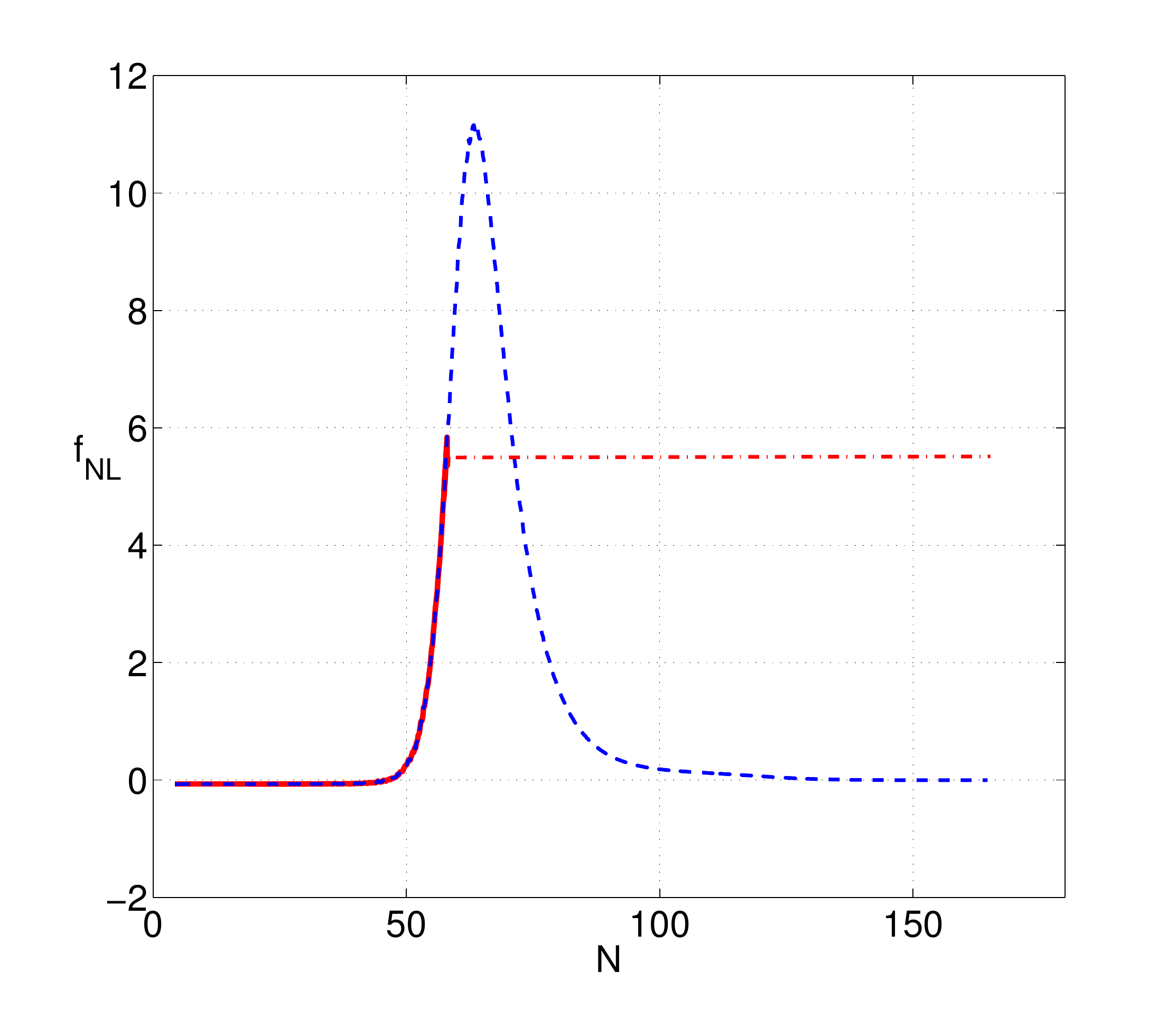}}
\caption[Slow-roll evolution of $\fnl$ in hybrid inflation]{Evolution of $\fnl$ for the hybrid inflation model defined in eq.~\eqref{eq:hybrid}. The blue dashed line represents the analytical approximation with no hybrid transition included. The solid red line is the numerical calculation, including the hybrid transition, for the parameter values in the text. The red dot-dashed line is added to illustrate the constant final level of $\fnl$.}
\label{figHy}
\end{figure}

\begin{figure}[h] 
\center{\includegraphics[width = 0.7\textwidth]{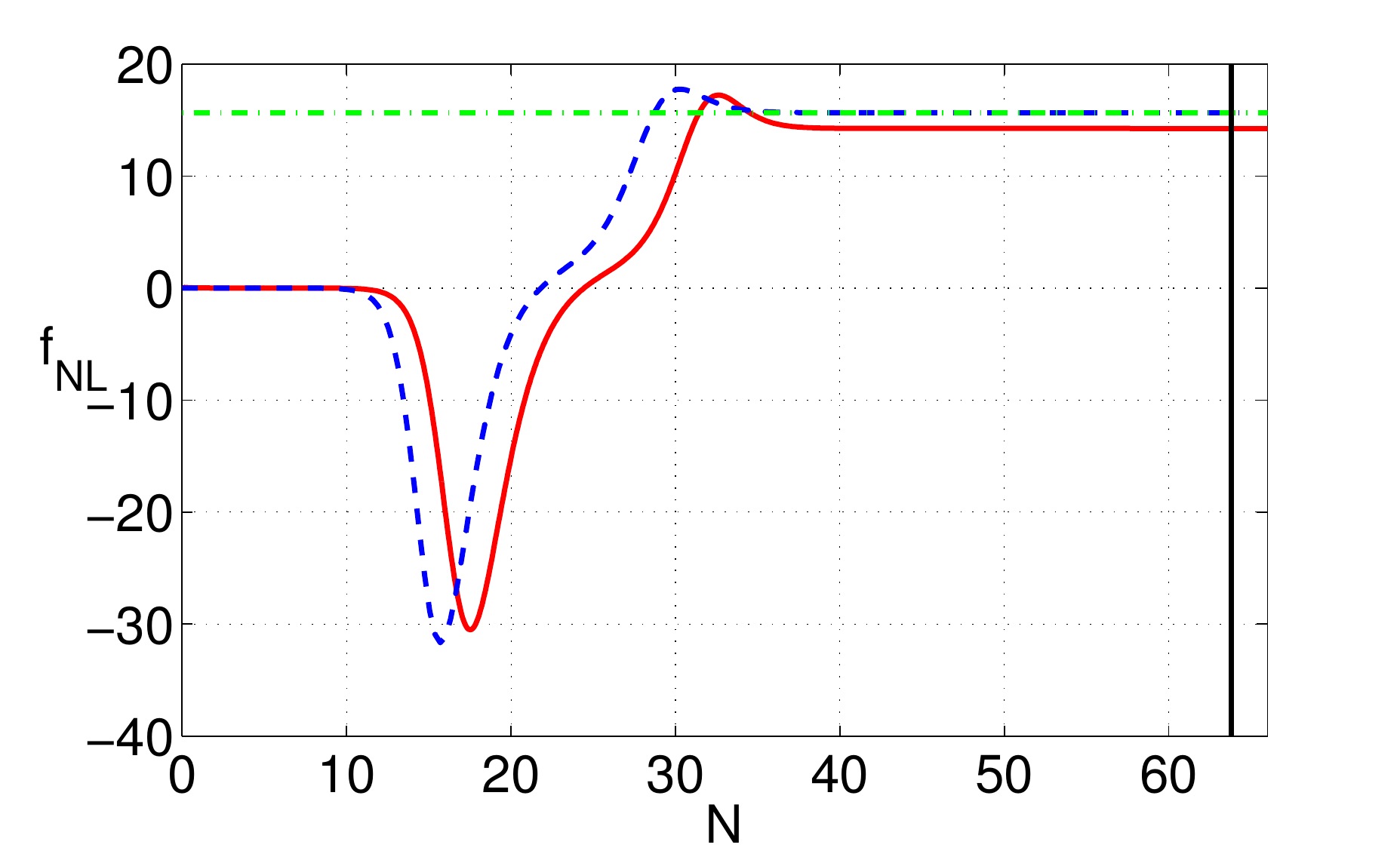}}
\caption[Slow-roll evolution of $\fnl$ for the quadratic-axion model]{Evolution of $\fnl$ for the quadratic axion model defined in eq.~\eqref{eq:2-field-axion} with the parameter choices as stated in the text. The solid red line is the numerical calculation, the blue dashed line is the slow-roll analytic approximation and the horizontal green dashed line represents the \hca~value. The vertical black line shortly after $N=60$ efolds denotes the end of inflation where it is clear that the slow-roll approximation must fail. Since the interesting dynamics occur significantly to the left of this line, slow-roll is a valid approximation and therefore it is unsurprising that the analytic calculation is a good approximation of the numerical evolution.}
\label{figAx1}        
\end{figure}

In figure~\ref{figHy} we plot the evolution of $\fnl$ for $m_\phi/m_\chi = 5$, $\eta_\phi = 4 \Mpl^2 m_\phi^2/ (\lambda v^4) = 0.08$ and initial conditions $\chi_* = 0.001~\Mpl$, $\phi_*=0.5~\Mpl$. We adjust the remaining parameters so that the waterfall occurs when $|\fnl| > 1$ and the waterfall takes much less than a Hubble time to complete. The blue dotted line represents the $\fnl$ calculated using the analytic techniques developed in \S\ref{sec:analytics}. This calculation considers the slow-roll fields only, ignoring the waterfall, and so inflation does not end and the bispectrum eventually decays towards negligible values. 

The solid red line represents a numerical evolution, terminated by the waterfall transition on the growing arm of the spike. We adopt the simple finite difference scheme to compute the derivatives of $N$, as discussed in \S\ref{sec:computation_techniques}. This requires the slow-roll approximation at horizon crossing, where initial conditions are set, but not subsequently. We have verified that our results are insensitive to changes in the step size of the finite difference scheme. Although slower than other approaches \cite{Mulryne:2009kh,Mulryne:2010rp}, the finite difference method has the advantage of straightforward comparison with analytic $\delta N$ methods. Moreover, it requires only the evolution  of an unperturbed universe, making a simple description of reheating---assuming thermal equilibrium and a single radiation fluid---easy to implement. These assumptions are at best quasi-realistic, but serve to model a plausible phenomenology.

As can be seen from figure~\ref{figHy}, before the hybrid transition, the numerical results follow the analytic prediction as expected. Inflation ends when the waterfall field $\psi$ becomes operative. Care must be taken in modelling this transition. We provide the waterfall field with a small value consistent with the root-mean-squared ({\sc rms}) value expected from quantum mechanical excitations of a massive field in de Sitter, $\psi_{\textsc{rms}} \approx H^3/M $, where $M$ is representative of the waterfall mass before the transition. The results are extremely insensitive to its precise value.\footnote{In reality, this {\sc rms} value is made up of many inhomogeneous short scale modes. Their collective evolution approximates that of a homogeneous mode, at least in the initial stages before the minimum is reached \cite{GarciaBellido:1996qt}. We expect this approximation captures at least some of the physics which occurs at the hybrid transition.}

The waterfall field is heavy at horizon crossing and is therefore unperturbed. Hence, we need not differentiate $N$ with respect to $\psi$. Using these assumptions, we find that $\fnl$ appears to be conserved through the hybrid transition (see figure~\ref{figHy}). The numerical evolution is only continued for a fraction of an efold after the transition, during which time $\fnl$ \emph{does} evolve due to the oscillating scalar fields. However, the resulting oscillations in $\fnl$ are decaying and are centred around a fixed value. This behaviour appears to be generic for a range of parameter values, provided the transition happens sufficiently rapidly (in less than an efold). One must already impose this `rapid transition' condition to avoid issues with primordial black holes~\cite{GarciaBellido:1996qt}.

This model illustrates a very simple point: If there is a transient growth of non-Gaussianity then it is often possible to impose some auxiliary mechanism to terminate inflation and allow this non-Gaussianity to be attained in the imposed adiabatic limit. Since the evolution up to this point is well-approximated by slow-roll, we are able to reliably employ analytic tools to predict the non-Gaussianity.

\ssec{Large non-Gaussianity at a `natural' adiabatic limit}
\label{sec:large-natural}

We now consider a two-field inflationary model in which an adiabatic limit may be attained during slow-roll inflation without any need to employ a secondary mechanism to terminate the dynamics.

This case can be illustrated using a model closely related to the $N$-axion model of Kim et al.~\cite{Kim:2010ud}, in which the potential assumes the form
\be
\label{eq:naxion}
V=\sum_I \Lambda_I^4 \, \left[ 1-\cos \left(\frac{2\pi \phi_I}{f_I} \right) \right].
\ee
The sum is taken over a large number of uncoupled axions $\phi_I$ with decay constants $f_I$. Each axion potential has a hilltop at $\phi_I = f_I/2$ and the field rolls to the minimum at $\phi_I = 0$. In ref.~\cite{Kim:2010ud} the initial conditions were chosen randomly, with only a small number accessing the hilltop region. The axions collectively contribute to a phase of assisted inflation, which ends gracefully when the final field to roll exits from slow-roll. The asymptotic $\fnl$ can be moderate or large. We will reconsider this potential in \S\ref{sec:model2}. 

In this section we illustrate this case with the help of a related effective model consisting of two fields. Dynamically, the large number of axions which begin away from the hilltop region serve only to source the Hubble rate and so these may be effectively replaced with a single quadratic field. We retain a single cosine field near the hilltop that will source the non-Gaussianity. (This model has some similarity to the scenario of Boubekeur and Lyth \cite{Boubekeur:2005fj}.) Our potential therefore takes the form
\begin{equation}
\label{eq:2-field-axion}
V= \frac{1}{2} m^2 \phi^2 + \Lambda^4 \left (
1-\cos \frac{2 \pi \chi}{f}
\right) ,
\end{equation}
where $\Lambda$ and $f$ are constants.

In figure~\ref{figAx1} we show a numerical evolution of $\fnl$ for this model with parameters $f= \Mpl$ and $\Lambda^4 = 25 m^2 f^2/(4\pi^2)$, which makes the mass of the axion five times greater than the mass of $\phi$. The initial conditions are $\phi_*= 16~\Mpl$ and $\chi_*=(f/2-0.001)~\Mpl$. The cosine and quadratic together define a potential with adjacent ridge and valley features. Later, in \S\ref{sec:shapes}, we shall discuss and explain why ridge shapes lead to {\it negative} spikes in $\fnl$. As a consequence of its large mass, the axion rolls off the ridge quite early and starts rolling down the bottom of the valley whilst slow-roll is still a good approximation. The positive isocurvature mass in the valley focusses the bundle towards a {\sc 1d} line and so enforces an adiabatic limit. In this case the difference between the slow-roll calculation and the full numerical evolution is at the level of a few percent, consistent with the accuracy of the slow-roll approximation.

\sec{Non-slow-roll adiabatic limit}
\label{sec:model2}

This class of models reach an adiabatic limit during the epoch in which the scalar fields dominate, but only after the slow-roll approximation fails. 

\begin{figure}[h]
\center{\includegraphics[width = 0.7\textwidth]{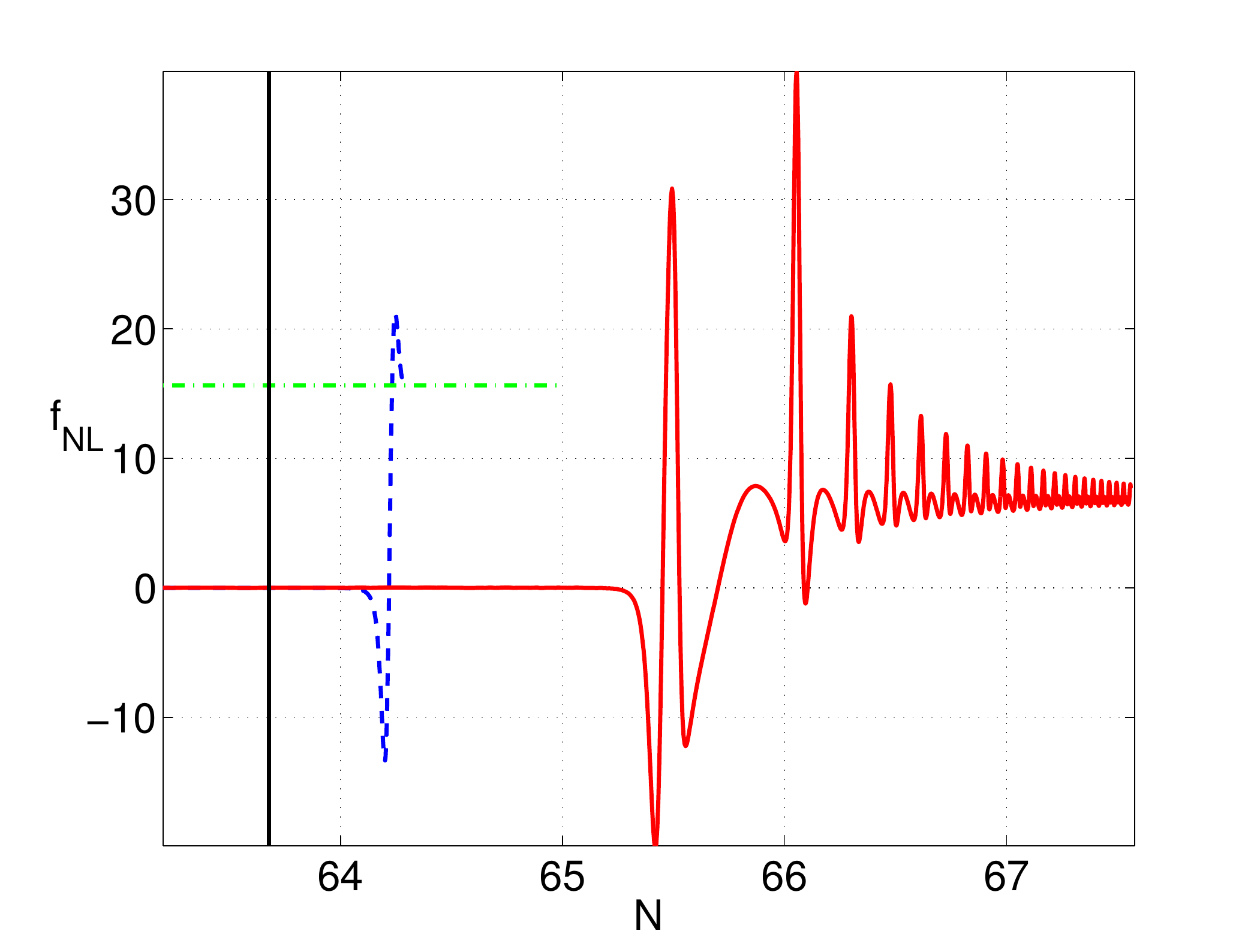}}
\caption[Non-slow-roll evolution of $\fnl$ with adiabatic limit]{Evolution of $\fnl$ for the potential~\eqref{eq:2-field-axion} with the parameter choices given in \S\ref{sec:model2}. The solid red line is the numerical evolution and the blue dashed line is the analytic calculation. Only the final few e-folds are shown. From the vertical black line---which identifies the end of inflation	computed using the exact equations of motion---one can see that the interesting dynamics occur outside of the slow-roll regime and so it is unsurprising that the analytic approximation is quantitatively unreliable. The horizontal green dashed line represents the \hca~value.}
\label{figAx2}        
\end{figure}

The quadratic-axion potential \eqref{eq:2-field-axion} may exhibit this behaviour if the parameter choices are suitably chosen. Figure~\ref{figAx2} compares the analytic and numerical evolution of $\fnl$ for $f= \Mpl$ and $\Lambda^4 =  m^2 f^2/(4 \pi^2)$, giving both fields the same mass. In this case the axion starts to evolve only near the end of inflation, when $\phi$ is approaching the minimum. Indeed, much of the axion's evolution takes place while $\phi$ is oscillating. In these circumstances the adiabatic limit cannot be calculated analytically. Nevertheless, the important features can still be understood. While $\phi$ is oscillating near the minimum, its potential energy contributes to the energy density in a way not accounted for by the slow-roll approximation. If we suppose the $\phi$ oscillations do not lead to rapid reheating or preheating, we may expect $\zeta$ to approach a constant as the trajectories settle in the minimum and Hubble friction drains their energy. In this simple example, $\fnl$ oscillates around an asymptotic value which is lower than would be expected if the fields obeyed slow-roll evolution throughout. In more sophisticated examples, where complex dynamical behaviour can occur during the oscillating phase, it would be necessary to follow their decay in precise detail~\cite{Traschen:1990sw,Kofman:1994rk,Shtanov:1994ce,
Chambers:2007se,Chambers:2008gu,Bond:2009xx,Kohri:2009ac}.

\ssec{$N$-axion model}
\label{sec:n-field-models}

We now briefly consider the $N$-axion model of Kim et al.~\cite{Kim:2010ud} with the potential~\eqref{eq:naxion}. Where the parameters $\Lambda_I$ and $f_I$ take common values $\Lambda$ and $f$ for each species, and $f \lesssim \Mpl$, this generates a large $\fnl$ at the adiabatic limit. Although larger $\fnl$ can na\"{\i}vely be obtained by decreasing $f$, it is necessary to simultaneously increase the number of fields in order to obtain sufficient inflation. This presents computational challenges, especially for the finite difference scheme. 

\begin{figure}[htb] 
\center{\includegraphics[width = \textwidth]{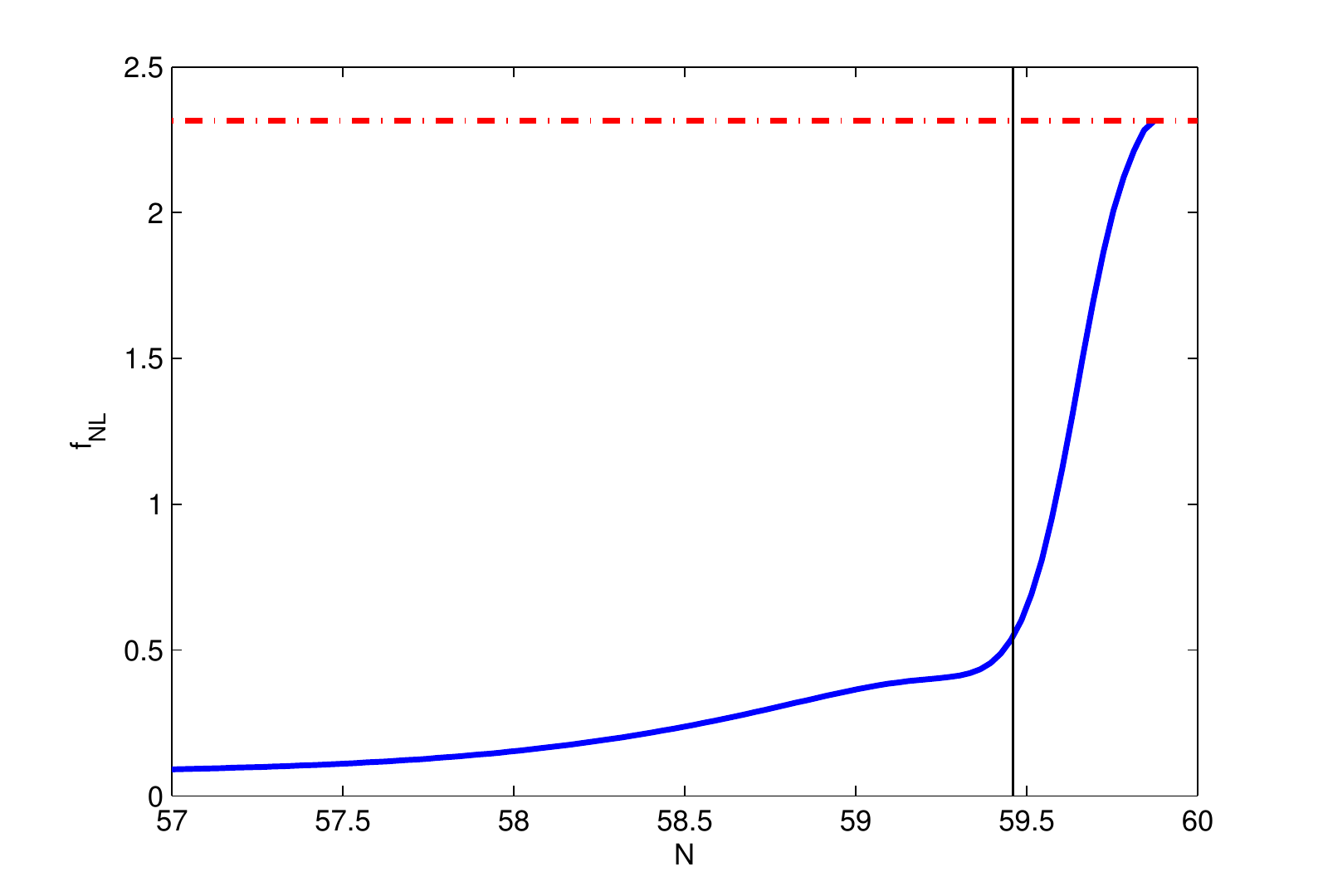}}
\caption[Evolution of $\fnl$ for the $N$-axion model]{Evolution of $\fnl$ for the $N$-axion model. The solid blue line shows the analytic calculation which employs the the slow-roll approximation. The horizontal dot-dashed red line is its asymptotic \hca~value. The vertical black line corresponds to $\epsilon = 1$. Since $\fnl$ has not reached the adiabatic limit in advance of this point, a numerical analysis is required to obtain a reliable value for $\fnl$.}
\label{fig:naxion}    
\end{figure}

There is also another difficulty. As $f$ decreases, the approach of $\fnl$ to its asymptotic limit occurs later in the evolution. In figure~\ref{fig:naxion} we show one realisation of this behaviour for 1800 fields and $f = \Mpl$, with initial conditions for the fields randomly distributed in the range $0 < \phi_I < \pi \Mpl$. The slow-roll phase ends, at the latest, when $\epsilon = 1$ which is marked by the vertical black line. The evolution near to and rightwards of this line is not trustworthy and should be replaced by a numerical calculation. 

There is a specific case where this model can be related exactly to the two-field axion-quadratic model of eq.~\eqref{eq:2-field-axion}: One can place all but one of the scalar fields to be initially close to the minimum of the axion potential, $\phi_I\ll f/2$, with identical initial conditions. These fields then act like a large number of fields with a quadratic potential. Since they all evolve from an identical initial condition, the dynamics of the many fields is completely identical to the dynamics of a \emph{single} field $\Phi^2=\sum_I \phi_I^2$, with a quadratic potential of the same mass as the individual $\phi_I$ fields. The remaining field can be initially placed close to the hilltop. For $f=\Mpl$, this reduces identically to the two-field axion-quadratic model studied in figures~\ref{figAx1} and~\ref{figAx2}.

Because of the computational challenge related to the large number of fields, Kim et al.~\cite{Kim:2010ud} employed the \hca. As we have discussed, this will represent a valid prescription for attaining the adiabatic value of $\fnl$, provided that the adiabatic limit is reached before the assumption of slow-roll is violated. Our analysis has shown that it is possible that slow-roll is violated before $\fnl$ reaches the adiabatic value. In this case, we still expect the $N$-axion model to produce a large adiabatic value of $\fnl$, but the quantitatively accurate value of $\fnl$ can only be determined numerically. 

Finally, it is interesting to note that the $N$-axion model produces a large $\fnl$ without any need to fine-tune the initial conditions on the field space. It is interesting to study whether or not the fine-tuning of initial conditions required to give large $\fnl$ in two-field models may be reduced in models with many fields. 

\sec{No inflationary adiabatic limit}
\label{sec:model3}

Finally we consider a model that does not reach an adiabatic limit in the phase space described by purely scalar field dynamics. We consider the two-field model of Byrnes et al.~\cite{Byrnes:2008wi}
\begin{equation}
	\label{eq:no-reconverge}
	V= V_0 \, \phi^2 e^{-\lambda \chi^2} .
\end{equation}
In ref.~\cite{Byrnes:2008wi} the dynamics were followed only until the end of slow-roll, at which time the non-Gaussianity is large. However, at this point, the isocurvature modes are not exhausted and the curvature perturbation is still evolving. To study this model, we make the the same parameter choices as Byrnes et al, setting $\lambda=0.05\, \Mpl^{-2}$, $\phi_*=16\, \Mpl$ and $\chi_*=0.001\, \Mpl$.

\begin{figure}[h]
\center{\includegraphics[width = 0.7\textwidth]{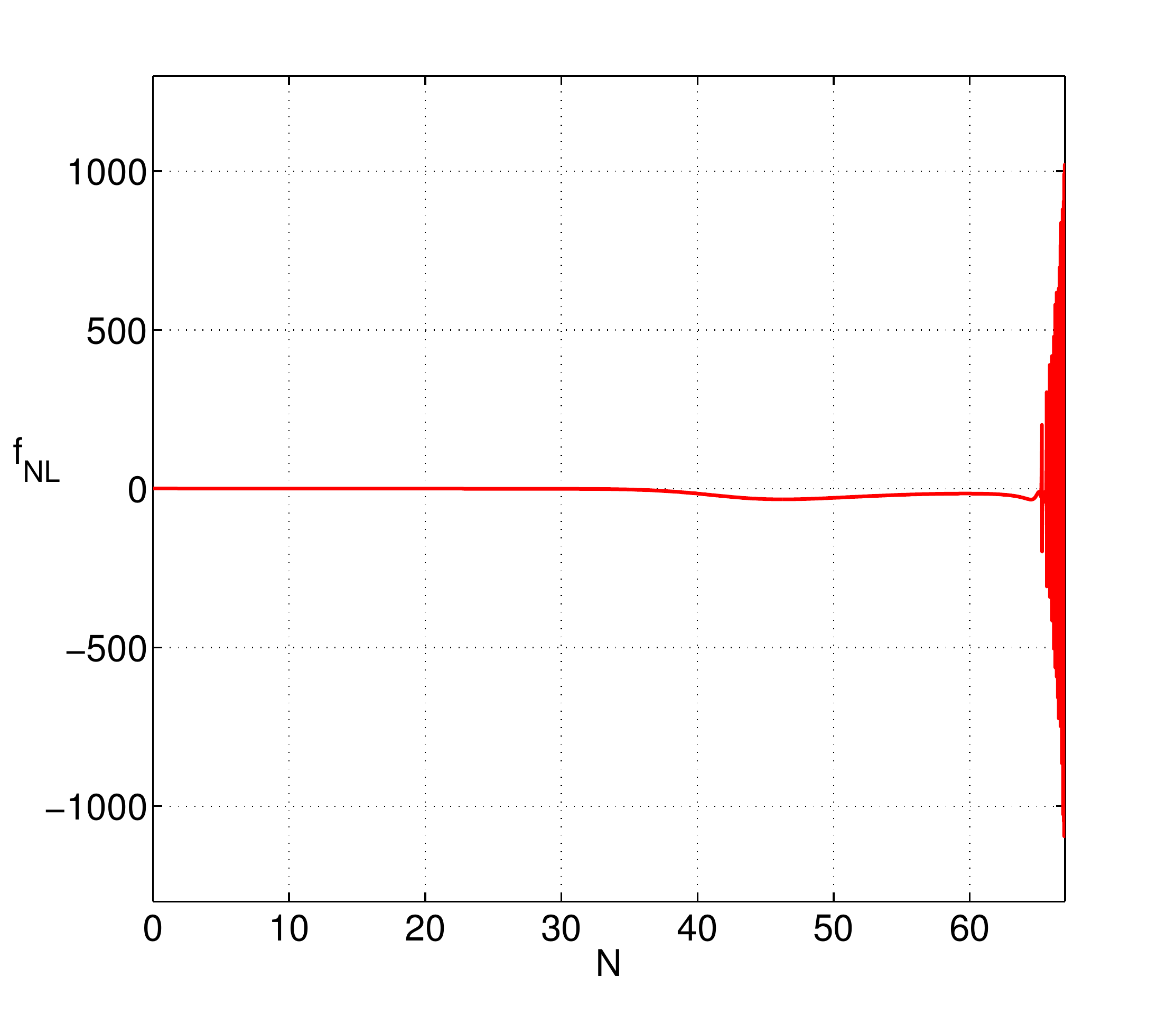}}
\caption[Evolution of $\fnl$ without an adiabatic limit]{Numerical evolution of $\fnl$ for the potential~\eqref{eq:no-reconverge} without reheating. Oscillations about the degenerate vacuum lead to growing oscillatory non-Gaussianity and the model is not predictive.}
\label{figChris2}  
\end{figure}

The initial evolution involves descent from a ridge in the potential landscape which leads to a negative `spike' of $\fnl$. The fields then evolve into and begin oscillations about the degenerate vacuum defined by the line $\phi=0$. This does not represent an attractor. This behaviour fails to focus the bundle either before or after the slow-roll approximation fails. Consequently, $\fnl$ continues to evolve and subsequently oscillates wildly. We illustrate this evolution in figure~\ref{figChris2}.

\begin{figure}[h]
\begin{center}
\includegraphics[width=0.7\textwidth]{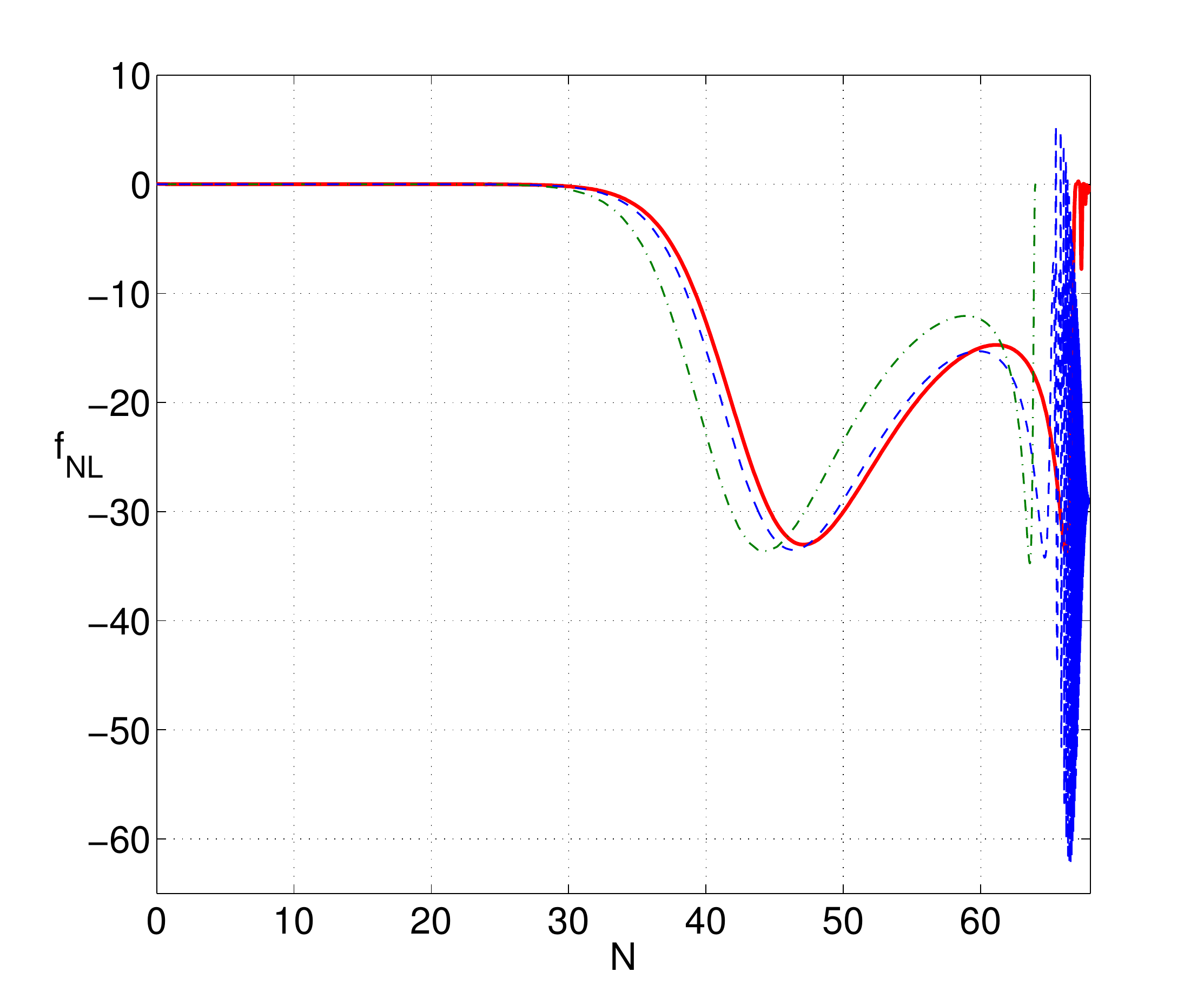}
\includegraphics[width=0.7\textwidth]{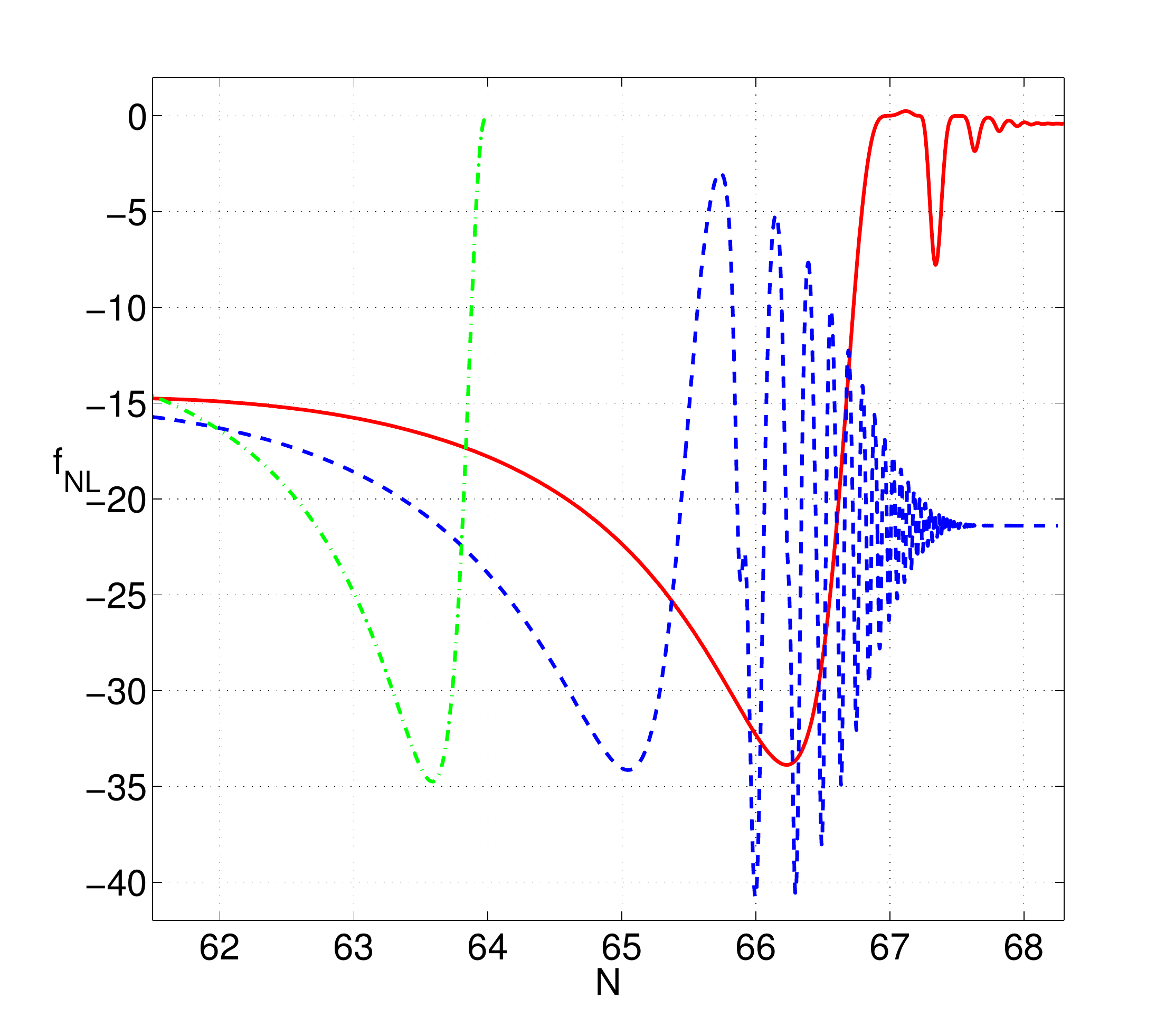}
\caption[Evolution of $\fnl$ to an adiabatic limit imposed by reheating]{Top panel: The solid red line is the numerical evolution of $\fnl$ for the parameter values quoted in the text and $\Gamma = (V_0/10)^{1/2}\, \Mpl^{-1}$. The blue dashed line represents the corresponding plot with $\Gamma = (V_0/100)^{1/2}\, \Mpl^{-1}$, and the green dot-dashed line represents the analytical evolution. The analytic evolution terminates when the $\chi$ field reaches zero, because the slow-roll expressions is unable to evolve past this point. \\
Bottom panel: Magnification of the evolution in the vicinity of the end of inflation. The asymptotic value of $\fnl$ depends on $\Gamma$,	and therefore on microphysical details of the reheating phase.}
\label{fig:Chris1}
\end{center}
\end{figure}

To reach an adiabatic limit we must apply a prescription for reheating. As we shall see, the details of such mechanisms can have important consequences on the final value of cosmological observables. Here, we adopt a very simple perturbative model in which energy is transferred from the field into a single radiation component. The dynamical equations are in this case given by
\begin{align}
	\ddot{\phi}_I + 3 H \dot{\phi_I} & = - \Gamma_I \dot{\phi_I}
		- \frac{\partial V}{\partial \phi_I}, \\
	\dot{\rho} & = - 4 H \rho +\sum_I \Gamma_I \dot\phi_I^2 ,
\end{align}
where $\rho$ is energy density of radiation, and the $\Gamma_I$ represent the decay rate from species $I$. We illustrate the effect of reheating in figure~\ref{fig:Chris1}. The final value of $\fnl$ is sensitive to our choice of $\Gamma_I$, and hence the time-scale of reheating. We take $\Gamma_I = \Gamma$ for both fields, making reheating begin approximately when $H=\Gamma$ and take place on a uniform density hypersurface. After reheating, if the radiation is the only contribution to the energy density, then the statistics of $\zeta$ at this time will be the ones relevant for observation. Figure~\ref{fig:Chris1} indicates that these will depend on microphysical details of the reheating phase, at least through $\Gamma$. This analysis appeared in Elliston et al.~\cite{Elliston:2011dr}. The dependence on reheating parameters has since been studied in greater detail by Leung et al.~\cite{Leung:2012ve}.

Our aim here has not been to present a realistic inflationary model. Rather, we wished to demonstrate that, if no attractor exists within the inflationary regime, we must follow the dynamics until all observable quantities stop evolving at the adiabatic limit. We can expect the asymptotic value of each observable to be sensitive to this evolution, including the time scale and details of reheating.

\paragraph{Summary.} The discussions on adiabaticity in this chapter inform us of the domain within which the analytic non-Gaussianity formulae derived in \S\ref{sec:analytics} may be legitimately applied. This chapter therefore provides an essential contextual basis for the subsequent analysis which we now turn to in chapter~\ref{ch:heatmaps}. 

\clearpage{\pagestyle{empty}\cleardoublepage}
\chapter{Heatmaps and potential shapes}
\label{ch:heatmaps}

\begin{addmargin}[0.05\textwidth]{0.05\textwidth}
This chapter analyses and interprets the analytic results of \S\ref{sec:analytics}, with a focus on non-Gaussianity in two-field inflation. For an arbitrary separable inflationary potential, the bispectrum and trispectrum are  respectively analysed in \S\S\ref{sec:bi_heatmaps} and \ref{sec:tri_heatmaps} in terms of graphical {\it heatmaps}. This work follows Elliston et al.~\cite{Elliston:2012wm} and Elliston~\cite{Elliston:2013uga}, where we apply, simplify and extend the methodology of Byrnes et al.~\cite{Byrnes:2008wi}. Throughout, we shall give due consideration to the implications of adiabaticity, following the discussion in chapter \ref{ch:adiabatic}. Our analysis motivates further investigation of how different {\it shapes} in the inflationary potential source non-Gaussianity, which we consider in~\S\ref{sec:shapes}, following Elliston et al.~\cite{Elliston:2012wm,Elliston:2011dr,Elliston:2011et}. Finally we validate our conclusions by comparing to illustratory examples in~\S\ref{sec:models}.
\end{addmargin}

\begin{center}
\partialhrule
\end{center}
\vspace{-3em}
\begin{quote}
\list{}{\leftmargin 1.7cm \rightmargin\leftmargin} \endlist
\begin{center}
{\it ``Great simplicity is only won by an intense moment or by years of intelligent effort.''}
\flushright{\shifttext{1.5cm}{---T.S. Eliot.}}
\end{center}
\end{quote}
\vspace{-1em}
\begin{center}
\partialhrule
\end{center}

\sec{Bispectrum heatmap analysis}
\label{sec:bi_heatmaps}

We now reformulate the analytic results for non-Gaussianity parameters derived in \S\ref{sec:analytics} in eqs.~\eqref{eq:ss_params_fnl}--\eqref{eq:ss_params_gnl} and also in eq.~\eqref{eq:ps_params_fnl}--\eqref{eq:ps_params_gnl}. Our method is an extension of the graphical approach employed by Byrnes et al.~\cite{Byrnes:2008wi} in their study of the $\fnl$ parameter. This approach does not require any particular form of the separable inflationary potential to be specified and so we can obtain the spectrum of possible modes of behaviour which therefore makes it a powerful tool. We begin by reviewing the analysis of the bispectrum of Byrnes et al.~\cite{Byrnes:2008wi}. Usefully we show that it is possible to greatly simplify the analytic expressions they obtained, which allows us to draw stronger conclusions. We then extend the analysis to the trispectrum  in \S\ref{sec:tri_heatmaps} to consider the parameters $\tnl$ and $\gnl$. 

Recently, Peterson and Tegmark~\cite{Peterson:2010np,Peterson:2010mv,Peterson:2011yt} have also undertaken a study of the bi and tri-spectrum parameters in the setting of slow-roll inflation, arriving at compact relationships between $\fnl$, $\tnl$, $\gnl$, and the tilts of the curvature and isocurvature power spectra. Our approach is complimentary to their study, and provides new insights. The key advantage of our method is that our analytic expressions for $\fnl$, $\tnl$ and $\gnl$ are written directly in terms of potential parameters. This means that our results make it easy to understand the evolution of non-Gaussianity in terms of the shape of the inflationary potential. 

\sssec{Variables}

In order to rewrite the non-Gaussianity formulae in eqs.~\eqref{eq:ss_params_fnl}--\eqref{eq:ss_params_gnl} and \eqref{eq:ps_params_fnl}--\eqref{eq:ps_params_gnl} in a simple form it is convenient to define two new variables. The first variable is the angle $\theta$, defined in eq.~\eqref{eq:rotation}, that prescribes the angle of phase space flow. This can be written in terms of the slow-roll parameters $\epp$ and $\epc$, defined in eq.~\eqref{eq:sr_parameters}, as
\be
\frac{\epp}{\ep} =\cos^2 \theta \,, \qquad
\frac{\epc}{\ep} = \sin^2 \theta.
\ee
Since we assume that both fields are monotonically decreasing (which follows from our use of the slow-roll equations of motion), $\theta$ is constrained to lie in the range $0\leq \theta \leq \pi/2$. We note that $\epp$ can be written in terms of the angle $\theta$, and that $\epp^*$ is written in terms of the different angle $\theta^*$. 

Eqs.~\eqref{eq:ss_params_fnl}--\eqref{eq:ss_params_gnl} and \eqref{eq:ps_params_fnl}--\eqref{eq:ps_params_gnl} also involve the quantities $u$ and $w$, defined in eqs.~\eqref{eq:ss_u}--\eqref{eq:ss_v} and \eqref{eq:ps_u}--\eqref{eq:ps_v}, which lie in the range zero to one. It proves convenient to define a second parameter, $\alpha$, to replace these variables as 
\be
u = \cos^2 \alpha \,, \qquad
w = \sin^2 \alpha.
\label{eq:alpha}
\ee
We note that for product-separable potentials, $\alpha = \theta$. The situation is not so simple in the sum-separable case, as we shall discuss in \S\ref{sec:fnl_heatmap_evo}.

Substituting these definitions into the non-Gaussianity parameters in eqs.~\eqref{eq:ss_params_fnl}--\eqref{eq:ss_params_gnl} and \eqref{eq:ps_params_fnl}--\eqref{eq:ps_params_gnl}, we can eliminate the variables $u,w,\epp$ and $\epc$ in favour of $\alpha,\theta$ and $\ep$. One then finds $\fnl$, $\tnl$ and $\gnl$ are only functions of $\alpha,\theta$ and $\theta^*$, as well as $\ep$ and the other slow-roll parameters. Some of these slow-roll parameters are evaluated at horizon crossing, whilst others are evaluated on a later uniform energy density hypersurface, usually labelled `$c$', but we have dropped this label for simplicity. Quantities without a `$*$' label are therefore assumed to be calculated on the later-time uniform density hypersurface.

\ssec{Simplifying the bispectrum}

For the $\fnl$ parameter, the procedure outlined above leads to the expressions 
\be
\bal{3}
\ds{\frac{6}{5} \fnl} &= f_1 \ep^* - f_2 \eta_{ss}^* + f_3
\eta_{\sigma s}^* + 2 f \, \Omega \, (\eta_{ss}-\ep) \qquad && \mbox{---for sum-separable} \,, \vspace{2mm}\\
\ds{\frac{6}{5} \fnl} &= - f_2 \eta_{ss}^* + f_3 \eta_{\sigma s}^* +
2 f  \eta_{ss} &&\mbox{---for product-separable} \,,
\eal
\label{eq:fnl_full}
\ee
where, with some overlap with ref.~\cite{Byrnes:2008wi}, we have defined the functions $f (\alpha,\theta^*)$, $f_i (\alpha,\theta^*)$, $\Lambda$ and $\Omega$ as
\begin{align}
f &= \ds{\frac{\sin^2 2\alpha}{4 \Lambda^2} (\cos^2 \alpha - \cos^2 \theta^*)^2 \,,} \qquad \qquad & f_2 &= \ds{\frac{1}{\Lambda^2} \left( \cos^6 \alpha \sin^4 \theta^* + \sin^6 \alpha \cos^4 \theta^* \right)} \,, \nonumber \vspace{2mm}\\
f_1 &= \ds{\frac{\sin^2 2\theta^*}{2\Lambda}}  \,, &
f_3 &= \ds{\frac{\sin 2\theta^*}{2\Lambda^2} \left( \cos^6 \alpha \sin^2 \theta^* - \sin^6 \alpha \cos^2 \theta^* \right)}  \,, \nonumber \vspace{2mm} \\
\Lambda &= \ds{\cos^4 \alpha \sin^2 \theta^* + \sin^4 \alpha \cos^2 \theta^* \,,} &
\Omega &= \ds{\frac{V^2}{V_*^2} \frac{\sin^2 2 \theta}{\sin^2 2 \alpha}} \,. 
\end{align}

We have introduced the useful parameter $\Omega$ which is bounded as $0 \leq \Omega \leq 1$. The non-trivial upper limit follows from eqs.~\eqref{eq:ss_u} and \eqref{eq:ss_v} and the associated definitions found in \S\ref{sec:form_sum} to give
\be
\bal{2}
\Omega &= \frac{V^2 \epp \epc}{(U^* \ep +W \epp - U \epc)(W^* \ep - W \epp + U \epc)} \\
&\leq \frac{V^2 \epp \epc}{(U \ep +W \epp - U \epc)(W \ep - W \epp + U \epc)} = 1 \,,
\eal
\label{eq:nifty_relation}
\ee
where the second inequality follows using $U \leq U^*$ and  $W \leq W^*$.

The functions $f_{1 \to 3}$ and $f$ all multiply quantities of $\O (\ep)$ or smaller, and so a necessary, though not sufficient, condition for $\fnl$ to be large is that the magnitude of one or more of these functions is large. We will not specify the value of slow-roll parameters since this would require specialisation to particular models. Our analysis, therefore, identifies only the conditions for which it is \emph{possible} to produce a large $\fnl$ during slow-roll inflation.

Because we are only interested in cases where $\fnl$ can be large, the expressions \eqref{eq:fnl_full} may be further simplified by noting that $|f_1|$ is bounded by order of unity and so the term $f_1 \epsilon^*$ is negligible. Similarly, the term $f_3 \eta_{\sigma s}^*$ is negligible and can be dropped. This latter result follows by noting that we have the freedom to interchange between $\eta_{\sigma \sigma}$, $\eta_{\sigma s}$ and $\eta_{ss}$ via the relations\footnote{These relations only exist because of our assumption of separability; eq.~\eqref{eq:eta_relationship} may be understood as the kinematic frame representation of the equation $\eta_{\phi \chi}=0$ for sum-separable potentials or $\eta_{\phi \chi}=2\sqrt{\epp \epc}$ for product-separable potentials.}
\be
\bal{3}
\eta_{\sigma s} &= \frac{1}{2} \tan 2 \theta \, (\eta_{ss} - \eta_{\sigma
\sigma}) &&\mbox{---for sum-separable} \,, \vspace{2mm}\\
\eta_{\sigma s} &= \frac{1}{2} \tan 2 \theta \, (\eta_{ss} - \eta_{\sigma
\sigma}+2\ep) \qquad \qquad &&\mbox{---for product-separable} \,.
\eal
\label{eq:eta_relationship}
\ee
Considering first the sum-separable case and expanding $f_3 \eta_{\sigma s}^*$ as
\be
 f_3 \eta_{\sigma s}^* = \Big[\sin 2 \theta^* f_3 \Big] \eta_{\sigma s}^* 
+\Big[\frac{1}{2} (1-\sin 2 \theta^*) \tan 2\theta^* f_3  \Big] (\eta_{ss}^*- \eta_{\sigma \sigma}^*),
\label{eq:trick1}
\ee 
then the negligibility of $ f_3 \eta_{\sigma s}^*$ follows from the fact that the terms in square brackets can never become larger than order of unity. A way of understanding how this procedure can work is that $\eta_{\sigma s}^*$ tends to zero in the limits of $\theta^* \to 0, \pi/2$ which counters the divergence in $f_3$. Since eq.~\eqref{eq:eta_relationship} has the same structure for both sum and product-separable potentials, we may neglect the term $f_3 \eta_{\sigma s}^*$ in both cases.

A final simplification follows by noting that $f_2 = 1 + f$ (this follows from a non-trivial proof involving standard trigonometric relations). We arrive at the extremely simple approximate expressions for $\fnl$ 
\be
\bal{3}
\displaystyle{\frac{6}{5} \fnl} &\simeq \displaystyle{f\, \Big[- \eta_{ss}^* +
2 \Omega \, (\eta_{ss} - \ep) \Big]} \qquad \qquad &&\mbox{---for sum-separable}
\,,\vspace{2mm}\\
\displaystyle{\frac{6}{5} \fnl} &\simeq \displaystyle{ 
f\, \Big[- \eta_{ss}^* + 2 \eta_{ss} \Big]} &&\mbox{---for product-separable} \,,
\eal
\label{eq:fnl_direction}
\ee
where $f$ is positive definite. These simpler expressions make it transparent that the condition for $\fnl$ to be large is that $f \gg 1$. We emphasise that these approximate expressions will be highly accurate when $|\fnl|>1$. 

\ssec{Bispectrum heatmap}

\begin{figure}[h]
\centering
\includegraphics[width=0.7\textwidth]{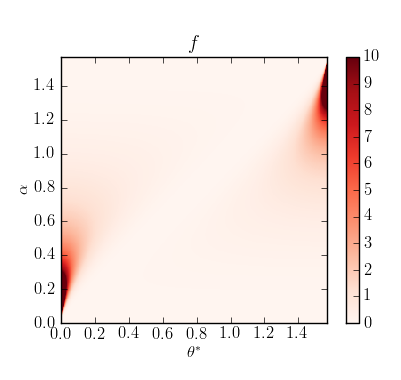}
\caption[Bispectrum heatmap]{Heatmap of the function $f$. Since $\fnl$, $\eta_{ss}$ and $\ep$ are symmetric under field exchange $\phi \leftrightarrow \chi$, the function $f$ must have the same symmetry. We can see this by inverting the heatmap though the point $(\theta^*,\alpha) = (\pi/4 , \pi/4)$ which leaves $f$ invariant.}
\label{fig:f}
\end{figure}

The expressions \eqref{eq:fnl_direction} imply that a necessary condition for $|\fnl| > 1$ is that $f \gg 1$. We can see when this occurs by plotting $f(\alpha, \theta^*)$, as shown in figure~\ref{fig:f}, which we refer to as a {\it heatmap}. One can immediately see that the function is only significant in a small region of the {\sc 2d} $(\alpha, \theta^*)$ space. In some sense this demonstrates the fine-tuning required in order to obtain a large value of $\fnl$ in two-field slow-roll inflationary models. However, to better understand this assertion requires a more thorough understanding of the parameter $\alpha$ which we shall now discuss. 

\sssec{Time-dependence of $\alpha$}
\label{sec:fnl_heatmap_evo}

For product-separable potentials, $\alpha$ and $\theta$ are trivially related as $\alpha = \theta$. The initial (horizon crossing) value of $\alpha$ for product-separable potentials is therefore $\alpha_{\rm init} = \theta^*$. For sum-separable potentials the same initial condition is found. This can be seen by taking the limit $c \to *$ in eqs.~\eqref{eq:ss_u} and \eqref{eq:ss_v}. This means that all separable inflationary models start on the diagonal of the heatmaps at horizon crossing. Since $f=0$ on the diagonal, we know that initially $\fnl$ will be given by the various negligible terms that we have dropped in eq.~\eqref{eq:fnl_direction}, and so $|\fnl| \ll 1$ initially, as we expect~\cite{Seery:2005gb}.

As the model evolves from a given $\theta^*$, $\alpha$ varies such that the model traces a vertical line on the heatmap. For this trajectory to ever intercept one of the regions in which $f \gg 1$, the initial conditions must be such that the initial phase-space velocity is dominated by one of the two fields $\phi$ or $\chi$. Up to an arbitrary field-relabelling, this equates to the demand that $0 < \theta^* \ll \pi/2$. 

Since $\alpha$ has different interpretations in the two different separable scenarios, we consider them separately.

\paragraph{Product-separable.} As an example, let us take a product-separable model with $\theta^* \ll 1$, such that the horizon crossing conditions correspond to a position on the diagonal of the heatmap of $f$ near the lower left-hand corner. If the angle $\theta$ ($=\alpha$) increases towards $\pi/2$, there can be a transient `spike' in $\fnl$ as the model passes through the region in which $f$ is large. If the trajectory turns back, so that $\theta$ decreases, it may well pass back through this region again and another transient signal in $\fnl$ can be produced. Of course, whether or not a significant spike will occur is also dependent on the slow-roll parameters that define the model. The magnitude of this spike in $\fnl$ increases as $\theta^*$ decreases towards zero. We must be careful with the interpretation, however, since our analysis presumes separable potentials and this places strong constraints on the possible modes of behaviour. If the initial field velocity is \emph{exactly} aligned with one of the field axes then it will (classically) remain so indefinitely, since this implies $\theta = \theta^*$ {\it always}. Furthermore, for neighbouring initial conditions with sufficiently small values of $\theta^*$, one may find that $\theta$ will take longer than the duration of observable inflation to grow sufficiently for there to be a significant enhancement of $\fnl$. Consequently it follows that there will be an upper bound on the value $\fnl$ achievable by any such potential.\footnote{There will also be another upper bound on $|\fnl|$ imposed by the quantum scatter of the field near horizon crossing which will prohibit the limit $\theta^* \to 0$ being physically realisable.}

\paragraph{Sum-separable.} To make progress with understanding the dynamics in the sum-separable case, it is necessary to understand the relationship between $\alpha$ and $\theta$. To do this we differentiate eq.~\eqref{eq:alpha} to find 
\be
\alpha' = \frac{V}{V^*} \frac{\sin^2 2 \theta}{\sin^2 2 \alpha} \theta' \,,
\label{eq:alphadash}
\ee
where a prime represents a derivative with respect to an arbitrary time variable. The fractions in eq.~\eqref{eq:alphadash} are positive definite and so $\alpha$
increases as $\theta$ increases and vice versa.

\paragraph{Vacuum dominated sum-separable.} More progress is possible by considering the vacuum-dominated limit, where $V \simeq V^*$. This is a good approximation for some models of inflation such as hybrid inflation. Eliminating the ratio $V/V^*$ from eq.~\eqref{eq:alphadash}, we see that $\alpha \simeq \theta$. Moreover, one also finds $\Omega = 1$ in this limit. Consequently, for vacuum-dominated sum-separable potentials, one may use the heatmaps in the same way as for product-separable potentials. Furthermore, we see that in this case the sum and product-separable formulae for $\fnl$ are identical, apart from the presence of the slow-roll parameter $\ep$ in the sum-separable case. Thus for models with $\eta_{ss} \gg \ep$, there is a very near equivalence between product-separable potentials and vacuum-dominated sum-separable potentials in terms of their contribution to $\fnl$.

\paragraph{General sum-separable.} We now consider eq.~\eqref{eq:alphadash} for general sum-separable models without vacuum-domination. The ratio $V/V^*$ is initially unity and decreases towards zero. This ensures that, whilst the angle is monotonically varying, $\alpha$ lags behind $\theta$. Furthermore, the difference between $\alpha$ and $\theta$ will become more pronounced the smaller the ratio $V/V^*$. When this ratio goes to zero, we see that $\alpha$ remains constant despite any subsequent turning of the trajectory in phase space. Similarly, $\alpha$ will cease to evolve if either of the limits $\theta \to 0$ or $\theta \to \pi/2$ are reached. Physically, these limits correspond to straight lines in the field phase space under which conditions it is well known that $\zeta$ does not evolve. We note that $\Omega$ is zero in any of these three limits and so we are only left with the $\eta_{ss}^*$ term in $\fnl$. This is the approximate \hca~formula for $\fnl$. We note that since $\alpha \neq \theta$, it is quite possible for $f$ to be large when $\alpha$ becomes a constant, and so produce an observationally relevant constant $\fnl$. For a given evolution, the final value of $\alpha$ is readily extracted once the initial and final field values are known, and hence one can check whether the correct value of $\alpha$ is reached in order to give a large non-Gaussianity.

The final case to consider is when the trajectory turns back on itself during its evolution. There is no barrier to constructing sum-separable potentials which exhibit this behaviour for particular evolutions. This will mean that the model moves up and then down a vertical line on the heatmap, perhaps many times. To fully and quantitatively understand how this movement occurs requires a knowledge of how the potential $V$ varies with $\theta$, which is necessarily model-specific. However, it is possible to gain some more detailed intuition for how $\alpha$ varies with $\theta$ by rewriting eq.~\eqref{eq:alphadash} as
\be
h'(\alpha) = \frac{V}{V^*} h'(\theta)\,,
\label{eq:alphadash2}
\ee
where we have defined the function $h(x) = 4x - \sin 4x$, which is monotonic in $x$. One sees that $h(\alpha)$ and $h(\theta)$ increase and decrease simultaneously and hence the velocity $h'(\alpha)$ is always smaller than the velocity $h'(\theta)$. Since $V/V^*$ is constantly decreasing then so is the range of values of $h(\alpha)$ which the evolution can reach. Ultimately $h'(\alpha) \to 0$ and $h(\alpha)$ takes a constant value. Since $h(\alpha)$ is monotonic in $\alpha$ we see that restricting the range of $h(\alpha)$ translates into introducing `excluded regions' at the top and bottom of the heatmaps, the size of which will grow as the potential drops, and ultimately the whole of the heatmap will be excluded except for the final value of $\alpha$. Once again, for a given evolution, the final value of $\alpha$ can be readily calculated, and one can check if it is in the regime $f \gg 1$.

\paragraph{The role of $\Omega$}

The value of $\Omega$ determines which of the terms $\eta_{ss}^*$ or $\eta_{ss}$ dominates in the sum-separable formula for $\fnl$~\eqref{eq:fnl_direction}. At horizon crossing it takes a value of unity. For product-separable or vacuum-dominated sum-separable potentials, $\Omega$ is effectively fixed to be unity for subsequent times, and so which term dominates depends on how $\eta_{ss}$ evolves during the evolution. On the other hand, if $\Omega \to 0$ then the $\eta_{ss}^*$ term will dominate. It is instructive, therefore, to think of sum-separable inflationary models belonging to one of two limiting classes, those for which $\Omega \simeq 1$ throughout and those for which $\Omega \to 0$ at some point. We now briefly consider each of these cases in turn.

Evolutions for which $\Omega \simeq 1$ have vacuum-dominated sum-separable potentials. In the simplest cases of interest, such as falling from a potential ridge or rolling into a vacuum dominated valley~\cite{Elliston:2011dr}, the absolute value of the potential does not change significantly during this phase of evolution. If $\eta_{ss} \simeq \eta_{ss}^*$ then one finds $\frac{6}{5}\fnl \simeq f \eta_{ss}^*$. Since $f$ is positive definite, we see that the sign of $\fnl$ is the same as the sign of the isocurvature mass. Thus a ridge shape ($\eta_{ss} < 0$) leads to a negative $\fnl$ and a valley shape ($\eta_{ss} > 0$) leads to a positive $\fnl$. If inflation does not end abruptly, for example by a hybrid transition, but the fields continue their evolution, then $\fnl$ will continue to evolve until $\theta \to 0$ or $\theta \to \pi/2$. In either of these limits $f\to 0$ and so $\fnl$ is much smaller than unity.

Evolutions for which $\Omega \to 0$ naturally reach a limit in which $\alpha$ becomes constant, and the \hca~becomes a good approximation for the limiting value of observable statistics. This may or may not coincide with a scenario in which the isocurvature perturbations decay and an adiabatic limit is reached. If an adiabatic limit is arrived at, then we can be certain that there will be no further evolution of $\zeta$ and its statistics. When $\Omega \to 0$ we find $\frac{6}{5}\fnl \simeq -f \eta_{ss}^*$. Due to the minus sign, if we begin in a region with a negative isocurvature mass, for example on a ridge, then such a model will eventually produce a positive $\fnl$, whereas if we begin in a region with a positive isocurvature mass, then $\fnl$ will ultimately reach a negative limiting value.

In summary, in the context of these models, the positive $\fnl$ that is marginally preferred by \wmap~data~\cite{Komatsu:2010fb} can be generated in two simple ways: The evolution can begin with a small $\theta^*$ and an initially \emph{negative} isocurvature mass, and then evolve until $\alpha$ naturally takes a small constant value for which $f$ is large. Alternatively, the evolution can evolve from a small $\theta^*$ during which time the isocurvature mass is \emph{positive}, and some mechanism may then interrupt the dynamics whilst $f$ is large.

We now turn to analyse the trispectrum. Our aims are twofold. First, to understand the types of inflationary models and the initial and final conditions for which the observational non-Gaussianity parameters $\tnl$ and $\gnl$ can be large. Secondly, to infer if it is possible to relate these non-Gaussianity parameters to one another, perhaps for specific classes of models. Such relations are potentially important in order to observationally exclude classes of models.

\sec{Trispectrum heatmap analysis}
\label{sec:tri_heatmaps}

The material in \S\ref{sec:bi_heatmaps} is similar to that appearing in Byrnes et al.~\cite{Byrnes:2008wi}, although our analysis is somewhat simpler. We now extend our method to consider the trispectrum which has not been considered before. The trispectrum parameters for sum-separable potentials follow from eqs.~\eqref{eq:ss_params_tnl}--\eqref{eq:ss_params_gnl} and may be written as~\cite{Elliston:2012wm}
\begin{align}
\tnl &=
\tau_1 {\eta_{ss}^* }^2
- \tau_2 \eta_{ss}^* \eta_{\sigma s}^*
+ \tau_3 {\eta_{\sigma s}^*}^2 
- \tau_4 \ep^* \eta_{ss}^*
+ \tau_5 \ep^* \eta_{\sigma s}^* 
+ \tau_6 {\ep^*}^2 
- \tau_7 \, \Omega \, \eta_{ss}^* (\eta_{ss}-\ep)
\nonumber \\ & \qquad 
- \tau_8 \, \Omega \, \eta_{\sigma s}^* (\eta_{ss}-\ep) 
+ \tau_9 \, \Omega \, \ep^* (\eta_{ss}-\ep)
+ 4 \tau \, \Omega^2 \, (\eta_{ss}-\ep)^2 \,, 
\label{eq:ss_tnl_full}\\
\frac{27}{25} \gnl &=
\tau_1 {\eta_{ss}^* }^2
- \tau_2 \eta_{ss}^* \eta_{\sigma s}^*
+ \tau_3 {\eta_{\sigma s}^*}^2 
- \frac{1}{4} \tau_4 \ep^* \eta_{ss}^*
+ \frac{1}{4} \tau_5 \ep^* \eta_{\sigma s}^* 
+\frac{1}{4} \tau_2 {\xi_{sss}^*}^2 \nonumber \\
& \qquad 
-\frac{1}{2} \tau_3 {\xi_{\sigma ss}^*}^2
- \frac{3}{4} \tau_7 \, \Omega \, \eta_{ss}^* (\eta_{ss}-\ep)
- \frac{3}{4} \tau_8 \, \Omega \, \eta_{\sigma s}^* (\eta_{ss}-\ep) \nonumber \\
&\qquad +g_1 \, \Omega^{3/2} \left( \xi_{sss}^2 - 2 \eta_{\sigma s} (\eta_{ss} + \ep) \right) 
+4 g_3 \, \Omega \, \frac{V}{V^*} \cos 2 \theta \, \eta_{ss} (\eta_{ss}-\ep) \,,
\label{eq:ss_gnl_full}
\end{align}
and, for the product-separable potentials, eqs~\eqref{eq:ps_params_tnl}--\eqref{eq:ps_params_gnl} may be written as
\begin{align}
\tnl &=
\tau_1 {\eta_{ss}^* }^2
- \tau_2 \eta_{ss}^* \eta_{\sigma s}^*
+ \tau_3 {\eta_{\sigma s}^*}^2 
- \tau_7 \eta_{ss}^* \eta_{ss}
- \tau_8 \eta_{\sigma s}^* \eta_{ss}
+ 4 \tau \eta_{ss}^2  \,,
\label{eq:ps_tnl_full}\\
\frac{27}{25} \gnl &=
\tau_1 {\eta_{ss}^* }^2
- \tau_2 \eta_{ss}^* \eta_{\sigma s}^*
+ \tau_3 {\eta_{\sigma s}^*}^2 
+ \tau_3 \ep^* \eta_{ss}^*
+\frac{1}{4} \tau_2 {\xi_{sss}^*}^2
-\frac{1}{2} \tau_3 {\xi_{\sigma ss}^*}^2 \nonumber \\
& \qquad 
- \frac{3}{4} \tau_7 \eta_{ss}^* \eta_{ss}
- \frac{3}{4} \tau_8 \eta_{\sigma s}^* \eta_{ss}
+g_1 \left( \xi_{sss}^2 - 2 \eta_{\sigma s} \eta_{ss}\right) 
+4 g_2 \eta_{ss}^2 \,,
\label{eq:ps_gnl_full}
\end{align}
where the various functions occurring in these expressions are defined by
\begin{align}
\tau_1 &= \displaystyle{\frac{1}{\Lambda^3}(\cos ^8 \alpha \sin ^6 \theta^*+\sin ^8 \alpha \cos ^6 \theta^*)}\,, &
\tau_4 &= 2 f_1 f_2 \vspace{2mm}\,, \nonumber \\
\tau_2 &= \displaystyle{\frac{\sin 2 \theta^*}{\Lambda^3}(\cos ^8 \alpha \sin ^4 \theta^*-\sin ^8 \alpha \cos ^4 \theta^*)}\,, \qquad &
\tau_5 &= 2 f_1 f_3\vspace{2mm}\,, \nonumber \\
\tau_3 &= \displaystyle{\frac{f_1}{2 \Lambda^2}(\cos ^8 \alpha \sin ^2 \theta^*+\sin ^8 \alpha \cos ^2 \theta^*)}\,,  &
\tau_6 &= f_1^2\vspace{2mm}\,, \nonumber \\
\tau_7 &= \displaystyle{\frac{4 f}{\Lambda} \left(\cos^2 \alpha \sin^2 \theta^* + \sin^2 \alpha \cos^2 \theta^*\right)}\,,  &
\tau_9 &= 4 f_1 f \vspace{2mm}\,, \nonumber \\
\tau_8 &= \displaystyle{-\frac{ \sin 2 \theta^* \sin^2 2 \alpha}{2\Lambda^2} \left(\cos^2 \alpha - \cos^2 \theta^* \right)}\,,  &
\tau &= \displaystyle{\frac{f \sin^2 2 \alpha}{4 \Lambda}}\vspace{2mm}\,, \nonumber \\
g_1 &= \displaystyle{g_3\sin 2 \alpha }\,, \qquad\qquad
g_2 = \displaystyle{g_3\cos 2 \alpha }\,, &
g_3 &= \displaystyle{-\frac{f}{2 \Lambda}(\cos^2 \alpha - \cos^2 \theta^*)}\,.
\label{eq:ss_tnl_para}
\end{align}

\ssec{Simplifying the trispectrum}

We can now follow similar methods to those we employed previously in our simplification of the bispectrum. A number of the terms in eqs.~\eqref{eq:ss_tnl_full}--\eqref{eq:ps_gnl_full} are negligible if the trispectrum parameters are large enough to detect. Since the trispectrum functions pre-multiply quantities that are second order in slow-roll, in this case we neglect any functions that are never larger than $10$, rather than order of unity. The manipulations involved are found in appendix~\ref{sec:trispectrum_simplification}, and we now quote the results.

\subsubsection{$\tnl$ simplified} 

After discarding terms that are unobservably small, we find the remarkably simple forms for $\tnl$ as
\be 
\bal{3}
\displaystyle{\tnl} &\simeq \displaystyle{{\cal C} \left( \frac{6}{5} \fnl
\right)^2 - \frac{12}{5} \fnl (\eta_{ss}^* - f_1 \ep^*)} \qquad && \mbox{---for sum-separable} \,,\vspace{2mm}\\
\displaystyle{\tnl} &\simeq \displaystyle{{\cal C} \left( \frac{6}{5} \fnl
\right)^2 - \frac{12}{5} \fnl \eta_{ss}^*} &&\mbox{---for product-separable} \,,
\eal
\label{eq:tnl_pre_final}
\ee
where
\be
{\cal C} = \frac{\tau}{f^2} = \frac{\Lambda}{(\cos^2 \alpha - \cos ^2 \theta^*)^2} \,.
\ee
A further simplification can be made by comparing the relative magnitude of the terms in these equations. Noting that ${\cal C} \geq 1$ and that $f_1$ is at most of order unity. The second terms in eqs.~\eqref{eq:tnl_pre_final} will therefore be suppressed relative to the first by at least $\O(\ep^*)$. We thus find
\be
\tnl \simeq {\cal C} \left( \frac{6}{5} \fnl \right)^2 \,,
\label{eq:tnl}
\ee
which is valid for both sum and product-separable potentials.

In ref.~\cite{Smidt:2010ra}, the ratio between $\tnl$ and $\left( \frac{6}{5} \fnl \right)^2$ was parametrized by $A_\mathrm{NL}$. Peterson et al.~\cite{Peterson:2010mv} subsequently showed that $A_\mathrm{NL} = 1/r_c^2$ for two-field models under slow-roll, where $r_c$ determines the fraction of the curvature perturbation which is sourced by the horizon crossing isocurvature mode. Our result, eq.~\eqref{eq:tnl}, is complementary to this analysis, explicitly showing the form of $r_c$ in terms of the dynamics of inflation for separable potentials.

\subsubsection{$\gnl$ simplified} 

Unsurprisingly, $\gnl$ does not simplify so neatly. After discarding terms that are unobservably small, $\gnl$ in the product-separable case may be written as
\be
\frac{27}{25} \gnl \simeq
\tnl \left( \frac{\eta_{ss}^* -\eta_{ss}}{\eta_{ss}^* -2\eta_{ss}} \right) 
-\frac{6}{5} \fnl (2 \eta_{ss}^* + \eta_{ss})-g_4 {\xi_{sss}^*}^2 
+ g_1 \Big[\xi_{sss}^2 - 2 \eta_{\sigma s} \eta_{ss}\Big],
\label{eq:ps_gnl}
\ee
where $g_4 = \frac{1}{4} \left(\tau_3 \sin 2 \theta^* \cos 2 \theta^* -\tau_2 \right)$. For sum-separable potentials we obtain two additional terms, and the expression takes the form
\begin{align}
\frac{27}{25} \gnl &\simeq 
\tnl \left( \frac{\eta_{ss}^* -\Omega \, (\eta_{ss}-\ep)}{\eta_{ss}^* -2 \, \Omega \, (\eta_{ss}-\ep)} \right) 
-\frac{6}{5} \fnl (2 \eta_{ss}^* + \Omega \, (\eta_{ss}-\ep))
-g_4 {\xi_{sss}^*}^2 \nonumber \\
& \qquad + g_1\, \Omega^{3/2} \, \Big[\xi_{sss}^2 - 2 \eta_{\sigma s}
(\eta_{ss}+\ep)\Big] -\frac{1}{2} f_1 f \ep^* \eta_{ss}^* \nonumber \\
& \qquad + 4 g_3 \, \Omega \, (\eta_{ss}-\ep) \left( \frac{V}{V^*}\cos 2 \theta
\eta_{ss} - \Omega \cos 2 \alpha (\eta_{ss} - \ep) \right).
\label{eq:ss_gnl}
\end{align}

These formulae for $\gnl$ are once again complimentary to those derived by Peterson et al.~\cite{Peterson:2010mv}. One advantage of our results is that, because of their explicit nature, we can easily consider which shapes in the inflationary potential will cause the different terms in eqs.~\eqref{eq:ps_gnl} and \eqref{eq:ss_gnl} to become large. We shall return to this theme in \S\ref{sec:shapes}. Another advantage of our formulae is that the \hca~is easily implemented.

\ssec{$\tnl$ heatmaps}

\begin{figure}[h]
\begin{minipage}{0.5\linewidth}
\centering
\includegraphics[width=\textwidth]{6heatmaps/f}
\end{minipage}
\begin{minipage}{0.5\linewidth}
\centering
\includegraphics[width=\textwidth]{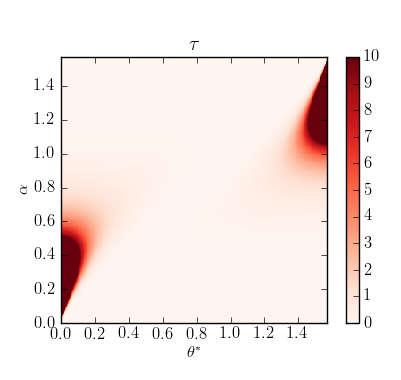}
\end{minipage}
\caption[Heatmaps of $f$ and $\tau$]{Heatmaps of $f$ and $\tau$ on the same scale. We note that both are large in the same regions of the parameter space. We also note that heatmaps of $\tau$ and $f^2$ are visually almost indistinguishable.}
\label{fig:ft}
\end{figure}

Referring to eqs. \eqref{eq:tnl_pre_final} we see that $\tnl$ can only be large when $\tau$ or $f$ are large. We plot these functions side-by-side in figure~\ref{fig:ft}. The fact that both functions peak in similar regions of the parameter space means that the classes of models that are capable of producing a large $\tnl$ are the same as the classes of models that can produce a large $\fnl$. 

\sssec{Simplifying the Suyama-Yamaguchi consistency relation}

\begin{figure}[h]
\centering
\includegraphics[width=0.7\textwidth]{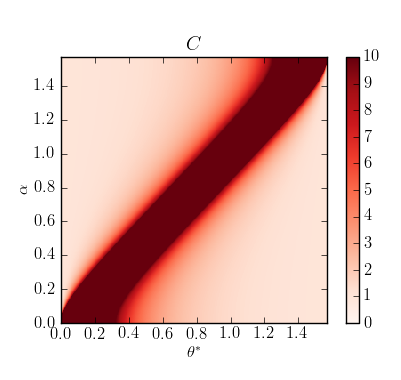}
\caption[Suyama-Yamaguchi heatmap]{Heatmap of the function $\C$ illustrating the conditions required for a model to deviate from the single field consistency result for $\tnl$. This is only physically interesting in the regions for which $f$ is  large which are close to the sides of the heatmap. The region where $\C\gg1$ overlaps with part---but not all of---the region where $f$ is large and one can see that this overlapping region is displaced from the sides of the heatmap.}
\label{fig:C}
\end{figure}

For canonical scalar field inflation, $\tnl$ and $\fnl$ satisfy the condition \cite{Suyama:2007bg}
\be
\tnl \geq \left( \frac{6}{5} \fnl \right)^2 \,,
\label{eq:sf_consistency}
\ee
where equality occurs for single field inflation.\footnote{See also the recent work by Sugiyama \cite{Sugiyama:2012tr} claiming that this equality is broken when contributions are included from all loops.} This is fully consistent with our result \eqref{eq:tnl}, once we recall that $\C\geq1$. We plot $\C$ in figure~\ref{fig:C}. The interesting regions of this plot are those for which $f$ is not small, as shown in figure~\ref{fig:ft}. We see that if a model has $\theta^* \ll 1$ and $\theta$ subsequently increases, then such a model will first enter the region in which $\C\gg1$ and so $\tnl$ will grow whilst $\fnl$ remains small. It is unsurprising that $\tnl$ evolves first, since being associated with a higher order moment, it will be more sensitive to outliers of the $\delta N$ distribution and it is these that will evolve first. For larger $\alpha$, one can see that $\C \simeq 1$ to a very good degree of accuracy, and the single field relation becomes a good approximation. We note that $\tnl$ deviates more from the single field limit $\C = 1$ when $\theta^*$ is fine-tuned to be closer to the  $\theta^*=0$ axis.

\ssec{$\gnl$ heatmaps}
\label{sec:gnlheatmaps}

We now turn our attention to consider how $\gnl$ evolves, and so consider possible relations between $\gnl$ and $\tnl$.

\paragraph{The $\Omega=0$ case.}  First, we consider the case of non-vacuum dominated sum-separable potentials for evolutions which reach $\Omega=0$ such that all of the observational parameters have ceased to evolve. From eq.~\eqref{eq:ss_gnl} we see that 
\be
\frac{27}{25} \gnl \simeq \tnl - \frac{3}{5} \fnl \left(4\eta_{ss}^*- f_1
\ep^* \right) - g_4 {\xi_{sss}^*}^2 \,.
\label{eq:tnl_gnl_full}
\ee
The $\fnl$ term in eq.~\eqref{eq:tnl_gnl_full} is suppressed relative to the $\tnl$ term by a relative factor of $\fnl^{-1} \times {\cal O} (\ep^*)$ and so may be safely neglected. In cases where  ${\xi_{sss}^*}^2$ can be neglected, for example in the absence of significant terms in the potential beyond quadratic order at horizon crossing, we find a relation between $\tnl$ and $\gnl$ as
\be
\frac{27}{25} \gnl \simeq \tnl \,.
\label{eq:gnl_tnl}
\ee

\begin{figure}[h]
\centering
\includegraphics[width=0.7\textwidth]{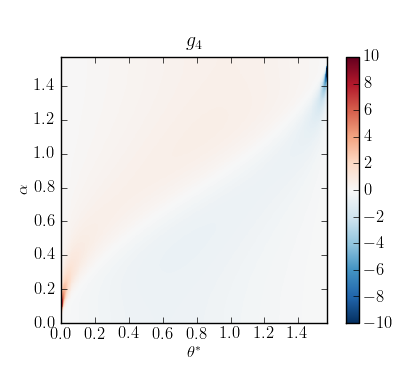} 
\caption[Heatmap of $g_4$]{Heatmap of the function $g_4$ which is antisymmetric about field exchange. The very small lobe in the bottom left hand corner has positive $g_4$, whilst the lobe in the top right hand corner has negative $g_4$. This pattern is repeated for $g_1$ and $g_3$ as seen in figures~\ref{fig:gg}.}
\label{fig:g4}
\end{figure}

Next, let us look at the term $g_4 {\xi_{sss}^*}^2$ and assess when it may be relevant. If an inflationary potential exists for which this term is important, and such a system reaches an adiabatic limit during slow-roll inflation, then it will have a signature such that $\gnl$ deviates from eq. \eqref{eq:gnl_tnl}. The heatmap for $g_4$ is plotted in figure~\ref{fig:g4} and we see that the areas in which $g_4$ is large are very small in comparison to the corresponding areas for the $\tau$ function. Practically speaking, this ensures that one has to tune the parameters of the model to a very high degree in order to access this region. In \S\ref{sec:shapes}, we show that, in the interesting limit where $\theta^*$ and $\alpha$ are small, we can accurately approximate $2 g_4 \simeq \theta^* \tau$. Thus a necessary (but not sufficient) condition for the ${\xi_{sss}^*}^2$ contribution to $|\gnl|$ to be large is that $\theta^* \tau \gg 1$. In addition, for $\gnl$ to deviate from eq. \eqref{eq:gnl_tnl} we require that $L = \theta^* {\xi_{sss}^*}^2 / 2{\eta_{ss}^*}^2$ is not small. Potentials with $L \gg 1$ therefore have the capacity to generate $|\gnl|\gg\tnl$.

Let us consider a sum-separable potential, in the case where $\theta^* \ll1$; to leading order in $\theta^*$ we find
\be
{\eta_{ss}^*}^2 = {\eta_{\chi}^*}^2\,,\qquad
{\xi_{sss}^*}^2 = {\xi_{\chi}^*}^2\,,\qquad
\theta^* = \sqrt{\epc^*/\epp^*}\,.
\ee
Expanding out the slow-roll parameters in terms of potential derivatives, one finds that $L \simeq V_{,\chi} V_{,\chi \chi \chi} / 2 V_{,\chi \chi}^2$.
If one considers a general power-law potential $U(\chi) = U_0 \chi^n$ then $L=0$ for $n=0,1,2$ and $L \leq 1/2$ for $n \geq 3$. On the other hand, for an exponential potential $U(\chi) = U_0 e^{\lambda \chi}$ then $L=1/2$. In these two cases we would therefore not expect to find a deviation from eq.~\eqref{eq:gnl_tnl} beyond a factor of 2. It is interesting to note that a potential of the form $U(\chi) = a \ln (\chi-b)$ for constants $a$ and $b$ has $L=1$ and so for such a potential the two leading order terms in $\gnl$ exactly cancel and so $\gnl$ is of order $\fnl \times {\cal O} (\ep^*) $.

We now ask if there are potentials with $L \gg 1$. Considering polynomial potentials  $V(\chi)$, the necessary condition is for the potentials to possess a linear term, a negligible quadratic term and at least one term beyond quadratic order. The simplest such potentials have a sloping inflection point of the form $U(\chi) = U_0 + h \chi + \frac{1}{6} \lambda \chi^3$. Nearby the inflection point one has $\eta_\chi \simeq 0$ whilst $\theta \, \xi_\chi^2 = \Mpl^3 h \lambda / U_0^2$ and so $L$ diverges. We shall examine such an inflection point further in \S \ref{sec:inflection}.

\paragraph{The $\Omega \neq 0$ case: }

\begin{figure}[h]
\begin{minipage}{0.5\linewidth}
\centering
\includegraphics[width=\textwidth]{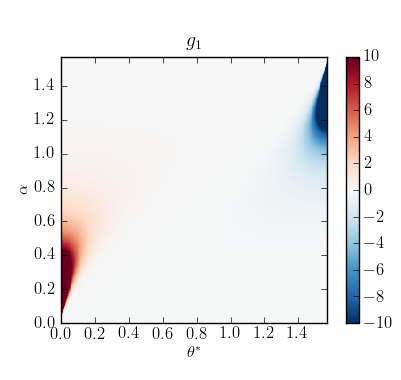}
\end{minipage}
\begin{minipage}{0.5\linewidth}
\centering
\includegraphics[width=\textwidth]{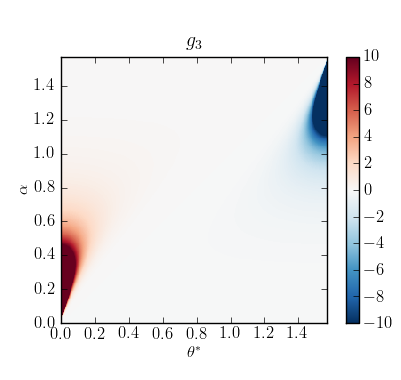}
\end{minipage}
\caption[Heatmaps of $g_1$ and $g_3$]{Plots of $g_1$ and $g_3$ on the same scale as used for the other heatmaps. We note that they are very similar and are anti-symmetric under field interchange. Both are large in the same regions of the parameter space as $\tau$.}
\label{fig:gg}
\end{figure}

It is considerably harder to make concrete statements about the value of the trispectrum parameter $\gnl$, and its relation to $\tnl$, when $\Omega \neq 0$. We can take a step in this direction by plotting the heatmaps for the remaining functions $g_1$ and $g_3$ which appear in the expressions \eqref{eq:ps_gnl} and \eqref{eq:ss_gnl}. These are shown in figure~\ref{fig:gg}. The similarity between these plots, and those of $f$ and $\tau$, tells us that the types of inflationary potential, and initial conditions, that can give rise to a large $\gnl$ are similar to those that may give rise to a large $\fnl$ and $\tnl$. No new regions of interest appear for $\gnl$ which were not present for $\fnl$.

\paragraph{The $\Omega \simeq 1$ case: } Finally, we consider the case $\Omega \simeq 1$. This arises for vacuum dominated potentials, and so can be relevant for models where inflation is terminated suddenly, perhaps through a waterfall transition. During a vacuum dominated phase of evolution, the fields evolve only very slowly, and likewise turns in field space progress slowly. 

It is instructive to consider how the non-Gaussianity parameters given by eqs.~\eqref{eq:fnl_direction}, \eqref{eq:tnl} and \eqref{eq:ss_gnl} may be approximated for $\Omega \simeq 1$, $\theta \ll 1$ and an isocurvature power-law potential of the form $V(\chi) \propto \chi^n$ where we presume $n$ to be a positive integer greater than or equal to $2$. In the regime of interest where $\theta \gg \theta^*$ (such that the functions $f$ and $\tau$ may be large) we require $|\chi| \gg |\chi^*|$. If $n=2$, such that the potential has a parabolic shape, then $\eta_{ss}$ is approximately constant and from eq. \eqref{eq:ss_gnl} for $\gnl$ we see that the first term will be negligible in this case. Furthermore, if the potential is well-described by an expansion to quadratic order we see that the other terms in $\gnl$ are also suppressed with respect to $\fnl$ and so $\gnl$ is not large. This was the case found by Byrnes et al.~\cite{Byrnes:2008zy} in their study of vacuum dominated quadratic potentials. 

However, for $n \geq 3$ the situation is different. The isocurvature mass $\eta_{ss}$ is now a function of $\chi$ and $|\eta_{ss}|$ will grow as $|\chi|$ grows. In the regime of interest, where $|\chi| \gg |\chi^*|$, we find that $\frac{27}{25} \gnl \simeq \frac{1}{2} \tnl$ from the first term in eq.~\eqref{eq:ss_gnl} and so it is clear that $\gnl$ is not generally a vanishing quantity when the potential is described by terms beyond quadratic order. We note that the remaining terms in eq.~\eqref{eq:ss_gnl} for $\gnl$ will mostly be small in this limit, but we will anticipate that the $g_1 \xi_{sss}^2$ term will contribute notably to the value of $\gnl$ during this transient evolution. We shall look at the specific case of $n=3$ in \S\ref{sec:inflectionpot}.

\sec{Shapes in the inflationary potential}
\label{sec:shapes}

An important feature of the results obtained in the previous section is that particular inflationary dynamics---mediated by particular shapes in the inflationary potential---lead to interesting evolution of non-Gaussianity. In this section we consider this correspondence in greater depth, demonstrating how our formulae for $\fnl$, $\tnl$ and $\gnl$ can be used to make both qualitative and quantitative predictions for the evolution of bi and tri-spectra as the fields evolve over generic features in the inflationary potential. Motivated by the analyses in the previous sections, as well as other work~\cite{Alabidi:2006hg,Byrnes:2008zy,Elliston:2011dr,Elliston:2011et}, we consider potentials with one of the following three general features: a ridge, a valley or an inflection point.

An important outcome of our analysis is that it provides simple expressions which give the sign and the peak values of the non-Gaussianity generated by these generic potential features. By Taylor expanding any given potential about the initial conditions, it is often possible to approximate the actual potential by one of these features over a range of field values. The dominant term in such a Taylor expansion will be a constant, justifying the subsequent analysis to presume $\alpha = \theta$. This method allows quantitative predictions to be obtained readily~\cite{Elliston:2011dr}, without the need for detailed calculations in each case.

Our heatmap analysis found that large non-Gaussianity parameters require small angles in the parameter space which correspond to potentials for which the motion is highly aligned to one of the field directions. We take this to be $\phi$ without loss of generality. We can then expand the functions such as $f$ in the limit where $\theta^*$ and $\alpha=\theta$ are small to find
\be
f \simeq \frac{\theta^6}{(\theta^4 + \theta_*^2)^2}\,, \qquad 
\tau \simeq \frac{\theta^8}{(\theta^4 + \theta_*^2)^3}\,,
\ee
and the other parameters are related as $g_1 \simeq \theta \tau$, $2 g_3 \simeq \tau$ and $2g_4 \simeq \theta^* \tau$. We now consider each feature in turn.

\ssec{Ridges}
\label{sec:ridge}

The simplest possible ridge that we can consider takes the form 
\be
V = V_0 + g \phi - \frac{1}{2} m_\chi^2 \chi^2 \,,
\ee
where $g$ and $m_\chi$ are taken to be positive and the $V_0$ term dominates. Since the potential is vacuum dominated and $\Omega \simeq 1$, the product-separable and sum-separable formulae for $\fnl$, $\tnl$ and $\gnl$ are identical if $\ep \ll \eta_{ss}$. To be consistent with our sign convention we stipulate that $\chi <0$ such that $V_{,I} >0$ and $0 \leq \theta \leq \pi/2$. The initial conditions are fine-tuned such that the field initially moves almost parallel to the top of the ridge, with $\theta \gtrsim 0$. The isocurvature direction is therefore almost precisely the $\chi$ direction, and $\eta_{ss} \simeq -m_{\chi}^2/W_0$ is a constant. This then gives
\be
\frac{6}{5}\fnl \simeq f \eta_\chi^* \,, \qquad
\tnl \simeq \tau {\eta_\chi^*}^2 \,, \qquad
\frac{27}{25} \gnl \simeq \frac{18}{5} \fnl \, \eta_\chi^* \,.
\label{eq:ridge_results1}
\ee
In this case $\gnl$ is subdominant, even to $\fnl$, and so we do not consider this further. The peak value of $\fnl$ and $\tnl$ can then be found by maximising the functions $f$ and $\tau$ as $\theta$ varies. They peak when $\theta^2 = \sqrt{3} \theta^*$ and  $\theta^2 = \sqrt{2} \theta^*$ respectively and so
\be
\frac{6}{5} \left. \fnl \right|_{\mathrm max} \simeq \frac{3 \sqrt{3}}{16} \frac{\eta_\chi^*}{\theta^*} \,, \qquad
 \left. \tnl\right|_{\mathrm max} \simeq \frac{4}{27} \frac{{\eta_\chi^*}^2}{\theta_*^2} \,.
\label{eq:ridge_results2}
\ee
The sign of $\fnl$ is negative due to the sign of $\eta_\chi^*$, and the amplitude of $\fnl$ reaches its peak after $\tnl$. The peak of $\tnl$ is only slightly larger than the square of $\frac{6}{5} \left. \fnl \right|_{\mathrm max}$, but due to the difference in peaking times it is quite possible to have $\tnl \gg |\fnl|$.

\sssec{Direct ridge calculation for $\fnl$}
\label{sec:direct_ridge}

We now show that these results can be approximately derived from a direct calculation~\cite{Elliston:2011dr} that does not require the analytic formalism developed in this chapter. The results work surprisingly well, even when the slow-roll approximation begins to break down. For simplicity we consider only the bispectrum. This provides a useful validation and helps to develop further qualitative intuition. 

\paragraph{Evolution of trajectories.}
Measuring length along each trajectory by the energy density, the slow-roll evolution equations are
\begin{align}
\frac{\d \phi}{\d V} 
&=
\frac{g}{g^2 + m_\chi^4 \chi^2}
\label{eq:phi-trajectory} \,, \\
\frac{\d \chi}{\d V}
&=
\frac{-m_\chi^2 \chi}{g^2 + m_\chi^4 \chi^2} \,.
\label{eq:chi-trajectory}
\end{align}
According to~\eqref{eq:chi-trajectory}, a trajectory emanating from $(V_*, \chi_*)$ 
and evolving to $(V_c, \chi_c)$ satisfies
\be
\frac{m_\chi^2}{2} ( \chi^2_c - \chi^2_* ) +
\frac{g^2}{m_\chi^2} \ln \frac{\chi_c}{\chi_*}
= V_* - V_c \,.
\label{eq:chi-dispersion}
\ee
Restricting our attention to the early stages of the turn for which 
$|\chi_c| \ll |g| / m_\chi^2$, then the logarithm dominates on the \lhs~of eq.~\eqref{eq:chi-dispersion} and the trajectories 
disperse exponentially as the potential drops. We note that $|\chi_c| = |g|/m_\chi^2$ defines the point at which the bundle is exactly halfway through the turn towards the $\chi$ direction.

\begin{figure}[h]
\centering
\includegraphics[width=\textwidth]{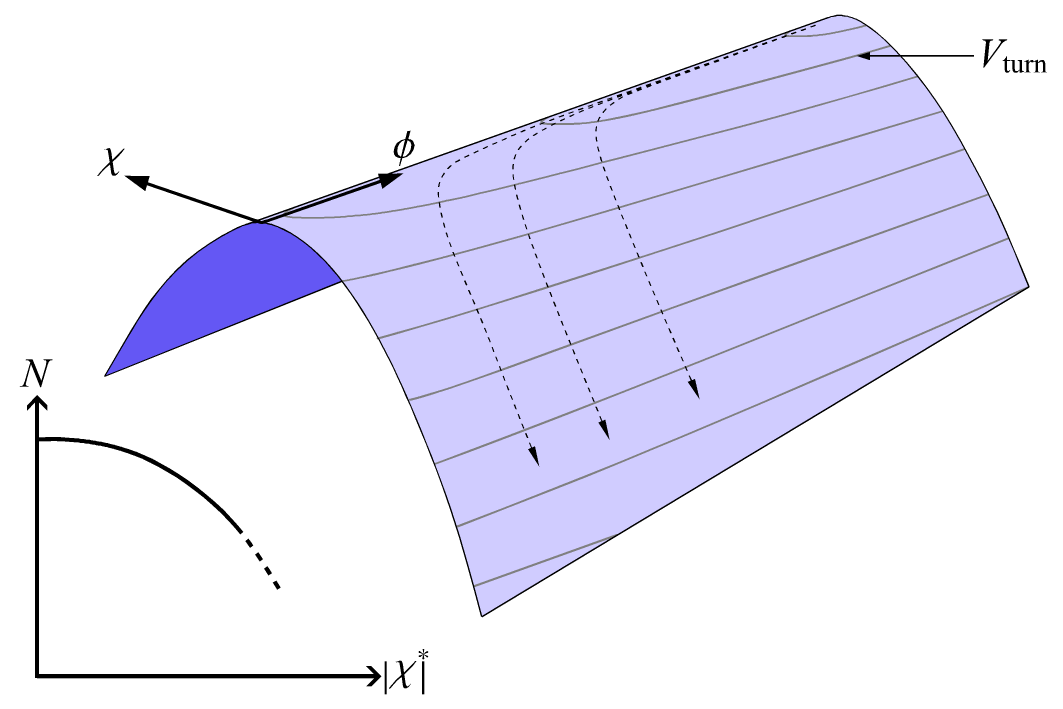}
\caption[$\delta N$ from the ridge mechanism]{Schematic demonstration of how a ridge may generate a large non-linear $\delta N$ distribution. Those bundle elements that start furthest from the ridge top (i.e. those with larger $|\chi^*|$) are the first to fall. They then reach a nearby uniform density hypersurface $V_{\rm turn}$ quickly and so $N$ decreases non-linearly with $\chi^*$. Noting $\chi < 0$, $N_{,\chi}$ is positive and $N_{,\chi \chi}$ is negative, and so $\fnl$ will be negative.}
\label{fig:ridge}
\end{figure}

This leads to the following physical picture: Trajectories begin close to the ridge with a Gaussian profile. Elements of the bundle furthest from the ridge top are the first to `turn' and fall from the ridge. Such bundle elements then rapidly hit subsequent uniform density hypersurfaces and so generate a large tail of negative $\delta N$ which will be measured by parameters such as $\fnl$. This picture is illustrated in figure~\ref{fig:ridge} from which it is physically clear that $\fnl$ will become negative as the bundle begins to fall from the ridge. 

One can further understand why it is possible for $\fnl$ to become large from this ridge mechanism. By its definition, a large $\fnl$ requires growth of the three point statistics, but without significant growth of the two point statistics. In the early stages of the turn, only the periphery of the bundle is affected, leading to a growth of $\langle \zeta \zeta \zeta \rangle$ but not $\langle \zeta \zeta \rangle$ and so the desired conditions are obtained. We note that this intuition leads us to hypothesise that higher-order moments of the $\delta N$ distribution will typically evolve first during turning in phase space, and therefore higher order non-Gaussianity parameters will also typically evolve first. We shall see examples of this behaviour in \S\ref{sec:models}.

\paragraph{$\delta N$ analysis.}
In order to obtain an alternative approximation of eq.~\eqref{eq:ridge_results2}, we now translate to $\zeta$ and repeat the above analysis in the language of the $\delta N$ method. Consider two trajectories rolling along the top of the ridge which are initially separated by a distance $(\delta \phi_*, \delta \chi_*)$. In this region, surfaces of constant energy density practically coincide with surfaces of constant $\phi$. Therefore, to bring this pair of trajectories to a common energy density $V = V_*$ requires a small excess expansion $\delta N \approx (2\epsilon_\phi^*)^{-1/2} \Mpl^{-1} \delta \phi_*$. The subsequent expansion between different uniform density hypersurfaces is then parametrized by the perturbation $\delta \chi_*$. 

Passing to the limit where $\delta \phi_*$ and $\delta \chi_*$ become infinitesimal and using eq.~\eqref{eq:chi-trajectory}, we conclude that on arrival at $V = V^c$ the trajectories have experienced expansion histories which differ by 
\begin{equation}
\d N \approx
\frac{1}{\sqrt{2 \ep_*} \Mpl} \, \d \phi_*
+ 2 \Mpl^{-2} m_\chi^4 \, \d \chi_*
\int_{V_*}^{V_c} \frac{V \; \d V}
{[ g^2 + m_\chi^4 \chi^2 ]^2}
\chi
\left( \frac{\partial \chi}{\partial \chi_*} \right)_{V},
\label{eq:ridge-delta-n}
\end{equation}
where the partial derivative is to be evaluated at constant $V$. Invoking the chain rule,~eq.~\eqref{eq:ridge-delta-n} determines all derivatives of $N$. We find
\begin{equation}
N_{,\chi \chi} =
2 \Mpl^{-2} m_\chi^4 
\int_{V_*}^{V_c} \frac{V \, \d V}
{[ g^2 + m_\chi^4 \chi^2 ]^2}
\left[
\frac{g^2 - 3 m_\chi^4 \chi^2}{g^2 + m_\chi^4 \chi^2}
\left( \frac{\partial \chi}{\partial \chi_*}\right)_{V}^2
+ \chi \left( \frac{\partial^2 \chi}{\partial \chi^2_*} \right)_{V} \right].	
\label{eq:n-chi-chi}
\end{equation}
So far our considerations have been general. 

Prior to the turn,~eq.~\eqref{eq:chi-dispersion} makes
$\partial^2 \chi / \partial \chi_*^2$ negligible whereas $\partial \chi / \partial \chi_* \approx \chi / \chi_*$ is exponentially growing. Substituting eq.~\eqref{eq:chi-dispersion} for $\chi(V)$ we can then integrate these expressions to find
\begin{align}
N_{,\chi} & \approx -
\frac{m_\chi^2 V_c}{\Mpl^2 g^2} \left( \frac{\chi_c^2}{\chi_*} \right),
	\label{eq:n-chi-estimate} \\
N_{,\chi\chi} & \approx 
 \frac{N_{,\chi}}{\chi_*} .
\label{eq:n-chi-chi-estimate}
\end{align}

Initially, $N_{,\chi}$ and $N_{,\chi\chi}$ are small in comparison with $N_{,\phi}$ 
and $N_{,\phi\phi}$. In addition, $N_{,\phi\chi} \approx -m_\chi^2 \chi^* / g \Mpl^2$ is constant and can safely be neglected. Therefore $\zeta$ is dominated by the fluctuation in $\phi$, which is practically Gaussian. Using~\eqref{eq:fnl}, we find
\begin{equation}
\frac{6}{5} \fnl \approx
\left[2 \epsilon_* 
- \left(\frac{N_{,\chi}}{N_{,\phi}}	\right)^3
\frac{g^2 \Mpl^2 }{V_* \chi_*} 
+ \O \Big(\frac{N_{,\chi}}{N_{,\phi}}\chi_* \Big)\right]
\left[	1 + \frac{N_{,\chi}^2}{N_{,\phi}^2}	\right]^{-2} .
\label{eq:fnl-peak-approach}
\end{equation}
While $|N_{,\chi}| \ll |N_{,\phi}|$, the first term dominates and eq.~\eqref{eq:fnl-peak-approach} gives $|\fnl| \sim \epsilon_* \ll 1$. As the trajectory moves away from the ridge, $N_{,\chi}$ becomes increasingly important whereas $N_{,\phi}$ is  constant. When $|N_{,\chi}|$ and $|N_{,\phi}|$ are comparable, $\fnl$ is dominated by the second term in~\eqref{eq:fnl-peak-approach} this causes a `spike' in $\fnl$. Estimating the peak to occur when $|N_{,\phi}| \approx |N_{,\chi}|$, we find
\begin{equation}
\left. \fnl \right|_{\text{peak}} \approx
-\sqrt{2 \epsilon_*} \frac{\Mpl}{|\chi_*|} ,
	\label{eq:fnl-peak}
\end{equation}
In this expression, and similar ones below, the numerical prefactor is uncertain by an $\O(1)$ quantity which depends on the precise balance between $N_{,\phi}$ and $N_{,\chi}$ at the peak. We can see that this analysis recovers the same scaling as eq.~\eqref{eq:ridge_results2}, as expected.

On approach to the spike, eq.~\eqref{eq:fnl-peak-approach} predicts that $\fnl$ is negative and growing like $(\chi/\chi_*)^6$. Subsequently, $\chi$ continues to increase and $|N_{,\chi}|$ eventually dominates $|N_{,\phi}|$. In this region $\zeta$ is composed almost entirely of the $\chi$ fluctuation. The non-Gaussianity then decays like $(\chi/\chi_*)^{-2}$. These estimates of the growth rate and decay rate are valid before the turn, where $\chi$ is growing exponentially.

Dropping numerical factors of order unity and using $\chi_{\rm turn} = -g/m_\chi^2$ to estimate $\fnl$ when the fiducial trajectory passes the turn, we find 
\begin{equation}
\left. \fnl \right|_{\text{turn}}
\sim \left. \eta_\chi \right|_{\text{turn}} .
\label{eq:fnl-turn}
\end{equation}
This is much less than eq.~\eqref{eq:fnl-peak} and corresponds to a time well after the `spike'. We can understand the smallness of eq.~\eqref{eq:fnl-turn} because this  is when the core of the bundle is part way through its turn and so the two-point statistics have grown to mask the growth of the three-point statistics.

\subsection{Valleys}
\label{sec:valley}

\begin{figure}[h]
\centering
\includegraphics[width=\textwidth]{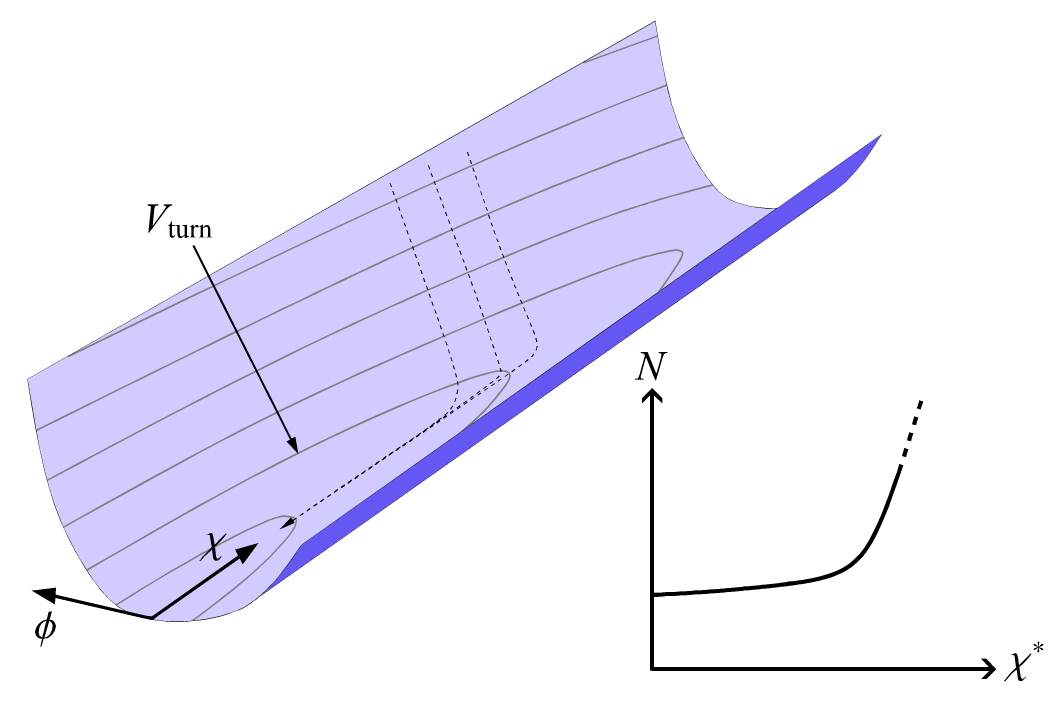}
\caption[$\delta N$ from the valley mechanism]{Schematic demonstration of how a valley may generate a large non-linear $\delta N$ distribution. The bundle initially evolves down the steep $\phi$ axis and those trajectories further from the $\chi=0$ axis experience a larger velocity in the $\chi$ direction. This leads to a non-linear compression of the bundle. At the valley bottom the bundle turns into the shallow $\chi$ direction. At the turning point, those bundle elements that started furthest from the $\phi$ axis have further to travel to the global minimum. The shallowness of the potential along the $\chi$ axis magnifies this into a large non-linear $\delta N$. One can see that $N_{,\chi}$ and $N_{,\chi \chi}$ are both positive and so $\fnl$ will be positive.}
\label{fig:valley}
\end{figure}

The simplest possible vacuum-dominated valley takes the form 
\be
\label{eq:valley_pot}
V = V_0 + \frac{1}{2} m_\phi^2 \phi^2 + \frac{1}{2} m_\chi^2 \chi^2 \,,
\ee
where $m_\phi$ and $m_\chi$ are taken to be positive and the $V_0$ term dominates. We again consider $\theta* \ll 1$, and assume that $m_\phi \gg m_\chi$, so that the initial motion is almost exactly parallel to the $\phi$ direction. Once again, therefore, the isocurvature direction is well approximated by the $\chi$ direction, and $\eta_{ss} = m_\chi^2/W_0$ is again constant. This then gives expressions for $\fnl$, $\tnl$ and $\gnl$ which are identical to those in eq.~\eqref{eq:ridge_results1}, but we note that the signs of $\fnl$ and $\gnl$ are now reversed. We can also understand the growth and positive sign of $\fnl$ intuitively as shown in figure~\ref{fig:valley}.

\ssec{Inflection points}
\label{sec:inflectionpot}

The analysis of \S \ref{sec:gnlheatmaps} with $\Omega \simeq 1$ illustrated how $\gnl$ is negligible for inflationary potentials that are approximately quadratic, but that $\gnl \sim \tnl$ for potentials with cubic or higher order shapes. We now consider the simplest such shape, with an inflection point of the form
\be
V = V_0 + g \phi + \frac{1}{6} \lambda \chi^3 \,.
\ee
Such a potential is also motivated by studies of d-brane inflation~\cite{Dias:2012nf}. The regime of interest is close to but below the inflection point with $|\chi| \ll 1$ such that $\theta \ll 1$. The non-Gaussianity parameters are only large in the regimes where $f$ or $\tau$ are large, which requires $|\chi| \gg |\chi^*|$. In this limit we find that the leading order non-Gaussianity is given as
\be
\frac{6}{5}\fnl \simeq 2 f \eta_{ss} \,, \qquad
\tnl \simeq 4 \tau \eta_{ss}^2 \,, \qquad
\frac{27}{25} \gnl \simeq \frac{5}{8} \tnl \,.
\ee
For a general power-law potential, the numerical relation between $\gnl$ and $\tnl$ will take a different value. Since $\gnl$ simply follows the evolution of $\tnl$, we need only calculate the peak values of $\fnl$ and $\tnl$ as
\be
\frac{6}{5} \left. \fnl \right|_{\mathrm max} \simeq -1.08 \frac{|\xi_{sss}^*|}{{\theta^*}^{3/4}} \,, \qquad
 \left. \tnl\right|_{\mathrm max} \simeq 1.48  \frac{{\xi_{sss}^*}^2}{{\theta^*}^{3/2}} \,.
\ee
where these peaks respectively occur when $\theta^2 = \sqrt{3}\, \theta^*$ and $\theta^2 = \sqrt{13/3}\, \theta^*$. 

\section{Concrete Models} \label{sec:models}

In \S\S\ref{sec:bi_heatmaps}--\ref{sec:shapes} we have developed a qualitative and quantitative understanding of non-Gaussian evolution in two-field inflationary models. In this section we demonstrate the usefulness of these tools, illustrating them with concrete examples of simple models which can produce large values of $\tnl$ and/or $\gnl$.

\subsection{Two-field hybrid inflation} 
\label{sec:hybrid}

Our first example is the two-field hybrid model studied in \S\ref{sec:model1} and extensively elsewhere~\cite{Lyth:2005qk,Alabidi:2006hg,Byrnes:2008wi,Byrnes:2008zy,Sasaki:2008uc, 
Naruko:2008sq,Huang:2009vk,Mulryne:2011ni,Abolhasani:2010kr,
Suyama:2010uj,Mulryne:2009ci,Choi:2012he,Lyth:2012yp}. We are returning to this example because it provides a simple scenario for demonstrating some of the intuition that we have developed. Until the time at which the waterfall field becomes operative, the effective potential may be written in the simple form 
\be
\label{eq:hybrid2}
V =V_0 \left( 1 + \frac{1}{2} \eta_\phi \phi^2 + \frac{1}{2} \eta_\chi \chi^2 \right),
\ee
Let us restrict our attention to the case in which $\eta_\phi$ and $\eta_\chi$ are both positive, and so the potential has a valley-like shape during inflation. This provides an illustrative example of our discussion in \S \ref{sec:valley}. If either $\eta_\phi$ or $\eta_\chi$ were negative, the potential would contain a ridge. We further assume that $\eta_\phi > \eta_\chi$, and that the initial conditions are such that $0 < \chi^* \ll \Mpl$, whilst $\phi^* \sim \O (\Mpl)$ is significantly displaced from zero. Hence, the trajectory initially rolls towards the minimum along the $\phi$ axis before turning and rolling very slowly towards the $\chi$ axis. This is precisely the scenario depicted in figure~\ref{fig:valley}.

These initial conditions imply a small $\theta^*$, and since the potential is vacuum-dominated, $\alpha \simeq \theta$ and the heatmaps are particularly simple to interpret. The initial (horizon crossing) condition corresponds to a position on the diagonal in the bottom left hand corner of any given heatmap. As the trajectory begins its slow turn into the valley, the point on the heatmap proceeds vertically upwards and we can see that a strong non-Gaussian signal is expected during the early stages of the turn. When the angle grows larger, the trajectory moves out of the regions of the heatmaps where $f$ and $\tau$ are large and so the non-Gaussianity decays. Since the large non-Gaussianity occurs during the early stages of the turn we see that $\eta_{ss} \approx \eta_\chi$ is a constant whilst all components of the $\xi_{ijk}^2$ parameter are zero.

\begin{figure}[t] 
\center{\includegraphics[width = 0.7\textwidth]{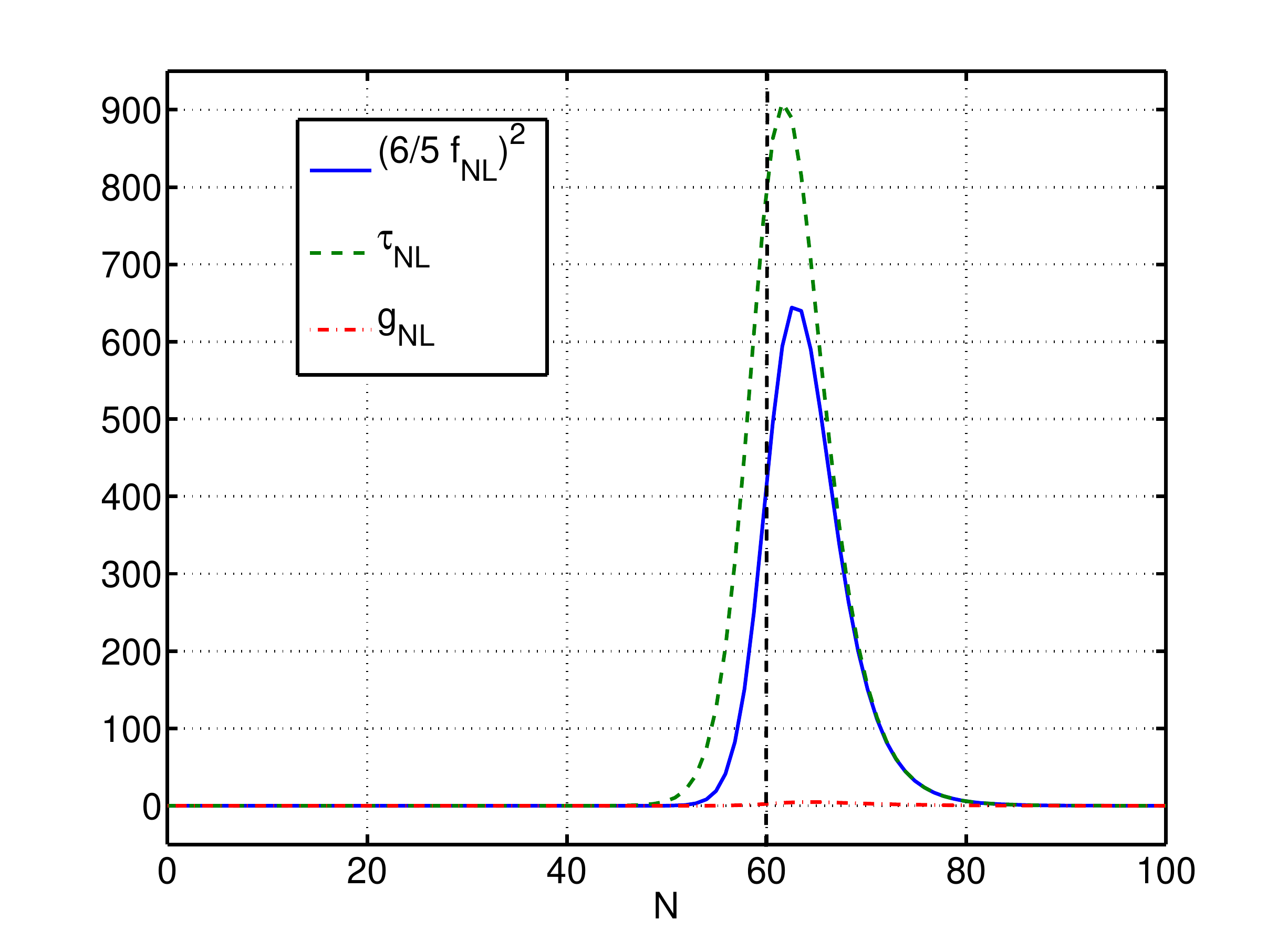}}
\caption[Analytic non-Gaussianity evolution in the two-field hybrid model]{Analytic evolution of non-Gaussianity for the two-field hybrid potential \eqref{eq:hybrid2} with parameter values as described in the text. The solid blue line shows the evolution of ${\rm sign} (\fnl) (6/5 \fnl)^2$, $\tnl$ is shown by the dashed green line, and $\gnl$ is negligible as shown by the dot-dashed red line.}
\label{fig:2fHy}
\end{figure}

In figure~\ref{fig:2fHy} we show how $\fnl$, $\tnl$ and $\gnl$ evolve for the potential \eqref{eq:hybrid2}. This evolution uses the full analytical formulae derived in \S\ref{sec:twofields} and so provides a test of our subsequent simplifications. The parameter values used are $\eta_\phi=0.09$ and $\etc=0.0025$ with the initial conditions taken as $\phi^*=0.9 \, \Mpl$ and $\chi^*=0.001 \, \Mpl$. The evolution of the trispectrum is exactly as we expect: $\fnl$ and $\tnl$ both begin at negligible values, grow to large positive peaks and then decay again to negligible values. We confirm that the peak amplitudes are as expected from eq.~\eqref{eq:ridge_results2} to within $10\%$. Furthermore, we see that $\tnl$ peaks before $\fnl$ in agreement with the discussion regarding eq.~\eqref{eq:tnl} and also in agreement with the results in \S\ref{sec:valley}. This quadratic shape of the potential also gives a small $\gnl$ as expected. 

After $60$ efolds, where we assume the evolution is terminated by the waterfall field, we find $n_s = 1.07$, which is clearly in violation of the \wmap~bounds on the spectral index, which demands a red spectral tilt for priors of zero running and zero tensor--scalar ratio. We do not therefore consider this model a viable candidate for inflation. Moreover, we know that there does not exist a choice of parameters or initial conditions (where both fields are less than the Planck scale) which can support a red tilt \cite{Alabidi:2006hg}. Nevertheless, we still consider the model a useful illustration of our analysis, and in particular as an example of a potential with a valley feature. 

\subsection{Axion-quartic potential} 

We now consider a modification of the axion-quadratic model of \S\ref{sec:model1}. The quadratic potential is replaced with a quartic with the effect that it is also possible to achieve a consistent spectral index, in addition to achieving a large non-Gaussianity. Whilst this is not necessary insofar as providing a test of our analytic results, it is comforting to know that it is possible for such models to be made observably consistent. The potential takes the form
\be
\label{eq:2AxQuart}
V = \frac{1}{4} g \phi^4 + \Lambda^4 \left( 1 - \cos \left( 2 \pi \chi / f \right) \right).
\ee
We take initial conditions such that $\chi^*$ is close to the hilltop region $\chi_* \lesssim f/2$, while $\phi$ starts its evolution at a large field value ($\phi^* \approx 23 \Mpl$ is required for roughly $60$ efolds), the velocity is initially almost entirely in the $\phi$ direction and rolls along the ridge defined by $\chi = f/2$. The initial angle $\theta^*$ is therefore close to zero, and as the trajectory slowly turns off the ridge, $\theta$ increases.

From the heatmaps, one can see that the initial growth of the angle $\theta$ will lead to an increase in the magnitude of $\fnl$ and $\tnl$ as $\alpha$ passes up through the `hot-spot' region where $f$ and $\tau$ peak. After the initial turn from the ridge-top, the trajectory will fall quickly down the steep side of the ridge, before turning back towards its original direction as it reaches the valley bottom. This leads to a decrease of $\alpha$ back through the hot-spot region once more. Importantly, the potential $V$ drops by a small (but non-vanishing) amount between these two turns. From eq.~\eqref{eq:alphadash2} this means that the initial growth of $\alpha$ is slightly larger than the subsequent decay and so $\alpha$ will not make it to zero as $\theta \to 0$. Instead, it will end up with a small positive value. We therefore expect the non-Gaussianity parameters may take large constant values asymptotically, where we also know that the phase space trajectory ends up in a valley and so the asymptotic formula $27 \gnl / 25 = \tnl$ in eq.~\eqref{eq:gnl_tnl} will be an accurate prediction.

\begin{figure}[htb] 
\center{\includegraphics[width = 0.7\textwidth]{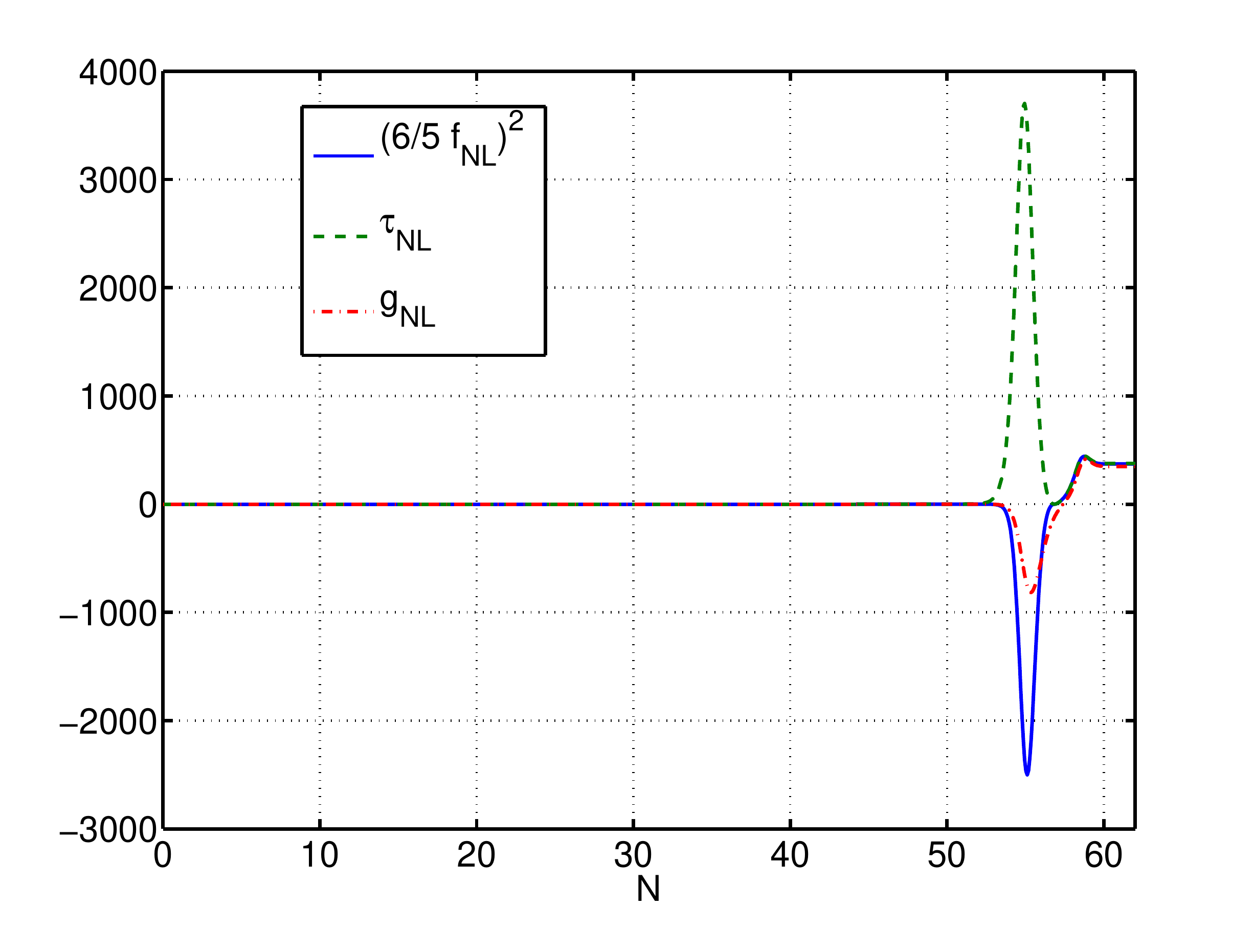}}
\caption[Non-Gaussianity evolution for the axion-quartic model]{Evolution of ${\rm sign} (\fnl) (6/5 \fnl)^2$ (solid blue line), $\tnl$ (dashed green line), and $\gnl$ (dot-dashed red line) for the axion-quartic model discussed in the text, calculated using the analytic formulae presented \S\ref{sec:form_sum}.}
\label{fig:axion}
\end{figure}

In figure~\ref{fig:axion} we give the full analytic evolution of $\fnl$, $\tnl$ and $\gnl$ for a specific realisation of this model. The parameter values chosen are $f=1\Mpl$ and $\Lambda^4/g = (25/2\pi)^2\Mpl^4$, with the overall normalisation fixed to agree with \wmap~7-year power spectrum amplitude. The initial conditions are $\phi^*=22.5 \Mpl$ and $\chi^*=f/2-0.001\Mpl$ which gives us the \hca~value of $\alpha$ as $0.022$ which is small in agreement with the above discussion. In this case, the final constant spectral index has a value of $n_s = 0.949$ which is within the \wmap~7-year $95\%$ contours. The evolution is exactly as we expect, with $\fnl$ and $\tnl$ both beginning with negligible values and $\fnl$ then evolving to a large negative peak, while $\tnl$ grows to a large positive one, before both peaks decay. Despite inflation ending soon after the axion rolls, the peak values of $\fnl$ and $\tnl$ are described by the formulae in \S \ref{sec:ridge} with about $30 \%$ accuracy. The subsequent evolution is also interesting. As the  trajectory evolves into the valley there is a positive spike in $\fnl$, typical of this evolution. We see that there is a delay between the evolution of $\fnl$ and $\tnl$ as we expect from our discussion in \S \ref{sec:ridge}. Finally, the constant asymptotic values of $\fnl$, $\tnl$ and $\gnl$ are reached where we find $27 \gnl / 25 = \tnl$ holds very well. In this case $\tnl \approx (6/5 \fnl)^2$, though this is not as a consequence of reaching an adiabatic regime, unlike the relation between $\tnl$ and $\gnl$.

\subsection{Inflection point model}
\label{sec:inflection}

The discussion in \S\ref{sec:gnlheatmaps} motivated sloping inflection features in the potential as a scenario in which the condition $\gnl \gg \tnl$ might be found in an adiabatic regime where $\Omega \to 0$. We now present such a model, but note that severe tuning is required to find parameters such that $\gnl$ is both large and dominant in the adiabatic limit. We expect, therefore, that the simple relation \eqref{eq:gnl_tnl} will hold for the vast majority of models.

We choose an inflection potential similar to that studied in ref.~\cite{Elliston:2011et} which takes the form 
\be           
V = V_0 \left( \frac{1}{4!} \phi^4 + U_0 + h \chi + \frac{1}{3!}\lambda \chi^3 +
\frac{1}{4!} \mu \chi^4\ \right)\,.
\ee                        
The value of $V_0$ is fixed by the \wmap~power spectrum, and there is an inflection point feature at $\chi = 0$. We further assume that $U_0$ and $\mu$ are fixed by the requirement that there is a minimum of the potential at $\chi_{\rm min} = -F$, where $F$ is taken to be positive. We note that beginning some way above/below the inflection point leads to a negative/positive asymptotic value of $\fnl$ as explored in ref.~\cite{Elliston:2011et}. 

From an exploration of the parameter space for this model we have found it very difficult to obtain a large and dominant $\gnl$ whilst satisfying the constraints of slow-roll. We present a compromise scenario where the $\chi$ field rolls when $\ep \sim 0.3$. Such a model has parameter values $h = 0.05\, \Mpl^3$, $\lambda = 10^{4}\, \Mpl$ and $F=0.1\, \Mpl$ and the evolution begins with $\phi^*=22.5\, \Mpl$ and $\chi^*=0$.\footnote{A realistic application of our formulae requires that the model is insensitive to changes in the horizon crossing conditions within the range of the quantum scatter. Such changes do not affect the predictions of this model.} We compare the evolution from our analytic expressions to the numerical non-slow-roll evolution calculated using a finite-difference code identical to that used in chapter \ref{ch:adiabatic}. When the deviation from slow-roll is not too large, the non-slow-roll evolution mimics a superposition of the slow-roll evolution and small rapid oscillations. We therefore find that our analytic results are surprisingly applicable when slow-roll begins to break. For this example, the analytic spectral index is $n_s=0.964$ and this is within $1\%$ of the non-slow-roll value. Our calculation gives $\gnl=-432$ which agrees with the non-slow-roll code to within a factor of two. In general, if the evolution of observables for a given model has not settled down to a constant value before the end of slow-roll inflation then one must use numerical methods to obtain an accurate result. However, a crude estimate is very easy to calculate through the slow-roll analysis. To conclude, in this final example we have tried to engineer a model to break from the simple result $27 \gnl / 25 = \tnl$ but we have found that this is very hard to achieve.

\paragraph{Summary.}
In this chapter we have developed intuition about the relation between inflationary dynamics and the evolution of primordial non-Gaussianity. We have followed the methodology of Byrnes et al.~\cite{Byrnes:2008wi} which has the great benefit of not requiring us to specify the form of the separable inflationary potential. Following Elliston et al.~\cite{Elliston:2012wm} and Elliston~\cite{Elliston:2013uga}, we have shown that the bispectrum analysis of Byrnes et al.~\cite{Byrnes:2008wi} may be simplified, leading to stronger conclusions. We also extend this method to the trispectrum, showing that the types of inflationary dynamics that can give rise to a large bispectrum are similar to those that generate a large trispectrum. We also derive new relations between the non-Gaussianity parameters, showing that the trispectrum parameters may be related as $27 \gnl / 25 = \tnl$ for the vast majority of \hca~scenarios. This analysis motivated us to consider the role of generic shapes in the inflationary potential as sources of non-Gaussianity, including ridges, valleys and inflection points. 
\clearpage{\pagestyle{empty}\cleardoublepage}
\chapter{Generalised single field inflation}
\label{ch:singlefield}

\begin{addmargin}[0.05\textwidth]{0.05\textwidth}
In this final chapter we consider the observational consequences of various generalisations to chaotic inflation that arise in low energy effective string theory. We begin in \S\ref{sec:barechaotic} with a review of standard chaotic inflation. The $\alpha'$ corrections from string theory, as motivated in \S\ref{sec:alphaprime}, are then added to the chaotic inflationary scenario in \S\ref{sec:includingalphaprime}. Subsequent sections then focus on particular $\alpha'$ terms: \S\S\ref{sec:single-transform}--\ref{sec:single-bd} consider non-minimally coupled Higgs inflation with a non-canonical kinetic term, as well as Brans-Dicke theories. Gauss-Bonnet assisted inflation is included in \S\ref{sec:single-gb} and Galileon inflation is considered in \S\ref{sec:single-g}. For each scenario we derive theoretical predictions which we then constrain using current observational data. This follows our work in ref.~\cite{DeFelice:2011jm}.
\end{addmargin}

\begin{center}
\partialhrule
\end{center}
\vspace{-3em}
\begin{quote}
\list{}{\leftmargin 2cm \rightmargin\leftmargin} \endlist
\begin{center}
{\it ``Everything should be made as simple as possible, but not simpler.''}
\flushright{\shifttext{1.5cm}{---Attributed to Albert Einstein.}}
\end{center}
\end{quote}
\vspace{-1em}
\begin{center}
\partialhrule
\end{center}

\sec{The predictions of standard chaotic inflation}
\label{sec:barechaotic}

Linde's chaotic inflation model~\cite{Linde:1983gd} is one of the simplest candidate models of inflation. It has an action of the form
\be
S = \int \d^4 x \, \sqrt{-g} \left[ \frac{\Mpl^2}{2} R + X - V_0 \vp^p \right],
\label{eq:chaoticaction}
\ee
where $V(\vp) = V_0 \vp^p$ is the inflationary potential with constants $\{V_0,p\}$ and the kinetic energy takes the canonical form $X = -g^{\mu \nu} \partial_\mu \vp \partial_\nu \vp/2$. Our interest will mainly be with the cases $p=2$ and $p=4$ which take the forms $V = m^2 \vp^2 / 2$ and $V=\lambda \vp^4 /4$ respectively. We note that the mass $m$ has dimension of $\Mpl$ whereas $\lambda$ is dimensionless.

The potential slow-roll parameters $\ep$ and $\eta$ can be calculated from eqs.~\eqref{eq:sf-ep} and~\eqref{eq:sf-eta} as
\be
\label{eq:single-sr}
\ep = \frac{\Mpl^2}{2} \frac{p^2}{\phi^2}, \qquad 
\eta = \Mpl^2 \frac{p(p-1)}{\phi^2}.
\ee
For reasonable positive values of $p$, slow-roll inflation occurs in the regime $\phi \gg \Mpl$. For the remainder of this chapter we restrict ourselves to $\phi>0$ without loss of generality. Note that we use $\vp$ for the fully-perturbed field and $\phi$ for the background field, as consistent with the rest of this thesis.

Since this is a single field model, it does not have appreciable infrared dynamics and so its predictions are robustly found a few efolds after horizon exit. These follow directly from the general formulae derived in chapter \ref{ch:formalisms} and their analytic approximations derived in \S\ref{sec:analytics}. We note that the horizon crossing approximation is valid in this scenario and so we may employ expressions such as eq.~\eqref{eq:sshca} to find the adiabatic values of the $\delta N$ coefficients. At lowest order in slow-roll these follow as
\be
N_{,\phi} = \frac{\phi^*}{\Mpl^2 \, p}, \qquad
N_{,\phi \phi} = \frac{1}{\Mpl^2 \, p} .
\ee
Using the formulae of \S\ref{sec:zetacorrelators} we therefore find adiabatic values for inflationary observables at leading order to be
\begin{align}
\P_\zeta &= \frac{H_*^2}{8 \pi^2 \Mpl^2 \ep^*},\\
n_\zeta - 1 &= -2 \ep^* \left( 1 + \frac{2}{p} \right),\\
\frac{6}{5} \fnl^{\rm local} &= \frac{2 \ep^*}{p},\\
r &= 16 \ep^* = -8 n_t .
\end{align}
We note that not all positive values of the parameter $p$ are compatible with observational constraints. The simplest possible model $V=m^2 \vp^2/2$---dubbed {\it vanilla inflation}---is compatible with observational data, whereas the {\it self-coupling potential} $V=\lambda \vp^4/4$ is not. We now explain briefly where the discrepancy with observational data arises.

The scalar spectral index $n_\zeta$ has the correct sign for all positive values of $p$; not all values of $p$ will be observationally viable but $p=2$ and $p=4$ are within the $2\sigma$ bounds. The tensor--scalar ratio $r$ is small, but with increasingly precise bounds on gravitational waves, larger values of $p$ are ruled out. One may see this by expressing the slow-roll parameter $\ep^*$ in terms of the number of efolds of inflation as $\ep^* \approx p/4N$. The tensor--scalar ratio then becomes $r = 4p/N$. Since we require about 55 efolds of observable inflation, we find $r \approx 0.15$ for $p=2$ and $r\approx0.29$ for $p=4$. Comparing to the data shown in figure~\ref{fig:constraint}, the self-coupling potential is disfavoured since it predicts a tensor--scalar ratio in excess of observational constraints. 

The self-coupling potential $V=\lambda \vp^4 /4$ also suffers from a problem of requiring very small parameter values. This arises from the scalar power spectrum amplitude $\P_\zeta$ which may be theoretically fixed by the constant $\lambda$. However, for 55 efolds of inflation, one recovers the observed value of $\P_\zeta$ (given in eq.~\eqref{eq:Pzeta_data}) for $\lambda \sim 10^{-13}$. Such a small coupling constant is not the generic prediction of particle physics where coefficients nearer to order of unity are generally expected.

Finally, we see that there is a negligible bispectrum for these simple chaotic inflationary models. This is unsurprising since the single field, canonical and slow-roll nature of chaotic inflation ensures that a negligible bispectrum follows immediately from Maldacena's theorem~\cite{Maldacena:2002vr}. 

\sec{Including $\alpha'$ corrections}
\label{sec:includingalphaprime}

In \S\ref{sec:alphaprime} we discussed the task of embedding chaotic inflation within string theory and how this could lead to $\alpha'$ modifications of the action \eqref{eq:chaoticaction} including:
\begin{itemize}
\item Non-minimal coupling of the inflaton $\vp$ to the Ricci scalar,
\item Non-canonical kinetic terms,
\item Higher derivative quantum gravity terms such as the Gauss-Bonnet (\gb) term $\G \equiv R^2 -4 R_{\alpha \beta}R^{\alpha \beta} + R_{\alpha \beta \gamma \delta}R^{\alpha \beta \gamma \delta}$,
\item Non-linear field interactions such as the Galileon term $J(\vp,X) \Box \vp$.
\end{itemize}
The \gb~term and the Galileon interaction $J(\vp,X) \Box \vp$ give rise to second-order field equations. This property is welcome since it avoids the propagation of extra degrees of freedom which may lead to negative kinetic energy states known as the Ostrogradski instabilities or {\it ghosts}. 

A key question is to determine the observational consequences of such $\alpha '$ terms for different models of inflation such as the chaotic inflationary model. For example, in \S\ref{sec:barechaotic} we illustrated that the `bare' self-coupling potential $V = \lambda \vp^4/4$ is disfavoured by observational constraints on the tensor--scalar ratio $r$ and furthermore there was a difficulty to explain the very small coupling constant required. It is interesting to see whether these problems may be addressed through the inclusion of additional ingredients into the action, in particular those motivated from string theory which may be naturally expected to be present.

For the self-coupling potential, it has been shown that multiplying the Ricci scalar by a non-minimal coupling of the form $1-\zeta\vp^{2}/\Mpl^2$ allows us to realise larger values of $\lambda$ (it should be clear from the context whether $\zeta$ is referring to the curvature perturbation or the non-minimal coupling). Furthermore, the self-coupling potential becomes observationally viable in the limit where $\zeta$ is negative and $|\zeta|\gg1$; the tensor--scalar ratio reduces to the order of $10^{-3}$ with the scalar spectral index $n_\zeta\approx0.96$ \cite{Makino:1991sg,Fakir:1992cg,Kaiser:1994vs,Komatsu:1997hv,
Komatsu:1999mt,Tsujikawa:2004my,Linde:2011nh}. Recently, there has been renewed interest in non-minimally coupled inflation models by identifying the inflaton as the Higgs boson appearing in the standard model of particle physics (for example see ref.~\cite{Bezrukov:2007ep}).

The self-coupling model may also be saved by the introduction of the nonlinear kinetic interaction $(1/M^{3})X\square\vp$. This interaction is invariant under Galilean symmetry $\partial_{\mu}\vp \to \partial_{\mu}\vp+b_{\mu}$ in Minkowski spacetime, where $b_\mu$ is a constant vector. For this type of model and a self-coupling potential, it was recently shown that the tensor--scalar ratio can reduce to $r \approx 0.18$~\cite{Kamada:2010qe}. Moreover, the coupling constant $\lambda$ may attain a more natural value of order $\lambda\sim 0.01$. We expect a modification of this result given our choice of a more general function $J(\vp,X)$ that is a differentiable function of both $\vp$ and $X$.

With these motivations in mind we incorporate the various generalisations into a general action of the form
\be
\hspace{-1.5cm}S = \int \! \d ^4 x \sqrt{-g} \left[ \frac{\Mpl^2}{2} F(\vp)R + \omega(\vp)X-V(\vp)-\xi(\vp){\cal G}-J(\vp,X)\Box \vp \right],
\label{eq:action}
\ee
where $F(\vp)$, $\omega(\vp)$ and $\xi(\vp)$ are differentiable functions of $\vp$. The field $\vp$ can be identified as a dilaton coupled to all the terms in the Lagrangian. De Felice and Tsujikawa~\cite{DeFelice:2011zh} considered an action of the form \eqref{eq:action}, deriving the background field equations and slow-roll parameters, as well as the second and third order perturbations from which they derived expressions for those observational quantities of interest in our study. We review these results in \S\ref{sec:single-background}. Our aim is to constrain the analysis of ref.~\cite{DeFelice:2011zh} by confronting various different inflationary scenarios with current observational data. 

We note that the bispectrum for the action \eqref{eq:action} was also studied in ref.~\cite{DeFelice:2011zh} although we shall not consider this here. This is because a large non-Gaussianity was shown to arise only if the sound speed $c_{\rm s}$ is small---and we shall show that $c_{\rm s} \approx 1$ for the action \eqref{eq:action} in the cases that we consider. The prediction of a negligible bispectrum is significant because the inflationary models considered in this chapter can be ruled out if observations detect a non-zero bispectrum.

\sssec{Routemap}

For the purposes of analysing the general action~\eqref{eq:action} we consider each of the additional terms (those not present in eq.~\eqref{eq:chaoticaction}) individually. This will allow us to assess how each of these $\alpha'$ corrections affect inflationary observables without the analysis becoming intractable. This is also pragmatic, since we do not presently have sufficient data to constrain every term in eq.~\eqref{eq:action} simultaneously. For non-minimally coupled models, we shall move to the Einstein frame, as discussed in \S\ref{sec:single-transform}. In this frame we shall consider non-minimally coupled Higgs inflation and a non-canonical kinetic term $\omega(\vp)X$ in \S\ref{sec:single-minimally}. In \S\ref{sec:single-bd} we consider Brans-Dicke theories which have explicit couplings $\vp R$ and $(\omega_{{\rm BD}}/\vp)X$. We consider \gb~coupling in \S\ref{sec:single-gb} and Galileon coupling in \S\ref{sec:single-g}. 

The theoretical predictions of each of these scenarios will be largely determined analytically, although we shall use some numerical methods where appropriate. We will calculate the scalar power spectrum $\P_\zeta$, the scalar spectral index $n_\zeta$ and the tensor--scalar ratio $r$. The scalar power spectrum will provide us with the value $V_0$ as discussed in \S\ref{sec:barechaotic}, leaving the scalar spectral index and the tensor--scalar ratio to constrain the remaining parameters. The value of the tensor power spectrum $\P_t$ is effectively constrained through the tensor--scalar ratio $r$. 

In some instances we shall compare the results to existing $1\sigma$ and $2\sigma$ likelihood contours, as produced from joint analysis of data from \wmap~7-year~\cite{Komatsu:2010fb}, \bao~\cite{Percival:2009xn} and \hst~\cite{Riess:2011yx}. For consideration of \gb~and Galileon corrections, we shall use the Cosmological Monte Carlo (\cosmomc) code~\cite{Lewis:2002ah}. In addition to \wmap, \bao~and \hst~data, we also use \lss~\cite{Reid:2009xm}, supernovae type {\sc ia} (\snia)~\cite{Kowalski:2008ez} and {\sc bbn}~\cite{Burles:1997ez}, by assuming a \lcdm~universe. In these cases we shall calculate the tensor spectral index $n_t$ which is a required constraint in \cosmomc. 

\section{Background theory}
\label{sec:single-background} 

This section briefly reviews the work of de Felice and Tsujikawa~\cite{DeFelice:2011zh} who provided a detailed theoretical analysis of the generalized action \eqref{eq:action}. Working with a metric of the spatially flat \frwl~form~\eqref{eq:frwl}, the background equations of motion are given by $\{E_1,E_2,E_3\}=0$ where
\begin{align}
E_{1} & \equiv 
3\Mpl^{2}FH^{2}+3\Mpl^{2}H\dot{F}-\omega X-V-24H^{3}\dot{\xi}-6H\dot{\phi}XJ_{,X}+2XJ_{,\phi}\,,
\label{E1eq} \\
E_{2} & \equiv 
3\Mpl^{2}FH^{2}+2\Mpl^{2}H\dot{F}+2\Mpl^{2}F\dot{H}+\Mpl^{2}\ddot{F}+\omega X-V-16H^{3}\dot{\xi} \nonumber \\ & \qquad
-16H\dot{H}\dot{\xi}-8H^{2}\ddot{\xi}-J_{,X}\dot{\phi}\dot{X}
-J_{,\phi}\dot{\phi}^{2} \,, \label{E2eq} \\
E_{3} & \equiv (\omega+6H\dot{\phi}J_{,X}+6H\dot{\phi}\, XJ_{,XX}
-2XJ_{,\phi X}-2J_{,\phi})\ddot{\phi}\nonumber \\ & \qquad
+(3\omega H+\dot{\phi}\,\omega_{,\phi}
+9H^{2}\dot{\phi}J_{,X}+3\dot{H}\dot{\phi}J_{,X}+3H\dot{\phi}^{2}J_{,\phi X}
-6HJ_{,\phi}-J_{,\phi\phi}\dot{\phi})\dot{\phi}\nonumber \\ & \qquad
-\omega_{,\phi}X+V_{,\phi}-6\Mpl^{2}H^{2}F_{,\phi}
-3\Mpl^{2}\dot{H}F_{,\phi}+24H^{4}\xi_{,\phi}
+24H^{2}\dot{H}\xi_{,\phi}\,.\label{E3eq}
\end{align}

Only two of the above equations are independent due to the Bianchi identities which prescribe that 
\be
\dot{\phi}E_{3}+\dot{E}_{1}+3H(E_{1}-E_{2})=0 \,.
\ee

The combined equation, $(E_{2}-E_{1})/(\Mpl^{2}H^{2}F)=0$, provides us with a formula for the slow-roll parameter $\epH$ as
\begin{align}
\epH \equiv
-\frac{\dot{H}}{H^{2}} &=
-\frac{\dot{F}}{2HF}
+\frac{\ddot{F}}{2H^{2}F}+\frac{\omega X}{\Mpl^{2}H^{2}F}
+\frac{4H\dot{\xi}}{\Mpl^{2}F}-\frac{8\dot{H}\dot{\xi}}{\Mpl^{2}HF}
\nonumber \\ & \qquad
\quad -\frac{4\ddot{\xi}}{\Mpl^{2}F} 
+\frac{3\dot{\phi}XJ_{,X}}{\Mpl^{2}HF}
-\frac{\ddot{\phi}XJ_{,X}}{\Mpl^{2}H^{2}F}
-\frac{2XJ_{,\phi}}{\Mpl^{2}H^{2}F}\,.\label{dotHeq}
\end{align}
Since $\epH \ll 1$ during inflation, the modulus of each term on the \rhs~of eq.~\eqref{dotHeq} is much smaller than unity. We do not consider the alternative scenario in which some of the terms on the \rhs~may be large but cancel in such a way that $\epH$ is still small. This leads us to introduce the following slow-roll parameters
\be
\bal{8}
\delta_{F} &\equiv \frac{\dot{F}}{HF}\,,\qquad
&\delta_{X} &\equiv\frac{\omega X}{\Mpl^{2}H^{2}F}\,,\qquad
&\delta_{\xi} &\equiv \frac{H\dot{\xi}}{\Mpl^{2}F}\,, \qquad
&\delta_{JX} &\equiv \frac{\dot{\phi}XJ_{,X}}{\Mpl^{2}HF}\,,\\ 
\delta_{\phi} &\equiv \frac{\ddot{\phi}}{H\dot{\phi}}\,,
&\delta_{J\phi} &\equiv\frac{XJ_{,\phi}}{\Mpl^{2}H^{2}F}\,,
&\eta_{F} &\equiv\frac{\dot{\delta}_{F}}{H\delta_{F}}\,,&\eta_{\xi} &\equiv\frac{\dot{\delta}_{\xi}}{H\delta_{\xi}}\,,
\label{slowvariation}
\eal
\ee
which lead to the relations
\begin{equation}
\frac{\ddot{F}}{H^{2}F}=\delta_{F}(\delta_{F}+\eta_{F}-\epH)\,,\qquad\frac{\ddot{\xi}}{\Mpl^{2}F}=\delta_{\xi}(\delta_{F}+\eta_{\xi}+\epH)\,.
\end{equation}
We may now substitute these slow-roll parameters into our expression for $\epH$ in eq.~\eqref{dotHeq} to obtain 
\begin{align}
\hspace{-2em}\epH &= 
\frac{2\delta_{X}-\delta_{F}+8\delta_{\xi}+6\delta_{JX}-4\delta_{J\phi}+\delta_{F}(\delta_{F}+\eta_{F})-8\delta_{\xi}(\delta_{F}+\eta_{\xi})-2\delta_{\phi}\delta_{JX}}{2+\delta_{F}-8\delta_{\xi}}, \\
&= \delta_{X}-\delta_{F}/2+4\delta_{\xi}+3\delta_{JX}
 -2\delta_{J\phi}+{\cal O}(\epH^{2})\,,\label{epap}
\end{align}
where in the latter step we have taken the leading-order contribution.

\ssec{Observable signatures}
\label{sec:single-perturb}

We now consider cosmological perturbations about a spatially flat \frwl~background, following ref.~\cite{DeFelice:2011zh}. As in chapter~\ref{ch:subhorizon}, it is convenient to employ the~\adm~metric~\cite{Arnowitt:1962hi}. Up to a gauge choice, we include perturbations in the scalar field $\delta \vp$ and both scalar and tensor perturbations of the metric.  However, since we are working with a single field model, it is convenient to use the curvature perturbation $\R$ and specialise to the uniform-field gauge where $\delta \vp=0$. We now quote the second and third order actions for this scenario as computed in ref.~\cite{DeFelice:2011zh}. 

After expanding the action \eqref{eq:action} up to second order, performing multiple integrations by parts and applying the energy and momentum constraints to eliminate the metric perturbations $\alpha_1$ and $\vartheta_1$ one obtains
\begin{equation}
S_{(2)}=\int \d t \, \d ^3 x \, a^{3} Z \left[\dot \R^2
-\frac{c_s^2}{a^2}\,\partial_i \R \partial_i \R \right],\label{eq:az2}
\end{equation}
where 
\begin{align}
Z & \equiv \frac{w_{1}(4w_{1}w_{3}+9w_{2}^{2})}{3w_{2}^{2}}\,,
\label{eq:defQ}\\
c_{s}^{2} & \equiv \frac{3(2w_{1}^{2}w_{2}H-w_{2}^{2}w_{4}
+4w_{1}\dot{w}_{1}w_{2}-2w_{1}^{2}\dot{w}_{2})}
{w_1(4w_{1}w_{3}+9w_{2}^{2})}\,,\label{eq:defc2s} 
\end{align}
and
\begin{align}
w_{1} & \equiv \Mpl^{2}\, F-8H\,\dot{\xi}\,,\\
w_{2} & \equiv \Mpl^{2}(2HF+\dot{F})-2\dot{\phi}XJ_{,X}-24H^{2}\dot{\xi}\,,\\
w_{3} & \equiv -9\Mpl^{2}F{H}^{2}-9\Mpl^{2}H\dot{F}+3\omega X+144H^{3}\dot{\xi} \nonumber \\ & \qquad
+18H\dot{\phi}(2XJ_{,X}+X^{2}J_{,XX})-6(XJ_{,\phi}+X^{2}J_{,\phi X})\,,\\
w_{4} & \equiv \Mpl^{2}F-8\ddot{\xi}\,.
\end{align}
In order to avoid the appearance of ghosts and Laplacian instabilities we respectively require $Z>0$ and $c_s^2>0$. One can express $w_{1\to4}$ in terms of the slow-roll parameters given in eq.~\eqref{slowvariation}. For example, one finds
\begin{equation}
w_{3}=-9\Mpl^{2}FH^{2}\left(1+\delta_{F}-\frac{1}{3}\delta_{X}-16\delta_{\xi}-4\delta_{JX}+\frac{2}{3}\delta_{J\phi}-2\delta_{JX}\lambda_{JX}+\frac{2}{3}\delta_{J\phi}\lambda_{J\phi}\right),
\end{equation}
where the quantities $\lambda_{JX}$ and $\lambda_{J\phi}$ are not necessarily small and are defined as
\begin{equation}
\lambda_{JX} \equiv \frac{XJ_{,XX}}{J_{,X}}\,,\qquad
\lambda_{J\phi} \equiv\frac{XJ_{,\phi X}}{J_{,\phi}}\,.
\end{equation}

We may now expand the speed of sound to leading order in slow-roll parameters viz
\be
c_s^2 \simeq
\frac{\delta_{X}+4\delta_{JX}-2\delta_{J\phi}+2\delta_{J\phi}\lambda_{J\phi}}
{\delta_{X}+6\delta_{JX}-2\delta_{J\phi}+6\delta_{JX}\lambda_{JX}
-2\delta_{J\phi}\lambda_{J\phi}}\,.
\label{cs}
\ee
For standard slow-roll inflation with $F=1$, $\omega=1$, $\xi=0$, and $J=0$, one obtains exactly $c_{s}^{2}=1$. Equation (\ref{cs}) shows that it is only the Galileon term that gives rise to contributions to $c_{\rm s}^{2}$ at linear order. Other terms, such as the \gb~term, only contribute to $c_{s}^{2}$ at the next order.

In order to express the scalar power spectrum analytically, it is helpful to define a new slow-roll parameter $\epsilon_s \equiv Z c_s^2 / (\Mpl^2 \, F)$, which may be expanded as
\begin{align}
\ep_s &= \delta_{X}+4\delta_{JX}-2\delta_{J\phi}
+2\delta_{J\phi}\lambda_{J\phi} \nonumber \\ 
& \qquad -2\delta_{J\phi}\delta_{F}\lambda_{J\phi}
+16\delta_{J\phi}\delta_{\xi}\lambda_{J\phi}
+2\delta_{\phi}\delta_{JX}\lambda_{JX}
+4\delta_{J\phi}\delta_{JX}\lambda_{J\phi} 
+3\delta_{F}^{2}/4 \nonumber \\ & \qquad
-12\delta_{\xi}\delta_{F} 
+2\delta_{JX}\delta_{\phi}-\delta_{F}\delta_{X}
-5\delta_{F}\delta_{JX}
+2\delta_{F}\delta_{J\phi}
+8\delta_{\xi}\delta_{X}
+40\delta_{\xi}\delta_{JX} \nonumber \\ & \qquad
-16\delta_{\xi}\delta_{J\phi}
-4\delta_{JX}\delta_{J\phi}
+2\delta_{JX}\delta_{X}
+48\delta_{\xi}^{2}
+7\delta_{JX}^{2} \nonumber \\ & \qquad \qquad
+\O(\epH^3)\,.
\label{eps}
\end{align}
In the standard slow-roll inflationary scenario one simply finds $\ep_s = Z/\Mpl^2=\delta_{X}=\epH$. Returning to the general case, the power spectrum of the curvature perturbation is given by~\cite{DeFelice:2011zh}
\begin{equation}
\P_\zeta = \frac{H^2}{8\pi^{2}Z c_s^3}
=\frac{H^2}{8\pi^2 \Mpl^2 F\ep_s c_s}\,,
\label{eq:power_orig}
\end{equation}
where it is assumed that these quantities are evaluated at the time of horizon crossing. The scalar spectral index follows, and may be expanded as
\begin{align}
n_\zeta -1 \equiv \frac{\d \ln \P_\zeta}{\d \ln k}
\bigg|_{c_s k = aH} 
& = -2\epH-\delta_Z-3s \,, \\
& \simeq -2\epH-\delta_{F}-\eta_s-s\,,
\label{nR}
\end{align}
where we have introduced three new slow-roll parameters 
\begin{equation}
\delta_{Z}\equiv\frac{\dot{Z}}{HZ}\,,\qquad 
s\equiv\frac{\dot{c}_{s}}{Hc_{s}}\,,
\qquad\eta_{s}\equiv\frac{\dot{\epsilon}_{s}}{H\epsilon_{s}}\,.
\label{etas}
\end{equation}
In writing these expressions we have assumed that both $H$ and $c_{s}$ vary slowly, such that $\d \ln k|_{c_s k=aH}$ may be approximated as $\d \ln a= H \d t$.

The tensor power spectrum is given by \cite{DeFelice:2011zh} 
\begin{align}
\P_t &= \frac{H^{2}}{2\pi^{2}Z_t c_t^3}\,, \\
Z_t &= \frac{w_1}{4} = \frac{\Mpl^2 F}{4} (1-8\delta_\xi)\,, \\
c_t^2 &= \frac{w_4}{w_1} = 1+8 \delta_\xi+\O(\epH^2)\,.
\end{align}
Taking the leading-order contribution in $\P_t$, it follows that $\P_t \simeq 2H^2/(\pi^2 \Mpl^2 F)$. The tensor spectral index is
\be
n_t \equiv \frac{\d \ln \P_t}{\d \ln k} \bigg|_{c_t k=aH} 
\simeq -2\epH-\delta_{F}\,,
\label{nT}
\ee
which is valid at first order in slow-roll. At times before the end of inflation when $\epH \ll 1$, both $\P_\zeta$ and $\P_t$ remain approximately constant and we can estimate the tensor--scalar ratio as 
\be
r \equiv \frac{\P_t}{\P_\zeta} \simeq
16 \frac{Z c_s^3}{\Mpl^2 F} = 16 c_s \ep_s \,.
\label{rgene}
\ee

\sssec{Non-Gaussianity}

The non-Gaussianity of scalar perturbations for the action \eqref{eq:action} follows from the third order perturbed action $S_{(3)}$. This has also been calculated in ref.~\cite{DeFelice:2011zh} (see also refs.~\cite{Maldacena:2002vr,Mizuno:2010ag,Kobayashi:2011pc} for related works). Under the slow-roll approximation the nonlinear parameter $\fnl^{\rm equil}$ was shown to be~\cite{DeFelice:2011zh}
\begin{align}
\fnl^{\rm equil} & \simeq \frac{85}{324}\left(1-\frac{1}{c_{s}^{2}}\right)-\frac{10}{81}\,\frac{\Lambda}{\Sigma}+\frac{55}{36}\,\frac{\epsilon_{s}}{c_{s}^{2}}+\frac{5}{12}\,\frac{\eta_{s}}{c_{s}^{2}}-\frac{85}{54}\,\frac{s}{c_{s}^{2}}
+\frac{5}{162}\,\delta_{F}\left(1-\frac{1}{c_{s}^{2}}\right)
\nonumber \\& \qquad 
-\frac{10}{81}\,\delta_{\xi}\left(2-\frac{29}{c_{s}^{2}}\right)+\delta_{JX}\left[\frac{20\,(1+\lambda_{JX})}{81\epsilon_{s}}+\frac{65}{162c_{s}^{2}\epsilon_{s}}\right]\,,\label{fnleq}
\end{align}
where 
\begin{align}
\Lambda & \equiv 
F^2 \left[\dot \phi H 
(XJ_{,X}+5X^2J_{,XX}+2X^3 J_{,XXX})
-\frac{2}{3}(2X^{2}J_{,\phi X}+X^{3}J_{,\phi XX})\right], \\
\Sigma & \equiv 
\frac{w_{1}(4w_{1}w_{3}+9w_{2}^{2})}{12\Mpl^{4}} \nonumber \\
&\simeq \Mpl^{2}F^{3}H^{2}
\big(\delta_{X}+6\delta_{JX}-2\delta_{J\phi}+6\delta_{JX}\lambda_{JX}
-2\delta_{J\phi}\lambda_{J\phi}\big). 
\end{align}

In the absence of the Galileon term (under which conditions $\delta_{JX}=0=\delta_{J\phi}$) one has $c_{s}^{2}\simeq1$ and 
$\ep_s \simeq \delta_{X}$ from eqs.~\eqref{cs} and~\eqref{eps} at linear order in slow-roll. In this case, the expansion of $c_{s}^{2}$ up to second order gives
\begin{equation}
c_{s}^{2}\simeq1-\frac{2\delta_{\xi}(\delta_{F}-8\delta_{\xi})
(3\delta_{F}-24\delta_{\xi}-4\delta_{X})}{\delta_{X}}\,,
\end{equation}
which shows that the \gb~contribution can only lead to minimal deviations from $c_s^2=1$. In this regime we can expand $\fnl^{\rm equil}$ in eq.~\eqref{fnleq} approximately as
\begin{equation}
\fnl^{\rm equil} \simeq \frac{55}{36}\epsilon_{s}
+\frac{5}{12}\eta_{s}+\frac{10}{3}\delta_{\xi}\,,
\end{equation}
from which it is evident that the bispectrum signal is negligible for theories with $J=0$. However, the presence of the Galileon term can potentially give rise to large non-Gaussianity.

\sec{Transformation to the Einstein Frame}
\label{sec:single-transform}

We start by considering non-minimally coupled theories which also include a coupling to the kinetic energy $X$, in the absence of the \gb~and Galileon terms ($\xi=0=J$). The action takes the form 
\begin{equation}
S=\int \d^{4}x\sqrt{- g}\left[\frac{M_{{\rm pl}}^{2}}{2}F(\vp) R
+\omega(\vp) X- V(\vp)\right].\label{actionsta}
\end{equation}
Models defined by the action \eqref{actionsta} have $c_{s}^{2}=1$ and $s=0$, such that 
\begin{align}
n_\zeta-1 & = -2\epH-\delta_{Z}\simeq
-2\epH-\delta_{F}-\eta_{s}\simeq-2\ep_{s}-\eta_{s}\,,\label{nRs}\\
n_t & \simeq -2\epH-\delta_{F}\simeq-2\epsilon_{s}\,,\label{nTs}\\
r & = \frac{16 Z}{\Mpl^{2}F} \simeq 16\epsilon_{s}\simeq-8n_{{\rm t}}\,,\label{rs} \\
Z &= \frac{F(2F\omega\dot{\phi}^{2}+3\Mpl^{2}\dot{F}^{2})}{(2HF+\dot{F})^{2}}\,.
\end{align}
In the last approximate equalities of eqs.~\eqref{nRs}--\eqref{rs} we have used the relation $\epsilon_{s}\simeq\epH+\delta_{F}/2$ which is valid at linear order in slow-roll. This follows from eqs.~\eqref{epap} and \eqref{eps}, which give $\epH \simeq \delta_{X}-\delta_{F}/2$ and $\epsilon_{s}\simeq\delta_{X}$ respectively. 

It makes for a simpler analysis if we transform the action \eqref{actionsta} into the Einstein frame in which the scalar field is minimally coupled. This may be achieved via a conformal transformation as described in \S\ref{sec:nonminimal} as 
\begin{equation}
\hat{g}_{\mu\nu}=F(\vp)g_{\mu\nu}\,,\label{ctrans}
\end{equation}
where we denote quantities calculated in the Einstein frame with circumflexes. It is then useful to define a new scalar field $\chi$ such that the kinetic term assumes canonical form (we note that this is only possible in general for single field models). The transformed action is given by (for example see ref.~\cite{Kaiser:2010ps})
\begin{equation}
\hat S=\int \d^{4}x\sqrt{-\hat{g}}\left[\frac{1}{2}\Mpl^{2}\hat{R}
-\frac{1}{2}\hat{g}^{\mu\nu}\partial_{\mu}\chi\partial_{\nu}\chi-\hat V(\chi)\right],\label{Eaction}
\end{equation}
where 
\begin{equation}
\hat V=\frac{V}{F^{2}}\,,\qquad\chi\equiv\int B(\vp)\, d\vp\,,\qquad 
B(\vp)\equiv\sqrt{\frac{3}{2}\left(\frac{\Mpl F_{,\vp}}{F}\right)^{2}
+\frac{\omega}{F}}\,.
\end{equation}

The following relations hold between the variables in the two frames:
\begin{equation}
\d \hat{t}=\sqrt{F}\, \d t\,,\qquad\hat{a}=\sqrt{F}\, a\,,\qquad
\hat{H}=\frac{1}{\sqrt{F}}\left(H+\frac{\dot{F}}{2F}\right).\label{Hre}
\end{equation}
We may define Einstein frame variables as
\begin{equation}
\hat{\epH}\equiv-\frac{1}{\hat{H}^{2}}\frac{\d\hat{H}}{\d\hat{t}}\,,
\qquad\hat{Z}\equiv\frac{1}{2\hat{H}^{2}}\left(\frac{\d\chi}{\d\hat{t}}
\right)^{2},\qquad
\hat{\delta}_{\hat{Z}}\equiv\frac{1}{\hat{H}
\hat{Z}}\frac{\d\hat{Z}}{\d\hat{t}},
\label{eq:einsteinvariables}
\end{equation}
which are related to the Jordan frame variables through the expressions~\cite{Komatsu:1999mt,Tsujikawa:2004my}
\begin{equation}
\hat{\epH}=\frac{\epH+\delta_{F}/2}{1+\delta_{F}/2}
-\frac{\dot{\delta}_{F}}{2H(1+\delta_{F}/2)^{2}}\,,\qquad
\hat{Z}=\frac{Z}{F}\,,\qquad
\hat{\delta}_{\hat{Z}}
=\frac{\delta_{Z}-\delta_{F}}{1+\delta_{F}/2}\,.
\end{equation}
Since $\hat{\epH}\simeq\epH+\delta_{F}/2$ and 
$\hat{\delta}_{\hat{Z}}\simeq\delta_{Z}-\delta_{F}$ at linear order in slow-roll, we find that eqs.~\eqref{nRs}--\eqref{rs} reduce to 
\begin{align}
n_\zeta-1 & \simeq -2\hat{\epH}-\hat{\delta}_{\hat{Z}}\,,\label{nRs2}\\
n_t & \simeq -2\hat{\epH}\,,\label{nTs2}\\
r & \simeq 16\frac{\hat{Z}}{\Mpl^{2}}=16\hat{\epH}\,.\label{rs2}
\end{align}
In the last equality of eq.~\eqref{rs2} we have used the relation 
$\hat{\epH}=\hat{Z}/\Mpl^{2}$, which follows from the background equation 
$\d \hat{H}/ \d \hat{t} = -(\d \chi /\d \hat{t})^2/(2 \Mpl^2)$. The results \eqref{nRs2}--\eqref{rs2} coincide with those derived from a purely Einstein frame calculation~\cite{Komatsu:1999mt,Tsujikawa:2004my}. This equivalence is a consequence of the fact that both the scalar and tensor spectra are unchanged under a conformal transformation~\cite{Makino:1991sg,Fakir:1992cg,Kaiser:1994vs,Komatsu:1997hv}.

Under the slow-roll conditions $|\d ^2\chi/\d \hat{t}^2| \ll 
|3 \hat{H} \d \chi/ \d \hat{t}|$ and $(\d \chi/ \d \hat{t})^2 /2 \ll \hat{V}$, the background equations of motion are approximately given by
\begin{equation}
3\Mpl^{2}\hat{H}^{2}\simeq \hat{V}\,,\qquad
3\hat{H}\frac{\d \chi}{\d \hat{t}} \simeq -\hat{V}_{,\chi}\,.
\label{Einback}
\end{equation}
We then find 
\begin{equation}
\hat{\epH}=\frac{\hat{Z}}{\Mpl^{2}}\simeq\frac{\Mpl^{2}}{2}
\left(\frac{\hat{V}_{,\chi}}{\hat{V}}\right)^{2},\qquad
\hat{\delta}_{\hat{Z}}\simeq
2\Mpl^{2}\left[\left(\frac{\hat{V}_{,\chi}}{\hat{V}}\right)^{2}-\frac{\hat{V}_{,\chi\chi}}{\hat{V}}\right].
\end{equation}
The expressions for $n_\zeta$, $n_t$ and $r$ in eqs.~\eqref{nRs2}--\eqref{rs2} can thus be explicitly written in terms of Jordan frame quantities as 
\begin{align}
n_\zeta-1 & \simeq 
-3\Mpl^{2}\left(\frac{\hat{V}_{,\chi}}{\hat{V}}\right)^{2}+2\Mpl^{2}\frac{\hat{V}_{,\chi\chi}}{\hat{V}}, \nonumber \\ &\simeq
\frac{\Mpl^{2}}{B^{2}}\left[2\frac{V_{,\phi\phi}}{V}-3\frac{V_{,\phi}^{2}}{V^{2}}-4\frac{F_{,\phi\phi}}{F}+4\frac{V_{,\phi}}{V}\frac{F_{,\phi}}{F}-2\frac{B_{,\phi}}{B}\left(\frac{V_{,\phi}}{V}-2\frac{F_{,\phi}}{F}\right)\right],
\label{nRex} \\
r & \simeq -8 n_t \simeq 8\Mpl^{2} \left(\frac{\hat{V}_{,\chi}}{\hat{V}}\right)^{2} 
\simeq 8\frac{\Mpl^{2}}{B^{2}}\left(\frac{V_{,\phi}}{V}-2\frac{F_{,\phi}}{F}\right)^{2}.\label{rex}
\end{align}
In the Jordan frame the number of efoldings from the horizon exit time $t$ (with the field value $\phi$) to the time $t_f$ at the end of inflation (with the field value $\phi_f$) is given by 
\begin{equation}
N=\int_{t}^{t_{f}} \!H \, \d t=\int_{\hat{t}}^{\hat{t}_{f}}
\!\hat{H}\, \d \hat{t}+\frac{1}{2}\ln\frac{F}{F_{f}}\,,\label{efold0}
\end{equation}
where $F_f \equiv F(\phi_f)$. Note that in the last equality we have used eq.~\eqref{Hre}. The scales relevant to the \cmbr~temperature anisotropy corresponds to $50 \lesssim N \lesssim 60$. The number of efoldings in the Einstein frame should be equivalent to that in the Jordan frame \cite{Catena:2006bd,Deruelle:2010ht}. Applying the slow-roll approximation in the Einstein frame, the frame-independent quantity \eqref{efold0} can be written as 
\begin{equation}
N\simeq \int_{\chi_{f}}^{\chi}\frac{\hat{V}}{\Mpl^{2}\hat{V}_{,\chi}} \, \d \chi
+\frac{1}{2}\ln\frac{F}{F_{f}}\,,\label{efold2}
\end{equation}
which we will use in the following sections.

\section{Inflation with non-minimal coupling and non-canonical kinetic terms}
\label{sec:single-minimally}

In this section we ignore the effects of the \gb~and the Galileon terms (i.e. let $\xi=J=0$) and instead focus on the non-minimal coupling $F(\vp)$ and non-canonical kinetic term $\omega(\vp) X$. The action is precisely of the form \eqref{actionsta} which the last section showed can be written in the Einstein frame. We take the function $F(\vp)$ to have the form 
\begin{equation}
F(\vp)=1-\zeta \frac{\vp^2}{\Mpl^2} \,.
\end{equation}
For the canonical field with $\omega(\vp)=1$, observational constraints for this non-minimal coupling have been studied for the chaotic potential $V = V_0 (\vp/\Mpl)^p$ using a combination of \wmap~1-year and \lss~data \cite{Tsujikawa:2004my}. Recently, the observational compatibility of this type of potential, as well as the different potential $V(\vp)=\lambda (\vp^2 -v^2)^2/4$, were examined in ref.~\cite{Linde:2011nh} using \wmap~7-year data. The latter potential appears in the context of Higgs inflation with the electroweak scale $v\sim10^{3}$~GeV \cite{Bezrukov:2007ep}. If the non-minimal coupling is negative with $|\zeta|\gg1$, it is possible to use the Higgs field as an inflaton because the self-coupling $\lambda$ can be of the order of $10^{-2}$ or $10^{-1}$ from the \wmap~normalization~\cite{Bezrukov:2007ep}. Since the field $\vp$ is much larger than the electroweak scale during inflation, the observational prediction of the potential $V(\vp)=\lambda (\vp^{2}-v^{2})^{2}/4$ is very similar to that of the potential $V = \lambda \vp^4 /4$. 

In addition to the Higgs non-minimal coupling described above, we shall consider the non-canonical kinetic term $\omega(\vp) X$. We will provide general formulae for $n_\zeta$, $r$, and $n_t$ in terms of the dimensionless background field $x=\phi/\Mpl$ and then apply these for two different choices of $\omega(\vp)$: Firstly, when $\omega(\vp)= {\rm constant}$, and secondly, the exponential coupling $\omega (\vp)=e^{\mu \vp/\Mpl}$, which is of the dilatonic form. 

The Jordan frame potential $V=V_0 x^p$, when written in the Einstein frame, takes the form
\begin{equation}
\hat{V}=\frac{V_{0} \, x^p}{(1-\zeta x^2)^2}\,.\label{Einspoten}
\end{equation}
For $p<4$ this has a local maximum at $x=\sqrt{p/[(4-p)|\zeta|]}$ and hence the non-minimal coupling makes it more difficult to realise inflation. On the other hand, if $p=4$, the potential \eqref{Einspoten} is asymptotically flat in the region $x \gg 1$. If $p>4$ the potential does not possess a local maximum and for $p>5+\sqrt{13}$ inflation does not occur.

From eqs.~\eqref{nRex} and \eqref{rex} it follows that 
\begin{align}
n_\zeta-1 & \simeq 
-\frac{1}{[\omega+(6\zeta-\omega)\zeta x^{2}]^{2}x^{2}}
\biggl\{(p-4)^{2}(6\zeta-\omega)(\zeta x^{2})^{3} \nonumber \\ &
\qquad +\big(24\omega-14p\omega+3p^{2}\omega+24p\zeta-12p^{2}\zeta \big)
(\zeta x^{2})^{2}\nonumber \\ &
\qquad +\big(-8\omega+4p\omega-3p^{2}\omega+24p\zeta+6p^{2}\zeta\big) \zeta x^{2} \nonumber \\ &
\qquad +p\omega(p+2)-\mu\omega x(1-\zeta x^{2})^{2}[(p-4)\zeta x^{2}-p]\biggr\}\,,\label{nRex2}\\
r & \simeq  -8n_{{\rm t}}\simeq\frac{8[p+(4-p)\zeta x^{2}]^{2}}{x^{2}[\omega+(6\zeta-\omega)\zeta x^{2}]}\,,\label{rex2}
\end{align}
where the variation of $\omega$ is parametrized by $\mu \equiv \Mpl \, \omega_{,\phi} / \omega$. We note that in these expressions, and the subsequent formula for $\P_\zeta$, the values of $\omega$ and $x$ are implicitly evaluated at horizon crossing. For the two cases $\omega(\vp)= {\rm constant}$ and $\omega(\vp)=e^{\mu\vp/\Mpl}$, the parameter $\mu$ is zero and constant respectively. Combining the slow-roll equations of motion \eqref{Einback} and the definitions \eqref{eq:einsteinvariables} into the power spectrum expression~\eqref{eq:power_orig} yields
\begin{equation}
\P_\zeta \simeq \frac{\hat{V}^3}{12\pi^2 \Mpl^6 \hat{V}_{,\chi}^2}
=\frac{V_0}{12 \pi^2 \Mpl^4} \frac{x^{p+2} [6 \zeta^2 x^2+\omega
(1-\zeta x^2)]}{(1-\zeta x^2)^2 [p+(4-p)\zeta x^2]^2}\,.
\label{Psam}
\end{equation}
This is constrained by the \wmap~normalization given in eq.~\eqref{eq:Pzeta_data}. We first consider the non-minimally coupled theories with $\omega = {\rm constant}$ and $\mu=0$, before proceeding to discuss the dilatonic case $\omega = e^{\mu \vp/\Mpl}$ for which $\mu$ is a non-zero constant.

\subsection{Non-minimal coupling with constant $\omega$}

Models with constant $\omega$ are found from the previous analysis by setting $\mu = 0$. Introducing a new field $\psi=\sqrt{\omega}\vp$, the kinetic term $\omega X$ reduces to the canonical form $-g^{\mu \nu} \partial_{\mu} \psi \partial_{\nu} \psi/2$. In this case the non-minimal coupling $\zeta \vp^2 R/2$ can be written as $\tilde{\zeta} \psi^2 R/2$ with $\tilde{\zeta}=\zeta/\omega$. The potential $V(\vp)=V_0(\vp/\Mpl)^p$ takes the power-law form $V=\tilde V_0(\psi/\Mpl)^p$, where $\tilde V_0=V_0/\omega^{p/2}$. This means that these theories reduce to non-minimally coupled theories with $\omega=1$ in terms of the field $\psi$. The ratio $\tilde{\zeta}=\zeta/\omega$ characterizes the effect of the non-minimal coupling on the inflationary observables $n_\zeta$, $n_t$, and $r$, and $\tilde V_0=V_0/\omega^{p/2}$ normalizes the scalar power spectrum.

From eq.~\eqref{efold2} the number of efoldings is given by 
\begin{align}
N &\simeq -\frac{1}{4\zeta}\ln\left|\frac{(p-4)\zeta x_{f}^{2}-p}
{(p-4)\zeta x^{2}-p}\right|^{\frac{3p\zeta-2\omega}{p-4}}
-\frac{1}{4}\ln\left|\frac{1-\zeta x^{2}}{1-\zeta x_{f}^{2}}\right|,
\qquad &\mbox{---}\textrm{for~}p\neq4 \,,\label{Nap0}\\
N &\simeq \frac{\omega-6\zeta}{8}(x^{2}-x_{f}^{2})
-\frac{1}{4}\ln\left|\frac{1-\zeta x^{2}}{1-\zeta x_{f}^{2}}\right|,
&\mbox{---}\textrm{for~}p=4\,,\label{Nap}
\end{align}
where $x_f \equiv \phi_f/\Mpl$. The result \eqref{Nap} can also be reproduced by taking the limit $p\to4$ in eq.~\eqref{Nap0}. We identify the end of inflation by the condition
$\hat{\epH}=1$, which gives  
\begin{equation}
x_{f}^{2}=\frac{\omega-\zeta p(4-p)-\sqrt{(\omega-2p\zeta)
(\omega-6p\zeta)}}{\zeta \big[\zeta(4-p)^{2}+2(\omega-6\zeta)\big]}\,.
\label{psif}
\end{equation}

Let us consider the limit where $|\zeta/\omega| \ll 1$. We implicitly assume that $\omega$ is not different from $\O(1)$. Expanding the \rhs~of eqs.~\eqref{Nap0} and \eqref{Nap} up to first order in $\zeta$ and then solving for $x$ using eq.~\eqref{psif} gives 
\begin{equation}
x^{2} \simeq \frac{p(p+4N)}{2\omega} \left[ 1
-\frac{8(p-4)N^2+4p (p-6)N+p^2(p-8)}{2(p+4N)}
\frac{\zeta}{\omega} \right],
\label{psi1}
\end{equation}
which is valid for both $p \neq 4$ and $p=4$. The spectral index \eqref{nRex2} and the tensor--scalar ratio \eqref{rex2} are then approximately given by 
\begin{align}
n_\zeta-1 & \simeq -{\frac{2(p+2)}{p+4N}}
\left[1-{\frac{4(p-2)(p-12){N}^{2}+2p({p}^{2}-12\,{p}+28)N+p^{2}(12-p)}
{\left(p+4\, N\right)\left(p+2\right)}} \frac{\zeta}{\omega}\right], \label{nRap} \\ 
r &\simeq {\frac{16p}{p+4\, N}}\left[1-{\frac{2N\bigl(2(p-12)N+p(p-10)\bigr)}
{p+4N}} \frac{\zeta}{\omega}\right].
\label{rap}
\end{align}
We see that the effect of the non-minimal coupling appears only in terms of the ratio $\zeta/\omega$.

Substituting eq.~\eqref{psi1} into the formula for $\P_\zeta$~\eqref{Psam} and expanding it up to first order in $\zeta$, it follows that 
\begin{align}
\P_\zeta &\simeq \frac{{\tilde V}_0}{\Mpl^4} \frac{p+4N}{24 \pi^2 p} 
\left[ \frac{p(p+4N)}{2} \right]^{p/2} \times \nonumber \\ & \qquad
\left[ 1-\frac{p^4+(4N-12)p^3+(8N^2-64N)p^2+(80N-112N^2)p+192N^2}
{4(p+4N)} \frac{\zeta}{\omega} \right] .\nonumber \\
\label{Pscon}
\end{align}
We can now equate this to the \wmap~power spectrum at $N=55$ efolds. In the absence of the non-minimal coupling ($\zeta=0$) one finds $m \simeq 6.8 \times 10^{-6}\Mpl$ for $p=2$ (where $\tilde V_0=m^2 \Mpl^2/2$) and $\lambda \simeq 2.0 \times 10^{-13} $ for $p=4$ (where $\tilde V_0=\lambda \Mpl^4/4$). If $\zeta \neq 0$, then the inside of the last parenthesis in eq.~\eqref{Pscon} is approximately given by $1+4\zeta/\omega$ for $p=2$ and $1+460 \zeta/\omega$ for $p=4$. As long as $|\zeta/\omega| \ll 1$, we conclude that the order of ${\tilde V}_0$ is not subject to change by the presence of the non-minimal coupling.

In the following we derive the numerical values of $n_\zeta$ and $r$ for $p=2$ and $p=4$ separately to compare the models with observations.

\subsubsection{Models with $p=2$}

In order to obtain the theoretical values of $n_\zeta$ and $r$ for $p=2$, we numerically solve the background equations of motion in the Jordan frame by identifying the end of inflation under the condition given in eq.~\eqref{psif}. We derive the numerical values of $x$ corresponding to the number of efoldings $N=55$ and then evaluate $n_\zeta$ and $r$ using the formulae \eqref{nRex2} and \eqref{rex2}.

\begin{figure}[h]
\begin{centering}
\includegraphics[width=0.7\textwidth]{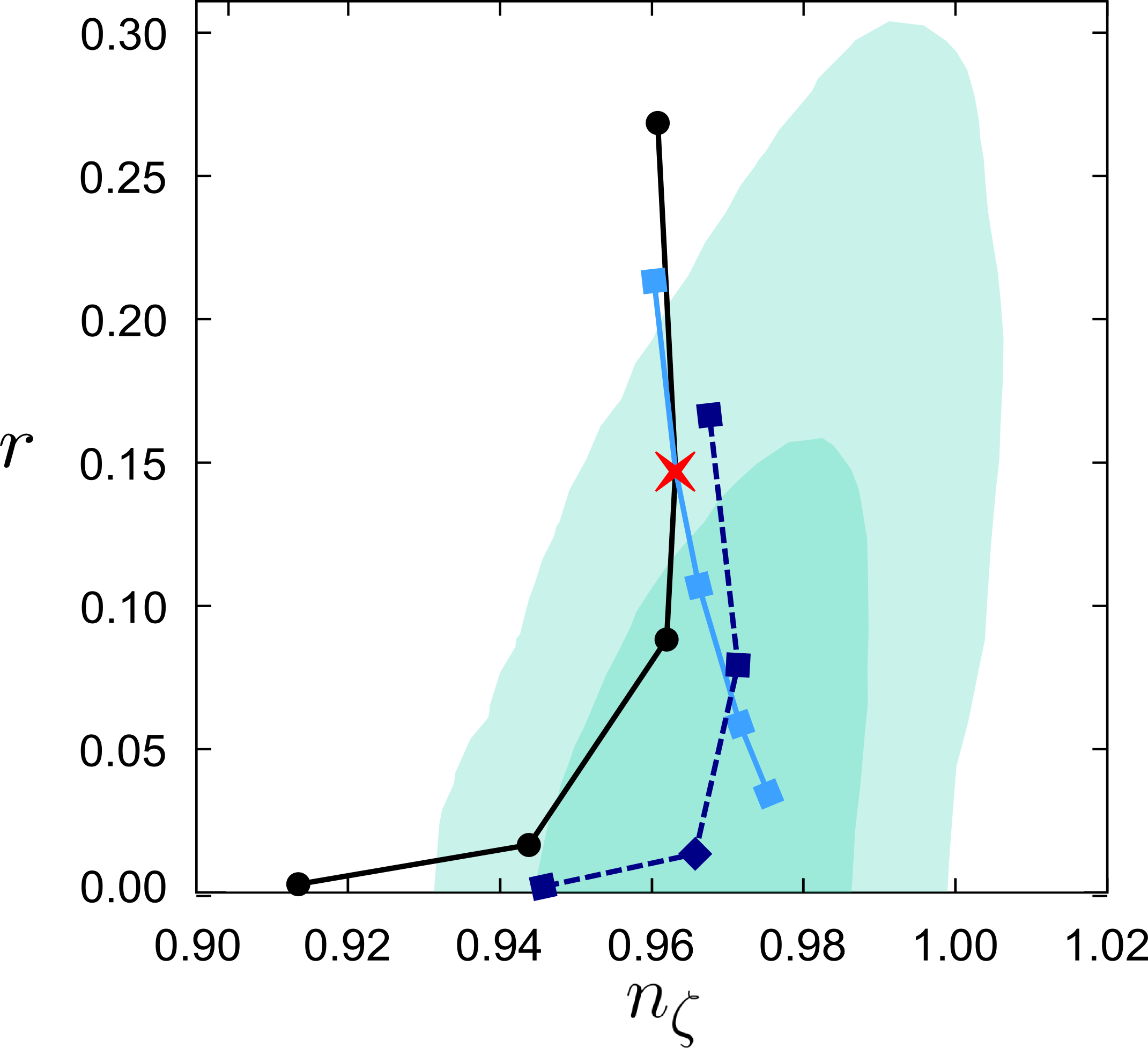} 
\par\end{centering}
\caption[Constraints on the model $V=m^2 \vp^2/2$ for non-minimal coupling and non-canonical kinetic terms]
{1$\sigma$ and 2$\sigma$ likelihood contours are shown in shades of green in the $(n_\zeta,r)$ plane, generated by the joint data analysis of \wmap~$7$-year, \bao, and \hst, with the pivot scale $k_{0}=0.002$~Mpc$^{-1}$. The red cross is the theoretical prediction for the canonical and minimally coupled potential $V(\vp)=m^2 \vp^2/2$ at $N=55$. Non-minimal and non-canonical couplings are illustrated by three lines where we label points from top to bottom as follows: \\
$\bullet$ {\it Solid black line with circles}---constant $\omega$ (i.e.~$\mu=0$) in the presence of the non-minimal coupling $\zeta \vp^2 R/2$ with $\zeta/\omega=0.001,0,-0.001,-0.005,-0.01$. \\
$\bullet$ {\it Solid light blue line with squares}---exponential coupling $e^{\mu\vp/\Mpl}X$ with $\mu=-0.05,0, 0.1,1,10$, in the absence of the non-minimal coupling. \\ 
$\bullet$ {\it Dashed dark blue line with squares}---exponential coupling $e^{\vp/\Mpl}X$ (i.e. $\mu=1$) in the presence of the non-minimal coupling with $\zeta=0.03, 0.01, -0.05,-0.1$.}
\centering{}\label{fig1} 
\end{figure}

In figure~\ref{fig1} we show the $1\sigma$ and $2\sigma$ observational contours constrained by the joint analysis of data from \wmap~7-year~\cite{Komatsu:2010fb}, \bao~\cite{Percival:2009xn} and \hst~\cite{Riess:2011yx}. This is derived by varying the two parameters $n_\zeta$ and $r$ with the consistency relation $r=-8 n_t$ (as derived in eq.~\eqref{rs}). Since the runnings of scalar and tensor spectral indices are suppressed as $\O(\epH^2)$, they are set to zero in the likelihood analysis. These results are valid for the theories with $\xi=0=J$.

In the limit $|\zeta|\ll1$, eqs.~(\ref{nRap}) and (\ref{rap}) give
\begin{align}
n_\zeta-1 &\simeq -\frac{4}{2N+1} \left[1
-\frac{4N+5}{2N+1} \frac{\zeta}{\omega}
+\frac { 2\left(104 {N}^{4}+160 {N}^{3}+84 {N}^{2}-30\,N-9 \right)}
{3 \left( 2N+1 \right)^{2}}\frac{\zeta^2}{\omega^2} \right], \label{nsp=2}\\
r &\simeq \frac{16}{2N+1}\left[1+{\frac {4N 
\left( 5N+4 \right)}{2N+1}}\,\frac\zeta\omega\right].
\label{rp=2}
\end{align}
For the scalar spectral index we have included the second-order correction in $\zeta/\omega$ because the dominant contribution to the first-order term in $\zeta/\omega$ in eq.~\eqref{nRap} vanishes for $p=2$. In the absence of the non-minimal coupling ($\zeta=0$), we calculate $n_\zeta=0.964$ and $r=0.144$ for $N=55$, which is inside the $2\sigma$ observational bound shown in figure~\ref{fig1}. A positive non-minimal coupling leads to an increase of $r$ relative to the case $\zeta=0$. Since $r$ is observationally bounded from above, this puts an upper bound on the positive value of $\zeta$. The parameter $\zeta$ is also bounded from below since a negative non-minimal coupling pushes the scalar spectral index too far from the scale-invariant limit. Together, $n_\zeta$ and $r$ constrain the non-minimal coupling ratio $\zeta/\omega$ as
\begin{equation}
-7.0\times10^{-3}<\zeta/\omega<7.0 \times 10^{-4} ~{\rm at}~95\%~\textsc{cl.}
\label{p=2con}
\end{equation}
This agrees with the range derived in ref.~\cite{Linde:2011nh} for $\omega=1$. The lower bound in eq.~\eqref{p=2con} is slightly tighter than the constraint $\zeta>-1.1\times10^{-2}$ (with $\omega=1$) \cite{Tsujikawa:2004my} obtained by using \wmap~1-year and \lss~data.

\subsubsection{Models with $p=4$}

\begin{figure}[h]
\begin{centering}
\includegraphics[width=0.7\textwidth]{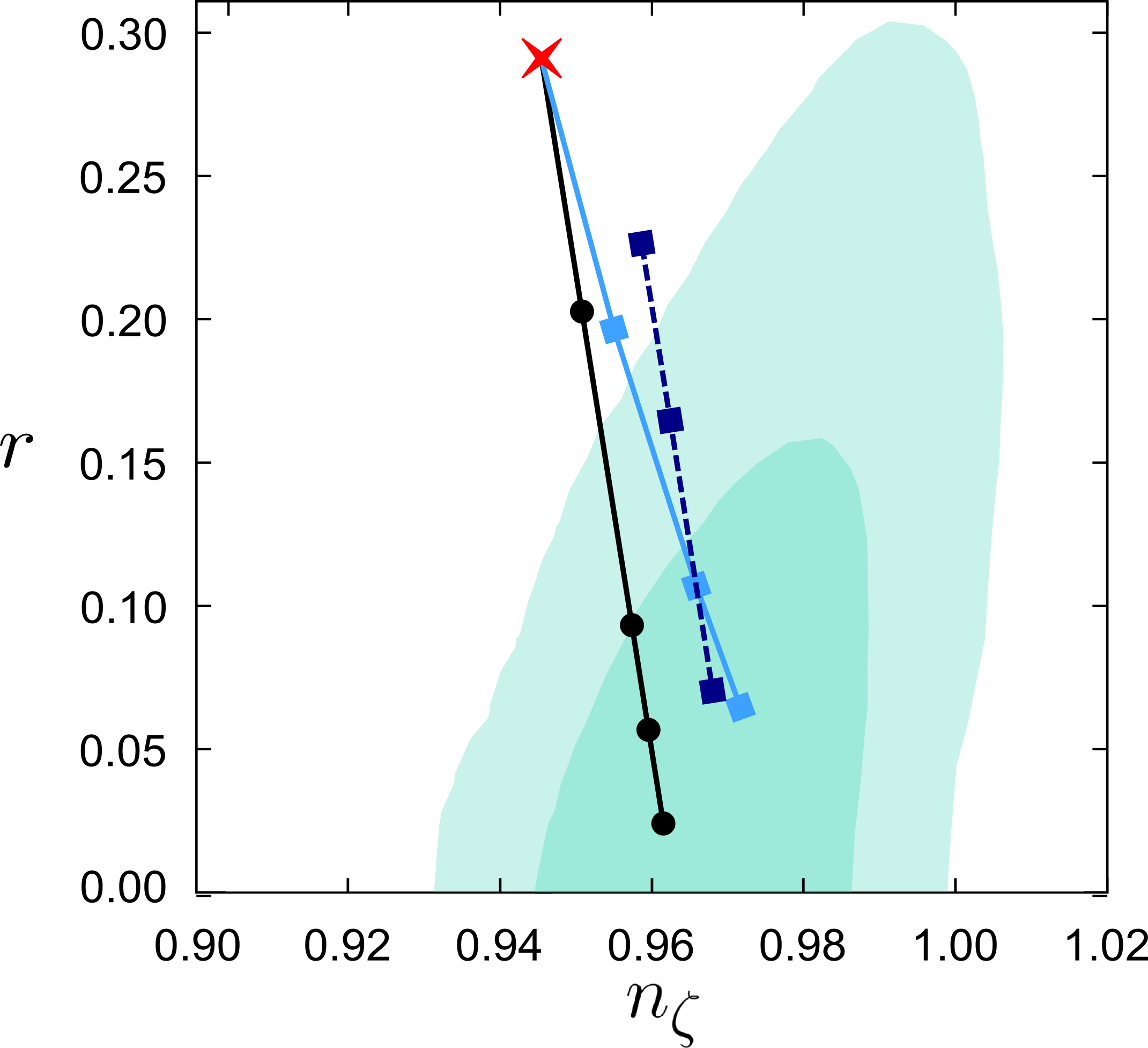} 
\par\end{centering}
\caption[Constraints on the model $V=\lambda \vp^4/4$ for non-minimal coupling and non-canonical kinetic terms]{The green regions are the same observational constraints as shown in figure~\ref{fig1}. The red cross is the theoretical prediction for the canonical and minimally coupled potential $V(\vp)=\lambda \vp^4/4$ at $N=55$. Non-minimal and non-canonical couplings are illustrated by three lines, where we label points from top to bottom as follows: \\
$\bullet$ {\it Solid black line with circles}---constant $\omega$ (i.e.~$\mu=0$) in the presence of a non-minimal coupling $\zeta \vp^2 R/2$ with $\zeta/\omega=0,-0.001,-0.005,-0.01,-0.03$. \\
$\bullet$ {\it Solid light blue line with squares}---exponential coupling $e^{\mu\vp/\Mpl}X$ with $\mu=0.1,1,10$ in the absence of non-minimal coupling. \\ 
$\bullet$ {\it Dashed dark blue line with squares}---exponential coupling $e^{\vp/\Mpl}X$ (i.e. $\mu=1$) in the presence of a non-minimal coupling with $\zeta =0.03, 0.02, -0.03$.}
\centering{}\label{fig2} 
\end{figure}

We proceed to the case of the self-coupling inflaton potential $V(\vp)=\lambda \vp^4/4$. In the regime $|\zeta/\omega|\ll1$, eqs.~(\ref{nRap}) and (\ref{rap}) give 
\begin{align}
n_\zeta-1 & \simeq -\frac{3}{N+1}\left[1+{\frac { 4\left( 2{N}^{2}+N-4
 \right)}{3(N+1)}}\,\frac\zeta\omega\right],\\
r & \simeq \frac{16}{N+1}\left[1+{\frac {4N \left( 2N+3 \right) 
}{N+1}}\,\frac\zeta\omega\right].
\end{align}
In the absence of non-minimal coupling one has $n_\zeta=0.946$ and $r=0.286$ for $N=55$, which is outside the $2\sigma$ observational bound, as shown in figure~\ref{fig2}. 

The presence of a negative non-minimal coupling leads to the increase of $n_\zeta$, whereas $r$ gets smaller. Hence it is possible for this mechanism to make the self-coupling potential consistent with observations. From the joint data analysis of \wmap~7-year~\cite{Komatsu:2010fb}, \bao~\cite{Percival:2009xn} and \hst~\cite{Riess:2011yx}, the non-minimal coupling is constrained to be
\begin{equation}
\zeta/\omega<-2.0 \times 10^{-3} ~{\rm at}~95\%~\textsc{cl.}
\end{equation}
This is tighter than the bound $\zeta<-3.0 \times 10^{-4}$ (with $\omega=1$) derived in ref.~\cite{Tsujikawa:2004my} obtained by using \wmap~1-year and \lss~data.

It is also interesting to consider the alternative limit of $\zeta < 0$ and $|\zeta/\omega| \to \infty$. In this case, inflation is realised by the flat potential $\hat V$ in the Einstein frame in the regime $x \gg 1$. One obtains $N\simeq-3\zeta x^{2}/4$ and $x_{f}^2\simeq-2\sqrt{3}/(3\zeta)$ from eqs.~(\ref{Nap}) and (\ref{psif}) respectively. The leading contributions to $n_\zeta$ and $r$ in the regime $N\gg1$ then follow from eqs.~(\ref{nRex2}) and (\ref{rex2}) as
\begin{align}
n_\zeta-1 & \simeq -2/N\,,\label{nslimit}\\
r & \simeq 12/N^{2}\,.\label{rlimit}
\end{align}
As long as $|\zeta|$ is sufficiently large relative to $\omega$, the effect of the term $\omega$ appears only as the next order corrections to (\ref{nslimit}) and (\ref{rlimit}) with the order of $\omega/(\zeta N^{2})$. For $N=55$, one finds $n_\zeta=0.964$ and $r=0.004$ which is well within the $1\sigma$ observational bound. These predictions, formally valid as $|\zeta| \to \infty$, define the Fakir-Unruh point in the $(n_\zeta,r)$ space.

In the regime $|\zeta/\omega| \gg 1$, the power spectrum (\ref{Psam}) reduces to $\P_\zeta \simeq \lambda N^2/(72 \pi^2 \zeta^2)$, so that the \wmap~normalization 
$\P_\zeta \simeq 2.4 \times 10^{-9}$ at $N=55$ gives
\begin{equation}
\lambda/\zeta^2 \simeq 5.6 \times 10^{-10}\,.
\end{equation}
For large negative non-minimal couplings, such as $\zeta \sim -10^{4}$, the self-coupling $\lambda$ can be of the order of $10^{-2}$. This was the property used in previous work on Higgs inflation~\cite{Bezrukov:2007ep}.

\subsection{Non-canonical kinetic coupling with $\zeta=0$}

Let us now consider a minimally coupled scenario ($\zeta=0$) with a dilatonic coupling $\omega(\vp)=e^{\mu\vp/\Mpl}$ to the kinetic energy $X$. After the field settles down to the potential minimum where $\vp=0$, the coupling $\omega(\vp)$ tends to unity and one recovers the standard kinetic energy $X$. The number of efoldings (\ref{efold2}) is given by 
\begin{equation}
N=\frac{1}{p\mu^{2}}\left[(\mu x-1)e^{\mu x}-(\mu x_{f}-1)e^{\mu x_{f}}\right].
\label{Nderi}
\end{equation}
Since $\hat{\epH} \simeq p^{2}/(2x^{2}\omega)$, we can estimate $x_{f}$ by setting $\hat{\epH}=1$ as
\begin{equation}
x_{f}^{2}\, e^{\mu x_{f}}=p^{2}/2\,,\qquad{\rm or}
\qquad x_{f}=\frac{2}{\mu} W\left(\frac{\sqrt{2} p |\mu|}{4} \right),\label{xf}
\end{equation}
where $W$ is Lambert's function~\cite{Corless:1996zz}. The scalar spectral index and the tensor--scalar ratio are found from eqs.~\eqref{nRex2} and \eqref{rex2} as 
\begin{align}
n_\zeta-1 & \simeq -\frac{p}{x^{2}e^{\mu x}}\left(p+2+\mu x\right),\label{nski}\\
r & \simeq \frac{8p^{2}}{x^{2}e^{\mu x}}\,.
\label{rki}
\end{align}

In the limit $|\mu|\ll1$, eq.~(\ref{xf}) allows us to rewrite eq.~(\ref{Nderi}) in the form 
\begin{equation}
N\simeq{\frac{2x^{2}-{p}^{2}}{4p}}+\frac{\left(8x^{3}
+\sqrt{2}p^{3}\right)\mu}{24p}\,.
\end{equation}
This can be solved for $x$ to yield
\begin{equation}
x^{2}\simeq\frac{p^{2}}{2}+2pN-\frac{\mu}{12}
\left[\sqrt{2}\, p^{3}+(2p^{2}+8pN)^{3/2}\right].
\end{equation}
We may now substitute this relation into the formulae for $n_\zeta$ and $r$ in eqs.~(\ref{nski}) and (\ref{rki}) to obtain
\begin{align}
n_\zeta -1 & \simeq -\frac{p+2}{2N}
\left[1-\frac{(p-1)\mu\sqrt{2pN}}{3(p+2)}\right],\label{nski2}\\
r & \simeq \frac{4p}{N}\left(1-\frac{\mu\sqrt{2pN}}{3}\right),
\end{align}
which are valid up to the first order in $\mu$. The presence of a positive value of $\mu$ results in a spectral index that is closer to the scale-invariant spectrum, meanwhile $r$ is attenuated. In figures~\ref{fig1} and \ref{fig2} we plot the theoretical values of $n_\zeta$ and $r$ in the ($n_\zeta,r$) plane for $p=2$ and $p=4$ respectively, with several different values of $\mu$. These are derived numerically by integrating the background equations and are therefore valid in the regime $\mu \gtrsim 1$. Interestingly, the models with large positive values of $\mu$ can be favoured observationally. On the other hand, the models with negative $\mu$ push the predictions away from the observationally allowed region. The joint observational constraints from \wmap~7-year~\cite{Komatsu:2010fb}, \bao~\cite{Percival:2009xn} and \hst~\cite{Riess:2011yx} give the following bounds on $\mu$: 
\begin{alignat}{2}
\mu &> -0.04~{\rm at}~95\%~\textsc{cl,} \qquad & &\mbox{---}{\rm for~} p=2\,,\\
\mu &> 0.2~{\rm at}~95\%~\textsc{cl,} & &\mbox{---}{\rm for~} p=4 \,.
\end{alignat}

Let us now consider the limit where $\mu\gg1$. In this regime the condition $\mu x\gg1$ is satisfied, so that eq.~\eqref{Nderi} yields $N\simeq xe^{\mu x}/(p\mu)$. The scalar spectral index and tensor--scalar ratio in eqs.~\eqref{nski} and \eqref{rki} then reduce to 
\begin{align}
n_\zeta-1 & \simeq -\frac{1}{N}\,,\label{nski3}\\
r & \simeq \frac{8p}{N}\frac{1}{\mu x}\,.
\label{rki3}
\end{align}
For a given $N$, $\mu x$ increases for larger $\mu$. This means that, in the limit $\mu\gg1$, one finds $n_\zeta-1 \to -1/N \simeq 0.982$ (for $N=55$) and $r\to0$. This limit is inside the $1\sigma$ observational bound. We have confirmed numerically that inflation is followed by oscillations of $\vp$ as required for a successful reheating phase.

\subsection{Combined non-minimal and non-canonical couplings}

To complete this section we consider the case in which both the non-minimal coupling $\zeta \vp^2 R/2$ and the non-canonical kinetic term $e^{\mu \vp/\Mpl}X$ are present. Since it is difficult to derive an analytic form for the number of efoldings $N$, we solve the background equations numerically to identify the values $x$ corresponding to $N=55$ before $x=x_f$, which happens when $\hat{\epH}=1$. We then use the formulae (\ref{nRex}) and (\ref{rex}) to evaluate $n_\zeta$ and $r$ for given values of $p$, $\mu$, and $\zeta$.

In figure~\ref{fig1} we plot the numerical values of $n_\zeta$ and $r$ in the two-dimensional plane for $p=2$ and $\mu=1$ with $\zeta=0.03, 0.01, -0.05, -0.1$. The presence of the term $e^{\mu \vp/\Mpl}X$ with $\mu>0$ leads to compatibility of non-minimally coupled models with larger values of $|\zeta|$ than is the case for $\mu=0$ and $\omega=1$. For instance, when $\mu=1$, we find that the non-minimal coupling is constrained to be 
\begin{equation}
 -0.12< \zeta<0.035~{\rm at}~95\%~\textsc{cl.}
 \label{p2com}
\end{equation}
This is wider than the range (\ref{p=2con}) which we found for $\mu=0$ and $\omega=1$. If $|\zeta|$ is larger than the bounds given by (\ref{p2com}), the effect of the non-minimal coupling is more important than that of the non-canonical kinetic term.

For $p=4$ and $\mu=0$, a positive non-minimal coupling is not allowed observationally because both $|n_\zeta-1|$ and $r$ tend to be larger than those for $\zeta=0$. However, the non-canonical kinetic term with $\mu>0$ allows the compatibility of the positive non-minimally coupled models with observations, as shown in figure~\eqref{fig2}. For example, if $\mu=1$ then $\zeta$ is constrained as
\begin{equation}
\zeta<0.025 ~{\rm at}~95\%~\textsc{cl.}
\label{p4com}
\end{equation}
For $\mu=1$, models with $\zeta<0$ are within the $1\sigma$ observational bound. In the limit of the large negative non-minimal coupling ($|\zeta| \gg 1$), the scalar spectral index and the tensor--scalar ratio are again given by eqs.~(\ref{nslimit}) and (\ref{rlimit}).

For $\mu$ larger than $\O(1)$, the effect of the non-canonical kinetic term tends to be more important. In the limit that $\mu \gg 1$, with a finite value of $\zeta$ (where $|\zeta| \lesssim 1$), $n_\zeta$ and $r$ approach the values given by eqs.~(\ref{nski3}) and (\ref{rki3}).

\sec{Brans-Dicke theories}
\label{sec:single-bd}

We now proceed to Brans-Dicke (\bd) theory \cite{Brans:1961sx} with the action
\begin{equation}
S=\int \d ^4 x\sqrt{-g}\left[\frac{1}{2} \Mpl \, \vp R
+\frac{\Mpl}{\vp} \wbd X-V(\vp)\right],
\label{Jframe}
\end{equation}
where $\wbd$ is the \bd~parameter. Under the conformal transformation (\ref{ctrans}) 
we obtain the Einstein frame action (\ref{Eaction}) with
\begin{align}
F &= \frac{\vp}{\Mpl}=e^{\mu\chi/\Mpl}\,, \\
\hat{V} &= e^{-2\mu\chi/\Mpl}\, V, \\
\mu &= \frac{1}{\sqrt{3/2+\wbd}}\,.\label{mudef}
\end{align}
We choose the integration constant for the field $\chi$ such that $\chi=0$ corresponds to $\vp=\Mpl$. 

\subsection{Case for a power-law potential}

So far our specific interest has been with inflationary models where the Jordan frame potential assumes the power-law form $V(\vp) = V_0 \vp^p$. For \bd~theory, the Einstein frame potential is then given by 
\begin{equation}
\hat{V} (\chi)= V_0 e^{\nu \chi/\Mpl}\,,\qquad
\nu \equiv \frac{p-2}{\sqrt{3/2+\omega_{{\rm BD}}}}\,.\label{exponential}
\end{equation}
For $\wbd = \O(1)$, inflation does not occur unless $p$ is close to 2. However, for $\wbd \gg 1$, it is possible to realise $|\nu| \ll 1$ even if $p$ is deviant from $2$.

We may now apply the potential (\ref{exponential}) to our general expressions for $n_\zeta$ and $r$ in eqs.~(\ref{nRex}) and (\ref{rex}) to find
\begin{align}
n_\zeta-1 & \simeq -\nu^{2}\,,\\
r & = -8n_{\rm t} \simeq 8\nu^{2}\,.
\end{align}
The \cmbr~likelihood analysis using data from \wmap~7-year~\cite{Komatsu:2010fb}, \bao~\cite{Percival:2009xn} and \hst~\cite{Riess:2011yx} provides bounds on $\nu$ of \cite{Ohashi:2011na}
\begin{equation}
0.09<\nu<0.23 ~{\rm at}~95 \% ~\textsc{cl.}
\end{equation}
This translates into a constraint on $\wbd$ as 
\begin{equation}
19(p-2)^{2}-3/2<\omega_{{\rm BD}}<123(p-2)^{2}-3/2\,.\label{omecon}
\end{equation}
The reason why the $p=2$ case (i.e. $\nu=0$) is disfavoured is that it recovers the Harrison-Zel'dovich scale-invariant spectrum which does not agree with the observed scalar spectral index $n_\zeta=1$. On the other hand, if $p=4$ then eq.~\eqref{omecon} gives the bound $75<\wbd<491$. Note that such bounds only apply around the time of horizon crossing where we are approximating $\wbd$ to be constant. 

A problem with the power law potential $V(\vp) = V_0 \vp^p$ is that it yields an exponential potential in the Einstein frame which does not naturally lead to an oscillatory reheating regime. We shall now consider Starobinsky's $f(R)$ model \cite{Starobinsky:1980te} which is a suitable potential that addresses this issue.

\subsection{Models including Starobinsky's $f(R)$ scenario}

$f(R)$ theory is a modification of the Einstein Hilbert action of the form
\begin{equation}
S=\int \d ^4 x\sqrt{-g} \, \frac{\Mpl^{2}}{2} \, f(R)\,.\label{actionfR}
\end{equation}
This theory is equivalent to \bd~theory with $\omega_{{\rm BD}}=0$ \cite{O'Hanlon:1972ya,Chiba:2006jp}. In fact the action (\ref{actionfR}) can be written as 
\begin{equation}
S=\int \d ^4 x\sqrt{-g}\left[\frac{\Mpl^{2}}{2}F(\vp) R - V(\vp)\right],
\label{actionfR2}
\end{equation}
where 
\begin{equation}
F=\frac{\vp}{\Mpl}=\frac{\partial f}{\partial R}\,,\qquad V(\vp)=\frac{\Mpl^{2}}{2}\left(R\frac{\partial f}{\partial R}-f\right).
\end{equation}
Starobinsky's model has $f(R)=R+R^{2}/(6M^{2})$ from which we find the Ricci curvature as $R=3M^{2}(\vp/\Mpl-1)$ and the inflationary potential as
\begin{equation}
V(\vp)=\frac{3M^{2}}{4}(\vp-\Mpl)^{2}\,.\label{Stapo}
\end{equation}
We consider the more general potential 
\begin{equation}
V(\vp)=V_{0}(\vp-\Mpl)^{p}\,,
\end{equation}
for arbitrary values of $\wbd$; Starobinsky's model is recovered as a special case by setting $p \to 2$ and $\wbd \to 0$. The potential in the Einstein frame reads
\begin{equation}
\hat{V} = V_{0}{\Mpl}^{p}e^{(p-2)\mu\chi/\Mpl}\left(1-e^{-\mu\chi/\Mpl}\right)^{p},
\label{poEin}
\end{equation}
where $\mu$ is defined in eq.~(\ref{mudef}). For $|\wbd| \sim \O(1)$ (which prescribes $\mu \sim \O(1)$), inflation occurs in the regime $\chi\gg\Mpl$. The behaviour of the potential (\ref{poEin}) depends on the values of $p$: 
\begin{itemize}
\item $p=2$: The potential (\ref{poEin}) becomes constant for $\chi\gg\Mpl$. We note that it is not necessary for $p$ to exactly equal $2$ for this behaviour to occur. Since $\hat{V}$ is approximated as $\hat{V} \propto\chi^{2}$ in the regime $\chi\ll\Mpl$, inflation is followed by a successful reheating. 
\item $p>2$: The field rolls down the potential towards $\chi=0$. 
\item $p<2$: The field rolls to $\chi=+\infty$ or $\chi=0$ depending on the initial condition. In the latter case the potential does not have a minimum at $\vp=0$ and so reheating is problematic.
\end{itemize}
From eqs.~(\ref{nRex}) and (\ref{rex}) we may calculate the observational quantities
\begin{align}
n_\zeta-1 & = -\frac{\mu^{2}\left[4+2(3p-4)F+(p-2)^{2}F^{2}
\right]}{(F-1)^{2}}\,,\label{nRBD}\\
r & = \frac{8\mu^{2}[2+(p-2)F]^{2}}{(F-1)^{2}}\,.\label{rBD}
\end{align}
The number of efoldings (\ref{efold2}) reads 
\begin{align}
N & = \frac{1}{2\mu^{2}}\left(F-F_{f}\right)+\frac{1}{2}\left(1-\frac{1}{\mu^{2}}\right)
\ln\left(\frac{F}{F_{f}}\right), & &\mbox{---}{\rm for~} p=2,\label{N1}\\
N & = \frac{p}{2\mu^{2}(p-2)}\ln\left(\frac{2+(p-2)F}{2+(p-2)F_{f}}\right)+\frac{1}{2}\left(1-\frac{1}{\mu^{2}}\right)\ln\left(\frac{F}{F_{f}}\right),\quad & & \mbox{---}{\rm for~} p\neq 2, \label{N2}
\end{align}
where $F_{f}$ is the value of $F$ at the end of inflation. Using the criterion $\hat{\epH}=1$ for the end of inflation, we have
\begin{equation}
F_{f}=\frac{1+\sqrt{2}\mu}{1-(p-2)\mu/\sqrt{2}}\,.
\end{equation}
We split the remaining analysis into two cases, models with $p=2$ and those with $p \neq 2$.

\subsubsection{Models with $p=2$}

For the theories with $|\wbd| \sim \O(1)$ (and thus $\mu \sim \O (1)$) one has $F_{f}=1+\sqrt{2}\mu={\cal O}(1)$ and $N\simeq F/(2\mu^{2})$, which means that $F\gg1$ for $N\gg1$. From eqs.~\eqref{nRBD} and \eqref{rBD} it follows that 
\begin{align}
\label{eq:cbb1}
n_\zeta-1 & \simeq -\frac{4\mu^{2}}{F}\simeq-\frac{2}{N}\,,\\
r & \simeq \frac{32\mu^{2}}{F^{2}}\simeq\frac{8}{\mu^{2}N^{2}}=\frac{4(3+2\wbd)}{N^{2}}\,, \label{eq:cbb2}
\end{align}
which are valid for $-3/2<\omega_{{\rm BD}}< \O(1)$. The metric $f(R)$ gravity corresponds to $\wbd=0$, whereas the limit $\wbd \to -3/2$ corresponds to Palatini $f(R)$ gravity \cite{Sotiriou:2008rp,DeFelice:2010aj}. In this latter case the tensor--scalar ratio vanishes and a separate analysis is required~\cite{Tamanini:2010uq}.

If $\wbd \gg 1$, then one has $\mu\ll1$ and hence $F$ is close to unity even during inflation. The end of inflation is characterized by the condition $\hat{\epH}=1$, which gives $F_{f}=1+\sqrt{2}\mu\simeq1$. In this case the number of efoldings (\ref{N1}) is approximately given by $N\simeq(\chi/\Mpl)^{2}/4$. The scalar spectral index and the tensor--scalar ratio then follow as
\begin{align}
n_\zeta-1 & \simeq -8\frac{\Mpl^{2}}{\chi^{2}}\simeq-\frac{2}{N}\,,\\
r & \simeq 32\frac{\Mpl^{2}}{\chi^{2}}\simeq\frac{8}{N}\,,
\end{align}
which match with those for the chaotic inflation model with the potential $V(\vp)=m^{2}\vp^{2}/2$ \cite{Stewart:1993bc}.

\begin{figure}[h]
\begin{centering}
\includegraphics[width=0.7\textwidth]{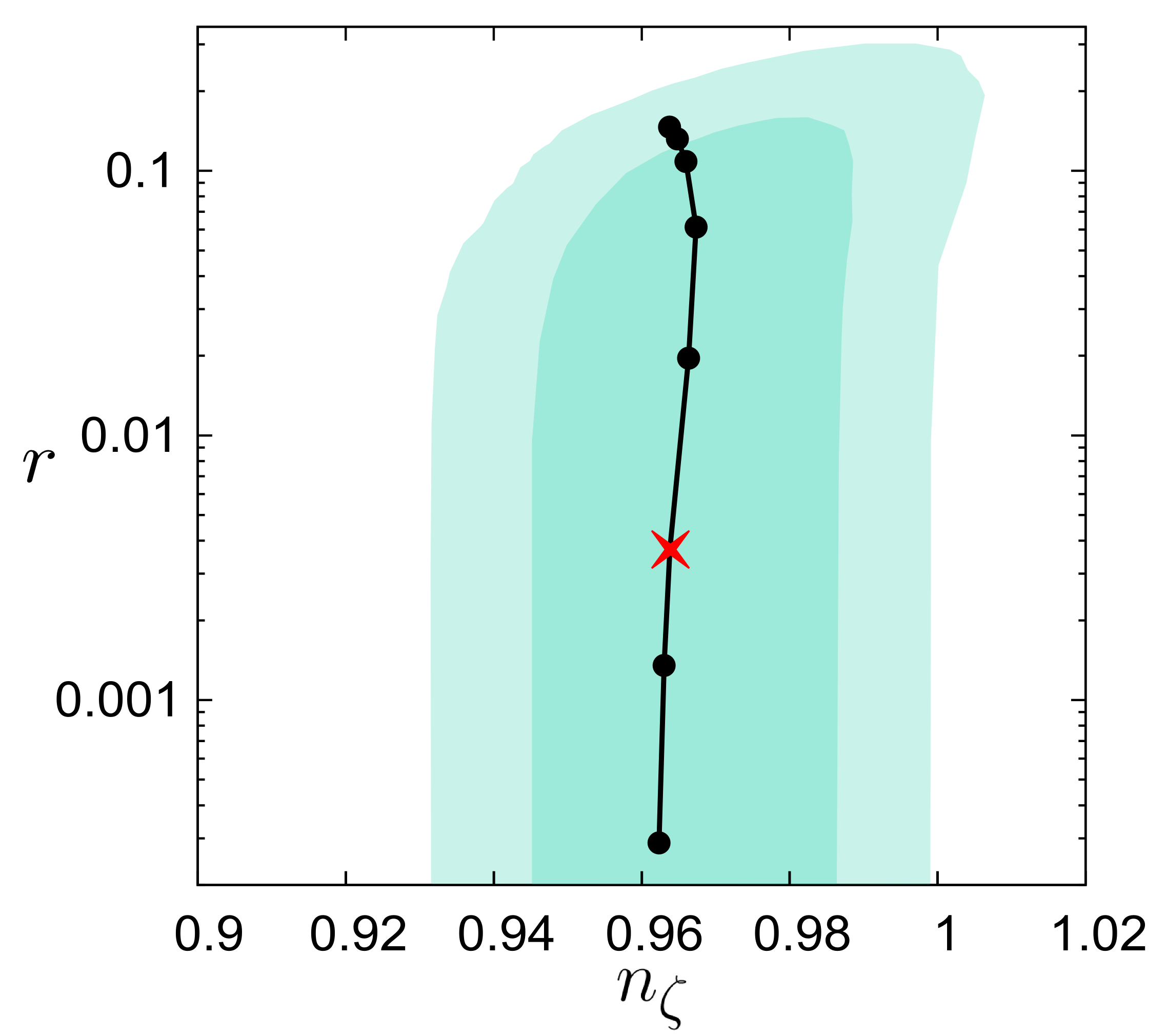} 
\par\end{centering}
\caption[Constraints on Brans-Dicke theory]
{In green are the 1$\sigma$ and 2$\sigma$ observational contours in the logarithmic $(n_\zeta,r)$ plane as discussed in figure \ref{fig1}. The dotted points show the theoretical predictions for \bd~theories with the potential $V(\vp)=V_{0}(\vp-\Mpl)^{2}$ and $N=55$. From bottom to top the points correspond to $\wbd=-1.4, -1, 0, 10, 10^2, 10^3, 10^4$ and $\omega_{{\rm BD}}\to\infty$. The red cross indicates $\wbd=0$ and represents Starobinsky's model $f(R)=R+R^{2}/(6M^{2})$. For larger $\wbd$ the two observables $n_\zeta$ and $r$ approach those for chaotic inflation with the quadratic potential $m^{2}\vp^2/2$.}
\centering{}\label{fig3} 
\end{figure}

From the general results \eqref{eq:cbb1} and \eqref{eq:cbb2} we see that the tensor--scalar ratio depends on the parameter $\wbd$, while the scalar spectral index is practically independent of $\wbd$. In figure~\ref{fig3} we plot the theoretical predictions of $n_\zeta$ and $r$ for several different values of $\wbd$ by fixing $N=55$. Shown also are the 1$\sigma$ and 2$\sigma$ observational contours constrained by the joint data analysis of \wmap~7-year~\cite{Komatsu:2010fb}, \bao~\cite{Percival:2009xn} and \hst~\cite{Riess:2011yx}. The $f(R)$ model $f(R)=R+R^{2}/(6M^{2})$, which corresponds to $\wbd=0$, is well within the 1$\sigma$ observational contour. While present observational data is compatible with $\wbd \gg 1$, it will be of interest to see how the Planck satellite~\cite{Planck:2006aa} can provide an upper bound on $\wbd$.

Using the approximate relation $F \simeq 2\mu^2 N+F_f$, which follows from eq.~(\ref{N1}), the \wmap~normalization for the scalar power spectrum $\P_\zeta = \hat{V}^3/(12 \pi^2 \Mpl^6 \hat{V}_{,\chi}^2)$ is given by 
\begin{equation}
\P_\zeta \simeq \frac{V_0}{12 \pi^2 \Mpl^2}
\frac{(\sqrt{2}\mu N+1)^4 \mu^2}
{(2\mu^2 N+1+\sqrt{2} \mu)^2} =2.4 \times 10^{-9}\,,
\end{equation}
around $N=55$. Since $\wbd=0$ and $\mu=1/\sqrt{3/2}$ for the model $f(R)=R+R^2/(6M^2)$, the mass scale $M$ is constrained to be $M \simeq 3 \times 10^{13}$~GeV. More generally, the energy scale $V_0$ varies with the parameter $\wbd$.

\subsubsection{Models with $p\ne2$}

In this case, for the parameter $\wbd$ of the order of unity, the number of efoldings (\ref{N2}) cannot be much greater than unity unless $F$ is much larger than $F_{f} \sim \O(1)$. If $F\gg1$, then eqs.~(\ref{nRBD}) and (\ref{rBD}) give 
\begin{align}
n_\zeta-1 & \simeq -\mu^{2}(p-2)^{2}\,,
\label{nsrneq1} \\
r & \simeq 8\mu^{2}(p-2)^{2}\,.
\label{nsrneq2}
\end{align}
Since $\mu\sim{\cal O}(1)$ in this case, the results (\ref{nsrneq1}) and (\ref{nsrneq2}) mean that for small $\wbd$ both the scalar spectral index and the tensor--scalar ratio are incompatible with observations apart from the case where $p$ is close to 2. On reflection this is unsurprising since only for $p\approx2$ is there a flat region of the potential that will give slow-roll along with its signatures of near scale invariance and suppressed tensor modes.

When $\wbd \gg 1$ one has $\mu\ll1$ and hence $F=e^{\mu\chi/\Mpl}$ is close to 1. In this case eqs.~(\ref{nRBD}) and (\ref{rBD}) give 
\begin{align}
n_\zeta-1 & \simeq -p(p+2)\frac{\Mpl^{2}}{\chi^{2}}\simeq-\frac{p+2}{2N}\,,\\
r & \simeq 8p^{2}\frac{\Mpl^{2}}{\chi^{2}}\simeq\frac{4p}{N}\,,
\end{align}
where we have used the approximate relation $N\simeq\chi^{2}/(2p\Mpl^{2})$. These results match with those of chaotic inflation with the potential $V=V_0 \vp^p$. The self-coupling potential $p=4$ is excluded observationally, both in the regimes $\wbd \gg 1$ and $\wbd = \O(1)$. Even for other values of $\wbd$ it is difficult to satisfy observational constraints unless $p$ is close to 2.

\sec{Gauss-Bonnet corrections}
\label{sec:single-gb}

In this section we study the effects of the \gb~term on the chaotic inflationary scenario, described by the action 
\begin{equation}
S=\int \d ^4 x\sqrt{-g}\left[\frac{\Mpl^2}{2}R
+X-V(\vp)-\xi(\vp) \G \right].
\end{equation}
In order to confront this model with observations, it is convenient to rewrite inflationary observables in terms of the potential slow-roll parameters \eqref{eq:single-sr}. The background equations are
\begin{align}
3\Mpl^{2}H^{2} - \dot{\phi}^{2}/2 - V - 24H^{3}\dot{\xi} &= 0 \,,\label{sta1}\\
\ddot{\phi}+3H\dot{\phi}+V_{,\phi}+24H^{2}\xi_{,\phi}(H^{2}+\dot{H}) &= 0\,.\label{sta2}
\end{align}
The parameters $\epH$ and $\ep_s$ follow from eqs.~(\ref{epap}) and (\ref{eps}) and at linear order take the simple forms
\begin{equation}
\epH=\epsilon_{s}+4\delta_{\xi}\,,\qquad
\epsilon_{s}=\delta_{X}\,.
\end{equation}

The background equations of motion~(\ref{sta1}) and (\ref{sta2}) allow us to express the potential $V$ and
its derivative $V_{,\phi}$ as 
\begin{align}
V & = 3\Mpl^{2}H^{2}\left(1-\frac{1}{3}\ep_{s}-8\delta_{\xi}\right),
\label{Veq}\\
V_{,\phi} & = -H\dot{\phi}\left[3-\epH+\frac{1}{2}\eta_{s}+12\frac{\delta_{\xi}}{\epsilon_{s}}(1-\epH)\right].\label{Vphi}
\end{align}
At leading order, eq.~(\ref{Vphi}) yields
\begin{align}
V_{,\phi} & \simeq -3H\dot{\phi}\left(1+\frac{4\delta_{\xi}}{\epsilon_{s}}\right),\label{dVphi}\\
V_{,\phi\phi} & \simeq -3H^{2}\left[\frac{1}{2}\eta_{s}-2\epsilon_{s}-16\delta_{\xi}-\frac{4\delta_{\xi}}{\epsilon_{s}}\left(8\delta_{\xi}+\frac{1}{2}\eta_{s}-\eta_{\xi}\right)\right].
\end{align}
These results allow us to express the potential slow-roll parameters $\ep$ and $\eta$ in terms of the other slow-roll parameters viz
\begin{align}
\ep & \simeq \epsilon_{s}\left(1+\frac{4\delta_{\xi}}{\epsilon_{s}}\right)^{2},\label{epVGB}\\
\eta & \simeq -\frac{1}{2}\eta_{s}\left(1-\frac{4\delta_{\xi}}{\epsilon_{s}}\right)+2\epsilon_{s}+4\delta_{\xi}\left[4+\frac{1}{\epsilon_{s}}(8\delta_{\xi}-\eta_{\xi})\right].
\end{align}
This enables us to obtain the inversion formulae
\begin{align}
\epsilon_{s} & \simeq \frac{1}{2}
\left[\ep-8\delta_{\xi}+\sqrt{\ep^{2}-16\ep \delta_{\xi}}\right],\label{epsin}\\
\eta_{s} & \simeq -\frac{2}{1-4\delta_{\xi}/\epsilon_{s}}
\left( \eta -2\epsilon_{s}-4\delta_{\xi}
\left[4+\frac{8\delta_{\xi}-\eta_{\xi}}{\epsilon_{s}}\right]\right),\label{etasGB}
\end{align}
where we have taken the positive sign in eq.~(\ref{epsin}) to reproduce $\epsilon_{s} \to \ep$ for $\delta_{\xi} \to 0$.

From eqs.~(\ref{nR}), (\ref{nT}), and (\ref{rgene}) the inflationary observables are given by 
\begin{align}
n_\zeta-1 & = -2\epsilon_{s}-\eta_{s}-8\delta_{\xi}\,,\label{nsGB0}\\
n_t & = -2\epsilon_{s}-8\delta_{\xi}\,,\label{ntGB0}\\
r & = 16\epsilon_{s}\,.\label{rGB0}
\end{align}
Upon substitution of eqs.~\eqref{epsin} and \eqref{etasGB}, the above expressions for inflationary observables are dependent on the four variables $\{\ep, \eta , \delta_{\xi} , \eta_{\xi}\}$. By specifying the functional forms of $V(\vp)$ and $\xi(\vp)$, we can reduce this number of variables. For the chaotic inflation potential $V(\vp) = V_0 \vp^p$, the potential slow-roll parameters \eqref{eq:single-sr} are related via 
\begin{equation}
\eta =\frac{2(p-1)}{p}\epsilon\,.
\label{etaepsilonpowerlaw}
\end{equation}
For the \gb~coupling we prescribe the function $\xi(\vp)$ to be of the dilatonic form
\begin{equation}
\xi(\vp)=\xi_{0}e^{\mu\vp/\Mpl},
\end{equation}
where $\xi_{0}$ and $\mu$ are constants. It then follows that 
\begin{equation}
\eta_{\xi}=-2\epsilon_{s}+\eta_{s}/2-8\delta_{\xi} - \mu\sqrt{2\epsilon_{s}}\,.\label{etaxi}
\end{equation}
Combining eq.~(\ref{etaxi}) with eq.~(\ref{etasGB}), we obtain 
\begin{equation}
\eta_{s}\simeq 2\left[2\epsilon_{s} - \eta 
+4\delta_{\xi}\left(\frac{\mu\sqrt{2\epsilon_{s}}+16\delta_{\xi}}{\epsilon_{s}}+6\right)\right].\label{etasap}
\end{equation}
Substituting eq.~(\ref{etasap}) into eq.~(\ref{nsGB0}), the scalar spectral index can be written as 
\be
n_\zeta-1 \simeq -6\epsilon_{s}+2\eta -8\delta_{\xi}\left(7+\frac{\mu\sqrt{2\epsilon_{s}}+16\delta_{\xi}}{\epsilon_{s}}\right),\label{nsGB}
\ee
where 
\begin{equation}
\eta=\frac{2(p-1)}{p}\ep\simeq\frac{2(p-1)}{p}\epsilon_{s}\left(1+\frac{4\delta_{\xi}}{\epsilon_{s}}\right)^{2}.\label{etaV}
\end{equation}
For fixed values of $p$ and $\mu$ one can carry out a \cmbr~likelihood analysis in terms of $n_\zeta$, $r$, and $n_t$ by varying the two parameters $\epsilon_{s}$ and $\delta_{\xi}$.

\begin{figure}
\begin{centering}
\includegraphics[width=0.7\textwidth]{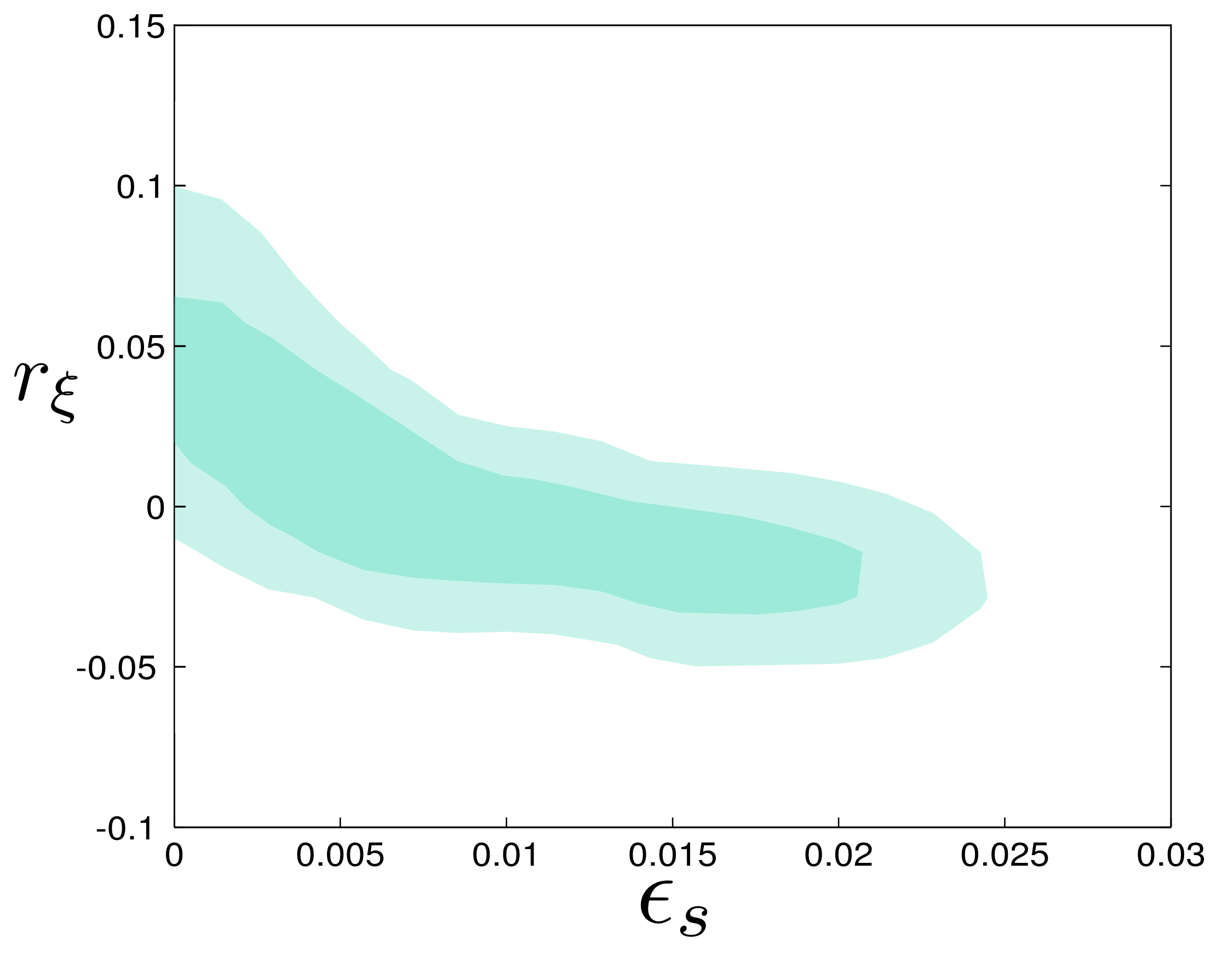} 
\par\end{centering}
\caption[Constraints on Gauss-Bonnet assisted inflation]{\cosmomc~likelihood contours at 1$\sigma$ and $2\sigma$ in the $(\ep_s,r_{\xi})$ plane for the potential $V(\vp)=m^{2}\vp^{2}/2$ with $\mu=1$ ($r_{\xi}=\delta_{\xi}/\epsilon_{s}$). We use data from \wmap~7-year, \lss~(including \bao), \hst, \snia~and \bbn~with the pivot scale $k_{0}=0.002$ Mpc$^{-1}$.}
\centering{}\label{fig4} 
\end{figure}

In figure~\ref{fig4} the observational constraints on the parameters $\epsilon_{s}$ and $r_{\xi}\equiv\delta_{\xi}/\epsilon_{s}$ are plotted for $p=2$ and $\mu=1$. This is produced using the \cosmomc~code~\cite{Lewis:2002ah} with input data from \wmap~7-year~\cite{Komatsu:2010fb}, \lss~\cite{Reid:2009xm}, \bao~\cite{Percival:2009xn}, \hst~\cite{Riess:2011yx}, \snia~\cite{Kowalski:2008ez} and {\sc bbn}~\cite{Burles:1997ez}, by assuming a \lcdm~universe. The ratio $r_{\xi}$ is constrained to be $|r_{\xi}|<0.1$ at 95\% \cl, which means that the effect of the \gb~term needs to be suppressed for observational compatibility. Consequently, the energy scale $V_0$ is similar to that of standard chaotic inflation.

From figure~\ref{fig4} we find that the slow-roll parameter $\epsilon_{s}$ is bounded to be $\epsilon_{s}<0.025$ at 95\% \cl. In the presence of the \gb~term, small values of $\epsilon_{s}$ can give rise to a scalar spectral index close to $n_\zeta=0.96$. For example, when $\epsilon_{s}=0.002$, $r_{\xi}=0.05$, $\mu=1$, and $p=2$, one has $n_\zeta = 0.962$ from eq.~(\ref{nsGB}). This is different from standard chaotic inflation in which small values of $\epsilon_{s}$ push the spectrum close to the Harrison-Zel'dovich one (which is not favoured observationally). Hence we find that the allowed range of $\epsilon_{s}$ tends to be wider in the presence of the \gb~coupling.

Let us now estimate the two observables $n_\zeta$ and $r$ in terms of the number of efoldings $N$, under the condition $|r_{\xi}|\ll1$. Since $\epsilon_{s}\simeq \ep-8\delta_{\xi}$ from eq.~(\ref{epsin}), we find eqs.~(\ref{nsGB}) and (\ref{rGB0}) reduce to 
\begin{align}
n_\zeta -1 & \simeq -\left(2+\frac{4}{p}\right)\ep -8\delta_{\xi}\left(1+\mu\sqrt{\frac{2}{\ep}}\right),\label{nsGB2}\\
r & \simeq 16\ep \left(1-\frac{8\delta_{\xi}}{\ep}\right).\label{rGB2}
\end{align}
From eqs.~(\ref{Veq}) and (\ref{dVphi}) one has $H/\dot{\phi}\simeq-(1+4\delta_{\xi}/\epsilon_{s})V/(\Mpl^{2}V_{,\phi})$. Using the relation $\epsilon_{s}=\delta_{X}$ and the definition of $\delta_{\xi}$, it follows that $H/\dot{\phi}=-\phi/(p\Mpl^{2}+8H^{2}\xi_{,\phi}\phi)$. Since we are considering the case where $H^{2}\xi_{,\phi}\phi/\Mpl^{2}\ll1$, the number of efoldings for the potential $V=V_0 \vp^p$ is 
\begin{equation}
N=\int_{\phi}^{\phi_{f}}\frac{H}{\dot{\phi}}\, \d \phi\simeq\frac{x^{2}-x_{f}^{2}}{2p}+\bar{N}_{p}\,,\qquad{\rm where}\qquad\bar{N}_{p}\equiv-\frac{8\xi_{0}\mu}{3p^{2}}\frac{V_{0}}{\Mpl^{4}}\int_{x_{f}}^{x}e^{\mu x}x^{p+2} \, \d x\,.\label{efoldap}
\end{equation}
Here $x_{f}$ is the value of $x=\phi/\Mpl$ at the end of inflation. We identify the end of inflation by the condition $\ep=1$, when $x_{f}=p/\sqrt{2}$. For the specific potentials $p=2$ and $p=4$, eq.~(\ref{efoldap}) may be integrated to give 
\begin{align}
\bar{N}_{2} = &-\frac{m^{2}}{3\Mpl^{2}}\frac{\xi_{0}}{\mu^{4}} \Big\{ e^{\mu x}[\mu^{4}x^{4}+4(-\mu^{3}x^{3}+3\mu^{2}x^{2}-6\mu x+6)] \nonumber \\
&-4e^{\sqrt{2}\mu}\,[\mu^{4}-2\sqrt{2}\mu^{3}+6(\mu^{2}-\sqrt{2}\mu+1)] \Big\} ,\label{N2d}
\end{align}
\begin{align}
\!\!\!\!\!\! \bar{N}_{4} = &-\frac{\lambda\xi_{0}}{24\mu^{6}}\,\biggl\{ e^{\mu x}(\mu^{6}x^{6}-6\mu^{5}x^{5}+30\mu^{4}x^{4}-120\mu^{3}x^{3}+360\mu^{2}x^{2}-720\mu x+720)\nonumber \\
 & -16e^{2\sqrt{2}\mu}\left[32\mu^{6}-48\sqrt{2}\mu^{5}+120\mu^{3}(\mu-\sqrt{2})+90\mu(2\mu-\sqrt{2})+45\right]\biggr\}, \label{N4d}
\end{align}
where we have set $V_{0} \to m^{2}\Mpl^{2}/2$ for $p=2$, and $V_{0} \to \lambda\Mpl^{4}/4$ for $p=4$. 

For positive $\mu$ of the order of unity, the dominant contributions to $\bar{N}_{p}$ come from the first terms in eqs.~(\ref{N2d}) and (\ref{N4d}), i.e. 
$\bar{N}_{2}\simeq-m^{2}\xi_{0} x^{4} e^{\mu x} /(3\Mpl^{2})\,$
and $\bar{N}_{4}\simeq-\lambda\xi_{0} x^{6} e^{\mu x}/24$, 
for the scales relevant to the \cmbr~($\mu x\gg1$). In this case the number of efoldings (\ref{efoldap}) is approximately given by 
\begin{alignat}{2}
N & \simeq \frac{1}{4}x^{2}\left[1-\frac{4}{3}\left(\frac{m}{\Mpl}\right)^{2}\xi x^{2}\right]-\frac{1}{2} \,, \qquad & &\mbox{---}{\rm for~} p=2\,,\label{Np2}\\
N & \simeq \frac{1}{8}x^{2}\left(1-\frac{1}{3}\lambda\xi x^{4}\right)-1\,, & &\mbox{---}{\rm for~} p=4\,.\label{Np4}
\end{alignat}
Since $\delta_{\xi}\simeq-\mu V_{0}p \, x^{p-1}\xi /(3\Mpl^{4})$ and
$\ep = p^{2}/(2x^{2})$, one can express the scalar spectral index (\ref{nsGB2}) and the tensor--scalar ratio (\ref{rGB2}) in terms of $x$. By treating the $\xi$-dependent terms in eqs.~(\ref{Np2}) and (\ref{Np4}) as small corrections, $n_\zeta$ and $r$ can be written in terms of $N$. For $p=2$ one finds
\begin{align}
n_\zeta-1 & \simeq -\frac{2}{N}\left[1-\frac{16}{3}N^{2}\left(\frac{m}{\Mpl}\right)^{2}\mu^{2}\xi\right], \\
r & \simeq \frac{8}{N}\left[1+\frac{32}{3}N^{3/2}\left(\frac{m}{\Mpl}\right)^{2}\mu\,\xi\right], 
\end{align}
whereas for $p=4$ one finds
\begin{align}
n_\zeta-1 & \simeq -\frac{3}{N}\left[1-\frac{256}{9}N^{3}\lambda\mu^{2}\xi\right], \\
r &\simeq \frac{16}{N}\left[1+\frac{128\sqrt{2}}{3}N^{5/2}\lambda\mu\,\xi\right].
\end{align}
These expressions are valid for positive $\mu$ of the order of unity. If $\xi>0$ (in which case $\delta_{\xi}<0$ under the assumption $\dot{\phi}<0$), then the effect of the \gb~coupling leads to the approach to the scale-invariant spectrum, while $r$ gets larger. Alternatively, negative values of $\xi$ lead to a decease of $r$, but $n_\zeta$ deviates further from unity. Since $\epsilon_{s}$ is approximately given by $\epsilon_{s}\approx\epsilon\approx p^{2}/(8N)$, the scales relevant to the \cmbr~anisotropy ($50<N<60$) correspond to $0.008<\epsilon_{s}<0.01$ for $p=2$. It is therefore unsurprising that figure~\ref{fig4} constrains the ratio $r_{\xi}$ to the tight bounds of $-0.04<r_{\xi}<0.03$ (at 95\% \cl) for this range of $\epsilon_{s}$. The self-coupling potential $V(\vp)=\lambda\vp^{4}/4$ is not saved by taking into account the \gb~term with positive $\mu$, because the \gb~coupling does not simultaneously lead to an increase of $n_\zeta$ and a decrease of $r$.

For negative $\mu$ with $|\mu|={\cal O}(1)$, the exponential term $e^{\mu x}$ in eqs.~(\ref{N2d}) and (\ref{N4d}) is much smaller than unity for \cmbr~scales ($x\gg1$). In this case we have 
\begin{align}
\!\!\!\! \bar{N}_{2} & \simeq \frac{4m^{2}}{3\Mpl^{2}}\frac{\xi_{0}e^{\sqrt{2}\mu}}{\mu^{4}}\left[\mu^{4}-2\sqrt{2}\mu^{3}+6(\mu^{2}-\sqrt{2}\mu+1)\right],\label{N2ne}\\
\!\!\!\! \bar{N}_{4} & \simeq \frac{2\lambda\xi_{0}e^{2\sqrt{2}\mu}}{3\mu^{6}}\left[32\mu^{6}-48\sqrt{2}\mu^{5}+120\mu^{3}(\mu-\sqrt{2})+90\mu(2\mu-\sqrt{2})+45\right]. \label{N4ne}
\end{align}
The scalar spectral index and the tensor--scalar ratio are then approximately given by 
\be
n_\zeta-1 \simeq -\frac{2}{N}\left(1+\frac{\bar{N}_{2}-1/2}{N}\right),\qquad r\simeq\frac{8}{N}\left(1+\frac{\bar{N}_{2}-1/2}{N}\right),
\ee
for $p=2$, whilst for $p=4$ one finds
\be
n_\zeta-1 \simeq -\frac{3}{N}\left(1+\frac{\bar{N}_{4}-1}{N}\right),\qquad~~~r\simeq\frac{16}{N}\left(1+\frac{\bar{N}_{4}-1}{N}\right).
\ee
To reach these expressions we have assumed $N\gg\bar{N}_{p}$ and ignored the exponential term $e^{\mu x}$. When $\mu<0$, one can show that $\bar{N}_{2}$ and $\bar{N}_{4}$ in eqs.~(\ref{N2ne}) and (\ref{N4ne}) are positive for $\xi_{0}>0$ and negative for $\xi_{0}<0$. In the latter case the presence of the \gb~term leads to the approach to the Harrison Zel'dovich spectrum. Such a scenario was discussed in ref.~\cite{Satoh:2008ck}. Since $m/M_{{\rm pl}}$ and $\lambda$ are much smaller than unity by the \wmap~normalization ($m/M_{{\rm pl}} \simeq 6.8 \times 10^{-6}$ and $\lambda \simeq 2.0 \times10^{-13}$), one has $|\bar{N}_{p}|\ll1$ for $|\xi_{0}| < \O(1)$. For $\mu<0$ the effect of the \gb~term on inflationary observables appears only for very large values of $\xi_0$ such as $|\xi_0| \sim 10^{10}$ \cite{Satoh:2008ck}.

\paragraph{Gauss-Bonnet section summary.} The presence of a \gb~coupling with positive $\mu$ does not save the self-coupling potential. Our \cmbr~liklihood analysis have shown that, for the quadratic potential, the \gb~coupling needs to be suppressed as 
$|\delta_{\xi}/\epsilon_s|<0.1$. If $\mu$ is negative then both $|n_\zeta-1|$ and $r$ can decrease for negative $\xi_0$, but this requires a large coupling constant. 

\sec{Galileon inflation}
\label{sec:single-g}

Finally we study chaotic inflation in the presence of the Galileon-like self-interaction $J(\vp,X)\square\vp$. We specify the functional form of $J(\vp,X)$, as 
\begin{equation}
J(\vp,X)=\Phi(\vp)X^{n},\qquad\Phi(\vp)=
\frac{\theta}{M^{4n-1}}e^{\mu\vp/\Mpl},\label{Phiex}
\end{equation}
where $n$ and $\mu$ are constants and $\theta=\pm1$. The constant $M$ has a dimension of mass with $M>0$. We have introduced the exponential form for $\Phi$ motivated by the dilaton coupling in low-energy effective bosonic string theory. In fact the low-energy effective Lagrangian at the next to leading order in the Regge parameter includes a term of the form $\alpha' f(\vp) X \square \vp$ (see the third term in eq.~(2) of ref.~\cite{Cartier:2001is}). Here we consider the power-law function $\Phi X^n$ by generalizing previous studies~\cite{Kamada:2010qe}. Note that, for $n=1$ and $\mu=0$, we recover the Galileon term which satisfies the Galilean symmetry in the limit of Minkowski spacetime \cite{Nicolis:2008in}.

The background equations of motion (\ref{E1eq}) and (\ref{E3eq}) can be written as 
\begin{align}
V &= 3\Mpl^{2}H^{2}\left(1-\frac{1}{3}\delta_{X}-2\delta_{JX}
+\frac{2}{3}\delta_{J\phi}\right),\label{bega1}\\
V_{,\phi} &= -3H\dot{\phi}\bigg\{ 1+(3-\epH)
\frac{\delta_{JX}}{\delta_{X}}-\frac{\mu^{2}}{3n}\delta_{JX}
+2(n-1)\frac{\delta_{J\phi}}{\delta_{X}}+ \nonumber \\ & \qquad \qquad
\frac{\delta_{\phi}}{3}
\left[1+6n\frac{\delta_{JX}}{\delta_{X}}-2(n+1)
\frac{\delta_{J\phi}}{\delta_{X}}\right]\bigg\}.\label{bega2}
\end{align}
To compare with observations, we seek an expression for $n_\zeta-1$ in terms of a minimal set of independent slow-roll parameters. Since $\P_\zeta = H^{2}/(8\pi^{2}Z c_{s}^{3})$, it is important to find the expressions for $Z$ and $c_{s}$. From eqs.~(\ref{eq:defQ}) and (\ref{eq:defc2s}) it follows that 
\begin{align}
\frac{Z}{\Mpl^{2}} & =  \frac{\delta_{X}+6n\delta_{JX}-2(n+1)\delta_{J\phi}+3\delta_{JX}^{2}}{(1-\delta_{JX})^{2}}\,,\label{eq:QG1}\\
c_{s}^{2} & =  \frac{\delta_{X}+2(2+n\delta_{\phi})\delta_{JX}+2(n-1)\delta_{J\phi}-\delta_{JX}^{2}}{\delta_{X}+6n\delta_{JX}-2(n+1)\delta_{J\phi}+3\delta_{JX}^{2}}\,,\label{eq:cG1}
\end{align}
where we have used the relations $\lambda_{JX}=n-1$ and $\lambda_{J\phi}=n$. Hence, we can derive an exact expression for $n_\zeta-1$ in terms of the slow-roll parameters entering eqs.~(\ref{eq:QG1}) and (\ref{eq:cG1}) and their first derivatives; these derivatives introduce other slow-roll parameters but these are not all independent as we now discuss.

For the choice of the function $\Phi$ in eq.~(\ref{Phiex}) we have
\begin{equation}
\delta_{J\phi}=\pm\frac{\mu}{\sqrt{2}n}\,\delta_{JX}\sqrt{\delta_{X}}\,,\label{eq:dGX1}
\end{equation}
where the $\pm$ signs in this expression are compatible with those in the expression $\dot{\phi}=\pm\sqrt{2}\Mpl H\sqrt{\delta_{X}}$. Equation (\ref{eq:dGX1}) shows that $\delta_{J\phi}$ is in general suppressed relative to $\delta_{JX}$ and so it is appropriate to consider $\delta_{J\phi} = \O(\epH^{3/2})$. This relation also implies
that 
\begin{equation}
\eta_{J\phi}\equiv\frac{\dot{\delta}_{J\phi}}{H\delta_{J\phi}}
=\frac{\eta_{X}}{2}+\eta_{JX}\,,\label{etaGphi}
\end{equation}
where 
\begin{equation}
\eta_{X}\equiv\frac{\dot{\delta}_{X}}{H\delta_{X}}\,,\qquad{\rm and}
\qquad\eta_{JX}=\frac{\dot{\delta}_{JX}}{H\delta_{JX}}\,.
\end{equation}
From the definition of $\delta_{X}$ and $\delta_{JX}$ we obtain
\begin{align}
\eta_{X} & =  2(1-\delta_{JX})\delta_{\phi}+2\delta_{X}+6\delta_{JX}
-4\delta_{J\phi}\,,\label{etaX}\\
\eta_{JX} & =  (2n+1-\delta_{JX})\delta_{\phi}\pm\mu\sqrt{2\delta_{X}}
+\delta_{X}-2\delta_{J\phi}+3\delta_{JX}\,,\label{eq:eGX1}
\end{align}
where we have used the relation 
\begin{equation}
\epH = \delta_{X}+3\delta_{JX}-2\delta_{J\phi}-\delta_{\phi}\delta_{JX}\,.\label{epGa}
\end{equation}
It should be noted that the relations~(\ref{eq:dGX1})-(\ref{epGa}) are all exact. We may use these to find $\{\eta_{J\phi} , \eta_{X} , \eta_{JX} \}$ in terms of the three slow-roll parameters $\{ \delta_{\phi} , \delta_{X} , \delta_{JX} \}$.

We finally use a last constraint coming from the fact that we have chosen a power-law form $V=V_0 \vp^p$ for the inflationary potential. The relation (\ref{etaepsilonpowerlaw}) between the potential slow-roll parameters $\ep$ and $\eta$ leads to 
\begin{equation}
\frac{\dot{V}_{,\phi}}{HV_{,\phi}}=\frac{p-1}{p}\frac{V_{,\phi}}{HV}\,\dot{\phi}\,.
\end{equation}
This equation can be used to set the last constraint on the slow-roll variables. At lowest order we have 
\begin{align}
\delta_{\phi} &= \frac{(\delta_{X}+3\delta_{JX})[(2-p)\delta_{X}+6\delta_{JX}]}
{p(\delta_{X}+6n\delta_{JX})} \mp \frac{3\sqrt{2}\delta_{JX}\sqrt{\delta_{X}}}
{\delta_{X}+6n\delta_{JX}}\mu \nonumber \\ & \qquad
-\frac{2(n-1)\delta_{X}\delta_{JX}(\delta_{X}-3\delta_{JX})}
{n(\delta_{X}+6n\delta_{JX})^{2}}\mu^{2}
+{\cal O}(\epH^{3/2}).\label{eq:d2pF}
\end{align}
Using this relation we can express $\eta_{J\phi}$, $\eta_{X}$, and $\eta_{JX}$ in terms of two slow-roll parameters $\delta_{X}$ and $\delta_{JX}$.

We are now ready to explicitly calculate the scalar index $n_\zeta-1=-2\epH-\delta_{Z}-3s$, where $\delta_{Z}=\dot{Z}/(HZ)$ and $s=\dot{c}_{s}/(Hc_{s})$ are evaluated by taking the time derivatives of eqs.~(\ref{eq:QG1}) and (\ref{eq:cG1}). This yields
\begin{align}
n_\zeta-1 &\simeq -\frac{2\left(\delta_{{X}}+3\,\delta_{{JX}}\right)}{p\left(\delta_{{X}}+4\,\delta_{{JX}}\right)\left(\delta_{{X}}+6n\,\delta_{{JX}}\right)^{2}}\bigg[
{\delta_{{X}}^{3}}(p+2) \nonumber \\ &
+{\delta_{{X}}^{2}}\delta_{{JX}}[(3p-6)n^{2}+(12p+27)n+4p+8]\nonumber \\
 & +\delta_{{X}}{\delta_{{JX}}^{2}}[(57p+30)n^{2}+(54p+105)n+6]+72n{\delta_{{JX}}^{3}}(3np+2n+1)\bigg]\nonumber \\
 & \pm {\frac{3\sqrt{2}\,\delta_{{JX}}\sqrt{\delta_{{X}}}\,[(7n+2)\,\delta_{{X}}\delta_{{JX}}+n{\delta_{{X}}^{2}}+24n\,{\delta_{{JX}}^{2}}]}{\left(\delta_{{X}}+6n\,\delta_{{JX}}\right)^{2}\left(\delta_{{X}}+4\,\delta_{{JX}}\right)}}\mu\nonumber \\
 & -\frac{2\delta_{{\it JX}}\delta_{{X}}}{n\left(\delta_{{X}}+6\, n\delta_{{\it JX}}\right)^{3}\left(4\,\delta_{{\it JX}}+\delta_{{X}}\right)^{2}}\,\bigg[{\delta_{{X}}}^{4}+\left(9\, n+8+6\,{n}^{3}-24\,{n}^{2}\right){\delta_{{X}}^{3}}\delta_{{\it JX}}\nonumber \\
 & +\left(4-99\,{n}^{2}+54\, n-42\,{n}^{3}\right){\delta_{{X}}^{2}}{\delta_{{\it JX}}^{2}}+\left(132\,{n}^{2}-282\,{n}^{3}-24-132\, n\right)\delta_{{X}}{\delta_{{\it JX}}^{3}} \nonumber \\ & 
-72n\left(n^{2}-3n+6\right){\delta_{{\it JX}}^{4}}\bigg]\mu^{2},
\label{nscom}
\end{align}
where, in order to derive this result, we have also included the terms of order $\O(\epH^{3/2})$ not shown in eq.~(\ref{eq:d2pF}). The factor of $\pm$ is compatible with those in the expression for $\dot{\phi}$ in terms of $\delta_{X}$. The tensor--scalar ratio (\ref{rgene}) and the tensor spectral index (\ref{nT}) are approximately given by 
\begin{align}
r & \simeq 16\frac{(\delta_{X}+4\delta_{JX})^{3/2}}
{(\delta_{X}+6n\delta_{JX})^{1/2}}\,,\label{rcom}\\
n_{{\rm t}} & \simeq -2(\delta_{X}+3\delta_{JX})\,,\label{ntcom}
\end{align}
and the scalar propagation speed squared is 
\begin{equation}
c_{s}^{2}\simeq\frac{\delta_{X}+4\delta_{JX}}
{\delta_{X}+6n\delta_{JX}}\,.\label{scalarga}
\end{equation}

If $|\delta_{JX}|\ll\delta_{X}$ then these observables reduce to
\begin{align}
n_\zeta-1 & \simeq -\frac{2(p+2)}{p}\delta_{X}
\pm 3\sqrt{2}n\mu\frac{\delta_{JX}}{\sqrt{\delta_{X}}}\,,
\label{nslimi1}\\
r & \simeq 16\delta_{X}\simeq-8n_{{\rm t}}\,.\label{nsrga}
\end{align}
On the other hand, in the limit where $\delta_{JX}\gg\delta_{X}$, one finds
\begin{align}
n_\zeta-1 & \simeq -\frac{3(3np+2n+1)}{pn}\delta_{JX}
 \pm \frac{\mu}{\sqrt{2}n}\,
\sqrt{\delta_{X}}\,,
\label{nsrga0} \\
r & \simeq \frac{64}{3}\sqrt{\frac{6}{n}}\delta_{JX}\simeq
-\frac{32}{9}\sqrt{\frac{6}{n}}n_{{\rm t}}\simeq
-\frac{8.7}{\sqrt{n}}\, n_{{\rm t}}\,,\label{nsgali}
\end{align}
which agree with the results in ref.~\cite{Kamada:2010qe} derived for $n=1$ and $\mu=0$.

To be concrete, in the following discussion we focus on the theories with $n=1$, $\mu\neq 0$, and $\theta=-1$. In this scenario $\delta_{JX}>0$ for $\dot{\phi}<0$, so that the conditions for the avoidance of ghosts and Laplacian instabilities ($Z>0$ and $c_{s}^{2}>0$) are always satisfied. In this case we need to take the minus sign for the terms proportional to $\mu$ in eqs.~(\ref{nscom}), (\ref{nslimi1}), and (\ref{nsrga0}). Since 
$V_{,\phi}\simeq-3H\dot{\phi}(1+3H\dot{\phi}\Phi)$ from eq.~(\ref{bega2}), the field velocity corresponding to $\dot{\phi}<0$ is 
\begin{equation}
\dot{\phi}\simeq\frac{\sqrt{1-4\Phi V_{,\phi}}-1}{6H\Phi}\,,
\end{equation}
where we have used the fact that $\Phi<0$. Employing the approximate relation $V\simeq3H^{2}\Mpl^{2}$, the two slow-roll parameters $\delta_{X}$ and $\delta_{JX}$ can be expressed in terms of $\phi$ as 
\begin{equation}
\delta_{X}\simeq\frac{\Mpl^{2}(\sqrt{1-4\Phi V_{,\phi}}-1)^{2}}
{8V^{2}\Phi^{2}}\,,\qquad\delta_{JX}\simeq\frac{\delta_{X}}{6}
(\sqrt{1-4\Phi V_{,\phi}}-1)\,.\label{delXGX}
\end{equation}
The number of efoldings is given by 
\begin{align}
N = \int_{\phi}^{\phi_{f}}\frac{H}{\dot{\phi}}\, \d \phi & \simeq
\frac{2}{\Mpl^{2}} \int_{\phi}^{\phi_{f}}\frac{\Phi V}{\sqrt{1-4\Phi V_{,\phi}}-1}\, 
\d \phi \nonumber \\ & =
2B^{4}\int_{x_{f}}^{x}\frac{x^{p}e^{\mu x}}{\sqrt{1
+4B^{4}px^{p-1}e^{\mu x}}-1} \, \d x\,,\label{efoldGa}
\end{align}
where 
\begin{equation}
B\equiv\left(\frac{V_{0}}{M^{3}\Mpl}\right)^{1/4},\qquad 
x\equiv\frac{\phi}{\Mpl}\,,\qquad x_{f}\equiv\frac{\phi_{f}}{\Mpl}\,.
\end{equation}
We determine the value of $x_{f}$ at the end of inflation using the condition $\epH \simeq \delta_{X}+3\delta_{JX}=1$.

In the limit $B \to 0$ (i.e. $\delta_{JX} \to 0$) we have
$\epH\simeq\delta_{X}\simeq p^{2}/(2x^{2})$ and $N\simeq x^{2}/(2p)-p/4$, so that eqs.~\eqref{nslimi1} and \eqref{nsrga} give
\begin{equation}
n_\zeta \simeq 1-\frac{2(p+2)}{4N+p}\,,\qquad 
r\simeq\frac{16p}{4N+p}\simeq-8 n_t\,.
\label{nrstan}
\end{equation}
In eq.~(\ref{nrstan}) we have not taken into account the contributions coming from the term $\mu$, because we do not have an analytic expression for general $p$. Numerical calculations show that in the regime $B \ll 1$ both $n_\zeta$ and $r$ become smaller for $\mu>0$. If $\mu<0$, then $n_\zeta$ decreases, whereas $r$ increases.

In the opposite limit where $B \gg 1$ it follows that 
\begin{equation}
\epH \simeq3\delta_{JX}\simeq\frac{p^{3/2}}{2B^{2}}x^{-(p+3)/2}
e^{-\mu x/2}\,,\qquad N\simeq\frac{B^{2}}{\sqrt{p}}\int_{x_{f}}^{x}x^{(p+1)/2}
e^{\mu x/2} \d x\,,
\end{equation}
and $\delta_{X} \simeq px^{-(p+1)}e^{-\mu x}/(2B^4)$. In order to have $N \approx 55$ for $B\gg1$, the integral inside the expression of $N$ needs to be much smaller than unity, which requires $x\ll1$ for $|\mu|={\cal O}(1)$. Using the approximation $|\mu x| \ll 1$, we have $x_{f}^{(p+3)/2}\simeq p^{3/2}/(2B^{2})$ and 
\begin{equation}
N \simeq \frac{B^2}{\sqrt{p}} \frac{2}{p+3} x^{(p+3)/2}
\left[ 1+\frac{p+3}{2(p+5)}\mu x \right]-\frac{p}{p+3}\,.
\end{equation}
From eqs.~(\ref{nsrga0}) and (\ref{nsgali}) it then follows that 
\begin{align}
n_\zeta &\simeq
1-\frac{3(p+1)}{(p+3)N+p}
\left[1-\frac{2(p-1)}{3(p+1)(p+5)} \mu x \right], 
\label{nrgalie0} \\
r &\simeq \frac{64\sqrt{6}}{9}\frac{p}{(p+3)N+p}
\left( 1-\frac{\mu x}{p+5} \right).
\label{nrgali}
\end{align}
For $N=55$, in the limit where $\mu\to0$, one finds $n_\zeta=0.9675$ and $r=0.1258$ for $p=2$, whereas for $p=4$ one finds $n_\zeta=0.9614$ and $r=0.1791$. In the regime $B\gg1$ the current observations can be consistent with both models. In the presence of the exponential coupling with positive $\mu$, the scalar spectral index gets larger for $p>1$, while the tensor--scalar ratio becomes smaller.

In the intermediate regime between $B \ll 1$ and $B \gg 1$, we evaluate $n_\zeta$ and $r$ as follows: For given values of $p$, $\mu$, and $B$ we identify the field value $x=\phi/\Mpl$ corresponding to $N=55$ by integrating eq.~(\ref{efoldGa}) numerically. We derive $\delta_{X}$ and $\delta_{JX}$ from eq.~(\ref{delXGX}) which allows us to obtain $n_\zeta$ and $r$ by using the formulae (\ref{nscom}) and (\ref{rcom}). We have also solved the background equations numerically to the end of inflation and confirmed that the above method provides an accurate estimation for $n_\zeta$ and $r$.

\begin{figure}[h]
\begin{centering}
\includegraphics[width=0.7\textwidth]{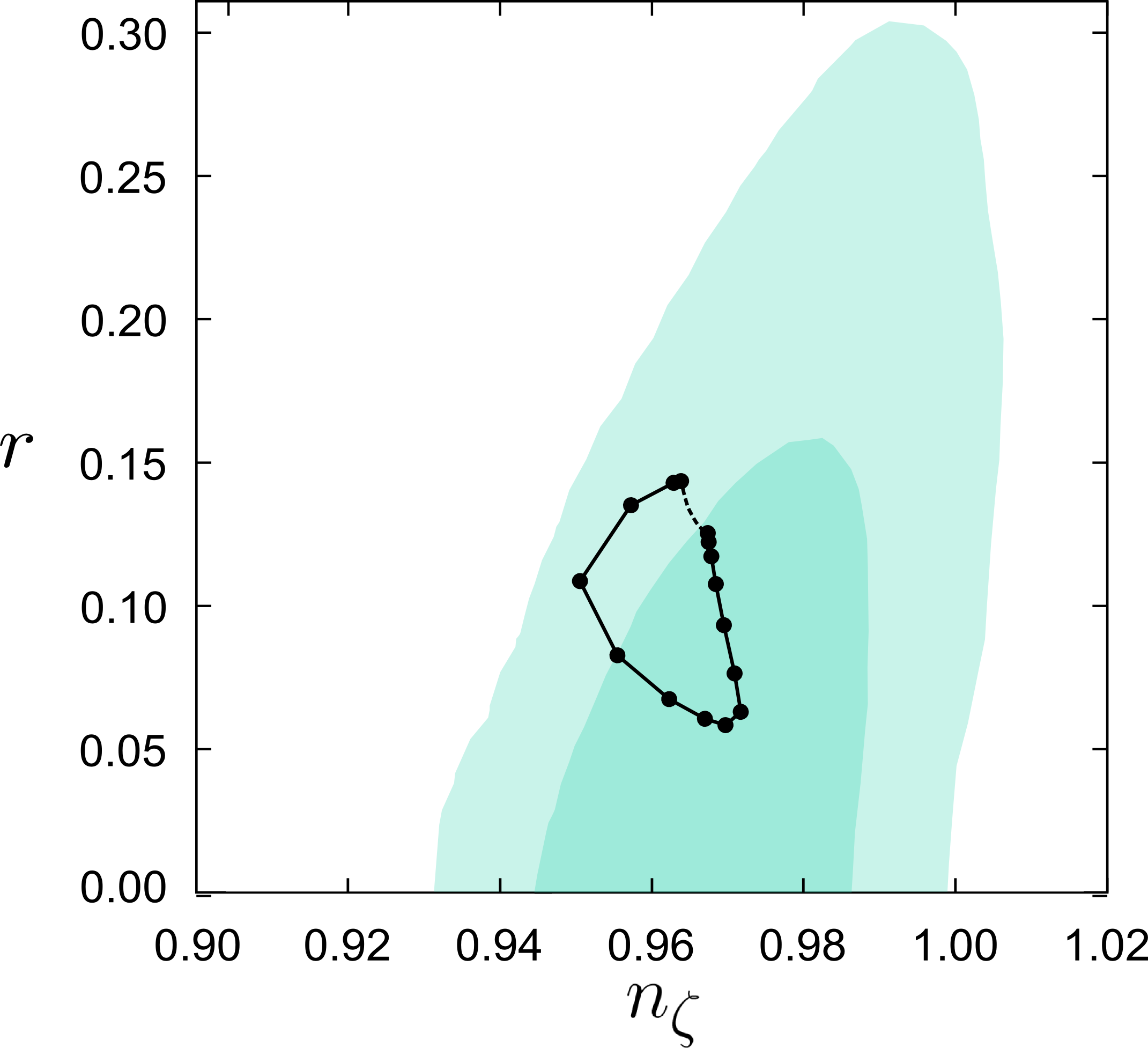} 
\par\end{centering}
\caption[Predictions for Galileon inflation with $V=m^2 \vp^2/2$]
{Theoretical values of $n_\zeta$ and $r$ for the standard chaotic inflationary potential $V(\vp)=m^{2}\vp^{2}/2$ in the presence of the dilatonic Galileon coupling $J=-(1/M^{3})e^{\mu\vp/\Mpl}X$ with $\mu=1$ (solid line). Proceeding anti-clockwise from the top of the loop, the points correspond to the cases with
$B=0, 10^{-5/2}, 10^{-9/4}, 10^{-2}, 10^{-7/4}, 10^{-3/2}, 10^{-5/4}, 0.1, 10^{-1/2}, 1, 10^{1/2}, 10, 10^{3/2}, 10^{2}$ with $N=55$. In the limit $B \to \infty$ one has $n_\zeta=0.9675$ and $r=0.1258$. The dotted curve corresponds to the case where $\mu=0$. We also show the $1\sigma$ and $2\sigma$ observational contours derived by the joint data analysis of \wmap~$7$-year, \bao~and \hst, with the consistency relation $r=-8n_t$.}
\centering{}\label{gap2} 
\end{figure}

\begin{figure}[h]
\begin{centering}
\includegraphics[width=0.7\textwidth]{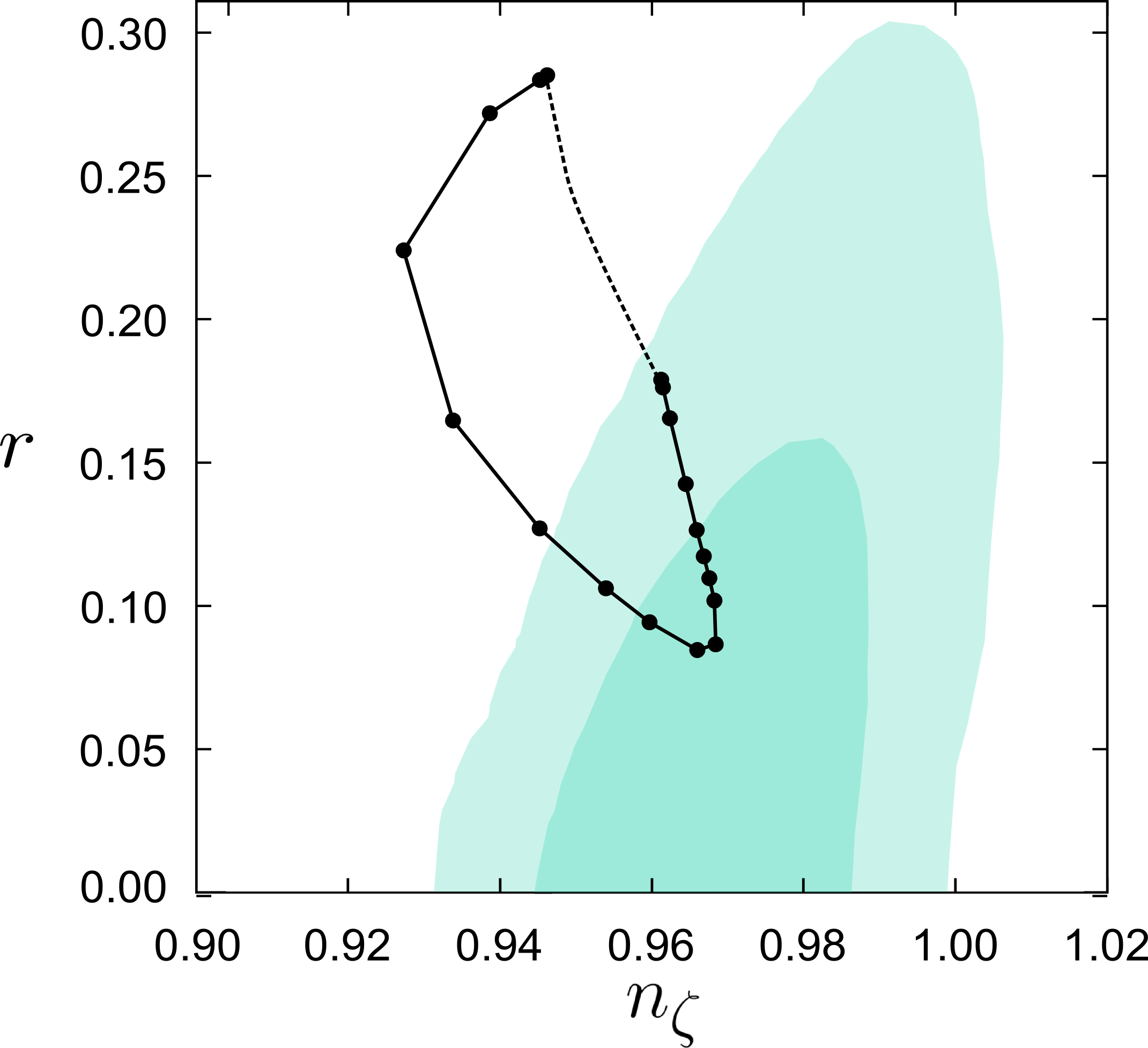} 
\par\end{centering}
\caption[Predictions for Galileon inflation with $V=\lambda \vp^4/4$]
{Similar to figure~\ref{gap2}, but for the potential $V(\vp)=\lambda\vp^{4}/4$ with $\mu=1$ (solid line). The points correspond to the cases with
$B=0, 10^{-9/2}\!, 10^{-17/4}\!, 10^{-4}\!, 10^{-15/4}\!, 10^{-7/2}, 10^{-13/4}, 10^{-3}, 10^{-11/4}\!, 10^{-5/2}\!, 10^{-2}, 10^{-3/2}\!, 10^{-3/4}\!,$ $10^{-1/2}, 10^{-1/4}, 1, 10^{1/2}, 10^{3/4}$, $10^{3/2}, 10^{3}$ with $N=55$. In the limit where $B\to\infty$ one finds $n_\zeta=0.9614$ and $r=0.1791$. The dotted curve corresponds to the case where $\mu=0$.}
\centering{}\label{gap4} 
\end{figure}

The theoretical values of $n_\zeta$ and $r$ for $\mu=1$ are plotted in figures~\ref{gap2} and \ref{gap4} (corresponding to $p=2$ and $p=4$ respectively) with several different values of $B$. Increasing $B$ from $B=0$ leads $n_\zeta$ to decrease initially, before it increases briefly and then finally decreases towards the point given by eq.~(\ref{nrgalie0}). Meanwhile $r$ decreases up to some value of $B$, with a minimum smaller than 0.1, before starting to increase towards the asymptotic value (\ref{nrgali}). The above peculiar curved trajectories in the $(n_\zeta, r)$ plane occur because of the presence of the exponential Galileon coupling with $\mu>0$. For $\mu=0$ the theoretical curve is well approximated by a straight line that connects the two asymptotic points corresponding to $B \to 0$ and $B \to \infty$. The actual line is shown as a dashed line in figures~\ref{gap2} and \ref{gap4}.


If $\mu<0$ and $B$ is increasing, then $r$ increases up to some value of $B$, whereas $n_\zeta$ decreases. The maximum values of $r$ for $\mu=-1$ are about 0.35 and 0.68 for $p=2$ and $p=4$, respectively. If $B$ is increased further, $r$ starts to decrease towards the point given by eq.~(\ref{nrgali}) (with $n_\zeta$ starting to increase at some value of $B$). Compared to the case $\mu>0$, this behaviour is not desirable to satisfy the observational bounds, especially for $p=4$. In the following discussion we shall therefore focus on the case of the positive $\mu$.

From eqs.~(\ref{rcom}) and (\ref{ntcom}) we may consider the ratio 
\begin{equation}
\frac{r}{n_t}=-8\frac{(1+4R_{J})^{3/2}}
{(1+6R_{J})^{1/2}(1+3R_{J})}\,,
\end{equation}
where $R_{J}\equiv\delta_{JX}/\delta_{X}$. For $0\le R_{J}<\infty$, the ratio $r/n_t$ is constrained to be in the narrow range $-8.71<r/n_t\le-8$. We carry out a \cmbr~likelihood analysis in terms of $n_\zeta$ and $r$ by using the two consistency relations $r=-8n_t$ and $r=-8.71 n_t$. We find that the observational constraints on $n_\zeta$ and $r$ are similar in both cases. Hence the constraints using the standard consistency relation $r=-8n_t$ should be trustworthy even in the intermediate regime. Figure \ref{gap2} shows that the quadratic inflaton potential is consistent with observations even in the presence of the exponential Galileon coupling with $\mu=1$. From figure~\ref{gap4} we find that the self-coupling inflaton potential can be saved by taking into account the exponential Galileon coupling.

\begin{figure}[h]
\begin{centering}
\includegraphics[width=0.7\textwidth]{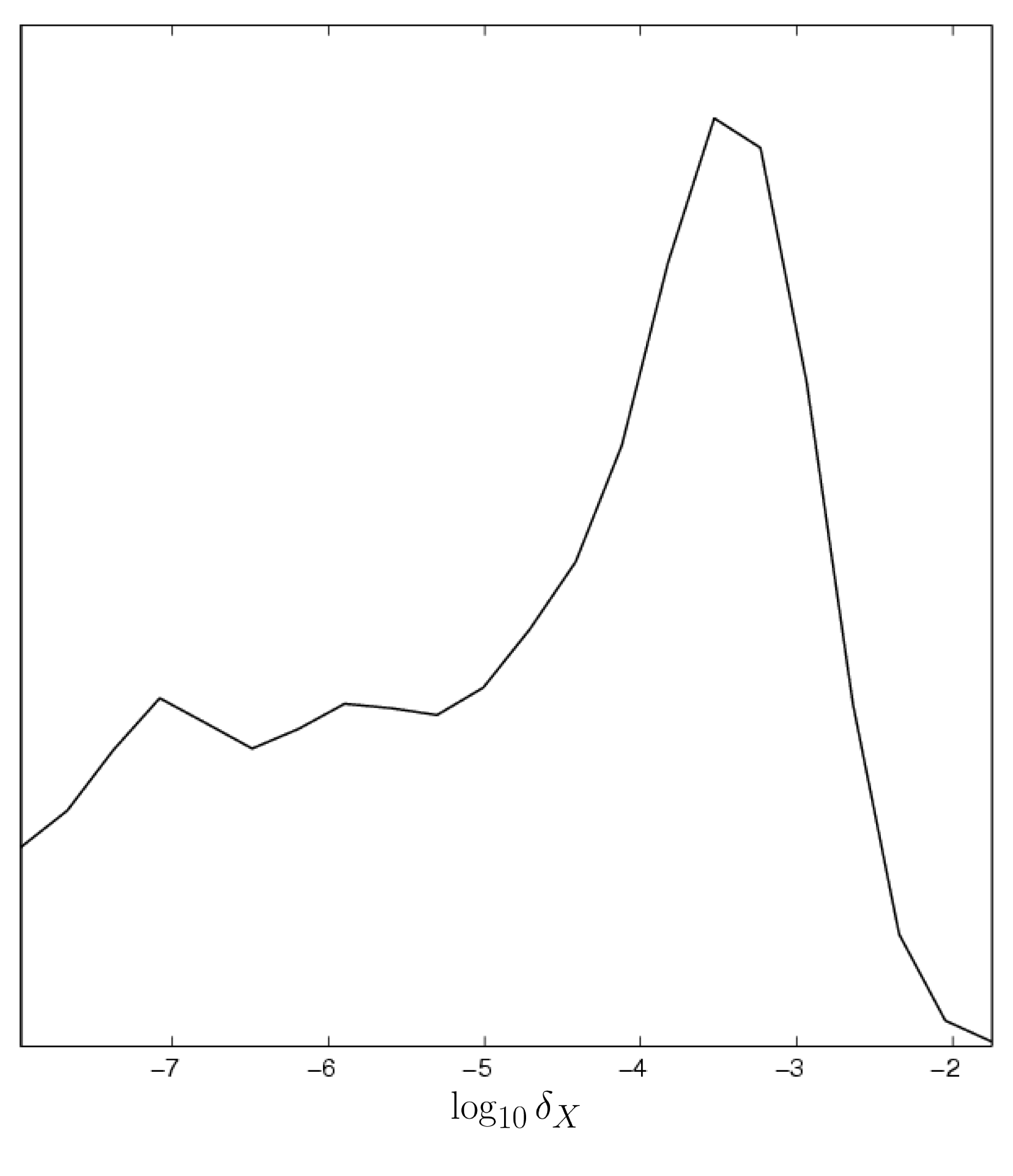} 
\par\end{centering}
\caption[Probability distribution for Galileon inflation]
{{\sc 1d} marginalized probability distribution of the parameter $\delta_X$ for Galileon modified inflation with the quartic potential $V(\vp)=\lambda \vp^{4}/4$ and $\mu=1$. We use the fitting function (described in \S\ref{sec:fitting}) that gives the relation between $\delta_X$ and $\delta_{JX}$ for $N=55$ in the regime $10^{-8}<\delta_X<0.018$. The parameter $\delta_X$ is constrained by the joint data analysis of \wmap~$7$-year, \lss~(including \bao), \hst, \snia~and \bbn, with the pivot scale $k_{0}=0.002$ Mpc$^{-1}$.}
\centering{}\label{constraintga} 
\end{figure}

For the theoretical points shown in figures~\ref{gap2} and \ref{gap4}, we can calculate the values of $\delta_X$ and $\delta_{JX}$ corresponding to $N=55$. It is then possible to derive fitting functions that relate $\delta_{JX}$ with $\delta_X$. The fitting function for $p=4$ and $\mu=1$ is given in eq.~(\ref{appeneq}) in \S\ref{sec:fitting}. This allows us to run the \cosmomc~code in terms of just one inflationary parameter $\delta_X$. In figure~\ref{constraintga} we show the {\sc 1d} marginalized probability distribution for $p=4$ and $\mu=1$, constrained by the joint data analysis of \wmap~7-year, \bao, \hst, \snia~and \bbn. In the absence of the Galileon coupling ($\delta_{JX}=0$), one has $\delta_{X}=p/(4N+p)\simeq 0.018$ for $N=55$, which is observationally excluded. In the opposite limit of large Galileon coupling such that $\delta_{JX}\gg\delta_{X}$, it follows that $\delta_{JX}\simeq p/\{3[(p+3)N+p]\}=3.4\times10^{-3}$ for $N=55$. Since this case is marginally inside the 2$\sigma$ observational contour in figure~\ref{gap4}, we find a suppressed probability distribution for smaller $\delta_X$ in figure~\ref{constraintga}. The intermediate regime such as $10^{-4} \lesssim \delta_X \lesssim 10^{-3}$ is most favoured observationally, because the corresponding theoretical points can be deep inside the 2$\sigma$ bound in figure~\ref{gap4}. In figure~\ref{gap4} the theoretical point for $B=10^{-3/2}$ gives $\delta_X=3.5 \times 10^{-4}$, which actually corresponds to the highest probability in figure~\ref{constraintga}. Hence the effect of the exponential Galileon coupling can work to save the self-coupling inflaton potential.

The scalar spectrum $\P_\zeta$ at the scale $k=0.002$ Mpc$^{-1}$ (for $n=1$) is subject to the \wmap~normalization: 
\begin{equation}
\P_\zeta=\frac{\sqrt{3}}{\pi^{2}}\left(\frac{\Mpl}{M}\right)^{6}
\left(\frac{V_{0}}{\Mpl^{4}}\right)^{3}\frac{y^{1/2}x^{3p}
e^{2\mu x}}{(y-1)^{2}(2y+1)^{3/2}}\simeq2.4\times10^{-9}\,,
\label{WMAPnor}
\end{equation}
where $y\equiv(1+4pB^{4}x^{p-1}e^{\mu x})^{1/2}$. In the limit that $B\gg1$ we obtain the result $m \approx 10^{16} (10^{12}\,{\rm GeV}/M)$ GeV for $p=2$ (with $V_{0}=m^{2}\Mpl^{2}/2$) and $\lambda\approx(10^{12}\,{\rm GeV}/M)^{4}$ for $p=4$ (with $V_{0}=\lambda\Mpl^{4}/4$), which agree with those obtained in ref.~\cite{Kamada:2010qe} for $\mu=0$. In order to have $B\gg1$ for $p=4$, we require that $\lambda\gg(M/\Mpl)^{3}$. Combining this with the \wmap~normalization, the mass scale $M$ is constrained to be $M\ll10^{-4}M_{{\rm pl}}$. If we demand that the coupling $\lambda$ is smaller than 1, this gives another constraint $M>4\times10^{-7}\Mpl$. In the intermediate regime between $B\gg1$ and $B\ll1$ we need to solve eq.~(\ref{WMAPnor}) to relate $M$ and $V_{0}$ after identifying the values of $x$ at $N=55$ numerically. In the regime $B\ll1$ we recover the standard mass scales of inflaton: $m/\Mpl \simeq 6.8 \times 10^{-6}$ for $p=2$ and $\lambda \simeq 2.0 \times 10^{-13}$ for $p=4$.

Since we regard $M$ to be a cut-off scale for the function $J(\vp,X)$, the effective theory can be trusted as long as $H\lesssim M$. This relation yields the constraint $B^{4}x^{p}\lesssim\Mpl/M$. For the case $B\gg1$ and $p=2$, we find that the effective theory can be trusted for $x \lesssim M/m \approx (M/10^{14}\,{\rm GeV})^2$. For $B\gg1$ and $p=4$, this constraint reduces to $x \lesssim \lambda^{-1/4} (M/\Mpl)^{1/2} \approx (M/10^{14}\,{\rm GeV})^{3/2}$.

The scalar propagation speed squared (\ref{scalarga}), in the regime 
$\delta_{JX}\gg\delta_{X}$, reduces to $c_{s}^{2} \simeq 2/(3n)$. 
Consequently, the non-Gaussianity parameter $f_{{\rm NL}}^{{\rm equil}}$
is constrained to be small for $n=1$ although it may be possible to have 
$|f_{{\rm NL}}^{{\rm equil}}|\gg1$ for $n\gg1$.

\paragraph{Galileon section summary.} We have shown that the Galileon-like self-interaction $\propto e^{\mu \vp/\Mpl} X^n \square \vp$ can have interesting effects on inflationary observables. Specifically, models with $\mu>0$ can lead to the compatibility of some chaotic inflationary potentials that are otherwise in tension with current observational bounds. We have confirmed this property for the self-coupling potential by carrying out a \cmbr~likelihood analysis. In addition, we have obtained analytic formulae for the regime where the Galileon term dominates over the standard kinetic term, $\delta_{JX} \gg \delta_X$. \\
\clearpage{\pagestyle{empty}\cleardoublepage}
\chapter{Conclusions}
\label{ch:conclusions}

Despite enormous observational and theoretical advances in recent years, the dynamics of our Universe are not fully understood. The focus of this thesis has been on the early epoch of {\it inflation}---a hypothesised period of accelerated cosmic expansion that is expected to be driven by the physics operating at energies vastly in excess of the standard model of high energy particle physics. As discussed in chapter \ref{ch:introduction}, the density perturbations observed in the \cmbr~provide excellent support for the inflationary hypothesis. Whilst inflation may easily be achieved by invoking simple phenomenological models, our understanding will remain incomplete until we can realise inflation in the context of fundamental theories of interactions. 

There are two independent ways in which progress may be made: Firstly, more precise cosmological data will allow us to constrain viable theories of gravitation, as well as constraining the potentially exotic contents of our Universe. Secondly, particle physics and higher energy theory, such as string theory, are hoped to develop an improved understanding of the small scale nature of physical law. Whilst these two efforts are valuable in isolation, in union they may collaboratively inform each others' development and dramatically enhance our understanding of our Universe. This thesis has taken a number of steps towards developing these vital links between high energy theory and cosmological observations.\\

Chapter \ref{ch:subhorizon} performed the first complete covariant computation of the bispectrum for multi-field inflation with a non-trivial field metric $G_{IJ}(\vp^K)$. Not only is this scenario well motivated in the context of fundamental theory, via a conformal transformation it includes non-minimally coupled models of multi-field inflation which themselves represent an important class of modified gravity theories. Employing the covariant perturbation scheme of Gong and Tanaka~\cite{Gong:2011uw}, we perturbed the action \eqref{eq:mf-action} to third order and provided the first full quantization. We have been the first to include the required factors of parallel propagation that ensure tensorial transformation properties are maintained. The two and three-point correlators of field perturbations $\langle Q^I_{\vect{k}_1} Q^J_{\vect{k}_2} \rangle$ and $\langle Q^I_{\vect{k}_1} Q^J_{\vect{k}_2} Q^K_{\vect{k}_3} \rangle$ were then found around the time of horizon exit in eqs.~\eqref{eq:power-spectrum} and \eqref{eq:bispectrum}. 

The correlators differ from the standard result of Seery and Lidsey~\cite{Seery:2005gb} in two ways: Firstly, we obtained the covariantized form of the standard result, by promotion of partial derivatives to covariant derivatives and contracting all indices with the field-space metric. Secondly, new terms were shown to arise, mediated by the Riemann curvature of the field-space metric. Such curvature terms only modify the power spectrum beyond leading order in slow-roll. However, they affect the three-point function at leading order, and so it is important to consider their effect when making bispectrum predictions. Furthermore, the new terms are {\it evolving} and may thus lead to interesting infrared dynamics as discussed in \S\ref{sec:ir-safety}.\\

The superhorizon evolution of covariant perturbations was discussed in chapter \ref{ch:formalisms}, where we made contact with observations in terms of correlators of the curvature perturbation $\zeta$. We developed a covariant version of the transport formalism of Mulryne, Seery and Wesley~\cite{Mulryne:2009kh,Mulryne:2010rp,Mulryne:2013uka} for this purpose. This is mediated by the covariant Jacobi equation~\eqref{eq:jacobi-eq} which automatically incorporates curvature contributions which influence the evolution of the two and three-point functions. Importantly, we have verified that this formalism correctly reproduces the time-dependent growing modes near horizon crossing as generated by the apparatus of quantum field theory. This matching agrees to subleading order in both slow-roll and time-dependent perturbation theory. 

The Jacobi approach leads to covariant {\it time evolution operators} ${\bigamma^I}_m$ and ${\bigamma^I}_{(mn)}$ which evolve the correlators of field perturbations as in eqs.~\eqref{eq:sep-univ-2pf}--\eqref{eq:sep-univ-3pf}. For contact with observations, the final step is to transform to the uniform density gauge as described in \S\ref{sec:gaugetransformation}. In \S\ref{sec:zetacorrelators} we show how these equations recover the covariant extension of the $\delta N$ formalism in terms of the separate universe coefficients $N_{,i}$ and $N_{;ij}$. In summary, we have provided a clear and economical framework enabling perturbations to be evolved in a slow-roll inflationary model with a non-trivial field-space metric.

Chapter \ref{ch:formalisms} closes with \S\ref{sec:analytics}, where we considered simpler scenarios of multi-field inflation in which the perturbations may be tracked analytically. Through a novel direct calculation of the path variation $\delta N$, we recover the standard results for sum and product-separable potentials. Specialising to the subclass of two-field inflation, we found analytic expressions for inflationary observables in eqs.~\eqref{eq:ss_params_ns}--\eqref{eq:ss_params_gnl} and \eqref{eq:ps_params_ns}--\eqref{eq:ps_params_gnl}. This includes new compact trispectrum results. These expressions involve a rotation to the kinemetic bases at both boundaries which leads to results in a suitable form for subsequent analysis in chapter \ref{ch:heatmaps}. \\

Before launching into an analytic analysis of superhorizon perturbation evolution, we considered the observational domain of applicability of our analytic results. Such a domain is intimately related to the broader notion of adiabaticity, which we discussed in chapter \ref{ch:adiabatic}. The occurrence of an adiabatic limit for an inflationary model ensures that its observational predictions do not subsequently evolve---a crucial requirement given our ignorance of the physics operating in the subsequent phase of reheating. After giving a general discussion, we illustrated the variety of ways in which simple phenomenological models of inflation may reach an adiabatic limit. If adiabaticity is obtained during the regime of slow-roll inflation, the analytic results of \S\ref{sec:analytics} represent a viable approximation and the asymptotic value of inflationary observables is accurately determined by the \hca. In other adiabatic scenarios, we showed how inflationary models may require numerical analysis to obtain quantitatively reliable predictions. We also discussed a non-adiabatic model and demonstrated how cosmic observables are then dependent on the details of subsequent phases such as reheating. \\

Employing this theoretical background, we proceeded in chapter \ref{ch:heatmaps} to study the types of inflationary potential and horizon exit conditions that may lead to the generation of large non-Gaussianity in two-field slow-roll inflation. We were able to achieve this for arbitrary separable potentials by extending and simplifying the {\it heatmap} analysis of Byrnes et al.~\cite{Byrnes:2008wi}. After some challenging algebraic manipulations, we arrived at remarkably simple results that then allowed us to draw new conclusions about the relationship between inflationary dynamics and their predictions for inflationary observables.

Specifically, we have shown that the same inflationary dynamics that are capable of producing large values of the bispectrum parameter $\fnl$ are also capable of producing large values of the trispectrum parameters $\tnl$ or $\gnl$. Our results confirmed that a necessary requirement for a large local non-Gaussianity is that the horizon crossing field velocities must be dominated by one of the two fields. We also provided an explicit realisation of the Suyama-Yamaguchi consistency relation, arriving at the very simple result~\eqref{eq:tnl} where we were able to relate $\tnl$ and $\fnl$ via an approximate equality rather than an inequality. We showed that in a slow-roll adiabatic limit, $\gnl$ is usually trivially related to $\tnl$ as $(27/25) \gnl \simeq \tnl$---with deviations from this result being possible, but are shown to be very hard to engineer. The heatmaps that we derived may also be interpreted in a dynamic way, enabling them to be used as qualitative tools to understand and predict the qualitative evolution of non-Gaussianity in different models of inflation.

Our heatmap analysis demonstrated that generic features in inflationary potentials such as ridges, valleys and inflection points can lead to large transient $\tnl$ and $\fnl$ parameters, with $\tnl$ being the first to peak. $\gnl$ was shown to be transiently significant near an inflection point, representing a specific realisation of a more general result that $\gnl$ can only be transiently large if the potential is described by terms beyond quadratic order. We then proceeded to study these potential shapes in greater quantitative detail in \S\ref{sec:shapes}. This allowed us to calculate the peak non-Gaussianity in terms of the parameters that define the local shape of the potential and these results may be exported to any canonical multi-field slow-roll model.\\

Finally, in chapter \ref{ch:singlefield} we considered the effects of low energy effective string theory corrections to the simple phenomenological scenario of single field chaotic inflation with the potential $V(\vp)=V_0 (\vp/\Mpl)^p$. The self-coupling potential $V(\vp)=\lambda \vp^4/4$ is excluded by \cmbr~temperature anisotropy data, while the quadratic potential $V(\vp)=m^2 \vp^2/2$ is within the $2\sigma$ observational contour. We have clarified how various field couplings present in low-energy effective string theory modify the resulting cosmological observables for these models.

In particular we have shown that the inclusion of a non-canonical kinetic term $e^{\mu \vp/\Mpl}X$ with $\mu>0$ allows chaotic inflation models that are in tension with observations to be made compatible with them. This study also included a non-minimal coupling of the form $1-\zeta \vp^2 / \Mpl^2$, as typically present in models of Higgs inflation. We verified recent work~\cite{Linde:2011nh} where it was shown that a large negative coupling $\zeta$ can lead  the self-coupling potential to be observationally viable. For Brans-Dicke theory, we have found that the field potential of the form $V(\vp)=V_0 (\vp-\Mpl)^p$, where $p \approx 2$, produces observationally viable inflation followed by a successful reheating for all values of the \bd~paramter $\wbd >-3/2$.

We also considered the presence of a Gauss-Bonnet coupling of the form $\xi_0 e^{\mu \vp/\Mpl} \G$. We found that the \gb~coupling with positive $\mu$ does not save the self-coupling potential. For the quadratic potential, we have shown that the \gb~coupling needs to be suppressed ($|\delta_{\xi}/\epsilon_s|<0.1$) from a \cmbr~likelihood analysis. If $\mu$ is negative then it is possible to lead to the decrease of both $|n_\zeta-1|$ and $r$ for negative $\xi_0$, but we require a large coupling constant, such as $|\xi_0| \sim 10^{10}$, in order to produce a sizeable effect on inflationary observables. 

The final correction that we considered is the presence of a Galileon-like self-interaction $\propto e^{\mu \vp/\Mpl} X^n \square \vp$ for which we expressed the three inflationary observables $n_\zeta$, $r$, and $n_t$ in terms of two slow-roll parameters $\delta_X$ and $\delta_{JX}$. In the regime where the Galileon term dominates over the standard kinetic term ($\delta_{JX} \gg \delta_X$) we have derived analytic formulae for $n_\zeta$ and $r$ in terms of the number of efoldings $N$. We have shown that, for $\mu>0$, the Galileon term can lead to the compatibility of chaotic inflationary potentials with current observations. We have confirmed this property for the self-coupling potential by carrying out a \cmbr~likelihood analysis. \\

In summary, this thesis has developed new tools that improve our intuition about more complex models of inflation. We have used a number of different approaches in our consideration of multi-field inflation, exploiting both analytic and numerical methods and contributing to the development of existing formalisms. To complement this work we have considered well-motivated generalisations of single field inflation. In combination, these represent two well-motivated methods for extending our understanding about the observational predictions of generalised inflationary scenarios.
\clearpage{\pagestyle{empty}\cleardoublepage}
\appendix
\chapter{Appendices}
\label{ch:appendices}

\sec{The amount of observable inflation}
\label{sec:appendix_quantifying_inflation}

This appendix calculates the number of efolds of inflation by evaluating the two terms appearing in eq.~\eqref{eq:Ninf}. We begin with the ratios of the scale factors. The entropy density $s_{\rm A}$ of a particle species `A' is calculated as~\cite{Weinberg:2008zzc}
\be
s_{\rm A} = \frac{g_{\rm A}}{2 \pi^2 T} \int_0^\infty \! \d q \,
\frac{q^2}{e^{E/T}\pm 1}\left( E + \frac{q^2}{3E}\right),
\ee
where $E$ and $q$ are the energy and momentum, $T$ is the temperature and $g_{\rm A}$ is the species degeneracy factor. The factor $\pm$ takes positive values for fermions and negative for bosons. We have presumed equilibrium conditions and so the chemical potential is absent. In the non-relativistic limit, $E \approx m + q/2m$ and both bosons and fermions have a distribution function of the form $e^{-m/T}$. This yields a suppressed entropy as
\be
s_{\rm A}^{\rm non-rel} = \frac{g_{\rm A} m}{8 T} \left( \frac{2 m T}{\pi} \right)^{3/2} e^{-m/T}.
\ee
However, in the relativistic limit one obtains
\be
s^{\rm rel} = g \frac{2 \pi^2}{45} T^3 ,
\label{eq:srel}
\ee
where 
\be 
\label{eq:degeneracy}
g = \sum_{\rm bosons} \! g_{\rm A} + \frac{7}{8} \sum_{\rm fermions} \! g_{\rm A}.
\ee
The entropy is totally dominated by the relativistic species and so it is only relativistic species that contribute to $g$. There is no heat transfer in a homogeneous and isotropic universe and so the total entropy $S = a^3 s$ is constant. From eq.~\eqref{eq:srel} one therefore obtains the useful formula
\be
\label{eq:ced}
g(T) a^3 T^3 = {\rm constant},
\ee
where we have written $g$ as a function of temperature to emphasise that it varies over the thermal history of the Universe. 

There is an important subtlety should any particle species fall out of thermal equilibrium whilst still relativistic. This happened as the temperature dropped through $T\sim 1$ MeV at the time of neutrino decoupling, taking some of the entropy density with it. Shortly after, the temperature dropped below the threshold for the electrons and positrons to become non-relativistic and they promptly annihilated into photons. In this annihilation process, $g$ drops from $11/2$ to $2$, and so eq.~\eqref{eq:ced} informs us that the photon temperature is enhanced by a factor of $(11/4)^{1/3}$. Because the neutrinos are decoupled, their temperature is not modified. Consequently, if we wish to employ eq.~\eqref{eq:ced} today, we need to account for the lower temperature of the neutrinos. 

For example, we may find the present density of relativistic species from the general density formula
\be
\label{eq:raddensity}
\rho = \frac{g \pi^2}{30} T^4 ,
\ee
by taking $T$ to be the \cmbr~temperature and replacing $g$ with $g_{\rm eff}$ where
\be
g_{\rm eff} = 2 + \frac{7}{8} \times 6 \times \left(\frac{4}{11}\right)^{4/3}.
\ee

It is useful to take the ratio of eq.~\eqref{eq:ced} with its present-day value. One finds
\be
\label{eq:ced1}
\left(\frac{a}{a_0}\right)^3 \left(\frac{T}{T_0} \right)^3 = \frac{g_0}{g}.
\ee
In order to use $T_0$ as the \cmbr~temperature, the lower neutrino temperature is accounted for by using $g_0 = 43/11$. This formula directly relates the expansion to the temperature difference. Given that $T_{\rm reh} > 1$ MeV, eq.~\eqref{eq:ced1} becomes
\be
\label{eq:aratio}
\ln \left( \frac{a_{\rm reh}}{a_0} \right) 
= -\frac{1}{3} \ln \left( \frac{g_{\rm reh}}{g_0} \right) - \ln \left( \frac{T_{\rm reh}}{T_0} \right).
\ee

The ratios of the Hubble rates in eq.~\eqref{eq:Ninf} may be calculated from the flat Friedmann equation \eqref{eq:general-friedmann} as
\be
\label{eq:Hratio}
\frac{1}{2}\ln \left(\frac{H^2_{\rm reh}}{H^2_0} \right)
= \frac{1}{2}\ln \left( \frac{g_{\rm reh} \pi^2 T_{\rm reh}^4/30}{\rho_{\rm crit,0}} \right) 
= \frac{1}{2}\ln \Omega_{{\rm rel},0}
+ \frac{1}{2} \ln \left( \frac{g_{\rm reh}}{g_0} \right)
+ 2 \ln \left( \frac{T_{\rm reh}}{T_0} \right),
\ee
where we have used eq.~\eqref{eq:raddensity} for the energy density at the time of reheating. Combining eqs.~\eqref{eq:aratio} and \eqref{eq:Hratio} gives
\be
N_{\rm inf} = \frac{1}{2}\ln \Omega_{{\rm rel},0} 
+ \frac{1}{6} \ln \left( \frac{g_{\rm reh}}{g_0}\right) 
+ \ln \left( \frac{T_{\rm reh}}{T_0} \right).
\ee
\newpage
\sec{Computing $N_{,I}$}
\label{sec:firstdeltaN}

\noindent \abshrulefill ~{\sc points of notation} \abshrulefill
\vspace{-1mm}
\begin{itemize}
\item Einstein summation convention is {\it not} employed in \S\ref{sec:firstdeltaN}.
\item To ease the complexity of the formulae derived in this section (\S\ref{sec:firstdeltaN}), we shall adopt units of $\Mpl=1$ for this section only.\footnote{This is particularly useful when dealing with separable inflationary potentials since the natural choice for sum-separable potentials is to have each term of dimension $\Mpl^4$, whereas for product-separable potentials the natural choice is to have each multiplicative term dimensionless. Temporarily taking $\Mpl=1$ facilitates easier comparison between these two calculations.}
\end{itemize}
\hrule
\vspace{5mm}

The lack of summation convention provides us with new ways of manipulating expressions such as $N_{,I}$ given in eq.~\eqref{eq:initial_expansion}. Terms which are proportional to a delta function $\delta_{IK}$ are only present when $I=K$ and therefore one can interchange indices $I$ and $K$ freely in expressions of this form. For example, in eq.~\eqref{eq:initial_expansion} the first term can be manipulated as
\be
\left. \frac{V}{V_{,K}} \right|_* \delta_{IK}=
\left. \frac{V}{V_{,I}} \right|_* \delta_{IK} ,
\ee
where we emphasise that the delta function is not removed during this index manipulation since there is no summation occurring.

The path term can be expanded in a field basis to give
\be
\label{eq:Ni}
N_{,I} = \left. \frac{V}{V_{,K}} \right|_* \delta_{IK} - 
\left. \frac{V}{V_{,K}} \right|_c \frac{\partial \phi_K^c}{\partial \phi_I^*}-
\sum_J \int_*^c \frac{\partial}{\partial \phi_J} \left( \frac{V}{V_{,K}} \right)_{\phi_K} 
\left( \frac{\partial \phi_J}{\partial \phi_I^*} \right)_{\phi_K} \d \phi_K.
\ee
To proceed we need to obtain expressions for the derivatives $(\partial \phi_J / \partial \phi_I^*)_{\phi_K}$ and $\partial \phi_J^c / \partial \phi_I^*$ in terms of the potential and its derivatives. We now proceed to show how this can be done for two general classes of separable potential.

\ssec{Separable potentials}

To evaluate the derivatives mentioned above, it is necessary to relate different field values to one another. For a general inflationary potential and non-slow-roll evolution, this has not been shown to be possible. However, if we confine ourselves to the slow-roll dynamics of eq.~\eqref{eq:exactSR} then these equations of motion provide the relation
\be 
\label{eq:sr_relation}
\frac{\d \phi_K}{V_{,K}} = \frac{\d \phi_J}{V_{,J}}.
\ee 
For this to uniquely relate fields to one another, we demand a potential $V$ such that eq.~\eqref{eq:sr_relation} is manipulable to make each side a function of just one field. For such a potential, we can take the functional derivative 
$\partial / \partial \phi_I^*$ of eq.~\eqref{eq:sr_relation} and obtain a path-independent relationship between different fields.

We therefore require that the potential obeys
\be
\frac{V_{,K}}{V_{,J}} = \frac{q_{K}(\phi_K)}{q_{J}(\phi_J)}.
\ee
where the $q_K$ are different functions for $K=1,2,\cdots$. This ratio requires that the dependence on every other field $L \ne \{K,J\}$ is either zero or cancels in the ratio. Using the shorthand $U_I = U_I (\phi_I)$, $S = \sum_I U_I$ and $P = \prod_I U_I$, there are two general potential forms that give this result:
\begin{align}
V &= F(S), \\
V &= G(P),
\end{align}
where $F$ and $G$ are arbitrary functions. We refer to these as sum and product-separable potentials respectively.

\ssec{Sum-separable potentials}

The sum-separable potential allows us to write eq.~\eqref{eq:sr_relation} as
\be 
\label{eq:gss1}
\frac{\d \phi_K}{U_K '} = \frac{\d \phi_J}{U_J '},
\ee
where $U_I ' = \partial U_I / \partial \phi_I$. Integrating from $t^*$ to some general time $t$ and taking the functional derivative with respect to horizon exit field values yields
\begin{align}
\frac{\partial}{\partial \phi_I^*} \int_*^{\phi_K} \frac{1}{U_K '} \, \d \phi_K &= 
\frac{\partial}{\partial \phi_I^*} \int_*^{\phi_J} \frac{1}{U_J '} \, \d \phi_J, \nonumber \\
\frac{1}{U_K '} \frac{\partial \phi_K}{\partial \phi_I^*} 
- \frac{1}{{U_I '}^*} \delta_{IK} &=
\frac{1}{U_J '} \frac{\partial \phi_J}{\partial \phi_I^*} 
- \frac{1}{{U_I '}^*} \delta_{IJ}, \nonumber \\
\frac{\partial \phi_J}{\partial \phi_I^*} &=
\frac{U_J'}{{U_I'}^*} (\delta_{IJ} - \delta_{IK}) 
+ \frac{U_J'}{U_K'} \frac{\partial \phi_K}{\partial \phi_I^*}.
\label{eq:gss4}
\end{align}
We note that the final boundary condition in the above expressions has not yet been fixed. This is because we wish to consider two different final boundary conditions which will yield the two derivatives that we require in the end and path terms of eq.~\eqref{eq:Ni}. 

The first boundary condition is to hold $\phi_K$ constant in eq.~\eqref{eq:gss4} which gives the path term derivative as
\be
\label{eq:ss_path_deriv}
\left( \frac{\partial \phi_J}{\partial \phi_I^*} \right)_{\phi_K} = 
\frac{U_J'}{{U_I'}^*} (\delta_{IJ} - \delta_{IK}) .
\ee
The second boundary condition is one of uniform density from which we may derive an expression for the derivative $\partial \phi_K^c / \partial \phi_I^*$. This is obtained using the defining relation for the uniform density condition,
\be 
0 = \frac{\partial V^c}{\partial \phi_I^*} 
= {F'}^c \sum_J {U_J'}^c \frac{\partial \phi_J^c}{\partial \phi_I^*}
\Rightarrow \sum_J {U_J'}^c \frac{\partial \phi_J^c}{\partial \phi_I^*} = 0,
\label{eq:ss_end_deriv1}
\ee
where we have ignored the trivial potential $V = {\rm constant}$, and employed the notation $F' = \d F / \d S$. Evaluating eq.~\eqref{eq:gss4} on a uniform density hypersurface `$c$' and then substituting into \eqref{eq:ss_end_deriv1} gives
\be
\bal{1}
\sum_J {U_J'}^c \left( \frac{{U_J'}^c}{{U_I'}^*} (\delta_{IJ} - \delta_{IK}) 
+ \frac{{U_J'}^c}{{U_K'}^c} \frac{\partial \phi_K^c}{\partial \phi_I^*} \right) &= 0 \,, \\
 \frac{1}{{U_I'}^*} \sum_J \left. {U_J'}^2 \right|_c (\delta_{IJ} - \delta_{IK}) +
 \frac{1}{{U_K'}^c} \frac{\partial \phi_K^c}{\partial \phi_I^*} \sum_J \left. {U_J'}^2 \right|_c &= 0 \,, 
\eal
\ee
and so we obtain
\be
\frac{\partial \phi_K^c}{\partial \phi_I^*} =
\frac{{U_K'}^c}{{U_I'}^*} \left(
\delta_{IK} - \frac{{U_I'}^2}{\sum_J {U_J'}^2} \right)_c .
\ee
We can put this into a more pleasant form by using the potential slow-roll parameters
\be
\sqrt{ 2\ep_I} = \frac{F' U_I'}{F},
\ee
and also defining $u_I = \ep_I / \ep$ we get
\be
\label{eq:ss_end_deriv}
\frac{\partial \phi_K^c}{\partial \phi_I^*} =
\frac{{U_K'}^c}{{U_I'}^*} \left( \delta_{IK} - u_I^c \right).
\ee
Taking eq.~\eqref{eq:ss_path_deriv}, we now work on the path term in eq.~\eqref{eq:Ni} as
\begin{align}
{\rm Path} &= -\sum_J \int_*^c \frac{\partial}{\partial \phi_J} \left( \frac{V}{V_{,K}} \right)_{\phi_K} 
\left( \frac{\partial \phi_J}{\partial \phi_I^*} \right)_{\phi_K} \d \phi_K, \nonumber \\
&= -\sum_J \int_*^c \frac{\partial}{\partial \phi_J} \left( \frac{V}{V_{,K}} \right) 
\frac{U_J'}{{U_I'}^*} (\delta_{IJ} - \delta_{IK}) \, \d \phi_K, \nonumber \\
&= -\frac{1}{{U_I'}^*} \sum_J \int_*^c \frac{\partial}{\partial \phi_J} \left( \frac{F }{F' U_K'} \right)
 U_J' (\delta_{IJ} - \delta_{IK}) \, \d \phi_K ,
\label{eq:app_logical_1}
\end{align}
where we have substituted in the form of the potential. We note a subtle point: The requirement that $\phi_K$ is constant in the first bracket has been removed. This follows because if $K \ne J$ then $\phi_K$ is already constant under $\frac{\partial}{\partial \phi_J}$, else if $K = J$, the integrand is zero by virtue of the $(\delta_{IJ} - \delta_{IK})$ factor. Thus, whether or not $\phi_K$ is constant in the derivative is irrelevant. By this same logic, we are at liberty to remove the $U_K'$ term from the differential as 
\be
\label{eq:app_logical_2}
{\rm Path} = -\frac{1}{{U_I'}^*} \sum_J \int_*^c \frac{\partial}{\partial \phi_J} \left( \frac{F}{F'} \right)
\frac{U_J'}{U_K'} (\delta_{IJ} - \delta_{IK}) \, \d \phi_K.
\ee
We can now use eq.~\eqref{eq:gss1} to change the integration variable to $\d \phi_J$ as
\be
{\rm Path} = -\frac{1}{{U_I'}^*} \sum_J \int_*^c \frac{\partial}{\partial \phi_J} \left( \frac{F}{F'} \right)
(\delta_{IJ} - \delta_{IK}) \, \d \phi_J.
\ee
Performing the summation gives two terms
\be
{\rm Path} = \frac{\delta_{IK}}{{U_I'}^*} \int_*^c \sum_J \frac{\partial}{\partial \phi_J} \left( \frac{F}{F'} \right) \d \phi_J
- \frac{1}{{U_I'}^*} \int_*^c \frac{\partial}{\partial \phi_I} \left( \frac{F}{F'} \right) \d \phi_I.
\ee
We notice that the first term is simply the expansion
\be
\d \left( \frac{F}{F'} \right) = \sum_J \frac{\partial}{\partial \phi_J} \left( \frac{F}{F'} \right) \d \phi_J ,
\ee
leading the path term to take the form
\begin{align}
{\rm Path} &= - \frac{\delta_{IK}}{{U_I'}^*} \left. \frac{F}{F'} \right|_* + 
 \frac{\delta_{IK}}{{U_I'}^*} \left. \frac{F}{F'} \right|_c
- \frac{1}{{U_I'}^*} \int_*^c \frac{\partial}{\partial \phi_I} \left( \frac{F}{F'} \right) \d \phi_I, \nonumber \\
\label{eq:ss_path}
&= -\delta_{IK} \left. \frac{V}{V_{,I}} \right|_* + 
\delta_{IK} \frac{U_I^c}{{U_I'}^*} \left. \frac{V}{V_{,I}} \right|_c
- \frac{1}{{U_I'}^*} \int_*^c \frac{\partial}{\partial \phi_I} \left( \frac{F}{F'} \right) \d \phi_I.
\end{align}
Substituting eqs.~\eqref{eq:ss_path} and \eqref{eq:ss_end_deriv} into the original expression for $N_{,I}$ in eq.~\eqref{eq:Ni} we find
\be
\bal{1}
N_{,I} &= \left. \frac{V}{V_{,K}} \right|_* \delta_{IK} - 
\left. \frac{V}{V_{,K}} \right|_c \frac{{U_K'}^c}{U_I^*} \left( \delta_{IK} - u_I^c \right)
- \left. \frac{V}{V_{,I}} \right|_* \delta_{IK} \\
& \qquad + \left. \frac{V}{V_{,I}} \right|_c \frac{{U_K'}^c}{U_I^*} \delta_{IK}
- \frac{1}{{U_I'}^*} \int_*^c \frac{\partial}{\partial \phi_I} \left( \frac{F}{F'} \right) \d \phi_I .
\eal
\ee

There are two cancellations arising from the interchange of $I$ and $K$ labels in terms proportional to $\delta_{IK}$. The simplified result is
\be
N_{,I} = \left. \frac{1}{U_I'}\right|_* \left( \left. \frac{F u_I}{F'} \right|_c 
- \int_*^c \frac{\partial}{\partial \phi_I} \left( \frac{F}{F'} \right) \d \phi_I \right).
\ee

This result applies for an arbitrary function $F$ and we have eliminated the free indices. However, if the ratio $F/F'$ maintains any $S$ dependence then to perform the integral requires us to know how each of the $U_J$ varies as we vary $\phi_I$. Without new tools to understand how the $U_J$ evolve we must restrict ourselves to special cases. There are precisely two cases when the $S$ dependence is lost: the integrand being either zero or a constant. If the integrand is zero then we have $F/F' = A$ for constant $A$.  This solves to give a general solution of
$F = Be^{S/A}$ where $B$ is also a constant. However, one can set these constants to unity by a suitable redefinition of the potentials $U_I$, leading to the result
\be
\label{eq:ss_Ni_1}
N_{,I} = \frac{u_I^c}{\sqrt{2 \ep_I^*}}, \qquad \mbox{---}{\rm for~} V = e^S.
\ee
Let us now consider the scenario where the integrand is not zero, but is constant. The integral manipulates as
\begin{align}
\int_*^c \frac{\partial}{\partial \phi_I} \left( \frac{F}{F'} \right) \d \phi_I &= 
\int_*^c \frac{\d}{\d S} \left( \frac{F}{F'} \right) U_I' \, \d \phi_I, \nonumber \\
&=  \int_*^c \frac{\d}{\d S}\left(\frac{F}{F'} \right) \d U_I .
\end{align}
We can find that the general form of $F$ such that the integrand is a constant is $F = (B S+D)^{1/A}$ where $A,B,D$ are constants. We can eliminate the constants $B$ and $D$ by redefining the potentials $U_I$, leaving a general potential form as $V = S^{1/A}$ such that
\be
\int_*^c \frac{\partial}{\partial \phi_I} \left( \frac{F}{F'} \right) \d \phi_I = \int_*^c A \, \d U_I = A U_I^c - A U_I^*.
\ee
This then gives us $N_{,I}$ for another class of potentials as
\be
\label{eq:ss_Ni_2}
N_{,I} = \frac{A}{{U_I'}^*} \left( U_I^* - U_I^c + V_c^A u_I^c \right), \qquad \mbox{---}{\rm for~} V = S^{1/A}.
\ee

\ssec{Product-separable potentials}
\label{sec:prod-separable}

The product-separable potential allows us to write eq. \eqref{eq:sr_relation} as
\be 
\label{eq:gps1}
\frac{U_K}{U_K '} \, \d \phi_K = \frac{U_J}{U_J '} \, \d \phi_J .
\ee
Integrating from $t^*$ to some general time $t$ and taking the functional derivative gives
\begin{align} 
\frac{\partial}{\partial \phi_I^*} \int_*^{\phi_K} \frac{U_K}{U_K '} \, \d \phi_K &= 
\frac{\partial}{\partial \phi_I^*} \int_*^{\phi_J} \frac{U_J}{U_J '} \, \d \phi_J, \nonumber \\
\frac{U_K}{U_K '} \frac{\partial \phi_K}{\partial \phi_I^*} 
- \frac{U_I^*}{{U_I '}^*} \delta_{IK} &=
\frac{U_J}{U_J '} \frac{\partial \phi_J}{\partial \phi_I^*} 
- \frac{U_I^*}{{U_I '}^*} \delta_{IJ}, \nonumber \\
\label{eq:gps4}
\frac{\partial \phi_J}{\partial \phi_I^*} &= 
\frac{U_I^*}{{U_I'}^*} \frac{U_J'}{U_J} (\delta_{IJ} - \delta_{IK}) 
+ \frac{U_J'}{U_J} \frac{U_K}{U_K'} \frac{\partial \phi_K}{\partial \phi_I^*}.
\end{align}
We now manipulate eq.~\eqref{eq:gps4} to find the derivatives in the end and path terms of eq. \eqref{eq:Ni}. Holding $\phi_K$ constant in eq.~\eqref{eq:gps4} gives the path term derivative as
\be
\label{eq:path_deriv}
\left( \frac{\partial \phi_J}{\partial \phi_I^*} \right)_{\phi_K} = 
\frac{U_I^*}{{U_I'}^*} \frac{U_J'}{U_J} (\delta_{IJ} - \delta_{IK}) .
\ee
For the end term of eq.~\eqref{eq:Ni} we require the derivative $\partial \phi_K^c / \partial \phi_I^*$. We use the defining relation
\be 
0 = \frac{\partial V^c}{\partial \phi_I^*} 
= (G' P)_c \sum_J \frac{{U_J'}^c}{U_J^c} \frac{\partial \phi_J^c}{\partial \phi_I^*} 
\Rightarrow
\sum_J \frac{{U_J'}^c}{U_J^c} \frac{\partial \phi_J^c}{\partial \phi_I^*} = 0,
\label{eq:end_deriv1}
\ee
where we have simply ignored the trivial potential $V = {\rm constant}$, and employed the notation $G' = \d G / \d P$. Evaluating eq.~\eqref{eq:gps4} on a uniform density hypersurface `$c$' and then substituting into eq.~\eqref{eq:end_deriv1} gives
\begin{align}
\sum_J \frac{{U_J'}^c}{U_J^c} \left( \frac{U_I^*}{{U_I'}^*} \frac{{U_J'}^c}{U_J^c} (\delta_{IJ} - \delta_{IK}) 
+ \frac{{U_J'}^c}{U_J^c} \frac{U_K^c}{{U_K'}^c} \frac{\partial \phi_K^c}{\partial \phi_I^*} \right) &= 0 \,, \nonumber \\
 \frac{U_I^*}{{U_I'}^*} \sum_J \left. \frac{{U_J'}^2}{U_J^2} \right|_c (\delta_{IJ} - \delta_{IK}) +
 \frac{U_K^c}{{U_K'}^c} \frac{\partial \phi_K^c}{\partial \phi_I^*} \sum_J \left. \frac{{U_J'}^2}{U_J^2} \right|_c &= 0 \,, \nonumber \\
\frac{\partial \phi_K^c}{\partial \phi_I^*} =
\frac{{U_K'}^c}{{U_K}^c} \frac{U_I^*}{{U_I'}^*} \left(
\delta_{IK} - \frac{{U_I'}^2 / {U_I}^2}{\sum_J {U_J'}^2 / {U_J}^2} \right)_c .&
\end{align}
We can put this into a more pleasant form by using the slow-roll parameters
\be
\sqrt{ 2\ep_I} = \frac{G'P U_I'}{G U_I} ,
\ee
and also defining $u_I = \ep_I / \ep$ to obtain
\be
\label{eq:end_deriv}
\frac{\partial \phi_K^c}{\partial \phi_I^*} =
\frac{{U_K'}^c}{{U_K}^c} \frac{U_I^*}{{U_I'}^*} \left( \delta_{IK} - u_I^c \right) .
\ee
Taking eq.~\eqref{eq:path_deriv}, we now work on the path term in eq.~\eqref{eq:Ni} given by
\begin{align}
{\rm Path} &= -\sum_J \int_*^c \frac{\partial}{\partial \phi_J} \left( \frac{V}{V_{,K}} \right)_{\phi_K} 
\left( \frac{\partial \phi_J}{\partial \phi_I^*} \right)_{\phi_K} \d \phi_K \nonumber \\
&= -\sum_J \int_*^c \frac{\partial}{\partial \phi_J} \left( \frac{V}{V_{,K}} \right) 
\frac{U_I^*}{{U_I'}^*} \frac{U_J'}{U_J} (\delta_{IJ} - \delta_{IK})\, \d \phi_K \nonumber \\
&= -\frac{U_I^*}{{U_I'}^*} \sum_J \int_*^c \frac{\partial}{\partial \phi_J} \left( \frac{G U_K}{G'P U_K'} \right)
 \frac{U_J'}{U_J} (\delta_{IJ} - \delta_{IK})\, \d \phi_K \,,
\end{align}
where we have substituted in the form of the potential. By the same logic that links eqs.~\eqref{eq:app_logical_1} and \eqref{eq:app_logical_2}, we are at liberty to remove the $U_K$ and $U_K'$ terms from the differential as 
\be
{\rm Path} = -\frac{U_I^*}{{U_I'}^*} \sum_J \int_*^c \frac{\partial}{\partial \phi_J} \left( \frac{G}{G'P} \right)
\frac{U_K}{U_K'} \frac{U_J'}{U_J} (\delta_{IJ} - \delta_{IK}) \, \d \phi_K.
\ee
We can now use eq.~\eqref{eq:gps1} to change the integration variable to $\d \phi_J$ as
\be
{\rm Path} = -\frac{U_I^*}{{U_I'}^*} \sum_J \int_*^c \frac{\partial}{\partial \phi_J} \left( \frac{G}{G'P} \right)
(\delta_{IJ} - \delta_{IK}) \, \d \phi_J.
\ee
Performing the summation gives two terms
\be
{\rm Path} =  \delta_{IK} \frac{U_I^*}{{U_I'}^*} \int_*^c \sum_J \frac{\partial}{\partial \phi_J} \left( \frac{G}{G'P} \right) \d \phi_J
- \frac{U_I^*}{{U_I'}^*} \int_*^c \frac{\partial}{\partial \phi_I} \left( \frac{G}{G'P} \right) \d \phi_I.
\ee
We notice that the first term is simply the expansion
\be
\d \left( \frac{G}{G'P} \right) = \sum_J \frac{\partial}{\partial \phi_J} \left( \frac{G}{G'P} \right) \d \phi_J \,,
\ee
and so we find
\begin{align}
{\rm Path} &= - \delta_{IK} \frac{U_I^*}{{U_I'}^*} \left. \frac{G}{G'P} \right|_* + 
\delta_{IK} \frac{U_I^*}{{U_I'}^*} \left. \frac{G}{G'P} \right|_c
- \frac{U_I^*}{{U_I'}^*} \int_*^c \frac{\partial}{\partial \phi_I} \left( \frac{G}{G'P} \right) \d \phi_I, \nonumber \\
\label{eq:path}
&= - \delta_{IK} \left. \frac{V}{V_{,I}} \right|_* + 
\delta_{IK} \frac{U_I^*}{{U_I'}^*} \frac{{U_I '}^c}{U_I^c} \left. \frac{V}{V_{,I}} \right|_c
- \frac{U_I^*}{{U_I'}^*} \int_*^c \frac{\partial}{\partial \phi_I} \left( \frac{G}{G'P} \right) \d \phi_I.
\end{align}
Substituting eqs. \eqref{eq:path} and \eqref{eq:end_deriv} into eq. \eqref{eq:Ni} we find
\be
\bal{1}
N_{,I} &= \left. \frac{V}{V_{,K}} \right|_* \delta_{IK} - 
\left. \frac{V}{V_{,K}} \right|_c \frac{{U_K'}^c}{{U_K}^c} \frac{U_I^*}{{U_I'}^*} \left( \delta_{IK} - u_I^c \right) -\left. \frac{V}{V_{,I}} \right|_* \delta_{IK} \\
&\qquad + \left. \frac{V}{V_{,I}} \right|_c \frac{{U_K'}^c}{{U_K}^c} \frac{U_I^*}{{U_I'}^*} \delta_{IK}
- \frac{U_I^*}{{U_I'}^*} \int_*^c \frac{\partial}{\partial \phi_I} \left( \frac{G}{G'P} \right) \d \phi_I,
\eal
\ee
where we finally note that there are two cancellations since we can interchange the $I$ and $K$ labels in terms proportional to $\delta_{IK}$. The simplified result is
\be
N_{,I} = \left. \frac{U_I}{U_I'}\right|_* \left( \left. \frac{G u_I}{G' P} \right|_c 
- \int_*^c \frac{\partial}{\partial \phi_I} \left( \frac{G}{G'P} \right) \d \phi_I \right).
\ee

This result works for an arbitrary function $G$. However, if the ratio $G/G'P$ maintains any $P$ dependence then to perform the integral requires us to know how each of the $U_J$ varies as we vary $\phi_I$. Without new tools to understand how the $U_J$ evolve we must restrict ourselves to special cases. There are precisely two cases when the $P$ dependence is lost: the integrand being either zero or a constant. If the integrand is zero then we have $G/G'P = A$ for constant $A$.  This solves to give a general solution of $G = BP^{1/A}$ where $B$ is also a constant. However, one can remove the constants under a redefinition of the potentials $U_I$, leading us to the result
\be
N_{,I} = \frac{u_I^c}{\sqrt{2 \ep_I^*}}, \qquad \mbox{---}{\rm for}~V = P.
\label{eq:ps_Ni_1}
\ee
Let us now consider the scenario where the integrand is not zero, but a constant. The integral manipulates as
\begin{align}
\int_*^c \frac{\partial}{\partial \phi_I} \left( \frac{G}{G'P} \right) \d \phi_I &= \int_*^c \frac{\d}{\d P} \left( \frac{G}{G'P} \right) \frac{P U_I'}{U_I} \, \d \phi_I \,, \nonumber \\
&=  \int_*^c P \frac{\d}{\d P}\left(\frac{G}{G'P} \right) \d (\ln U_I) \,,
\end{align}
and we find that the general form of $G$ such that the integrand is a constant is $G = (B \ln P+D)^{1/A}$, where $A,B,D$ are constants. We can eliminate the constants $B$ and $D$ by redefining the potentials $U_I$, leaving a general potential of the form $V = (\ln P)^{1/A}$, such that
\begin{align}
\int_*^c \frac{\partial}{\partial \phi_I} \left( \frac{G}{G'P} \right) \d \phi_I &= \int_*^c A \, \d (\ln U_I) \,, \nonumber \\
&= A \ln U_I^c - A \ln U_I^*.
\end{align}
This then gives us a result for another form of separable potential as
\be
N_{,I} = A \left. \frac{U_I}{U_I'}\right|_* \left( \ln U_I^* - \ln U_I^c + V_c^A u_I^c \right), \qquad \mbox{---}{\rm for}~V = (\ln P)^{1/A}.
\label{eq:ps_Ni_2}
\ee

\ssec{Correspondence between sum and product-separable potentials}
\label{sec:correspondence}

It is clear that there is a correspondence between eqs. \eqref{eq:ss_Ni_1} and \eqref{eq:ps_Ni_1} and also between eqs. \eqref{eq:ss_Ni_2} and \eqref{eq:ps_Ni_2}. Essentially, redefining $\ln U_I \to U_I$ turns a product-separable potential into a sum-separable potential, and redefining $e^{U_I} \to U_I$ performs the opposite transformation. This correspondence was first discussed by Wang~\cite{Wang:2010si}. It is therefore not necessary for us to proceed with all four classes of potentials. We choose the pair of independent potentials that appear most natural, $V=P$ and $V=S^{1/A}$. Since we only need these two potentials, it is useful at this point to list the four key formulae that we need for subsequent calculations:

For the potential $V=P$:
\begin{align}
\label{eq:ps_Ni}
N_{,I} &= \frac{u_I^c}{\sqrt{2 \ep_I^*}}, \\
\label{eq:ps_deriv}
\frac{\partial \phi_K^c}{\partial \phi_I^*} &=
\sqrt{\frac{\ep_K^c}{\ep_I^*}} \left( \delta_{IK} - u_I^c \right).
\end{align}

For the potential $V=S^{1/A}$:
\begin{align}
\label{eq:ss_Ni}
N_{,I} &= \frac{1}{V_*^A \sqrt{2 \ep_I^*}} \left( U_I^* - U_I^c + V_c^A u_I^c \right), \displaybreak[0]\\
\label{eq:ss_deriv}
\frac{\partial \phi_K^c}{\partial \phi_I^*} &=
\left( \frac{V^c}{V^*} \right)^A \sqrt{\frac{\ep_K^c}{\ep_I^*}} \left( \delta_{IK} - u_I^c \right). 
\end{align}

We can put these results into a more standard form by using eqs.~\eqref{eq:ps_deriv} and \eqref{eq:ss_deriv} to substitute for $u_I$ in eqs.~\eqref{eq:ps_Ni} and \eqref{eq:ss_Ni} and rewriting as
\begin{align}
\label{eq:ps_Ni_hca}
N_{,I} &= \frac{U_I^*}{{U_I'}^*} \delta_{IK}
- \frac{U_K^c}{{U_K'}^c} \frac{\partial \phi_K^c}{\partial \phi_I^*}, \\ 
N_{,I} &= A \frac{U_I^* - U_I^c + V_c^A \delta_{IK}}{{U_I '}^*} 
- A \frac{\sum_J U_J^c}{{U_K'}^c} \frac{\partial \phi_K^c}{\partial \phi_I^*},
\label{eq:ss_Ni_hca}
\end{align}
for the potentials $V=P$ and $V=S^{1/A}$ respectively. One may then relabel $K=J$ in eq.~\eqref{eq:ps_Ni_hca}. Summing over $J$ in the last line of eq.~\eqref{eq:gss4} and then substituting into eq.~\eqref{eq:ss_Ni_hca} yields the other standard result. Together these are
\begin{align}
N_{,I} &= \frac{U_I^*}{{U_I '}^*} \delta_{IJ} - \frac{U_J^c}{{U_J'}^c} \frac{\partial \phi_J^c}{\partial \phi_I^*}, \\ 
N_{,I} &= A \frac{U_I^*}{{U_I '}^*} - A \sum_J \frac{U_J^c}{{U_J'}^c} \frac{\partial \phi_J^c}{\partial \phi_I^*},
\end{align}
for the potentials $V=P$ and $V=S^{1/A}$ respectively. 

\sec{Analytic expressions for product-separable potentials}
\label{sec:form_prod}

For a two-field potential with the product-separable form $V = \Mpl^4 \, U(\phi)W(\chi)$, 
the first derivatives $N_{,I}$ are
\begin{align}
\Mpl N_{,\phi} &= \frac{u}{\sqrt{2 \epp^*}}\,, \qquad u = \frac{\epp}{\ep}\,, \label{eq:ps_u} \\
\Mpl N_{,\chi} &= \frac{w}{\sqrt{2 \epc^*}} \,, \qquad w = \frac{\epc}{\ep} \,.
\label{eq:ps_v}
\end{align}

In the $\{ \phi,\chi \}$ frame the potential is product-separable which means that any of the parameters $\eta_{IJ}$ or $\xi_{IJK}^2$ with mixed derivatives can be written in terms of lower-order slow-roll parameters that do not have mixed derivatives. This allows us to use the single-index notation $\etp$ and $\xi_\phi^2$ for the remaining terms. In the kinematic basis the three $\eta$ components are
\begin{align}
\eta_{\sigma \sigma} &= \frac{\epp \etp + \epc \etc + 4 \epp \epc}{\ep} \,, \\
\eta_{\sigma s} &= \frac{\sqrt{\epp \epc}}{\ep} \big[(\etc -2\epc) - (\etp - 2\epp) \big]\,, \\
\eta_{ss} &= \frac{\epc \etp + \epp \etc - 4 \epp \epc}{\ep} \,.
\end{align}
The components of the $\xi_{IJK}^2$ tensor are
\begin{align}
\ep^{3/2}\, \xi^2_{\sigma \sigma \sigma} &= \epc^{3/2}\, \xic^2 + \epp^{3/2} \, \xip^2 +6\epp \epc \sqrt{\ep} (\etp + \etc)\,, \\
\ep^{3/2}\, \xi^2_{\sigma \sigma s} &= \epc \sqrt{\epp} \, \xic^2 - \epp \sqrt{\epc}\, \xip^2 + 2 \sqrt{\ep \epp \epc} \big[ (\epp-2\epc)\etp-(\epc-2\epp)\etc \big]\,, \\
\ep^{3/2}\, \xi^2_{\sigma ss} &= \epp \sqrt{\epc} \, \xic^2 + \epc \sqrt{\epp}\, \xip^2 + 2 \sqrt{\ep} \big[ (\epp-2\epc)\epp \etc+(\epc-2\epp) \epc \etp \big] \,, \\
\ep^{3/2}\, \xi^2_{sss} &= \epp^{3/2} \, \xic^2 - \epc^{3/2}\, \xip^2 +6 \sqrt{\ep \epp \epc} (\epc \etp - \epp \etc) \,.
\end{align}
The second derivatives of $N$ follow by differentiation of eqs.~\eqref{eq:ps_u} and~\eqref{eq:ps_v}:
\begin{align}
\Mpl^2 N_{,\phi \phi}&= u - \frac{u \etp^*}{2 \epp^*} + \frac{\A_P}{\epp^*},  \\
\Mpl^2 N_{,\phi \chi}&= - \frac{\A_P}{\sqrt{\epp^* \epc^*}},  \\
\Mpl^2 N_{,\chi \chi}&= w - \frac{w \etc^*}{2 \epc^*}  + \frac{\A_P}{\epc^*},
\end{align}
where we have substituted
\be
\A_P \equiv u w \, \eta_{ss}\,,
\label{eq:ps_A}
\ee
to put these equations into a form similar to that found for the sum-separable potential.
Taking the next derivative we find
\begin{align}
\hspace{-2em}\Mpl^3 N_{,\phi \phi \phi} &= \frac{1}{\epp^* \sqrt{2 \epp^*}}
	\left(
	- \frac{u}{2} \sqrt{\frac{\epp^*}{\ep^*}} {\xip^*}^2  
	- u \epp^* \etp^*
	+ u {\etp^*}^2
	- 3 (\etp^* - 2 \epp^*) \A_P 
	+ \B_P^2
	\right),\\
\hspace{-2em}\Mpl^3 N_{,\phi \phi \chi} &= \frac{1}{\epp^* \sqrt{2 \epc^*}}\Big( (\etp^* - 2 \epp^*) \A_P - \B_P^2 \Big), \\
\hspace{-2em}\Mpl^3 N_{,\phi \chi \chi} &= \frac{1}{\epc^* \sqrt{2 \epp^*}}\Big( (\etc^* - 2 \epc^*) \A_P + \B_P^2 \Big), \\
\hspace{-2em}\Mpl^3 N_{,\chi \chi \chi} &= \frac{1}{\epc^* \sqrt{2 \epc^*}}
	\left(
	- \frac{w}{2} \sqrt{\frac{\epc^*}{\ep^*}} {\xic^*}^2  
	- w \epc^* \etc^*
	+ w {\etc^*}^2
	- 3 (\etc^* - 2\epc^*) \A_P 
	- \B_P^2
	\right), \\
\!\!\!\!\!\! \!\!\!\!\! \!\!\!\!\!\! \!\! \!\! \Mpl \sqrt{2 \epp^*} \frac{\partial \A_P}{\partial \phi_*} &\equiv 
- \Mpl \sqrt{2 \epc^*} \frac{\partial \A_P}{\partial \chi_*}
\equiv \B_P^2 \,, \\
\B_P^2 &\equiv -\sqrt{uw}^3 \Big[
\xi_{sss}^2 +2 \frac{\epp-\epc}{\sqrt{\epp \epc}} \eta_{ss}^2 - 2 \eta_{\sigma s} \eta_{ss}
\Big]\,.
\label{eq:ps_B}
\end{align}
As with the sum-separable case, $\A_P$ is symmetric under field exchange whilst $\B_P^2$ is anti-symmetric. We take these results together to find
\begin{align}
\label{eq:ps_params1_ns}
\hspace{-2em}n_\zeta-1 &= - 4 \left(\frac{u^2}{\epp^*} + \frac{v^2}{\epc^*} \right)^{-1}
\sneg_space && \left[ 1 - 2 uw - \frac{u^2 \etp^*}{2 \epp^*} - \frac{w^2 \etc^*}{2 \epc^*} \right] - 2 \ep^*, \displaybreak[0]\\
\label{eq:ps_params1_fnl}
\hspace{-2em} \fnl &= \frac{5}{6} \left(\frac{u^2}{\epp^*} + \frac{w^2}{\epc^*} \right)^{-2} 
\jneg_space && \left[ 
	2 \left( \frac{u^3}{\epp^*} +  \frac{w^3}{\epc^*} \right)
	- \frac{u^3 \etp^*}{{\epp^*}^2 }
	- \frac{w^3 \etc^*}{{\epc^*}^2 } 
	+2 \left( \frac{u}{\epp^*} - \frac{w}{\epc^*} \right)^2 \A_P 
\right]\! \!\!\!\!\! \displaybreak[0]\\
\hspace{-2em}\tnl &= 4 \left(\frac{u^2}{\epp^*} + \frac{w^2}{\epc^*} \right)^{-3} \jneg_space && \Bigg[
	-\frac{u^4 \etp^*}{{\epp^*}^2}
	-\frac{w^4 \etc^*}{{\epc^*}^2}
	+\frac{u^4 {\etp^*}^2 }{4{\epp^*}^3 }
	+\frac{w^4 {\etc^*}^2 }{4{\epc^*}^3 } 
	+\frac{u^4}{\epp^*}
	+\frac{w^4}{\epc^*} \nonumber \\
	& && \left. - \frac{u^2}{ {\epp^*}^2} \left( \frac{u}{\epp^*} - \frac{w}{\epc^*} \right) \etp^* \A_P
	- \frac{w^2}{ {\epc^*}^2} \left( \frac{w}{\epc^*} - \frac{u}{\epp^*} \right) \etc^* \A_P \right. \nonumber \\
	& && + 2 \left( \frac{u^2}{\epp^*} - \frac{w^2}{\epc^*} \right) \left( \frac{u}{\epp^*} - \frac{w}{\epc^*} \right) \A_P  
	+\left( \frac{u}{\epp^*} - \frac{w}{\epc^*}\right)^2 \left( \frac{1}{\epp^*} + \frac{1}{\epc^*}\right) \A_P^2
\Bigg], \label{eq:ps_params1_tnl} \displaybreak[0]\\
\hspace{-2em}\gnl &= \frac{25}{27}\left(\frac{u^2}{\epp^*} + \frac{w^2}{\epc^*} \right)^{-3} 
\jneg_space && \Bigg[
	- \frac{u^4 \etp^*}{{\epp^*}^2}
	- \frac{w^4 \etc^*}{{\epc^*}^2}
	+ \frac{u^4 {\etp^*}^2}{{\epp^*}^3}
	+ \frac{w^4 {\etc^*}^2}{{\epc^*}^3} 
	- \frac{1}{2}\frac{u^4 {\xip^*}^2}{{\epp^*}^2 \sqrt{\ep^* \, \epp^*} }
	- \frac{1}{2}\frac{w^4 {\xic^*}^2}{{\epc^*}^2 \sqrt{\ep^* \, \epc^*}}	
	\nonumber \\ & && \left.
	-3 \frac{u^2}{{\epp^*}^2} \left(\frac{u}{\epp^*} - \frac{w}{\epc^*} \right) \etp^* \A_P 
	-3 \frac{w^2}{{\epc^*}^2} \left(\frac{w}{\epc^*} - \frac{u}{\epp^*} \right) \etc^* \A_P 	\right. \nonumber \\ & &&  
	+6 \left( \frac{u^2}{\epp^*} - \frac{w^2}{\epc^*}\right) \left( \frac{u}{\epp^*} - \frac{w}{\epc^*}\right) \A_P
	+ \left(\frac{u}{\epp^*} - \frac{w}{\epc^*} \right)^3 \B_P^2
\Bigg] . \label{eq:ps_params1_gnl}
\end{align}

One can rewrite the horizon crossing slow-roll parameters as
\begin{align}
\etp &= \eta_{ss} + 2 \epp - \sqrt{\frac{\epp}{\epc}} \eta_{\sigma s} \, , \qquad 
\label{eq:kinematic_subs_2a}
\etc = \eta_{ss} + 2 \epc + \sqrt{\frac{\epc}{\epp}} \eta_{\sigma s} \, , \\
\xip^2 &= \frac{\sqrt{\ep \epp}}{\epc}\xi_{\sigma ss}^2 - \sqrt{\frac{\ep}{\epc}} \xi_{sss}^2 
-2 \frac{\sqrt{\ep \epp}}{\epc}(\epp \etc - 2 \epc \etp)\, ,  \\
\xic^2 &= \frac{\sqrt{\ep \epc}}{\epp}\xi_{\sigma ss}^2 + \sqrt{\frac{\ep}{\epp}} \xi_{sss}^2 
-2 \frac{\sqrt{\ep \epc}}{\epp}(\epc \etp - 2 \epp \etc)\, .
\label{eq:kinematic_subs_2}
\end{align}
Substituting the relations~\eqref{eq:kinematic_subs_2a}--\eqref{eq:kinematic_subs_2} into eqs.~\eqref{eq:ps_params1_ns}--\eqref{eq:ps_params1_gnl} yields
\begin{align}
\hspace{-1em} n_\zeta-1 &= \left(\frac{u^2}{\epp^*} + \frac{w^2}{\epc^*} \right)^{-1} \jneg_space && \left[\frac{2w-2u}{\sqrt{\epp^* \epc^*}} \eta_{\sigma s}^* \right] + 2(\eta_{ss}^* - \ep^*) \,, \label{eq:ps_params_ns} \displaybreak[0]\\
\hspace{-1em} \frac{6}{5} \fnl &= \left(\frac{u^2}{\epp^*} + \frac{w^2}{\epc^*} \right)^{-2} \jneg_space &&
\left[ 
	- \left( \frac{u^3}{{\epp^*}^2} + \frac{w^3}{{\epc^*}^2} \right) \eta_{ss}^*  	+ \left( \frac{u^3}{\epp^*} - \frac{w^3}{\epc^*} \right) \frac{\eta_{\sigma s}^*}{\sqrt{\epp^* \epc^*}}
	+2 \left( \frac{u}{\epp^*} - \frac{w}{\epc^*} \right)^2 \A_P
\right]\!\!, \label{eq:ps_params_fnl} \displaybreak[0]\\
\hspace{-1em} \tnl &= \left(\frac{u^2}{\epp^*} + \frac{w^2}{\epc^*} \right)^{-3} \jneg_space &&
\Bigg[
	\left( \frac{u^4}{{\epp^*}^3} + \frac{w^4}{{\epc^*}^3} \right){\eta_{ss}^*}^2
	-2\left( \frac{u^4}{{\epp^*}^2} - \frac{w^4}{{\epc^*}^2} \right)\frac{\eta_{ss}^* \eta_{\sigma s}^*}{\sqrt{\epp^* \epc^*}} \nonumber \\
	& && \left. 
	+\left( \frac{u^4}{\epp^*} + \frac{w^4}{\epc^*} \right)\frac{{\eta_{\sigma s}^*}^2}{\epp^* \epc^*}
	- 4\left(\frac{u}{\epp^*}-\frac{w}{\epc^*}\right)^2 \left( \frac{u}{\epp^*} + \frac{w}{\epc^*} \right) \eta_{ss}^* \A_P 	
	\right. \nonumber \\
	& &&
	\hspace{-3.5em} + 4\left(\frac{u}{\epp^*}-\frac{w}{\epc^*}\right) \left( \frac{u^2}{\epp^*} + \frac{w^2}{\epc^*} \right) \frac{\eta_{\sigma s}^* \A_P}{\sqrt{\epp^* \epc^*}} 
	+4\left( \frac{u}{\epp^*} - \frac{w}{\epc^*}\right)^2 \frac{\ep^* \A_P^2}{\epp^* \epc^*}
\Bigg], \label{eq:ps_params_tnl} \displaybreak[0]\\
\hspace{-1em} \frac{27}{25} \gnl &= \left(\frac{u^2}{\epp^*} + \frac{w^2}{\epc^*} \right)^{-3} \jneg_space &&
\Bigg[
	\left( \frac{u^4}{{\epp^*}^3} + \frac{w^4}{{\epc^*}^3} \right){\eta_{ss}^*}^2
	-2 \left( \frac{u^4}{{\epp^*}^2} - \frac{w^4}{{\epc^*}^2} \right) \frac{\eta_{ss}^* \eta_{\sigma s}^*}{\sqrt{\epp^* \epc^*}} 
	+\left( \frac{u^4}{\epp^*} + \frac{w^4}{\epc^*} \right)\frac{{\eta_{\sigma s}^*}^2}{\epp^* \epc^*}	
	\nonumber \\
	& && \left.
	+ \left(\frac{u^4}{\epp^*} + \frac{w^4}{\epc^*} \right)\frac{\ep^* \eta_{ss}^*}{\epp^* \epc^*}  
+\frac{1}{2} \left(\frac{u^4}{{\epp^*}^2} - \frac{w^4}{{\epc^*}^2} \right) \frac{{\xi_{sss}^*}^2}{\sqrt{\epp^* \epc^*}}	
	\right. \nonumber \\
	& &&\left. 	
	-\frac{1}{2} \left(\frac{u^4}{\epp^*} + \frac{w^4}{\epc^*} \right) \frac{{\xi_{\sigma ss}^*}^2}{ \epp^* \epc^*}
	- 3\left(\frac{u}{\epp^*}-\frac{w}{\epc^*}\right)^2 \left( \frac{u}{\epp^*} + \frac{w}{\epc^*} \right) \eta_{ss}^* \A_P 
\right. \nonumber \\
	& &&
	\hspace{-1.5em} + 3\left(\frac{u}{\epp^*}-\frac{w}{\epc^*}\right) \left( \frac{u^2}{\epp^*} + \frac{w^2}{\epc^*} \right) \frac{\eta_{\sigma s}^* \A_P}{\sqrt{\epp^* \epc^*}}  
	+ \left( \frac{u}{\epp^*} - \frac{w}{\epc^*} \right)^3 \B_P^2  
\Bigg] . 
\label{eq:ps_params_gnl}
\end{align}

\sec{Simplification of trispectrum expressions}
\label{sec:trispectrum_simplification}

The function $|\tau_6|$, as defined in eq.~\eqref{eq:ss_tnl_para}, is bounded well within our limit of ten, and so this term may be immediately neglected. Furthermore, we can use eq.~\eqref{eq:eta_relationship} to manipulate various other terms. In the sum-separable case we find
\begin{align}
&\tau_2 \eta_{ss}^* \eta_{\sigma s}^* = \Big[\sin 2 \theta^*  \tau_2\Big] \eta_{ss}^* \eta_{\sigma s}^* 
+ \Big[(1-\sin 2\theta^*) \tau_2 \frac{1}{2} \tan 2 \theta^* \Big] \eta_{ss}^* (\eta_{ss}^* -\eta_{\sigma \sigma}^*)\,, \\
&\tau_3 {\eta_{\sigma s}^*}^2 = \Big[\sin^2 2\theta^* \tau_3\Big] {\eta_{\sigma s}^*}^2 
+ \Big[(1-\sin^2 2\theta^*) \tau_3 \frac{1}{4} \tan^2 2 \theta^* \Big]  (\eta_{ss}^* -\eta_{\sigma \sigma}^*)^2 \,, \\
&\tau_5 \ep^* \eta_{\sigma s}^* =  \Big[\sin 2 \theta^* \tau_5 \Big] \ep^* \eta_{\sigma s}^* 
+\Big[(1-\sin 2 \theta^*) \tau_5 \frac{1}{2} \tan 2\theta^* \Big] \ep^* (\eta_{ss}^*-\eta_{\sigma \sigma}^*)\,, \\
&- \tau_8 \, \Omega \, \eta_{\sigma s}^* (\eta_{ss}-\ep) =
-\Big[(1-\sin 2 \theta^*) \tau_8 \frac{1}{2} \tan 2\theta^* \Big] \, \Omega \, (\eta_{ss}^*-\eta_{\sigma \sigma}^*) (\eta_{ss}-\ep) \nonumber \\
& \qquad \qquad\qquad\qquad \qquad
-\Big[\sin^2 2 \theta^* \tau_8\Big] \, \Omega \, \eta_{\sigma s}^* (\eta_{ss}-\ep) \,,
\end{align}
and we find that all of the functions in square brackets never have a magnitude greater than ten and so these terms represent variations in the trispectrum that are significantly smaller than observables will ever probe. One can easily check that these results follow analogously for product-separable potentials and so these terms may be neglected for both types of separable potential.

Analogously to eq.~\eqref{eq:eta_relationship} there exist formulae relating the various $\xi_{ijk}^2$ components. For sum-separable potentials we have 
\begin{align}
\xi_{\sigma ss}^2 &= \frac{1}{2} \tan 2 \theta (\xi_{sss}^2-\xi_{\sigma \sigma s}^2)\,, \label{eq:ss_xi_relation1} \\
\xi_{\sigma \sigma s}^2 &= \frac{1}{2} \tan 2 \theta (\xi_{\sigma ss}^2-\xi_{\sigma \sigma \sigma}^2)\,,
\label{eq:ss_xi_relation2}
\end{align}
whereas for product-separable potentials these relations are of the form
\begin{align}
\xi_{\sigma ss}^2 &= \frac{1}{2} \tan 2 \theta \, (\xi_{sss}^2-\xi_{\sigma \sigma s}^2) + 2 \ep \eta_{ss} + 2 \tan 2 \theta \, \ep \, \eta_{\sigma s}\,,  \label{eq:ps_xi_relation1} \\
\xi_{\sigma \sigma s}^2 &= \frac{1}{2} \tan 2 \theta \, (\xi_{\sigma ss}^2-\xi_{\sigma \sigma \sigma}^2) - 2 \ep \, \eta_{\sigma s} 
+ 2 \tan 2 \theta \, \ep \, (\eta_{ss}+\ep)\,. \label{eq:ps_xi_relation2}
\end{align}
Using eqs.~\eqref{eq:ss_xi_relation1} and \eqref{eq:ss_xi_relation2} we can then simplify part of the sum-separable expression for $\gnl$ in eq.~\eqref{eq:ss_gnl_full} as
\begin{align}
\frac{1}{4} \tau_2 {\xi_{sss}^*}^2 -\frac{1}{2} \tau_3 {\xi_{\sigma ss}^*}^2 
&= \frac{1}{4} \tau_2 {\xi_{sss}^*}^2 - \sin^2 2 \theta^* \frac{1}{2} \tau_3 {\xi_{\sigma ss}^*}^2 
-(1-\sin^2 2 \theta^*) \frac{1}{2} \tau_3 \nonumber \\
& \qquad \times \left(\frac{1}{2} \tan 2 \theta^* {\xi_{sss}^*}^2 
- \frac{1}{4} \tan^2 2 \theta^* \left({\xi_{\sigma ss}^*}^2 
- {\xi_{\sigma \sigma \sigma}^*}^2\right) \right) \nonumber \\
&= -g_4{\xi_{sss}^*}^2 - \Big[ \frac{1}{2} \tau_3 \sin^2 2 \theta^* \Big]
\left(\frac{3}{4}{\xi_{\sigma ss}^*}^2 + \frac{1}{4}{\xi_{\sigma \sigma \sigma}^*}^2  \right), \nonumber \\
&\simeq -g_4{\xi_{sss}^*}^2 \,,
\end{align}
where we have defined $g_4 = \frac{1}{4} \left(\tau_3 \sin 2 \theta^* \cos 2 \theta^* -\tau_2 \right)$ in the expressions above. The term in square brackets in the penultimate line can never be large and so is neglected.

The product-separable case follows similarly, yielding the same answer, however the calculation is unsurprisingly more involved. Manipulating three of the terms in eq.~\eqref{eq:ps_gnl_full} for $\gnl$ by using eq.~\eqref{eq:ps_xi_relation1} we find
\begin{align}
\frac{1}{4} \tau_2 {\xi_{sss}^*}^2 + \tau_3 \ep^* \eta_{ss}^* -\frac{1}{2} \tau_3 {\xi_{\sigma ss}^*}^2 
&= -g_4{\xi_{sss}^*}^2 - \Big[ \frac{1}{2} \tau_3 \sin^2 2 \theta^* \Big]
\left( {\xi_{\sigma ss}^*}^2 - 2 \ep^* \eta_{ss}^*\right) \nonumber \\
& \qquad + \frac{1}{4} \tau_3 \sin 2 \theta^* \cos 2 \theta^* \left({\xi_{\sigma \sigma s}^*}^2 - 4 \ep^* \eta_{\sigma s}^* \right). 
\label{eq:ps_xi_working}
\end{align}
We now expand the last term of eq.~\eqref{eq:ps_xi_working} by substituting for ${\xi_{\sigma \sigma s}^*}^2$ using eq.~\eqref{eq:ps_xi_relation2}. We also use eq.~\eqref{eq:eta_relationship} to rewrite $\eta_{\sigma s}^*=\sin 2 \theta^* \eta_{\sigma s}^* + (1-\sin 2 \theta^*) \frac{1}{2} \tan 2 \theta^*(\eta_{ss}^* - \eta_{\sigma \sigma}^*+2\ep^*)$ and so we ultimately find 
\begin{align}
\frac{1}{4} \tau_2 {\xi_{sss}^*}^2 + \tau_3 \ep^* \eta_{ss}^* -\frac{1}{2} \tau_3 {\xi_{\sigma ss}^*}^2 
&=-g_4{\xi_{sss}^*}^2 
- \Big[ \frac{1}{2} \tau_3 \sin^2 2 \theta^* \Big]
\left( \frac{3}{4}{\xi_{\sigma ss}^*}^2 + \frac{1}{4} {\xi_{\sigma \sigma \sigma}^*}^2 -3 \ep^* \eta_{ss}^* \right. \nonumber \\
& \hspace{-2cm} \left. - {\ep^*}^2 +3 \ep^* \eta_{\sigma s}^* \cos 2 \theta^* + \frac{3}{2} (1-\sin 2\theta^*)(\eta_{ss}^* - \eta_{\sigma \sigma}^* + 2 \ep^*)\right).
\end{align}
The term in square brackets is always negligible and this multiplies terms no larger than ${\cal O} ({\ep^*}^2)$ and so may be ignored, leaving the same simple result that we found for sum-separable potentials.

After these various terms have been neglected from eqs. \eqref{eq:ss_tnl_full}
to \eqref{eq:ps_gnl_full}, we can simplify the remaining terms, rewriting them
by means of the following trigonometric relations (similarly to the bispectrum calculation, these simple formulae follow standard trigonometric identities, although the algebra is not itself trivial)
\be
\bal{2}
\tau_1 &= \tau + 2 f + 1 \,,  \\
\tau_4 &= 2 f_1 ( 1+ f)\,, \\
\tau_7 &= 4(\tau+f)\,, \\
2 g_2 &= \tau-f \,.
\eal
\ee

\sec{Fitting function for Galileon inflation}
\label{sec:fitting}

\begin{figure}[h]
\begin{centering}
\includegraphics{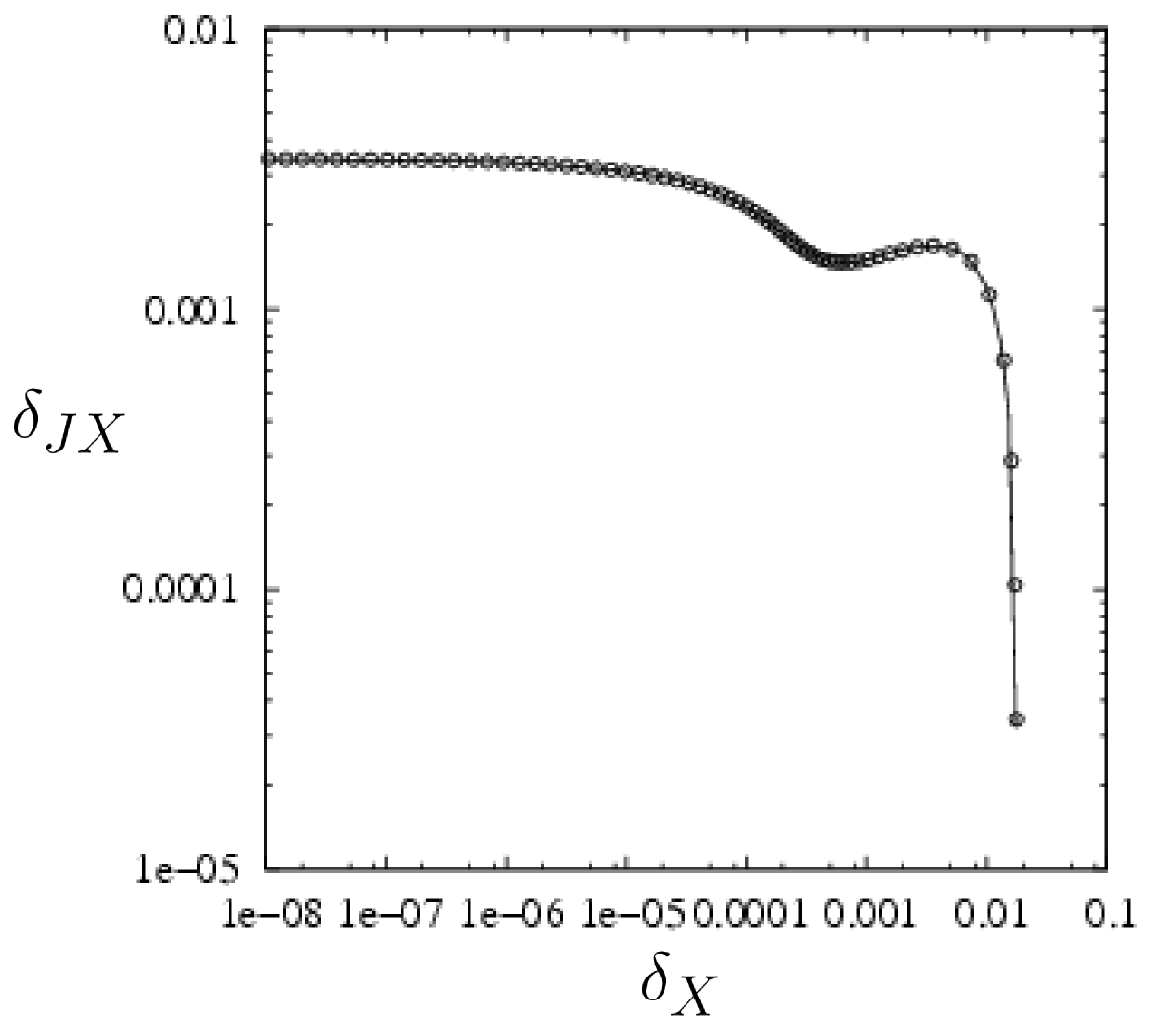} 
\par\end{centering}
\caption[Fitting function for Galileon inflation]{Numerical data points corresponding to the values of $\delta_{JX}$ and $\delta_X$ satisfying the constraint $N=55$. Each data point corresponds to a particular value $B=10^{i/8}$ with $i=-32,\dots,32$. The fitting function we use for \cmbr~likelihood analysis is also plotted.}
\centering{}\label{fittof} 
\end{figure}

We present a fitting function for the quartic potential $V(\vp)=\lambda \vp^4/4$ in the presence of the Galileon-type coupling $J=-(1/M^{3})e^{\mu\vp/\Mpl}X$ with $\mu=1$. We numerically find the field value $\phi$ giving $N=55$ before the end of inflation and evaluate $\delta_X$ and $\delta_{JX}$ for several different values of $B$ ($B=10^{i/8}$ with $i=-32,\dots,32$). These slow-roll parameters can be approximated by the following 
fitting function (found by using the method of least squares)
\begin{align}
\delta_{JX} &= 
-5.25192634579698+540.210808997015\,\delta_{{X}}^{1/2}-3509.55978587371\,\delta_{{X}}^{1/3}
\nonumber\\ & \qquad
+15290.159752272\,\delta_{{X}}^{1/4}-38509.9526724544\,\delta_{{X}}^{1/5}+53949.4042466374\,\delta_{{X}}^{1/6}
\nonumber\\ &\qquad
-38908.1718682253\,\delta_{{X}}^{1/7}\,+11232.4296410833\,\delta_{{X}}^{1/8}
-224.28358682764\,\delta_{{X}}
\nonumber\\ &\qquad
+6155.30047533836\,{\delta_{{X}}^{2}}\,-519243.629001884\,{\delta_{{X}}^{3}}\,+38227861.764318\,{\delta_{{X}}^{4}}
\nonumber\\ &\qquad
-1897688289.07932\,{\delta_{{X}}^{5}}\,+54200448383.7942\,{\delta_{{X}}^{6}}\,-665839723646.196\,{\delta_{{X}}^{7}}. \nonumber \\
\label{appeneq}
\end{align}
We have used this expression in the regime $10^{-8}<\delta_X<0.018$ for our \cmbr~likelihood analysis in figure~\ref{constraintga}. Finally, in figure~\ref{fittof}, we show both the numerical data and the fitting function $\delta_{JX}(\delta_X)$. Since its inverse function, on the whole interval, is multivalued, we have used $\delta_X$ as the independent slow-roll parameter.

\clearpage{\pagestyle{empty}\cleardoublepage}




\begin{singlespace}

\bibliography{thesis}
%
\bibliographystyle{./utphys-ih}

\end{singlespace}

\end{document}